\documentclass[11pt]{article} 
\newtheorem{definition}{Definition}  
\newtheorem{theorem}{Theorem}  
\newtheorem{proposition}{Proposition}  
\newtheorem{lemma}{Lemma}  
\usepackage{vmargin}   
\setmarginsrb{3.2cm}{2.5cm}{3.2cm}{2.5cm}{1cm}{1cm}{1cm}{1.6cm}   
\usepackage{amsmath}  
\usepackage{latexsym} 
\usepackage{amssymb}  
\usepackage[dvips]{graphicx}  
\usepackage{rotating}   
\usepackage{psfrag} 

\title{{ Hamiltonian Quantization of Chern-Simons theory with $SL(2,{\mathbb C}) $ Group}}  
  
\author{ E. BUFFENOIR, K. NOUI, Ph. ROCHE \\  
 Laboratoire de physique math\'ematique et th\'eorique  \\  
 Universit\'e Montpellier 2, 34000 Montpellier, France.}  
\date{\today}  
\begin{document}  
  
\def\kE#1#2#3#4#5#6{{\buildrel {#1} \over k}{}^{#2}_{\! #3} \!\otimes \!{\buildrel {#4} \over E}{}^{#5}_{#6}}    
    
\def\coefalambda#1#2#3{\Lambda^{#1#2}_{#3}}    
 \def\coefblambda#1#2#3#4{\Lambda^{#1#2}_{#3#4}}    
\def\coefclambda#1#2#3#4#5#6{\Lambda^{#1#2}_{#3#4}(#5,#6)}

\def\ontop#1#2{\buildrel {{}_{#1}}\over #2}    
\def\twoontop#1#2#3{\buildrel {{}_{(#1,#2)}}\over #3}    
    
\def\End{\rm End}    
\def\Hom{\rm Hom}    
\def\Irr{\rm Irr}    
\def\dim{\rm dim}    
\def\id{\rm id}    
\def\Mat{\rm Mat}    
\def\Im{\rm Im} 
    
\def\Rff#1#2#3#4{{\buildrel #1 #2 \over {\bf R}}_{{} \atop {\!\!\!#3#4}}}    
\def\Rppff#1#2#3#4#5#6{{\buildrel #1 #2 \over R}\!{}^{(+)}{}^{#3#4}_{#5#6}}    
\def\Rmmff#1#2#3#4#5#6{{\buildrel #1 #2 \over R}\!{}^{(-)}{}^{#3#4}_{#5#6}}    
\def\Rpff#1#2{\buildrel #1#2 \over {R'}}    
\def\Rmff#1#2{\buildrel #1#2 \over {R^{-1}}}    
    
\def\GPhin#1#2#3{\Phi^{{}_{(#1)}{}_{(#2)}}_{\;\;\;\;\;{}_{(#3)}}}    
\def\GPhic#1#2#3{{\hat{\Phi}}^{{}_{(#1)}{}_{(#2)}}_{\;\;\;\;\;{}_{(#3)}}}    
\def\GPhip#1#2#3{\Phi^{{}_{(#1)}{}_{(#2)}}_{{}_{[+]}\;\;{}_{(#3)}}}    
\def\GPhim#1#2#3{\Phi^{{}_{(#1)}{}_{(#2)}}_{{}_{[-]}\;\;{}_{(#3)}}}    
\def\GPhie#1#2#3{\Phi^{{}_{(#1)}{}_{(#2)}}_{{}_{[\epsilon]}\;\;{}_{(#3)}}}    
\def\GPhis#1#2#3{\Phi^{{}_{(#1)}{}_{(#2)}}_{{}_{[\sigma]}\;\;{}_{(#3)}}}    
    
\def\CGphi#1#2#3{\phi^{{#1}{#2}}_{#3}}    
\def\CGpsi#1#2#3{\psi_{{#1}{#2}}^{#3}}    
\def\Clebphi#1#2#3#4#5#6{\left(\begin{array}{ll}     
\!#4 & \!#5 \\ \!\!\!#1 & \!\!\!#2 \end{array}\vline \begin{array}{ll} \!#3 \\ #6 \end{array}    
 \!\!\right)}    
\def\Clebpsi#1#2#3#4#5#6{\left(\!\begin{array}{ll}     
#6 \\ \!#3 \end{array}\vline \begin{array}{ll} #1 & \!\!\!#2 \\ #4 & \!\!\!#5 \end{array}\!\!    
 \right)}    
\def\sixj#1#2#3#4#5#6{\left\{\!\!\begin{array}{ll}     
#4 & \!\!\!#5 \\ #1 & \!\!\!#2 \end{array}\!\vline \begin{array}{ll} #3 \\ #6 \end{array}    
 \!\!\right\}}    
\def\sixjn#1#2#3#4#5#6#7{\left\{\!\!\begin{array}{ll}     
#4 & \!\!\!#5 \\ #1 & \!\!\!#2 \end{array}\!\vline \begin{array}{ll} #3 \\ #6 \end{array}    
 \!\!\right\}_{\!\!{}_{(#7)}}}    
\def\valsixj#1#2#3#4#5#6{\left\{\!\!\left\{\!\!\begin{array}{ll}     
#4 & \!\!\!#5 \\ #1 & \!\!\!#2 \end{array}\!\vline \begin{array}{ll} #3 \\ #6 \end{array}    
 \!\!\right\}\!\!\right\}}    
\def\nor6j#1#2#3#4#5#6{{\cal N}\left(\!\!\begin{array}{ll}     
#4 & \!\!\!#5 \\ #1 & \!\!\!#2 \end{array}\!\vline \begin{array}{ll} #3 \\ #6 \end{array}    
 \!\!\right)}    
\def\noraskey6j#1#2#3#4#5#6{\Gamma\left(\!\!\begin{array}{ll}     
#4 & \!\!\!#5 \\ #1 & \!\!\!#2 \end{array}\!\vline \begin{array}{ll} #3 \\ #6 \end{array}    
 \!\!\right)}    
    
\def\ElemRedphi#1#2#3#4#5#6{\left[\!\!\begin{array}{ll} #1 & #2 \\     
#4 & #5 \end{array}    
\!\vline \!\begin{array}{ll} #3 \\  #6 \end{array}    
 \!\!\!\right]}    
 
\def\ElemRedpsi#1#2#3#4#5#6{\left[\!\! \!\begin{array}{ll} #1 \\  #4 \end{array} \!\vline\begin{array}{ll} #2 & #3 \\     
#5 & #6 \end{array}       
 \!\!\!\right]} 
 
\def\Entrela#1#2#3#4#5#6#7#8#9{\left[\!\!\begin{array}{lll}    
 \; \mbox{\small \it #1 } & \; \mbox{\small \it #2 } \\     
#7 & #8\\    
\;\mbox{\small \it #4} & \; \mbox{\small \it #5 } \end{array}\!\vline \!\begin{array}{lll}     
\; \mbox{\small \it #3 }  \\     
#9\\    
\;\mbox{\small \it #6}     
 \end{array} \!\!\!\right]}

\def\qphit43#1#2#3#4#5#6#7{{}_4{\tilde\Phi}_{3}{\left[\!\!\begin{array}{cccc}     
 #1 &  #2 &  #3& #4\\     
 #5 &  #6 &  #7& {} \end{array}    
 \!\!\right]}}    
    
\def\qphi43#1#2#3#4#5#6#7#8#9{{}_4{\Phi}_{3}{\left[\!\!\begin{array}{cccc}     
 #1 &  #2 &  #3& #4\\     
 #5 &  #6 &  #7& {} \end{array};#8,#9    
 \right]}}

\def\qn#1{[#1]}    
\def\qd#1{[d_{#1}]}

\def\halfinteger{\frac{1}{2}\mathbb{Z}^{+}}    
\def\onehalf{\frac{1}{2}}    
\def\ZZ{\mathbb Z}    
\def\NN{\mathbb N}    
\def\RR{\mathbb R}    
\def\CC{\mathbb C}    
    
\def\ZZ{\mathbb Z}  
\def\GG{\mathbb G}  
\def\LL{\mathbb L}    
\def\VV{\mathbb V}  
\def\HH{\mathbb H}  
\def\UU{\mathbb U}  
\def\SSP{{\mathbb S}_P}  
\def\SSF{{\mathbb S}_F}  
\def\SS+{{\mathbb S}^{+}}    
\def\SSc{{\mathbb S}_c}    
\def\XX{\mathbb X}    
\def\AA{\mathbb A}  
\def\BB{\mathbb B}  
\def\MM{\mathbb M}

\def\Proof{\underline{\sf Proof:}}    
\def\eoProof{$\square$}    
\maketitle  

\begin{abstract} 
We analyze   the hamiltonian quantization of  Chern-Simons theory associated to the  real group  $SL(2, \CC)_{\RR},$ universal covering of  
the Lorentz group $SO(3,1).$ 
The algebra of observables is generated  by finite dimensional spin networks drawn on a punctured
 topological surface.
Our main result is a construction of a unitary representation of this algebra.
  For  
this purpose we use the formalism of combinatorial quantization of Chern-Simons theory, i.e we quantize the algebra of polynomial functions on the space of  flat $SL(2, \CC)_{\RR}-$connections on a topological surface $\Sigma$ with punctures. This algebra, the so called moduli algebra,  is constructed along the lines of Fock-Rosly, Alekseev-Grosse-Schomerus, Buffenoir-Roche using only finite dimensional representations of $U_q(sl(2,\CC)_{\RR})$. It is shown that this algebra admits a unitary representation acting on an Hilbert space which consists in wave packets of spin-networks associated to   principal unitary representations of  $U_q(sl(2,\CC)_{\RR}).$  The representation of the moduli algebra is constructed using only Clebsch-Gordan decomposition of a tensor product of a finite dimensional representation with a principal unitary representation of $U_q(sl(2,\CC)_{\RR})$. The  proof of unitarity of this representation  is non trivial  and is a consequence of properties of  $U_q(sl(2,\CC)_{\RR})$ intertwiners which  are studied in depth.
We analyze  the relationship between the insertion of a puncture colored with a principal 
representation and the presence of a world-line of a massive spinning particle in de Sitter space. 
\end{abstract} 
 
\section*{I. Introduction} 
In the  pioneering work of   \cite{AT,Wi1},  it has been shown that there  is an ``equivalence''  between 2+1 dimensional gravity with cosmological constant $\Lambda$ and Chern-Simons theory with a non compact group of the type $SO(3,1)$,  $ISO(2,1)$ or $SO(2,2)$  (depending on the sign of the cosmological constant $\Lambda$). A  good review on this subject is \cite{Ca}. As a result    
 the project of quantization of Chern-Simons theory for these groups has spin-offs on the program of canonical quantization of 2+1 quantum gravity. However  one should be aware that the two theories, nor in the classical case nor in the quantum case, are not completely  equivalent. These discrepencies arize  from various reasons. One of them, fully understood  by Matschull \cite{Ma},  is that the Chern-Simons formulation  includes degenerate metrics, and the classical phase space of Chern-Simons  is therefore quite different from the classical phase space of 2+1 gravity.  Another one comes from the structure of boundary terms (horizon, observer, particles) which have to be carefully related in the two models.\\ 
In this work, we study Chern-Simons formulation of 2+1 gravity in the case where the cosmological constant is positive i.e    Chern-Simons theory on a 3-dimensional compact oriented manifold $M=\Sigma \times \RR$ with  the real non compact  group $SL(2,\CC)_{\RR}$ universal covering of $SO(3,1)$.  This theory has been the subject of numerous studies, the main contributions being the work of E.Witten \cite{Wi2} using geometric quantization and the work of Nelson-Regge \cite{NR} using representation of the algebra of observables.  We will extend the analysis of Nelson-Regge using the so-called ``combinatorial quantization of Chern-Simons theory'' developped in \cite{FR, AGS1,AGS2,  BR1, BR2}. We first give the idea of the construction when the group is compact. 
Let $\Sigma$ be an oriented topological compact surface of genus $n$ and let us denote by $G=SU(2)$ and $\mathfrak g$ its Lie algebra. The classical  phase space of SU(2)-Chern Simons theory is the symplectic manifold $Hom(\pi_1(\Sigma), G)/{\text{Ad} G}$. The algebra of functions on this manifold  is a Poisson algebra which admits a quantization $M_{q}(\Sigma, G)$, called Moduli algebra in \cite{AGS1}, which is an associative algebra with an involution $*$. Note that $q$ is taken here to be a root of unity $q=e^{i\frac{\pi}{k+2}}$ where $k\in \NN$ is the coupling constant in the Chern-Simons action.  This algebra is built in two stages. One first defines a quantization of the Poisson algebra  $F(G^{2n})$ endowed with the Fock-Rosly Poisson structure \cite{FR}. This algebra is denoted ${\cal L}_{n}$ and called the graph algebra \cite{AGS1}. It is   the algebra generated by the matrix elements of the $2n$ quantum holonomies around the non trivial cycles $a_i, b_i$. $U_q({\mathfrak g })$  acts on  ${\cal L}_{n}$ by gauge transformations. The space of invariant elements ${\cal L}_n^{ U_q({\mathfrak g })}$ is a subalgebra of  ${\cal L}_{n}$ whose vector space basis  is entirely described by spin networks drawn on $\Sigma.$ 
If we define  $U_C$ to be the quantum holonomy around the cycle  
$C=\prod_{i=1}^n[a_i,b_i^{-1}]$,  one defines an ideal ${ \cal I}_C  $ of ${\cal L}_n^{ U_q({\mathfrak g })}$ which, when modded out, enforces the relation $U_C=1.$ As a result the moduli algebra is  
  $M_{q}(\Sigma, G)= {\cal L}_{n}^{U_q({\mathfrak g})}/{\cal I}_C.$  In \cite{Al,AS}  Alekseev and Schomerus have  constructed   its   unique unitary irreducible representation acting on a finite dimensional Hilbert space $H$. This is done in two steps. They have shown that there exists a unique unitary representation $\rho$ ($*$-representation) of the loop algebra  ${\cal L}_{n}$ acting on a finite dimensional space ${\cal H}$. The algebra generated by the matrix elements of $U_C$ is isomorphic to $U_q({\mathfrak g}),$ therefore $U_q({\mathfrak g})$ acts on ${\cal H}.$  $\rho$ can be restricted to the subalgebra   $ {\cal L}_{n}^{U_q({\mathfrak g})}$ and acts on the subspace of invariants ${\cal H}^{U_q({\mathfrak g})}=H.$ The ideal ${\cal I}_C$ is shown to be annihilated, as a result one obtains by this procedure a unitary representation of the moduli algebra.  This representation can be shown to be unique up to equivalence. Note that this construction is however implicit in the sense that no explicit formulae for the action of an element of  $M_{q}(\Sigma, G)$ is given in a basis of  ${\cal H}^{U_q({\mathfrak g})}.$ 
 In this brief exposition we have oversimplified the picture:  $q$ being a root of unity the formalism of weak quasi-Hopf algebras has to be used.  
 
We will modify this construction in order to handle the $SL(2,\CC)_{\RR}$ case. 
The construction of the moduli algebra in this case is straightforward and is parallel to the construction in the compact case. One defines the graph algebra  ${\cal L}_{n}$, generated by the matrix elements of the $2n$ quantum $SL(2, \CC)_{\RR}$  holonomies around the non trivial cycles $a_i, b_i$. This is a non commutative algebra on which  
 $U_q(sl(2,\CC)_{\RR})$ acts. We have chosen $q$ real, in complete agreement with the choice of the real invariant bilinear form on  
$SL(2,\CC )_{\RR}$ used to represent the $2+1$ gravity action with positive cosmological constant as a  Chern-Simons action.   
One  defines similarly  $M_{q}(\Sigma, SL(2, \CC)_{\RR} )= {\cal L}_{n}^{U_q(sl(2,\CC)_{\RR})}/{\cal I}_C$ which is a non commutative $*$-algebra, quantization of  the space of  functions on the moduli space of flat-$SL(2, \CC)_{\RR}$ connections. Although one can generalize the first step of the construction  of \cite{AS}, i.e constructing unitary  representations of  ${\cal L}_{n}$ acting on an Hilbert space ${\cal H}$, it is not possible to construct a unitary representation of $M_{q}(\Sigma, SL(2, \CC)_{\RR})$ by acting on  
${\cal H}^{U_q(sl(2, \CC)_{\RR})}.$ Indeed, there is no vector (of finite norm), except 0,  in the Hilbert space ${\cal H}$ which is invariant under the action of $U_q(sl(2, \CC)_{\RR}).$ This is a typical example of the fact that the volume of the gauge group is infinite (here it comes from the non compactness of $SL(2,\CC)_{\RR}$). To circumvent this problem  we use and adapt  the formalism of \cite{AS}  to directly  construct a representation of   $M_{q}(\Sigma, SL(2, \CC)_{\RR} )$ by acting on a vector space $H$. In a nutshell, $M_{q}(\Sigma, SL(2, \CC)_{\RR} )$ is generated by spin network colored by finite dimensional representations, whereas vectors in $H$ are integral of spin networks colored by principal representations of $U_q(sl(2,\CC)_{\RR})$. We give explicit formulae for the action of $M_{q}(\Sigma, SL(2, \CC)_{\RR} )$ on $H$, we endow this space with a structure of Hilbert space and show that the representation is unitary. Our approach uses as central tools the harmonic analysis of $U_q(sl(2, \CC)_{\RR})$ and an explicit construction of Clebsch-Gordan coefficients of principal representations of $U_q(sl(2, \CC)_{\RR})$, which have been developped in \cite{BR3,BR4}. 

Note that Nelson and Regge have previously succeeded to construct unitary representation of the Moduli algebra in the case of genus one in \cite{NR} and in the genus 2 case in the $SL(2, \RR)$ case in \cite{NR2}. Our method  works for any punctured surface of arbitrary genus and, despite certain technical points which have been mastered, is very natural. It is a non trivial implementation of the concept of refined algebraic quantization developped in \cite{ALMMT}.

\section*{II. Summary of the Combinatorial Quantization Formalism: the compact group case.} 
Chern-Simons theory with gauge group $G=SU(2)$ is  defined on a 3-dimensional compact oriented manifold $M$ by the action 
\begin{eqnarray} 
S(A)=\frac{\lambda}{4\pi} \int_M \text{Tr}(A \wedge dA + \frac{2}{3} A \wedge A  \wedge A)\;\;,\label{ChernSimonsaction} 
\end{eqnarray} 
where the gauge field $A = A_{\mu} dx^{\mu}$  and Tr is the Killing form on ${\mathfrak g}=su(2)$. In the sequel we will investigate the case where   Chern-Simons theory  has an hamiltonian formulation. We will therefore assume that the manifold $M=\Sigma \times \RR,$ where the real line can be thought as  being  the time direction and $\Sigma$ is a compact oriented surface and we will write $A=A_0 dt+ A_1 dx^1 +A_2 dx^2.$  
In the action (\ref{ChernSimonsaction}) $A_0$ appears as a Lagrange multiplier. Preserving the gauge choice $A_0=0$ enforces the first class constraint  
 
\begin{eqnarray} 
F_{12}(A)=\partial_1 A_2-\partial_2 A_1 +[A_1,A_2]=0.\label{flatness} 
\end{eqnarray} 
 
The space ${\cal A}(\Sigma, G)$  of $G$-connections on $\Sigma$ is an infinite dimensional affine symplectic space with   
 Poisson bracket: 
\begin{eqnarray} 
\{ A_{i }(x)\;  \stackrel{  \otimes}{ ,} \; A_{j }(y) \} = \frac{2\pi}{\lambda}\delta (x-y) \epsilon_{ij} t \;\; ,\label{poisson}   
\end{eqnarray} 
where $t \in {\mathfrak g} \otimes {\mathfrak g}$ is the Casimir tensor associated to the non-degenerate bilinear form Tr defined by $t= \sum_{a,b}(\eta^{-1})^{ab} T_a\otimes T_b$ where $T_a$ is any basis of ${\mathfrak g}, \eta_{ab}= \text{Tr}(T_aT_b)$ and $i,j \in \{1,2\}.$  
 
The constraint 
 (\ref{flatness})  generates gauge transformations  
\begin{eqnarray} 
{}^g A \; = \; g \; A \; g^{-1} \; + \; dg \; g^{-1} \;\; ,\forall \; g \in C^{\infty}(\Sigma,G) \;\;. 
\end{eqnarray} 
As a result, the classical phase space of this theory consists of the moduli space of flat G-bundles on the surface $\Sigma$ modulo the gauge transformations and  has been studied in \cite{At}.  
 
In order that $exp(iS(A))$ is gauge invariant under large gauge transformations, $\lambda$ has to be an integer.

The moduli space of flat connections $M(\Sigma, G)$ is defined using an infinite dimensional version of Hamiltonian reduction, i.e  
$M(\Sigma, G)=\{A\in {\cal A}(\Sigma, G), F(A)=0\}/{\cal G}$ where the group 
 ${\cal G}$  is the group of gauge transformations. 
The quantization of this space can follow two paths: quantize before applying the constraints or quantize after applying the constraints. 
The approach of Nelson Regge aims at developping the latter but it is cumbersome. We can take advantage of the fact that $M(\Sigma, G)$ is finite dimensional to replace the gauge theory on $\Sigma$ by a lattice gauge theory on $\Sigma$ following  Fock-Rosly's idea \cite{FR}. This method aims at quantizing before applying the constraints but in a finite dimensional framework. 
 
This framework can be generalized to the case of a topological surface $\Sigma$ with punctures $P_1,..., P_p$. If $A$ is a flat connection on a punctured surface one denotes by $H_x(A)$ the conjugacy class of the holonomy around a small circle centered in $x$. One chooses $\sigma_1, ..., \sigma_p$ conjugacy classes in $G$ and defines $M(\Sigma, G;\sigma_1,...,\sigma_p)=\{A\in {\cal A}(\Sigma, G), F(A)=0,\, H_{P_i}(A)= \sigma_i\}/{\cal G}$ where the group 
 ${\cal G}$  is the group of gauge transformations.  
The symplectic structure on this  space is well analyzed in \cite{AM}.

\subsection*{II.1. Fock-Rosly description of the moduli space of flat connections.} 
Functions on $M(\Sigma, G),$  also called observables, are  gauge invariant functions on $\{A\in { \cal A}(\Sigma, G), F(A)=0\} $.  Wilson loops are examples of observables, and are particular examples of the following construction  which associates to any  spin-network an observable. 
 Let us consider  an oriented graph on   $\Sigma,$  this graph consists in  a set of  oriented edges (generically denoted by $l$) which meet at vertices (generically denoted by $x$).  
Let $<$ be a choice of an  order on the set of oriented edges.  
 
It is  convenient to introduce the notations $d(l)$ and $e(l)$ respectively for the departure  point and the end point of an oriented  edge $l$. 
 
A spin network associated to an oriented graph on $\Sigma$ consists in two data: 
\begin{itemize} 
\item  a  coloring of the set of oriented edges i.e    each oriented edge  $l$ is associated to a finite dimensional module  $V_l$ of the algebra ${\mathfrak g}.$ We denote by $\pi_l$ the representation associated to $V_l.$ For any $l$ and $x$ we define $V^{+}_{(l,x)}=V_l$ if $e(l)=x$ and $V^{+}_{(l,x)}=\CC$ elsewhere, as well as $V^{-}_{(l,x)}=V_l^*$ if $d(l)=x$ and $V^{-}_{(l,x)}=\CC$ elsewhere. 
\item a coloring of the vertices i.e each vertex is associated to an intertwiner 
$\phi_x \in Hom_{\mathfrak g} (\otimes^{<}_{l} (V^+_{(l,x)}\otimes V^-_{(l,x)}) , \mathbb C).$  
\end{itemize} 
 
To each   spin network ${\cal N}$ we can associate a function on  $M(\Sigma, G),$ as follows: 
 let $U_l(A) \; = \; \pi_{l}(\stackrel{\leftarrow}{P} \exp \int_l A)$ , and define  
\begin{equation} 
f_{\cal N}(A)= 
(\otimes_x \phi_x)(\otimes^<_l U_l(A))\label{observableclassique} 
\end{equation} 
 where we have identified $V_l\otimes V_l^{\star}$ with $End(V_l)$.

 The Poisson structure on  $M(\Sigma, G),$ can be neatly described in terms of the functions $f_{\cal N}$  as first understood by Goldman \cite{Go}. Given two spin-networks ${\cal N}, {\cal N}'$ such that their associated graphs are  in generic position,  we have 
\begin{eqnarray} 
\{f_{\cal N} \;,\; f_{{\cal N}^\prime}\} \; = \frac{2\pi}{\lambda}\; \sum_{x\in {\cal N} \cap {\cal N}^{\prime}} f_{{\cal N} \cup_x {\cal N}^{\prime}} \; \epsilon_x({\cal N}; {\cal N}^{\prime}) \;\; ,
\end{eqnarray} 
where the graph $ {\cal N}\cup_x {\cal N}^{\prime}$ is defined to be the union of ${\cal N}$ and ${\cal N}^{\prime}$ with the additional   vertex $x$ associated to the intertwiner $ P_{12} t_{12}: V\otimes V'\rightarrow V'\otimes V$ (which can be viewed as an element of $Hom(V\otimes V' \otimes V^{\star} \otimes V^{\prime \star} ; \CC)$) and where the sign  $\epsilon_x({\cal N}; {\cal N}^{\prime}) \;=\; \pm 1$ is the index of the intersection of the  two graphs at the vertex $x$.    
Quantizing directly this Poisson structure is too complicated (see however \cite{NR, Tu}). 
We will explain now Fock-Rosly's construction  and the definition of the combinatorial quantization of the moduli space $M_q(\Sigma, G).$ 
 
Finite dimensional representations of $G$ are classified by a positive half integer $I\in \onehalf\NN$, and we will denote $\stackrel{I}{V}$ the associated module with representation $\stackrel{I}{\pi}.$

Let $\Sigma$ be a surface of genus $n,$ with $p$ punctures associated to a conjugacy class $\sigma_i, i=1,...,p$ of $G$. Fock-Rosly's  idea amounts  to replace  the  surface by an oriented fat graph ${\cal T}$ drawn on it and the space of connections on $\Sigma$ by the space of holonomies on this fat  graph.  We assume that the surface is divided by the graph into plaquettes such that either this plaquette is contractible or contains a unique  puncture.   Let us denote ${\cal T}^0$ the set of vertices of the graph, ${\cal T}^1 $ the set of  edges  and  ${\cal T}^2 $ the set of faces. 
 
The orientation of the surface induces at each vertex $x$ a cyclic order on the  set of edges $L_{x}$ 
incident to $x$.

 We can now introduce the space of discrete connections, which is an equivalent  name for lattice gauge field on ${\cal T}$. The space of discrete connections ${\cal A}({\cal T})$ on the  surface  
$\Sigma $ is defined as 
\begin{equation} 
{\cal A}({\cal T}) = \{U(l) \in G \; ; \; l \in {\cal T}^1 \} \;\; 
\end{equation} 
and the group of gauge transformations $G^{{\cal T}^0}$ acts on the discrete connections as follows: 
\begin{equation} 
U(l)^g = g(e(l)) \; U(l) \;  g(d(l))^{-1} \;\; \; \; , \forall g \in G^{ {\cal T}^{0}} \; 
\; . 
\end{equation} 
 The discrete connections can be viewed  as functionals of the connection 
 $A\in {\cal A}(\Sigma, G)$   as ${U}(l)=\stackrel{\leftarrow}{P}exp\int_l A.$ 
If $f\in {\cal T}^2,$ let $U(f)$ be the conjugacy class of $\stackrel{\leftarrow}{\prod}_{l\in \partial f}U(l).$ For each  $f\in {\cal T}^2,$ we denote $\sigma_f=1$ if $f$ is contractible and $\sigma_f=\sigma_{i}$ if $f$ contains the puncture $P_i$. 
The group   $G^{{\cal T}^0}$ has a natural  Lie-Poisson structure: 
\begin{eqnarray} 
\{\stackrel{I}{g}_{1}\!\!(x)\;,\;\stackrel{J}{g}_{2}\!\!(x) \} & = &  \frac{2\pi}{\lambda}\lbrack \stackrel{IJ}{r}_{12}  \; , \;\stackrel{I}{g}_{1}\!\!(x) \; \stackrel{J}{g}_{2}\!\!(x) \rbrack \;\; ,\\ 
\{\stackrel{I}{g}_{1}\!\!(x)\;,\;\stackrel{J}{g}_{2}\!\!(y) \} & = & 0 \;\; if \; x \neq y \;\; , 
\end{eqnarray} 
where we have used the notations 
\begin{eqnarray} 
\stackrel{I}{g}_{1}\!\!(x) \;=\;\stackrel{I}{g}\!\!(x)\;\otimes \;\stackrel{J}{1} \;\;\; ,\; \stackrel{J}{g}_{2}\!\!(x)\;=\;\stackrel{I}{1} \;\otimes \; \stackrel{J}{g}\!\!(x) \;\; , 
\end{eqnarray} 
and $\stackrel{I}{g}\!\!(x) \; \in End(\stackrel{I}{V}) \; \otimes \; F(G)_x$ ($F(G)_x$ being the functions on the group at the vertex $x$), $r \in {\mathfrak g}^{\otimes 2}$ is a classical r-matrix which satisfies the classical Yang-Baxter equation and  $r_{12} + r_{21} = t_{12}.$  
 
Fock and Rosly \cite{FR} have introduced a Poisson structure on the functions on  ${\cal A}({\cal T})$ denoted $\{,\}_{FR}$ such that the gauge transformation map 
 \begin{equation} 
G^{{\cal T}^0} \times {\cal A}({\cal T})\rightarrow  {\cal A}({\cal T}) 
\end{equation} 
is a Poisson map. 
 
 Note however that this Poisson structure is not canonical and depends on an additional item (called in their paper a ciliation ), which is a linear   order $<_x$  compatible with the cyclic order defined on the set of edges incident to the vertex $x$. 
 
We shall give here the Poisson structure on the space of discrete connections in the case where ${\cal T}$ is a triangulation: 
 
\begin{eqnarray} 
\{ \stackrel{I}{U}_1\!\!(l) \; , \;  \stackrel{J}{U}_2\!\!(l') \}_{FR} & = & \frac{2\pi}{\lambda}(\stackrel{IJ} {r}_{12} \; \stackrel{I}{U}_1\!\!(l) \; \stackrel{J}{U}_2\!\!(l')) \;\; , 
\; \; \text{if} \;\; e(l)=e(l') = x \; and \; l <_x l' \;\; ,\label{FockRosly1}\\ 
\{ \stackrel{I}{U}_1\!\!(l) \; , \;  \stackrel{J}{U}_2\!\!(l) \}_{FR} & = & \frac{2\pi}{\lambda}(\stackrel{IJ} {r}_{12} \; \stackrel{I}{U}_1\!\!(l) \; \stackrel{J}{U}_2\!\!(l) \;+  \stackrel{I}{U}_1\!\!(l) \; \stackrel{J}{U}_2\!\!(l) \; \stackrel{IJ} {r}_{21} )\;\; ,\label{FockRosly2} \\ 
\{ \stackrel{I}{U}_1\!\!(l) \; , \;  \stackrel{J}{U}_2\!\!(l') \}_{FR} & =  & 0 \;\; if \; l \cap l' = \emptyset \label{FockRosly3} \;,\; 
\end{eqnarray} 
the other relations can be deduced from the previous ones using the relation $U(-l)U(l)=1.$\\ 
The moduli space can be described as: 
$M(\Sigma, G, \sigma_1,..., \sigma_p)=\{ {\cal A}({\cal T}), U(f)=\sigma_f, f\in {\cal T}^2 \}/G^{{\cal T}^0}.$ 
The major result of \cite{FR} is that the Poisson structure $\{,\}_{FR}$ descends to this quotient, is not degenerate, independent of the choice of the fat graph and on the ciliation  and is the Poisson structure associated to the canonical symplectic structure on  
$M(\Sigma, G; \sigma_1,...,\sigma_p).$

A  quantization of Fock-Rosly Poisson bracket   has been analyzed in \cite{AGS1,AGS2, BR1}. In order to give a sketch of this construction we  will first recall standard results on quantum groups.\\

\subsection*{II.2. Basic notions on quantum groups.}  
  
Basic definitions and properties of the quantum envelopping algebra $U_q({\mathfrak g})$ where ${\mathfrak g}=su(2)$  are recalled in the appendix A.1.  
 $U_q({\mathfrak g})$ is  a quasi-triangular ribbon Hopf-algebra with counit $\epsilon: \; U_q({\mathfrak g}) \longrightarrow \CC$, coproduct $\Delta: \; U_q({\mathfrak g}) \longrightarrow  U_q(\mathfrak g)^{ \otimes 2}$ and antipode $S:\; U_q({\mathfrak g}) \longrightarrow  U_q({\mathfrak g})$. For a review on quantum groups, see \cite{CP}. The universal R-matrix $R$ is an element of $U_q({\mathfrak g})^{\otimes 2}$ denoted by $R = \sum_i x_i \; \otimes \; y_i=R^{(+)} $.  
It is also convenient to introduce $R^{\prime} = \sum_i y_i \; \otimes \; x_i$ and $R^{(-)}=R'{}^{-1}$.  
 Let $u = \sum_i S(y_i)\;x_i$,  $uS(u)$ is in the center of $U_q(\mathfrak g{})$ and there exists a central element $v$ (the ribbon element) such that $v^2=uS(u).$  
We will define the group-like element $\mu=q^{2J_z}$.    
  
 Finite dimensional irreducible representations $\stackrel{ I}{\pi}$ of $U_q({\mathfrak g})$ are labelled by $ I \in \frac{1}{2} \NN$ and let us define $\stackrel{I}{V}$ the associated module.  The tensor product $\stackrel{ I}{\pi} \otimes \stackrel{ J}{\pi}$  of two representations is  decomposed into irreducible representations $\stackrel{ K}{\pi}$   
\begin{eqnarray}  
\stackrel{ I}{\pi} \otimes \stackrel{ J}{\pi}= \sum_{ K} N^{ IJ}_{ K} \; \stackrel{ K}{\pi} \;\; ,  
\end{eqnarray}  
where the integers $N^{ IJ}_{ K}\in\{0, 1\}$ are the multiplicities. For any representations $\stackrel{ I}{\pi}$, $\stackrel{ J}{\pi}$, $\stackrel{ K}{\pi}$, we define  the Clebsch-Gordan maps $\Psi^{ K}_{ IJ}$ (resp. $\Phi_{ K}^{ IJ}$) as a basis of $Hom_{U_q({\mathfrak g})}(\stackrel{ I}{V} \otimes \stackrel{J}{V}, \stackrel{ K}{V})$ (resp. $Hom_{U_q({\mathfrak g})}(\stackrel{ K}{V}, \stackrel{ I}{V} \otimes \stackrel{ J}{V})$). These basis can always be chosen such that:  
\begin{equation}  
 N_{ K}^{ IJ} \Psi^{ L}_{ IJ}\Phi_{ K}^{ IJ}= \; N_{ K}^{ IJ} \;  
 \; \delta^{ L}_{ K} \;   
id_{ \stackrel{ K}{V}}\;\;\;, \;\;\;   
    \sum_{{ K}}\Phi_{ K}^{ IJ}\Psi^{ K}_{ IJ}=  
id_{ \stackrel{ I}{V} \otimes \stackrel{ J}{V}}.  
\end{equation}

For any finite dimensional representation ${I}$, we will define the quantum trace of an element  
 $M\in End (\stackrel{ I}{V})$ as $tr_q(M) = tr_{\stackrel{ I}{V}} (\stackrel{ I}{\mu} M).$  
The element $c_{I}=tr_{q}((\stackrel{ I}{\pi}\otimes id )(RR')) $ is a central element of   
$ U_q({\mathfrak g}).$ For any finite dimensional representation $I$ and for any irreducible module $V$ associated to the representation $\pi$ (not necessarily of finite dimension),   we will denote by $\vartheta_{I\pi}$ the complex number defined by  $\pi(c_I)=\vartheta_{I\pi} \id_{V}.$ For ${\mathfrak g}=su(2)$, $\vartheta_{IJ}=\vartheta_{I\stackrel{ J}{\pi}}=\frac{[(2I+1)(2J+1)]}{[2I+1] [2J+1]}$ where $I,J \in \frac{1}{2} \NN$ label irreducible representations of $ U_q(su(2))$ and quantum numbers $[x]$ is defined in the appendix. 
  
 Let us denote by $\{\stackrel{ I}{e}_i \mid i = 1 \cdots dim \stackrel{ I}{V} \}$ a particular basis of $\stackrel{I}{V}$ and $\{\stackrel{ I}{e}{}\!\!^i \;\mid i = 1 \cdots dim \stackrel{ I}{V} \}$ its dual basis. By duality, the space $Pol_q(G)$ of polynomials  on the quantum group inherits a structure of Hopf-algebra. It is generated as a vector space by the coefficients of the representations $\stackrel{ I}{\pi}$, which will be denoted by $\stackrel{ I}{g}{}\!\!^a_{b} = \big <\stackrel{ I}{e}{}\!\!^a \mid \; \stackrel{ I}{\pi}\!(.) \;\mid \stackrel{ I}{e}_b \big >. $  To simplify the notations, we define $\stackrel{ I}{g} \; =\sum_{a,b} \; \stackrel{ I}{E}{}\!\!^b_{a}  \otimes  \stackrel{ I}{g}{}\!\!^a_{b}  \in End(\stackrel{ I}{V})  \otimes  Pol_q(G)$ where the elements $\{\stackrel{ I}{E}{}\!\!^b_{a}\}_{a,b}$ is the canonical basis of $End(\stackrel{ I}{V})$. By a direct application of the definitions,  we have the fusion relations  
\begin{eqnarray}  
\stackrel{ I}{g}_1 \; \stackrel{ J}{g}_2\; = \; \sum_{ { K}} \Phi_{{ K} }^{ IJ} \; \stackrel{ K}{g}  \; \Psi^{{ K}}_{IJ}\;\; ,  
\end{eqnarray}  
which imply the exchange relations $\stackrel{ IJ}{R}_{12} \; \stackrel{ I}{g}_1 \; \stackrel{ J}{g}_2\;  = \; \stackrel{ J}{g}_2 \; \stackrel{ I}{g}_1 \; \stackrel{ IJ}{R}_{12} \;\; ,$  
where $ \stackrel{ IJ}{R}_{12} = (\stackrel{ I}{\pi} \otimes  \stackrel{J}{\pi})(R) \; \in \; End(\stackrel{I}{V})  \otimes  End(\stackrel{ J}{V}).$   
  
The coproduct is  
\begin{eqnarray}  
\Delta(\stackrel{ I}{g}{}\!\!^a_b)\; = \; \sum_c \stackrel{ I}{g}{}\!\!^a_c \; \otimes \; \stackrel{ I}{g}{}\!\!^c_b\;\; .  
\end{eqnarray}  
Up to this point, it is possible to give a presentation, of FRT type,  of the  defining relations of $U_q({\mathfrak g})$. Let us introduce, for each representation $I$, the element $ \stackrel{ I}{L}{}\!\!^{(\pm)}\; \in \; End(\stackrel{I}{V} ) \otimes U_q({\mathfrak g})$ defined by $\stackrel{ I}{L}{}\!\!^{(\pm)}= (\stackrel{ I}{\pi} \otimes id) (R^{(\pm)})$.  
The duality bracket is given by  
\begin{eqnarray}  
{\Big <} \stackrel{ I}{L}_1{}\!\!^{(\pm)} \; ,\stackrel{ J}{g}_2 \;  {\Big >} =\stackrel{ IJ}{R}{}\!\!^{(\pm)}_{12} \;\;.  
\end{eqnarray}  
These matrices satisfy the relations:  
\begin{eqnarray}  
\stackrel{ I}{L}_1{}\!\!^{(\pm)} \; \stackrel{ J}{L}_2{}\!\!^{(\pm)}  & = & \sum_{{ K}} \Phi^{ IJ}_{{ K}} \; \stackrel{ K}{L}{}\!\!^{(\pm)} \Psi_{ IJ}^{{ K}}\;\; ,\\  
\stackrel{IJ}{R}_{12}{}\!\!^{(\pm)} \; \stackrel{ I}{L}_1{}\!\!^{(+)} \;  \stackrel{ J}{L}_2{}\!\!^{(-)} & = & \stackrel{ J}{L}_2{}\!\!^{(-)} \; \stackrel{ I}{L}_1{}\!\!^{(+)} \;  \stackrel{ IJ}{R}_{12}{}\!\!^{(\pm)} \; \; , \\  
\Delta (\stackrel{ I}{L}{}\!\!^{(\pm)a}_{\;\;\;\;\;b}) & = & \sum_c \stackrel{ I}{L}{}\!\!^{(\pm)c}_{\;\;\;\;\;b} \; \otimes \; \stackrel{ I}{L}{}\!\!^{(\pm)a}_{\;\;\;\;\;c}\;\; .  
\end{eqnarray}  
The first fusion equation implies the exchange relations  
\begin{eqnarray}  
 \stackrel{ IJ}{R}_{12} \; \stackrel{ I}{L}_1{}\!\!^{(\pm)} \; \stackrel{ J}{L}_2{}\!\!^{(\pm)} \; = \;  \stackrel{ J}{L}_2{}\!\!^{(\pm)} \; \stackrel{ I}{L}_1{}\!\!^{(\pm)} \;  \stackrel{ IJ}{R}_{12} \;\; .  
\end{eqnarray}

\subsection*{II.3. Combinatorial Quantization of the moduli space of flat connections.}  
  
\begin{figure} 
\psfrag{A(1)}{$A(1)$}  
\psfrag{B(1)}{$B(1)$}  
\psfrag{A(2)}{$A(2)$}  
\psfrag{B(2)}{$B(2)$}  
\psfrag{A(n)}[][]{$A(n)\;\;$}  
\psfrag{B(n)}[][]{$B(n)\;\;$}  
\psfrag{M(1)}{$M(n+1)$}  
\psfrag{M(p)}{$M(n+p)$}  
\psfrag{x}{$x$} 
\centering 
\includegraphics[scale=0.7]{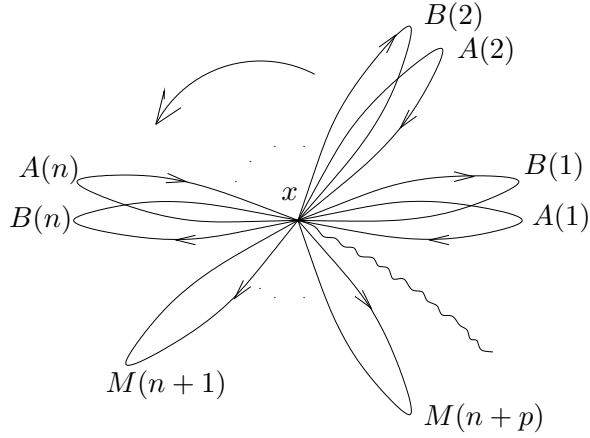} 
\caption{Standard Graph.}  
\label{Monodromy} 
\end{figure}

We are now ready to define a quantization of the space of  flat connections along the lines of \cite{AGS1}.  Because this construction can be shown to be independent of the choice of ciliated fat graph, we will choose a specific graph, called standard graph, which is shown in  figure \ref{Monodromy}. 
 
This graph consists in one vertex $x$ , $p+1$ 2-cells and $2n+p$ 1-cells. The $2n+p$ 1-cells are given with the orientation and the order $<$ of the picture.  
  
The space of discrete connections on this graph consists in   
  the holonomies $\{M(j) \mid j = n+1, \cdots, n+p\} $ around the punctures and the holonomies $\{A(i),\; B(i) \mid i = 1, \cdots, n\} $ around the handles. We can choose the associated curves in such a way that they have  the same base point $x$ on the surface.   
  
We associate to this graph ${\cal T}$ a   
quantization of the Fock-Rosly Poisson structure on  the space of discrete connections on ${\cal T}$ as follows:  
  
\begin{definition}{\rm (Alekseev-Grosse-Schomerus)\cite{AGS1}}  
The graph algebra ${\cal L}_{n,p}$ is an associative algebra generated  by the matrix elements of $(\stackrel{ I}{A}\!\!(i))_{i=1,\cdots,n}$, $(\stackrel{ I}{B}\!\!(i))_{i=1,\cdots,n}$, $(\stackrel{ I}{M}\!\!(i))_{i=n+1,\cdots,n+p}$ $\in End (\stackrel{ I}{V}) \; \otimes  \; {\cal L}_{n,p}$ and  satisfying the relations:  
\begin{eqnarray}  
\stackrel{ I}{U}_1\!\!(i) \;  \stackrel{ IJ}{R^{\prime}} \; \stackrel{ J}{U}_2\!\!(i) \; \stackrel{ IJ}{R^{(-)}}   & = & \sum_K \Phi_{ K}^{ IJ} \; \stackrel{ K}{U}\!\!(i) \; \Psi_{ IJ}^{ K} \;\;\text{(Loop Equation)}\;\; \forall \;i,\label{fusion}\\  
\stackrel{ IJ}{R} \; \stackrel{ I}{U}_1\!\!(i) \; \stackrel{ IJ}{R}{}\!\!^{-1} \;\stackrel{ J}{U}_2\!\!(j)    & = &  \stackrel{ J}{U}_2\!\!(j) \; \stackrel{ IJ}{R} \; \stackrel{ I}{U}_1\!\!(i) \; \stackrel{ IJ}{R}{}\!\!^{-1} \;\;\;\;\; \forall \;i<j,\\  
\stackrel{ IJ}{R} \; \stackrel{ I}{A}_1\!\!(i) \; \stackrel{ IJ}{R^{\prime}} \; \stackrel{ J}{B}_2\!\!(i)  & = & \stackrel{ J}{B}_2\!\!(i) \; \stackrel{ IJ}{R} \; \stackrel{ I}{A}_1\!\!(i) \; \stackrel{ IJ}{R}{}\!\!^{-1}  \;\;\;\;\; \forall \; i ,  
\end{eqnarray}  
where $U\!(i)$ is indifferently $A(i)$, $B(i)$ or $M(i)$.   
The relations are chosen in such a way that the co-action  $\delta$:  
\begin{eqnarray}  
\delta & : &  {\cal L}_{n,p} \;\;\; \rightarrow F_q(G) \; \otimes \; {\cal L}_{n,p} \nonumber\\  
       &   &   \stackrel{ I}{U}\!\!(i){}^a_b \;\; \mapsto \sum_{c,d}\stackrel{ I}{g}{}\!\!^a_c \;S( \stackrel{ I}{g}{}\!\!^d_b)  \; \otimes \; \stackrel{ I}{U}\!\!(i){}^c_d \; =\; \big( \stackrel{ I}{g} \;\stackrel{ I}{U}\!\!(i) \;S( \stackrel{ I}{g}) \big)^a_b\;\; \label{transfojauge}  
\end{eqnarray}  
is a morphism of algebra.  
\end{definition}  
This last property is the quantum version of the fact that the map (\ref{transfojauge}) is a Poisson map.  
Equivalently the coaction $\delta$ provides a right action of   $U_q({\mathfrak g})$  on ${\cal L}_{n,p}$ as follows:  
\begin{eqnarray}  
&&\forall a, b\in {\cal L}_{n,p},  
 \forall \xi\in  U_q({\mathfrak g}), (ab)^{\xi}=a^{\xi_{(1)}}b^{\xi_{(2)}},\\  
&&\stackrel{ I}{U}(i)^{\xi}= \stackrel{ I}{\pi}(\xi_{(1)})   
 \stackrel{ I}{U}(i)\stackrel{ I}{\pi}(S(\xi_{(2)})),\\  
&&\text{where} \;\;U\!(i) \;\;\text{is indifferently}\;\; A(i), B(i) \;\;\text{or} \;\; M(i).\nonumber   
\end{eqnarray}

Let us notice that, from (\ref{fusion}), $\stackrel{ I}{U}\!\!(i)$ admits an inverse matrix $\stackrel{ I}{U}\!\!(i)^{-1}$, see \cite{AGS1,BR1}.  \\

The space of gauge invariant elements is the subspace  of coinvariant elements of  ${\cal L}_{n,p}$ i.e   
 $ {\cal L}_{n,p}^{inv}=\{a \in{\cal L}_{n,p}, \delta(a)=1 \otimes a\}.$ This is an algebra because $\delta$ is a morphism of algebra.\\

In $ {\cal L}_{n,p}^{inv}$ we still have to divide out by the flatness condition, i.e the quantum version of the fact   
that the holonomies on contractible curves are trivial and that around punctures they belong to a fixed conjugacy class. Let us define   
\begin{eqnarray}  
\stackrel{ I}{C}&=&\stackrel{ I}{G}\!\!(1) \cdots \stackrel{ I}{G}\!\!(n) \cdot \stackrel{ I}{M}\!\!(n+1) \cdots \stackrel{ I}{M}\!\!(n+p)\label{defofC}  
\end{eqnarray}  
where $ \stackrel{ I}{G}\!\!(i) =v_{I}^2 \stackrel{ I}{A}\!\!(i) \; \stackrel{ I}{B}\!\!(i)^{-1} \stackrel{ I}{A}\!\!(i)^{-1} \stackrel{ I}{B}\!\!(i)$. The elements $\stackrel{ I}{C}$ satisfy the loop equation (\ref{fusion}).  
We will denote by ${\cal C}$ the subalgebra of ${\cal L}_{n,p}$ generated by the matrix elements of $\stackrel{ I}{C}, \forall \; I$. It can be shown that    
$tr_q(\stackrel{ I}{C})$ and $tr_q(\stackrel{ I}{M}(i)), i=n+1,...,n+p$ are central elements of the algebra  ${\cal L}_{n,p}^{inv}.$

We would like first to divide out by the relation $\stackrel{ I}{C}=1.$  
  
An annoying fact  is that the matrix elements of $\stackrel{ I}{C}-1$ do not belong to $ {\cal L}_{n,p}^{inv}.$ As a result in order to divide out by this relation we have to slightly modify the picture.   
  
Let $I$ be a finite dimensional representation of $U_q({\mathfrak g})$ and let   ${\cal J}_{I}\subset   {\cal L}_{n,p} \otimes  End(\stackrel{ I}{V})$ such that $X\in {\cal J}_{I}$ if and only if $X=\sum_{a,b}X^{a}_b \otimes E^{b}_a$ with $\delta(X^{a}_b)=\sum_{a',b'} g^{a}_{a'}X^{a'}_{b'}S(g^{b'}_{b})$.   For any $Y\in {\cal J}_{I}$ we define the invariant element $<Y(\stackrel{ I}{C}-1)>=  
\sum_{a,b}\stackrel{I}{\mu}{}^{bb}Y^{b}_a(\stackrel{ I}{C}{}^a_b-\delta^{a}_b).$   
Let $ {\cal I}_C$ be the ideal of  $ {\cal L}_{n,p}^{inv}$ generated by the elements $<Y(\stackrel{ I}{C}-1)>$ where $I$ is any finite dimensional representation of $U_q({\mathfrak g})$ and $Y$ is any element of ${\cal J}_{I}.$  
  
\begin{definition}  
We define ${\mathfrak M}_q(\Sigma, G,p)$ to be the algebra ${\cal L}_{n,p}^{inv}/ {\cal I}_C.$ When there is no puncture this is the Moduli algebra of \cite{AGS1}, and we will write in this case $M_q(\Sigma, G)={\mathfrak M}_q(\Sigma, G,p=0).$  
\end{definition}

In the case of punctures a quantization of the coadjoint orbits is necessary. This is also completely consistent with quantization of Chern-Simons theory, where punctures are associated to vertical lines colored by representations of the group $G$.   
  
\begin{definition}  
Let $\pi_1,..., \pi_p$ be the representations associated to the vertical lines coloring the punctures.  
We can define, following \cite{AGS1}, the moduli algebra   
$M_q(\Sigma, G, \pi_1,\cdots,\pi_p)={\mathfrak M}_q(\Sigma, G,p)/\{tr_q(\stackrel{ I}{M}(n+i))=\vartheta_{I\pi_{i}}, i=1,...,p,\forall I\}.$  
\end{definition}  
  
In order to introduce a generating family of  gauge invariant elements we have to define the notion of quantum spin-network. The definition of this object is the same as in the classical case except that the coloring of the edges are representations of $U_q({\mathfrak g})$ and that the coloring of the vertices are $U_q({\mathfrak g})$-intertwiners. To each quantum spin-network one associates an element of $M_q(\Sigma, G)$ by the same equation as (\ref{observableclassique}), the order $<$ is now essential because it orders non commutative holonomies in the tensor product.   
  
In the following proposition, we will construct an explicit basis of the vector space ${\cal L}_{n,p}^{inv}$ labelled by quantum spin networks. This will provide, after moding out by  the relations defining the moduli algebra, a generating family of this algebra.   
  
We will need the following notations: if $L=(L_1,..., L_r)$ and $L'=(L_1',...,L_s')$   
are sequences, we denote $LL'$ to be the sequence $(L_1,...,L_r,L_1',...,L_s').$  
If $L$ is a sequence we denote $L_{<j}=(L_1,..., L_{j-1}).$   
If $L=(L_1,..., L_r)$ is a finite sequence of irreducible representations of  $U_q(\mathfrak g)$ we denote $V(L)=\otimes_{j=1}^r \stackrel{L_j}{V}$ and we will denote   
$$\stackrel{{ N}{ L}}{R}{}\!\!^{(\pm)}  =  \stackrel{{ N}{ L}_1}{R}{}\!\!^{(\pm)} \cdots \stackrel{{ N}{ L}_r}{R}{}\!\!^{(\pm)}\;\;.$$

For $W$ an irreducible representation of $U_q({\mathfrak g})$ and $S=(S_3,\cdots,S_{r})$ a $(r-2)$-uplet of irreducible representations of  $U_q({\mathfrak g})$, we define the intertwiners: \\

  $ \Psi^{\;\;W}_{L}(S)  
 \in Hom_{U_q({\mathfrak g})}(V(L) , \stackrel{{ W}}{V})$ by   
\begin{equation}  
\Psi^{ W}_{L} (S)  =  \Psi^{\;\;\;\; W}_{ S_r L_r} \cdots \Psi^{\;\;{ S}_4}_{{S}_3 { L}_3} \Psi^{\;\;{ S}_3}_{{ L}_1 { L}_2}   
\end{equation}  
and $\Phi^{L}_{W}({ S}) \in Hom_{U_q({\mathfrak g})}(\stackrel{{ W}}{V} ,V(L))$ defined by   
\begin{equation}  
\Phi^{ L}_{\;\; W}({ S})   
=  \Phi^{{ L}_{1} { L}_2}_{\;\;{ S}_{3}}   
  \Phi^{{ S}_{3} { L}_3 }_{\;\;{ S}_{4}}   
 \cdots   
 \Phi^{{ S}_r { L}_{r} }_{\;\;W}.  
\end{equation}

\begin{definition}  
We will define a ``palette''   
as being   a  family   
$P = (I,J,N; K,L,U,T,W)$  where $I, J, K, L$ (resp. $N$) (resp. $U, T$) are $n$-uplets (resp.$p$-uplets) (resp. $n+p-2$-uplets ) of irreducible finite dimensional representations of $U_q(\mathfrak g)$ and $W$ is an irreducible finite dimensional representations of $U_q(\mathfrak g).$  
Any palette $P$ defines a unique quantum spin-network ${\cal N}_{P}$ associated to the standard graph, precisely:  
$(I,J,N)$ is coloring of the non contractible cycles $A(i), B(i), M(n+i)$ and   $(K,L,U,T,W)$ is associated to the intertwiner $\Psi^{\;\;W}_{IJN}(KU) \otimes \Phi_{\;\;W}^{IJN}(LT)$ coloring the vertex of this spin-network.
\end{definition}

We first define, for $i=1,..,n$ ,  
 $\theta(i) \in {\cal L}_{n,p}\otimes  Hom(\stackrel{{ L}_i}{V}, \stackrel{{ K}_i}{V})$ by:  
  
$$\theta(i)  =  \Psi^{\;{ K}_i}_{{ J}_i { I}_i} \stackrel{{ J}_i}{B}\!\!(i) \stackrel{{ J}_i { I}_i}{R'} \; \stackrel{{ I}_i}{A}\!\!(i) \stackrel{{ J}_i {I}_i}{R}{}\!\!^{(-)}  \; \Phi_{\;{ L}_i}^{{ J}_i { I}_i}.$$  
  
We can now associate to $I, J, K, L$ the element $\stackrel{ I,J}{\theta}{}\!\!^{(\pm)}_{n}({K,L})\in  {\cal L}_{n,p}\otimes  Hom(V(L),V(K)) $ by  
  
\begin{equation}  
\stackrel{ I,J}{\theta}{}\!\!^{(\pm)}_{n}({ K,L}) =   
\prod_{j=1}^n \big(\stackrel{K_{j}L_{<j}}{R^{(\pm)}}\theta(j)\big).  
\end{equation}

We associate to $N$ the element of ${\cal L}_{n,p}\otimes  Hom(V(N),V(N))$  
\begin{equation}   
\stackrel{N}{\theta}{}\!\!^{(\pm)}_{p}  =   
\prod_{j=1}^p \big( \stackrel{N_{j}N_{<j}}{R^{(\pm)}}  
 \stackrel{N_j}{M}(n+j)\big).   
\end{equation}  
We can introduce the elements of ${\cal L}_{n,p}\otimes  Hom(V(LN),V(KN)) $  
\begin{eqnarray}  
\stackrel{I,J,N}{\Omega}{}\!\!^{(\pm)}_{n,p}({ K,L}) \; = \; \stackrel{I,J}{\theta}{}\!\!^{(\pm)}_{n}({ K,L})\big(\prod_{j=1}^p \stackrel{{ N_j}{ L}}{R^{(\pm)}}\big) \stackrel{N}{\theta}{}\!\!^{(\pm)}_{p}  
(\prod_{j=1}^{n+p} \stackrel{{ (LN)_j}{(LN)_{<j}}}{R^{(\pm)}})^{-1}.  
\end{eqnarray}

\begin{proposition}  
Let $P$ be a palette labelling a quantum spin network ${\cal N}_P$ associated to the standard graph. We will define an element of ${\cal L}_{n,p}$  
\begin{eqnarray}  
\stackrel{ P}{\cal O}{}\!\!^{(\pm)}_{n,p} \; = \; \frac{v_{ K}^{1/2}}{v_{I}^{1/2}v_{J}^{1/2}} tr_q( \Psi^{\;\;W}_{ KN} ({U}) \; \stackrel{I,J,N}{\Omega}{}\!\!^{(\pm)}_{n,p}({K,L})  \; \Phi^{LN}_{\;\;W}({T}))  
\end{eqnarray}  
where we have defined   
$v_{ I}^{1/2} = v_{{I}_1}^{1/2} \cdots v_{{I}_n}^{1/2}.$ \\  
The elements $\stackrel{ P}{\cal O}{}\!\!_{n,p}^{(\pm)}$ are gauge invariant elements  and if $\epsilon \in\{ +,-\}$ is fixed the nonzero elements of the  family  $\stackrel{ P}{\cal O}{}\!\!_{n,p}^{(\epsilon )}$ is  a basis  of ${\cal L}_{n,p}^{U_q({\mathfrak g})}$  when $P$ runs over all the palettes.   
\end{proposition}  
  
  \Proof It is a simple consequence of    
\begin{eqnarray}  
\delta  (\stackrel{I,J,N}{\Omega}{}\!\!^{(\pm)}_{n,p}({ K,L}))  
 \; = \; g(KN)  
\stackrel{I,J,N}{\Omega}{}\!\!^{(\pm)}_{n,p}({K,L})S(g(LN)))  
\end{eqnarray}  
where $g(L)=\stackrel{{ L}_1}{g}_1\cdots \stackrel{{ L}_p}{g}_p$ and that $tr_q()$ is invariant under the adjoint action.  
$\Box $\\  
  
Remarks.  
  
1. The family  $\stackrel{ P}{\cal O}{}\!\!_{n,p}^{(+)}$ can be linearly expressed in term of the   
family $\stackrel{ P}{\cal O}{}\!\!_{n,p}^{(-)}$, and the coefficients of these linear transformations  
 can be exactly computed in terms of $6j$ coefficients.

2. The particular normalization of these  families  has been chosen in order to simplify the action of the star on these elements  
 (see next section).   
  
We will denote also by the same notation the image of $\stackrel{ P}{\cal O}{}\!\!_{n,p}^{(\pm)}$ in the quantum moduli space   
$M_q(\Sigma, G, \pi_1,\cdots,\pi_p).$  
  
Example.   
 
In the case where the surface is a  torus with no puncture ($n=1$ and $p=0$), the spin-networks are labelled by the colors $ IJ$ of the two non-contractible cycles  and the choice of the intertwiner is fixed by a finite dimensional representation $ W$. As a result   the vector space of gauge invariant functions ${\cal L}_{1,0}^{inv}$ is linearly generated by the following observables:  
\begin{eqnarray}  
\stackrel{ IJW}{\cal O}_{1,0} \; = \; \frac{v_{ W}^{1/2}}{v_{ I}^{1/2}v_{ J}^{1/2}} tr_q(\Psi^{\;\; W}_{ JI} \; \stackrel{ J}{B} \; \stackrel{ JI}{R'} \; \stackrel{ I}{A} \; \stackrel{ JI}{R}{}\!\!^{(-)} \; \Phi_{\;\; W}^{ JI}) \;\;,   
\end{eqnarray}  
for all finite dimensional representations $IJW$. The Moduli algebra  is generated as an algebra by the Wilson loops around the handles in the fundamental representation, i.e.  
\begin{eqnarray}  
W_A \; = \; tr_q(\stackrel{I}{A})\;, 
 W_B \; = \; tr_q(\stackrel{I}{B})\; \text{where } I=\onehalf. 
\end{eqnarray}

\subsection*{II.4 Alekseev's Isomorphisms Theorem}  
  
The construction of  the representation theory of ${\cal L}_{n,p}$  uses    Alekseev's method \cite{AGS1,AGS2, Al}:  
 we first build representations of ${\cal L}_{0,p}$ (the multi-loop algebra), then we build representations of  
 ${\cal L}_{n,0}$ (the multi-handle algebra) and we use these results to build representations of the graph   
algebra ${\cal L}_{n,p}$.  
  
\begin{lemma}  
The algebra ${\cal L}_{0,p}$ is isomorphic to the algebra $U_q({\mathfrak g})^{\otimes p}$.   
  
The algebra ${\cal L}_{0,p}$ is generated by the matrix elements of $\stackrel{{ I}}{M}\!\!(i)$, $i=1,...,p$, and the algebra $U_q({\mathfrak g})^{\otimes p}$ is generated by the matrix elements of $\stackrel{{ I}}{L}\!\!(j)^{(\pm)}$, $j=1,...,p$, where the label $j$ denotes one of the $p$ copies of $U_q({\mathfrak g})$. An  explicit isomorphism in term of these generators can be constructed as follows:  
\begin{eqnarray}  
{\cal L}_{0,p} \; \otimes \; End(\stackrel{{ I}}{V})   & \longrightarrow & U_q({\mathfrak g})^{\otimes p} \; \otimes  \; End(\stackrel{{ I}}{V}) \nonumber \\  
\stackrel{{ I}}{M}\!\!(i) & \mapsto & \stackrel{{ I}}{\mathfrak F}\!\!(i) \; \stackrel{{ I}}{\mathfrak M}\!\!(i) \; \stackrel{{ I}}{\mathfrak F}\!\!(i)^{-1} \; \; ,  
\end{eqnarray}  
where we have defined   
\begin{eqnarray}  
\stackrel{{ I}}{\mathfrak M}\!\!(i) \; = \; \stackrel{{ I}}{L}\!\!(i)^{(+)} \stackrel{{ I}}{L}\!\!(i)^{(-)-1} \;\;\;\;\; ,\;\;  
 \stackrel{{ I}}{\mathfrak F}\!\!(i)  \; = \; \stackrel{{ I}}{L}\!\!(1)^{(-)} \cdots \stackrel{{ I}}{L}\!\!(i-1)^{(-)} \;\; .  
\end{eqnarray}  
  
\end{lemma}  
\Proof See \cite{Al}   
$\Box$

As an immediate consequence, the representations of the loop algebra ${\cal L}_{0,p}$ are those of $U_q({\mathfrak g})^{\otimes p}$. A basis of the irreducible finite dimensional module labelled by $J$  is denoted as usual by $(\stackrel{ J}{e}_i \mid i= 1 \cdots dim(\stackrel{ J}{V}))$.  The action of the generators on this basis is given by:  
\begin{eqnarray}  
\stackrel{{ I}}{L}{}\!\!^{(\pm)a}_{\;\;\;\;\;b} \rhd \stackrel{J}{e}_i  \; = \;  \stackrel{ J}{e}_j \; \stackrel{ IJ}{R}{}\!\!^{(\pm) aj}_{\;\;\;\;\;bi} \;\;.  
\end{eqnarray}  
  
{}From this relation and the explicit isomorphism of the lemma 1, it is easy to find out  explicit expressions for the representations of ${\cal L}_{0,p}$ on the module $\stackrel{ I_1}{V} \otimes \cdots \otimes \stackrel{ I_p}{V}$. In particular, the action of $\stackrel{ I}{M}\!\!(i)$ on the basis $\stackrel{{ I}_1}{e}_{i_1} \otimes \cdots \otimes \stackrel{{ I}_p}{e}_{i_p}$ is given in term of product of R-matrices. \\

The previous theorem can be modified in order to  apply to the algebra ${\cal L}_{n,0}$. However, this algebra can not be represented as a direct product of several copies of $U_q(\mathfrak g)$. An easy way to understand this point is to consider, for example, the center of each algebras. The loop algebra ${\cal L}_{0,1}$ admits a subalgebra generated by the $\stackrel{ I}{W}\!\!(i) = tr_q(\stackrel{I}{M}\!\!(i))$ which are central elements.   
 One can show that the center of the handle algebra ${\cal L}_{1,0} $ is trivial \cite{AGS1,AGS2}.  
 To understand the representations of ${\cal L}_{n,0}$, we therefore have  to introduce one more object: the Heisenberg double.  
  
\begin{definition}  
Let $A$ be a Hopf algebra (typically $U_q(\mathfrak g)$) and $A^{\star}$ its dual. The Heisenberg double is an algebra defined as a vector space by  
\begin{eqnarray}  
H(A) = A \; \otimes \; A^{\star}\;\; ;\nonumber  
\end{eqnarray}  
the algebra law is defined by the following algebra morphisms   
\begin{eqnarray}  
A & \hookrightarrow & H(A)\;\;\;\;\; ; \;\;\;\;\; A^{\star}  \hookrightarrow  H(A) \nonumber \\  
x & \mapsto &  x\; \otimes \; 1 \;\;\;\;\; \;\; \;\;\;\;\; f  \mapsto 1 \; \otimes\; f \nonumber  
\end{eqnarray}  
and the exchange relations  
\begin{eqnarray}  
x  f & = &  (x \otimes 1)(1 \otimes f) =  x \otimes f = \sum_{(x),(f)} \big< x_{(1)}, f_{(2)} \big> \;  ( 1 \otimes f_{(1)})  (x_{(2)} \otimes 1) \;\; ,\label{echangeheisenberg}  
\end{eqnarray}  
where we have used Sweedler notation $\Delta(x) = \sum_{(x)} x_{(1)} \otimes x_{(2)}$.   
\end{definition}  
In the case where $A=U_q(\mathfrak g)$, the Heisenberg double may be seen as a quantization of $Fun(T^{\star} G)$. So, we can interpret the elements of $A^{\star}$ as functions and those of $A$ as derivations.   
  
\begin{proposition}  
$H(A)$ admits a unique irreducible representation $\Pi$ realized in the module $A^{\star}$ as follows:  
\begin{eqnarray}  
\Pi \;\; : \;\;  H(A) & \longrightarrow & End(A^{\star}), \nonumber \\  
   A^{\star} \ni f    & \mapsto &  m_f \;\;\;/\;\;\;m_f(g)  =  fg \;\;,\forall g \in A^*\nonumber\\  
   A \ni x  & \mapsto & \nabla_x \;\;\;/\;\;\; \nabla_x (g)  =  g_{(1)} \; \big< x, g_{(2)} \big > \;\; ,\forall g \in A^*.\nonumber  
\end{eqnarray}     
\end{proposition}

In the case where $A = U_q(\mathfrak g),$ $H(A)$ is generated as a vector space by $\stackrel{ I}{L}{}\!\!^{(+) a}_{\;\;\;\;\;b}  \stackrel{ J}{L}{}\!\!^{(-) c}_{\;\;\;\;\;d}  \otimes  \stackrel{ K}{g}{}\!\!^e_f$, and the exchange relations (\ref{echangeheisenberg}) take the simple form:  
\begin{eqnarray}  
 \stackrel{ I}{L}{}\!\!_1^{(\pm)}  \stackrel{ J}{g}_2 \; = \; \stackrel{ J}{g}_2 \; \stackrel{ I}{L}{}\!\!_1^{(\pm)}  \stackrel{ IJ}{R}{}\!\!_{12}^{(\pm)}\;\;.  
\end{eqnarray}

In order to understand the relation between the multi-handle algebra and the Heisenberg double $H(U_q(\mathfrak g))$, it is convenient to introduce left derivations $\stackrel{I}{\tilde {L}} \; \in End(\stackrel{I}{V}) \otimes H(U_q(\mathfrak g))$:  
\begin{eqnarray}  
\stackrel{I}{\tilde {L}} \;= \;v_{I}^2 \stackrel{I}{g} \; \stackrel{ I}{L}{}\!\!^{(+)-1}  \stackrel{ I}{L}{}\!\!^{(-)} \stackrel{I}{g}{}\!\!^{-1}   
\; = \; \stackrel{ I}{\tilde {L}}{}\!\!^{(+)} \stackrel{I}{\tilde {L}}{}\!\!^{(-)-1}\;\;,  
\end{eqnarray}  
where the last formula corresponds to the Gauss decomposition. As usual, left and right derivations commute with each other   
\begin{eqnarray}  
\stackrel{I}{\tilde {L}}{}\!\!_{1}^{(\epsilon)}  \stackrel{ J}{L}{}\!\!_{2}^{(\sigma)} \; = \; \stackrel{ J}{L}{}\!\!_{2}^{(\sigma)}  \stackrel{I}{\tilde {L}}{}\!\!_{1}^{(\epsilon)} \;\; , \forall \; (\epsilon \;,\;\sigma) \in \{+\;,\;-\} \;\;,  
\end{eqnarray}  
and realize two independent embeddings of $U_q({\mathfrak g})$ in $H(U_q(\mathfrak g))$.{} From the relations of the Heisenberg double, it is easy to show the following  relations:  
\begin{eqnarray}  
  \stackrel{ I}{\tilde {L}}{}\!\!_1^{(\pm)}  \stackrel{ J}{\tilde {L}}{}\!\!_2^{(\pm)}  & = & 
 \sum_K \Phi^{IJ}_{K}  \stackrel{ K}{\tilde {L}}{}\!\!{}^{(\pm)}  
  \Psi^{K}_{IJ} 
 \;\; , \\  
\stackrel{I J}{R}{}\!\!_{12}^{(\pm)}  \stackrel{ I}{\tilde {L}}{}\!\!_1^{(+)}  \stackrel{ J}{\tilde {L}}{}\!\!_2^{(-)} & =&  \stackrel{ J}{\tilde {L}}{}\!\!_2^{(-)}  \stackrel{ I}{\tilde {L}}{}\!\!_1^{(+)}  \stackrel{ I J}{R}{}\!\!_{12}^{(\pm)}, \;\; \\  
\stackrel{ I J}{R}{}\!\!_{12}^{(\pm)}  \stackrel{ I}{\tilde {L}}{}\!\!_{1}^{(\pm)}  \stackrel{ J}{g}_2 \;& = &\; \stackrel{ J}{g}_2 \; \stackrel{ I}{\tilde {L}}{}\!\!_{1}^{(\pm)} \;\; .  
\end{eqnarray}  
 
 The action of the elements $\stackrel{ I}{L}{}\!\!_1^{(\pm)},\stackrel{ I}{\tilde {L}}{}\!\!_1^{(\pm)},  \stackrel{ J}{g}$ through representation $\Pi$ are expressed as:  
\begin{eqnarray}  
&&\stackrel{ I}{L}{}\!\!_1^{(\pm)} \;\; \rhd \;\; \stackrel{ J}{g}_2 \; = \; \stackrel{J}{g}_2 \;  \stackrel{ I J}{R}{}\!\!_{12}^{(\pm)} \;\;\;\;,  
\stackrel{ I}{\tilde {L}}{}\!\!_1^{(\pm)} \;\; \rhd \;\; \stackrel{ J}{g}_2 \; = \; \stackrel{ I J}{R}{}\!\!_{12}^{(\pm)}{}^{-1} \; \stackrel{ J}{g}_2 \;\;,\\  
&& \stackrel{ I}{g}_1 \rhd\; \stackrel{ J}{g}_2 =  
 \stackrel{ I}{g}_1\stackrel{ J}{g}_2=\sum_K \Phi^{IJ}_{K}  
  \stackrel{ K}{g}\Psi^{K}_{IJ}.  
\end{eqnarray}  
  
The following lemma, due to Alekseev \cite{Al}, describes the structure of ${\cal 
L}_{n,0}:$     
\begin{lemma}  
The algebra ${\cal L}_{n,0}$ is isomorphic to the algebra $H(U_q({\mathfrak g}))^{\otimes n}$.  
\begin{eqnarray*}  
{\cal L}_{n,0} \; \otimes \; End(\stackrel{ I}{V}) & \longrightarrow &  H(U_q({\mathfrak g}))^{\otimes n} \; \otimes \; End(\stackrel{\bf I}{V}) \\  
\stackrel{ I}{A}\!\!(i) & \mapsto & \stackrel{ I}{\mathfrak H}\!\!(i) \; \stackrel{ I}{\mathfrak A}\!\!(i) \; \stackrel{ I}{\mathfrak H}\!\!(i)^{-1} \\  
\stackrel{ I}{B}\!\!(i) & \mapsto & \stackrel{ I}{\mathfrak H}\!\!(i) \; \stackrel{ I}{\mathfrak B}\!\!(i) \;  \stackrel{ I}{\mathfrak H}\!\!(i)^{-1}\;\; ,  
\end{eqnarray*}  
where  we have defined   
\begin{eqnarray}  
\stackrel{ I}{\mathfrak A}\!\!(i) & = & \stackrel{ I}{L}\!\!(i)^{(+)} \stackrel{ I}{g}\!\!(i) \stackrel{ I}{L}\!\!(i)^{(-)-1}\;\;\; , \;\;\; \stackrel{ I}{\mathfrak B}\!\!(i) \; = \; \stackrel{ I}{L}\!\!(i)^{(+)} \stackrel{I}{L}\!\!(i)^{(-)-1} \;\;\; ,\\  
\stackrel{ I}{\mathfrak H}\!\!(i) & = & (\stackrel{ I}{L}\!\!(1)^{(-)} \stackrel{ I}{\tilde {L}}\!\!(1)^{(-)}) \cdots (\stackrel{ I}{L}\!\!(i-1)^{(-)} \stackrel{ I}{\tilde {L}}\!\!(i-1)^{(-)})\;\;\; .  
\end{eqnarray}  
\end{lemma}  
  
Remark: 
this lemma can be used to build representations of the the multi-handle 
algebra  ${\cal L}_{n,0}.$ The two monodromies $\stackrel{I}{A}$ and 
$\stackrel{ I}{B}$ act on the vector space $F_q(G)$ as follows:   
\begin{eqnarray}  
\stackrel{ I}{A}_1 \rhd \stackrel{ J}{g}_2 \; = \; \sum_{i,K} 
\stackrel{ I}{x}_{i1} \Phi^{ IJ}_{K} \; \stackrel{ K}{g} \; \stackrel{ 
K}{y}_{i} \; \Psi^{ K}_{ IJ} \; \stackrel{ IJ}{R'}_{12} \;\;\;\; ,\;\;  
\stackrel{ I}{B}_1 \rhd \stackrel{ J}{g}_2 \; = \; \stackrel{ J}{g}_2 \; 
\stackrel{ IJ}{R}_{12} \; \stackrel{ IJ}{R'}_{12} \;\; ,  
\end{eqnarray}  
where $R=\sum_i x_i \otimes y_i$.  In the case of the multi-handle algebra, 
the action of the monodromies is given in term of product of R-matrices with 
Clebsh-Gordan maps.\\     
 
The following lemma shows that the graph algebra 
${\cal L}_{n,p}$ is isomorphic to ${\cal L}_{n,0} \otimes {\cal L}_{0,p}$. As 
a result, from the previous theorems, representations of the graph algebra 
${\cal L}_{n,p}$ is constructed from the representations of the multi-loop 
algebra and the multi-handle algebra.   
  
\begin{lemma}\label{lemmeAlekseev}  
The algebra ${\cal L}_{n,p}$ is isomorphic to the algebra $ H(U_q({\mathfrak 
g}))^{\otimes n}\otimes U_q({\mathfrak g}) ^{\otimes p} $, the isomorphism is 
given by  \begin{eqnarray*}  {\cal L}_{n,p} \; \otimes \; End(\stackrel{ 
I}{V}) & \longrightarrow &  H(U_q({\mathfrak g}))^{\otimes n} \otimes  
U_q({\mathfrak g})^{\otimes p} \otimes End(\stackrel{ I}{V})  \nonumber \\  \stackrel{ 
I}{A}\!\!(i) & \mapsto & \stackrel{ I}{\mathfrak K}\!\!(i) \; \stackrel{ 
I}{\mathfrak A}\!\!(i) \; \stackrel{ I}{\mathfrak K}\!\!(i)^{-1} \\  
\stackrel{ I}{B}\!\!(i) & \mapsto & \stackrel{ I}{\mathfrak K}\!\!(i) \; 
\stackrel{ I}{\mathfrak B}(i) \;  \stackrel{ I}{\mathfrak K}\!\!(i)^{-1}\;\; 
,\\  \stackrel{ I}{M}\!\!(n+i) & \mapsto & \stackrel{ I}{\mathfrak K}\!\!(n+i) 
\; \stackrel{ I}{\mathfrak M}\!\!(n+i) \;  \stackrel{ I}{\mathfrak 
K}\!\!(n+i)^{-1}\;\; ,  \end{eqnarray*}  where we have defined   
\begin{eqnarray}  \stackrel{ I}{\mathfrak A}\!\!(i)  & =  & \stackrel{ 
I}{L}\!\!(i)^{(+)}  \stackrel{ I}{g}\!\!(i)  \stackrel{ I}{L}\!\!(i)^{(-)-1} 
\;\; ,  \;\;\stackrel{ I}{\mathfrak B}\!\!(i)  =  \stackrel{ 
I}{L}\!\!(i)^{(+)} \; \stackrel{I}{L}\!\!(i)^{(-)-1} \;\; ,\\  \stackrel{ 
I}{\mathfrak M}\!\!(n+i) & = & \stackrel{ I}{L}\!\!(n+i)^{(+)}  \stackrel{ 
I}{L}\!\!(n+i)^{(-)-1} \; , \\  \stackrel{ I}{\mathfrak K}\!\!(i)  & = & 
\stackrel{ I}{\mathfrak H}\!\!(i) \; , \; \stackrel{ I}{\mathfrak K}\!\!(n+i) 
= \stackrel{ I}{\mathfrak H}\!\!(n+1)  \stackrel{ I}{\mathfrak F}\!\!(i) \;,  
\end{eqnarray}  where $\stackrel{ I}{\mathfrak H}\!\!(i)$ and $\stackrel{ 
I}{\mathfrak F}\!\!(i)$ have already been introduced.  \end{lemma}            
Let $\stackrel{I}{C}$ as defined by (\ref{defofC}), it can easily be shown 
that:  \begin{eqnarray}  
&&\stackrel{I}{C}=\stackrel{I}{C}{}^{(+)}\stackrel{I}{C}{}^{(-)-1}\;\;\text{with}\\  &&\stackrel{I}{C}{}^{(\pm)}=\prod_{j=1}^n\stackrel{I}{L}{}^{(\pm)}(j)\stackrel{I}{\tilde L}{}^{(\pm)}(j)  \prod_{k=1}^p\stackrel{I}{L}{}^{(\pm)}(n+k).  \end{eqnarray}    
{}From the relations (\ref{fusion}) satisfied by  $\stackrel{I}{C}$, one obtains that the algebra ${\cal C}$ is isomorphic to ${\cal L}_{0,1}$ and hence to $U_q({\mathfrak g}).$ Let us denote by   $i: U_q({\mathfrak g})\rightarrow {\cal C}$ the isomorphism of algebra defined by   $$i (\stackrel{I}{ L}{}^{(\pm)-1})=\stackrel{I}{C}{}^{(\pm)}.$$    
An important property is that the adjoint action of  ${\cal C}$ on the graph 
algebra is equivalent to the action of $U_q(\mathfrak{g})$, namely we have:  
\begin{equation}  i(\xi_{(1)})\;a\;i(S(\xi_{(2)}))=a^{\xi},\;\;\; 
\forall a\in {\cal L}_{n,p}, \forall \xi\in   U_q({\mathfrak 
g}).\label{xiaxi=axi}  \end{equation}  
This last property follows easily from the relation   
$$\stackrel{I}{C}{}^{(\pm)}_{1}\stackrel{J}{U}_{2}(i)\;\stackrel{I}{C}{}^{(\pm)-1}_{1}=  
\stackrel{IJ}{R^{(\pm)}_{12}}\;\stackrel{J}{U}_{2}(i)\;\stackrel{IJ}{R^{(\pm)}_{12}}{}^{-1}$$  
where $U(i)$ is indifferently $A(i), B(i), M(i).$
Note that the classical property that the constraint (\ref{flatness}) generates gauge transformation  is turned into (\ref{xiaxi=axi}) after quantization.

Finally, the representation theory of the graph-algebra ${\cal L}_{n,p}$ is obtained from those of the quantum group $U_q(\mathfrak g)$ and the Heisenberg double $H(U_q(\mathfrak g))$. If  $I=(I_1,...,I_p)$ are irreducible $U_q({\mathfrak g})$-modules,  
 ${\cal H}_{n,p}[I]=F_q(G)^{\otimes n}\otimes   
\stackrel{{I}_1}{V} \otimes \cdots \otimes \stackrel{{ I}_p}{V}$  are irreducible modules  of ${\cal L}_{n,p}$ defing  the  representation denoted   
 $\rho_{n,p}[I].$  
   
 ${\cal H}_{n,p}[I]$ is also a  $U_q(\mathfrak g)$-module associated to the representation   
$\rho_{n,p}[I]\circ i.$ As a result the subset of invariant elements is the vector space  ${\cal H}_{n,p}[I]^{U_q({\mathfrak g})}=\{v\in {\cal H}_{n,p}[I],\\
 (\stackrel{I}{C}{}^{(\pm)}-1)\rhd v=0\}$.

\begin{proposition}{\rm (Alekseev)} \label{prop:repofmoduli} 
The representation  $\rho_{n,p}[I]$  of ${\cal L}_{n,p}$ restricted to   
 ${\cal L}_{n,p}^{inv}$ leaves ${\cal H}_{n,p}[I]^{U_q({\mathfrak g})}$ invariant. As a result one obtains a representation  of  ${\cal L}_{n,p}^{inv}$ acting on ${\cal H}_{n,p}[I]^{U_q({\mathfrak g})}=  
H_{n,p}[I]$.  
This representation annihilates the ideal ${\cal I}_{C}$, therefore one obtains a representation of ${\mathfrak M}_q(\Sigma, G,p).$  
Moreover  ${ \rho}_{n,p}[I]$ annihilates the ideals generated by the relations $tr_q(\stackrel{J}{M}(n+i))=\vartheta_{J\pi_i}, i=1,...,p$, where $\pi_i=\stackrel{I_i}{\pi}$.  
As a result  ${ \rho}_{n,p}[I]$ descends to the quotient, defines a representation denoted  $\tilde{\rho}_{n,p}[I]$, of   
$M_q(\Sigma,G;\pi_1,...,\pi_p).$   
\end{proposition}  
  
This proposition is a direct consequence of the above constructions and the fact that $tr_q(\stackrel{J}{M}(n+i))$ is represented by   
$tr_q((\stackrel{J}{\pi}\otimes {\id})(RR')).$

To complete the construction of combinatorial quantization, the space of states $H_{n,p}[I]$  has to be endowed with a structure of Hilbert space and the algebra of observables   
$M_q(\Sigma,G;\pi_1,...,\pi_p)$ has to be endowed with a star structure such that the representation    
${\tilde \rho}_{n,p}[I]$ is unitary.  
  
 In the case of Chern-Simons theory with $G=SU(2),$ this last step of the construction  has been fully studied in \cite{AGS1, AGS2, AS}. We refer to these works for full details but let us put the emphasis on the following points:\newline
- $q$ is a root of unit which admits the following classical expansion (large $\lambda$  expansion) $q=1+i\frac{2\pi}{\lambda}+o(\frac{1}{\lambda})$;\newline
- $U_q(su(2))$ is endowed with a structure of star  weak-quasi Hopf: truncation on the spectrum of finite dimensional unitary irreducible representations holds and the representations ${\tilde \rho}_{n,p}[I]$ is the unique finite dimensional irreducible  representation of $M_q(\Sigma,G;\pi_1,...,\pi_p).$  
  
In the next section we will  modify the previous constructions and apply them  to the case of the group   
$SL(2,\CC)_{\RR}$.

\section*{III Combinatorial Quantization in the $SL(2,\CC)_{\RR}$ case.}	  
\subsection*{III.1. Chern-Simons theory with $SL(2,\CC)_{\RR}$ group.}  
Let $G=SU(2)$, we will denote by $G^{\CC}=SL(2,\CC)$ the complex group and by $SL(2,\CC)_{\RR}$ the   
realification of $SL(2,\CC)$. The real Lie algebra of $SL(2,\CC)_{\RR}$ denoted $sl(2,\CC)_{\RR}$ can equivalently be described by   
a star structure on its complexification $(sl(2,\CC)_{\RR})^{\CC}=sl(2,\CC)\oplus \overline{sl(2,\CC)}.$  
  
Chern-Simons theory with gauge group $SL(2,\CC)_{\RR}$ is  defined on a 3-dimensional compact oriented manifold $M$ by the action  
\begin{eqnarray}  
S(A)=\frac{\lambda}{4\pi} \int_M \text{Tr}(A \wedge dA + \frac{2}{3} A \wedge A  \wedge A)+  
\frac{\bar{\lambda}}{4\pi} \int_M \text{Tr}({\bar A} \wedge d{\bar A} + \frac{2}{3} {\bar A} \wedge {\bar A}  \wedge {\bar A})  
\;\;,\label{ChernSimonsactioncomplexe}  
\end{eqnarray}  
where the gauge field $A = A_{\mu} dx^{\mu}$ is a $sl(2,\CC)$ 1-form on $M$ and $Tr$ is the Killing form on $sl(2,\CC).$  
  
Following\cite{Wi2},  we can always write  $\lambda=k+is,$ with $s$ real and $k$ integer in order that $\exp(iS(A))$ is invariant under large gauge transformation.   
  
In this paper we will choose the case $k=0$ which is  selected when one expresses the action  
 of $2+1$ pure gravity with positive cosmological constant as a $SL(2,\CC)_{\RR}$ Chern Simons action \cite{Wi1}.   
  
We shall apply the program of combinatorial quantization in this case. {}From the expression of the Poisson structure on the space of flat connections, it is easy to see that $q$ has to satisfy $q=1+\frac{2\pi}{s}+o(\frac{1}{s})$ when $s$ is large. As a result we will develop the combinatorial quantization construction using the Hopf algebra   
$U_q(sl(2,\CC)_{\RR})$ with $q$ real.  
  
An introduction to the notion of complexification and realification in the Hopf algebra context can be found in the chapter 2 of \cite{BR3'}.  
  
\subsection*{III.2. Combinatorial Quantization Formalism in the $SL(2,\CC)_{\RR}$ case: the algebraic structures.}  
In this part we describe the modifications that have to be made to construct all the algebraic structures of the combinatorial quantization formalism in the $SL(2,\CC)_{\RR}$ case.  
  
In Fock-Rosly construction, we first change $G$ to $SL(2,\CC)_{\RR}$. The Lie algebra   
${\mathfrak g}$ is changed into $sl(2,\CC)_{\RR}$ which is equivalent to the Lie algebra $sl(2,\CC)\oplus  
 \overline{sl(2,\CC)}$ with star structure $\star$ defined by $ (a\oplus \bar{b})^{\star}=-(b\oplus \bar{a}).$  
Let $\dagger$ be the star structure on $sl(2,\CC)$ selecting the compact form,  $-\dagger$ identifies   
$sl(2,\CC)$ and $\overline{sl(2,\CC)}$ as $\CC$-Lie algebras.  
As a result we can equivalently describe $sl(2,\CC)_{\RR}$ as being the Lie algebra   
$sl(2,\CC)\oplus sl(2,\CC)$   
with star structure $(a\oplus  b)^{\star}=( b^{\dagger}\oplus a^{\dagger}).$    
We will denote by $\stackrel{I}{\pi}$ for $I\in \onehalf\ZZ^+$ the irreducible representations of dimension $2I+1$ of   
$su(2)$ which are also $\dagger$ representations of $ sl(2,\CC)$, and let $e_a$ be an orthonormal basis of this module.   
The contragredient  representation of $\stackrel{I}{\pi}$, denoted $\stackrel{\check{I}}{\pi}$ is  equivalent  to the conjugate  representation  because it is a $\dagger$ representation. In the $su(2)$ case it is moreover equivalent to the representation   $\stackrel{I}{\pi}$ through the intertwiner $\stackrel{I}{W}$: $\stackrel{\check{I}}{\pi}=\stackrel{I}{W}\stackrel{{I}}{\pi}  
\stackrel{I}{W}{}^{-1}$ where $\stackrel{I}{W}{}^a_b=(-1)^{I-a}\delta^{a}_{-b}.$

Finite dimensional irreducible representations of $sl(2,\CC)_{\RR}$ are labelled by a couple   
$I=(I^{l},I^{r})$ of   
positive half integers and we will denote by $\stackrel{I}{\VV}=\stackrel{I^l}{V}\otimes\stackrel{I^r}{V}$   
the $sl(2,\CC)\oplus sl(2,\CC)$  
 module labelled by the couple $I=(I^{l},I^{r})$ associated to the representation   
 $\stackrel{I}{\Pi}=\stackrel{I^l}{\pi}\otimes \stackrel{I^r}{\pi}.$   
These representations, except the trivial one,  are not $\star$-representations.  
  
{}From the action (\ref{ChernSimonsactioncomplexe}) the Poisson bracket on the space of $sl(2,\CC)\oplus sl(2,\CC)$-connections   
is expressed by   
\begin{eqnarray}  
\{A^l_i(x)\stackrel{\otimes}{,}A^l_j(y)\}&=&\frac{2\pi}{\lambda}\delta(x-y)\epsilon_{ij}t^{ll}\\  
\{A^r_i(x)\stackrel{\otimes}{,}A^r_j(y)\}&=&-\frac{2\pi}{\lambda}\delta(x-y)\epsilon_{ij}t^{rr}\\  
\{A^l_i(x)\stackrel{\otimes}{,}A^r_j(y)\}&=&0  
\end{eqnarray}  
where $t^{ll}$ (resp. $t^{rr}$ )is the embedding of $t$ in the $l\otimes l$ (resp. $r\otimes r$) component of   
$(sl(2,\CC)\oplus sl(2,\CC))^{\otimes 2}$.  
Note that  we have $A^l_i(x)^{\dagger}=-A^r_j(x)$ and $t^{\dagger\dagger}=t.$  
  
The spin-networks are defined analogously by replacing finite dimensional   
representations of ${\mathfrak g}$ by finite dimensional representations of $sl(2,\CC)\oplus sl(2,\CC).$  
  
For any representation $\stackrel{I}{\Pi}$ of $sl(2,\CC)\oplus sl(2,\CC)$ with $I=(I^l,I^r)$, we define   
$\stackrel{I}{\GG}\in \End(\stackrel{I}{\VV})\otimes Pol(SL(2,\CC)_{\RR})$ the matrix of coordinate functions on $SL(2,\CC)_{\RR}.$  
We have  
 \begin{equation}  
\stackrel{(I^l,I^r)}{\GG^{aa'}_{bb'}}{}^{\star}=  
\stackrel{(\check{I}{}^r,\check{I}{}^l)}{\GG^{a'a}_{b'b}}=\big((\stackrel{I^l}{W}\otimes \stackrel{I^r}{W} ) 
\stackrel{(I^l,I^r)}{\GG}(\stackrel{I^l}{W}{}^{-1}\otimes 
\stackrel{I^r}{W}{}^{-1})\big){}^{a'a}_{b'b}.\label{StaronGClassique}  
\end{equation}   
We denote  the holonomy of the $sl(2,\CC)\oplus sl(2,\CC)$ 
connection in the representation  $\stackrel{I}{\Pi}$ by   
$\stackrel{I}{\UU}(l)$, they satisfy the same relation   \begin{equation}  
\stackrel{(I^l,I^r)}{\UU^{aa'}_{bb'}}{}^{\star}  
=\big((\stackrel{I^l}{W}\otimes \stackrel{I^r}{W}) 
\stackrel{(I^l,I^r)}{\UU}(\stackrel{I^l}{W}{}^{-1}\otimes 
\stackrel{I^r}{W}{}^{-1})\big){}^{a'a}_{b'b}.  \label{StaronUClassique}  
\end{equation}  
We can define a Fock-Rosly structure on them, the Poisson bracket is the same as  
 (\ref{FockRosly1},\ref{FockRosly2},\ref{FockRosly3}) where the classical $r$ matrix of su(2) has been replaced by the $r$ matrix of $sl(2,\CC)_{\RR}$: $r_{sl(2,\CC)_{\RR}}=r^{ll}_{12}-r^{rr}_{21}.$  
  
We refer the reader to the article (\cite{BR3,BR3'}) for a thorough study of the quantum group $U_q(sl(2,\CC)_{\RR})$, see also the appendix (A.1) where basic definitions as well as fundamental results on harmonic analysis are described.  
It is important to stress that $U_q(sl(2,\CC)_{\RR})$ admits two equivalent definitions.    
  
The first one is  $U_q(sl(2,\CC)_{\RR})=U_q(sl (2,\CC))\otimes U_q(sl( 2,\CC))$ as an algebra with a suitable structure of coalgebra and $\star$ structure.  
  
 The second one,  suitable for the study of harmonic analysis, is    
$U_q(sl(2,\CC)_{\RR})=D(U_q(su(2)))$, the quantum double of $U_q(su(2))$, which is the quantum analog of   
Iwasawa decomposition.

$U_q(sl(2,\CC)_{\RR})$ is a quasi-triangular ribbon Hopf algebra endowed with a  
star structure (see appendix A.1).   
Finite dimensional representations of $U_q(sl(2,\CC)_{\RR})$ are labelled by a couple  
 ${ I} = (I^l,I^r)\in ({\onehalf }\ZZ^{+})^2={\SSF}$. The explicit description of these  
representations is  contained in the appendix A.1. The decomposition of the 
tensor product of these representations, and the explicit form of the 
Clebsch-Gordan maps, are described in the appendix A.2.    For any finite 
dimensional irreducible representation $\stackrel{I}{\Pi}$, we will define 
$\stackrel{I}{\VV}$ the associated module.       
Let us denote by $\lbrace 
\stackrel{A}{e}_i({ I}), \; i=-A, \cdots, A, \; A = \vert I^l-I^r \vert ,\cdots 
,I^l+I^r \rbrace$ an orthonormal basis of this vector space, and  $\lbrace 
\stackrel{A}{e}{}\!^i({ I}) \rbrace$ the dual basis.      
The algebra  
$Pol(SL_q(2,\CC)_{\RR})$ of polynomials on $SL_q(2,\CC)_{\RR}$ is generated by 
the matrix elements of the representations $\stackrel{ I}{\GG}{}\!\!^{Aa}_{Bb} 
= \big <\stackrel{A}{e}{}\!^a({ I}) \vert \stackrel{ I}{\Pi}(\cdot)   \vert 
\stackrel{B}{e}{}\!_b({ I}) \big>$.     
As in the previous section, let us 
introduce for each representation $ I$ the elements $\stackrel{ 
I}{\LL}{}\!\!^{(\pm)}{}^{Aa}_{Bb} \; \in \; End (\stackrel{I}{\VV}) \otimes 
U_q(sl(2,\CC)_{\RR})$ defined by $\stackrel{ I}{\LL}{}\!\!^{(\pm)}  =  
(\stackrel{ I}{\Pi}  \otimes  id) (\RR^{(\pm)})$ where $\RR$ is the 
$U_q(sl(2,\CC)_{\RR})$ R-matrix arising from the construction of the quantum double. 
Thanks to the factorisation theorem, i.e. 
$U_q(sl(2,\CC)_{\RR}) = U_q(sl(2)) \otimes_{R^{-1}} U_q(sl(2))$ as a Hopf 
algebra \cite{BR3,BR3'}, $\RR$ is expressed in term of $U_q(sl(2))$ 
R-matrices as $\RR^{(\pm)} = R_{14}^{(-)} R_{24}^{(\mp)} R_{13}^{(\pm)} 
R_{23}^{(+)}$.   It is therefore easy to obtain the ribbon 
elements of $U_q(sl(2,\CC)_{\RR})$ from those of $U_q(sl(2))$:  
\begin{eqnarray}  v_{ I} \; = \; v_{I^l} \; v_{I^r}^{-1} \; , \;\;\;\;\;\;\; 
\stackrel{ I}{\mu}\; = \;\stackrel{I^l}{\mu}\otimes  \stackrel{I^r}{\mu} \;.  
\end{eqnarray}    Let us now study the properties of the star structure on 
$U_q(sl(2,\CC)_{\RR})$ and, by duality, on $Pol(SL_q(2,\CC)_{\RR})$. In the 
case of $U_q(sl(2,\CC)_{\RR})$, the star structure, recalled in the appendix, 
is an antilinear involutive antimorphism satisfying in addition the condition  
\begin{eqnarray}  \forall \; a \in U_q(sl(2,\CC)_{\RR}),\;\; (\star \otimes 
\star) \Delta(a)\; = \; \Delta(a^{\star})  \label{starcoproduit}\;\;.  
\end{eqnarray}  It is easy to show the following relation between the antipode 
and the $\star$ :  \begin{eqnarray}  S \circ \star \; = \; \star \circ 
S^{-1}\;\; .  \end{eqnarray}   
The universal R-matrix of $U_q(sl(2,\CC)_{\RR})$ 
satisfies $\RR^{\star \otimes \star} = \RR^{-1}$ which is compatible with 
(\ref{starcoproduit}) and is a key property in order to build a star 
structure on the graph algebra associated to $U_q(sl(2,\CC)_{\RR})$.\\   

   By duality, 
$Pol(SL_q(2,\CC)_{\RR})$ is endowed with a star structure using the following 
definition:  \begin{eqnarray}  \alpha^{\star}(a) \; = \; 
\overline{\alpha(S^{-1}a^{\star})} \;\;\; \forall \; (\alpha, a) \; \in \; 
Pol(SL_q(2,\CC)_{\RR}) \times U_q(sl(2,\CC)_{\RR})\;\;.  \end{eqnarray}      
Let ${ I} =(I^l,I^r)\in {\SSF}$ labelling  a finite-dimensional representation 
of $U_q(sl(2,\CC)_{\RR})$ and let us define by $\tilde{I} = (I^r,I^l)$. The 
following properties of the action of the $\star$ and of the complex 
conjugation are proved in the appendix A.3.    The explicit action of the 
$\star$ involution on the generators $\stackrel{ I}{\GG}{}\!\!^{Aa}_{Bb}$ of 
$Pol(SL_q(2,\CC)_{\RR})$ is:  \begin{eqnarray}  \stackrel{ 
I}{\GG}{}\!\!^{\star} & = & \stackrel{\tilde{I} }{W} \;   \stackrel{\tilde{I} 
}{\GG} \; \stackrel{\tilde{I}  }{W}{}\!\!^{-1} \;   \end{eqnarray}  where we 
have defined $\stackrel{ I}{W}{}\!\!^{Aa}_{Bb} = 
\stackrel{\tilde{I}}{W}{}\!\!^{Aa}_{Bb}=e^{i \pi A} \; v_A^{-1/2} \; 
\stackrel{A}{w}_{ab} \; \delta^A_B$.      By duality,  the action of $\star$ 
on the generators     $\stackrel{I}{\LL}{}\!\!^{(\pm)}{}^{Aa}_{Bb}$ of 
$U_q(sl(2,\CC)_{\RR})$ is:  \begin{eqnarray}  \stackrel{I}{\LL}{}\!\!^{(\pm) 
\; \star} = \stackrel{\tilde{I} }{W} \; 
\stackrel{\tilde{I}}{\LL}{}\!\!^{(\pm)} \; \stackrel{\tilde{I} }{W}{}\!\!^{-1} 
\; .  \end{eqnarray}

We endow the graph-algebra with the following star 
structure:     
\begin{proposition}  The graph-algebra ${\cal L}_{n,p}$ is 
endowed with a star structure defined on the generators   $\stackrel{ 
I}{\AA}\!\!(i)$, $\stackrel{ I}{\BB}\!\!(i)$, $\stackrel{I}{\MM}\!\!(i)$ 
(denoted generically $\stackrel{ I}{\UU}\!\!(i))$, by  \begin{eqnarray}  
\stackrel{I}{\UU}{}\!\!(i)^{\star} \; = \; \sum_j v_{\tilde{ I}}^{-1} \; 
\stackrel{\tilde{ I}}{W} \; S^{-1}(\stackrel{\tilde{ I}}{x}_{j}) \; 
\stackrel{\tilde{ I}}{\UU}\!\!(i) \; \stackrel{\tilde{ I}}{y}_{j} \; 
\stackrel{\tilde{ I}}{\mu} \; \stackrel{\tilde{ I}}{W}{}\!\!^{-1} \;\; ,  
\end{eqnarray}  
where $\RR = \sum_j x_j \otimes y_j$. This star structure  is an  involutive antilinear automorphism 
which in the classical limit gives back the star properties 
(\ref{StaronUClassique}) on the holonomies. The definition of this  star structure is chosen in 
order that   the coaction $\delta$ is a star morphism.  \end{proposition}    
{\Proof}    We just have to prove that it is an involution and that it is 
compatible with the defining relations of the graph algebra. These two 
properties  are straightforward to verify. $\Box$ \\    The star structure on 
the graph algebra induces a star structure on the algebra ${\cal 
L}_{n,p}^{inv}.$  The action of this star structure on the generating family 
labelled by spin-network is described in the following proposition:    
\begin{proposition}  The  action of the star structure on ${\cal 
L}_{n,p}^{inv}$ satisfies:  \begin{eqnarray}  \stackrel{ 
P}{\cal O}{}\!\!^{(\pm)\star}_{n,p} \; = \; \stackrel{\tilde{ 
P}}{\cal O}{}\!\!^{(\mp)}_{n,p} \;\;  \end{eqnarray}  where $\tilde{ P}$ is 
the spin network deduced from $P$ by turning all the colors $I$ of $P$  into 
$\tilde{I}.$  \end{proposition}  
   {\Proof}  This follows from  the action of the star on the monodromies, the commutation 
relations of the monodromies and the properties with respect to the complex conjugation.        
 $\Box$   
 
 ${\mathfrak 
M}_q(\Sigma, SL(2,\CC)_{\RR},p)$ is  the  algebra defined by   ${\mathfrak 
M}_q(\Sigma, SL(2,\CC)_{\RR},p)={\cal L}_{n,p}/{\cal I}_{C}.$    Let 
$\stackrel{\alpha}{\Pi}$ be an irreducible unitary representation of   
$U_q(sl(2,\CC)_{\RR}),$  labelled by the couple $\alpha\in \SSP$ we can still 
define the complex numbers    
$\vartheta_{I\alpha} $ where $I\in \SSF$. The 
explicit formula, which is proved in the appendix A.3,  for $\vartheta_{I\alpha} $ is  \begin{equation}  
\vartheta_{I\alpha}=\frac{[(2I^l+1)(2\alpha^l+1)]}{[2\alpha^l+1]}\frac{[(2I^r+1)(2\alpha^r+1)]}{[2\alpha^r+1]}.  \end{equation}    Let 
 $\stackrel{\alpha_1}{\Pi},...,\stackrel{\alpha_p}{\Pi}$ be irreducible unitary representations of $U_q(sl(2,\CC)_{\RR})$ attached to the punctures of $\Sigma$, the moduli space   ${M}_q(\Sigma, SL(2,\CC)_{\RR};\alpha_1,...,\alpha_p )$ is defined by:  \begin{equation}  
 {M}_q(\Sigma, SL(2,\CC)_{\RR};\alpha_1,...,\alpha_p)=  
{\mathfrak M}_q(\Sigma, G,p)/\{tr_q(\stackrel{ 
I}{\MM}(n+i))=\vartheta_{I\alpha_i}, i=1,...,p,\forall I\in \SSF\}.  \end{equation}     
  
\begin{proposition}  
The star structure  on the graph algebra defines a natural star structure on the algebras $ {\mathfrak M}_q(\Sigma, SL(2,\CC)_{\RR},p)$ and  $ {M}_q(\Sigma, SL(2,\CC)_{\RR};\alpha_1,...,\alpha_p).$  
\end{proposition}  
{\Proof}  
This is a simple consequence of the fact that $\overline{\vartheta_{I\alpha}}=\vartheta_{\tilde{I}\alpha}$ for $\alpha\in \SSP$  and $I\in \SSF.$  
$\Box$  
  
The Heisenberg double of  $U_q(sl(2,\CC)_{\RR})$ is defined by  
$H(U_q(sl(2,\CC)_{\RR})=U_q(sl(2,\CC)_{\RR})\otimes Pol(SL_q(2,\CC)_{\RR}).$ 
  
In order to build unitary representations of the graph algebra and the moduli algebra in the  
next chapter we have to study the properties of Alekseev isomorphism with respect to the star structure.   
The already defined star structures on $U_q(sl(2,\CC)_{\RR})$ and $Pol(SL_q(2,\CC)_{\RR})$ naturally extend to  a star structure on $H(U_q(sl(2,\CC)_{\RR}),$ and therefore to   
$H(U_q(sl(2,\CC)_{\RR})^{\otimes n}\otimes U_q(sl(2,\CC)_{\RR}^{\otimes p}.$  
\begin{proposition}  
The Alekseev isomorphism  defined in lemma \ref{lemmeAlekseev}  
\begin{eqnarray}  
 {\cal L}_{n,p} \stackrel{\sim}{\longrightarrow} H(U_q(sl(2,\CC)_{\RR}))^{\otimes n} \otimes U_q(sl(2,\CC)_{\RR})^{\otimes p}\nonumber  
\end{eqnarray}  
is a star-isomorphism.  
\end{proposition}  
  
{\Proof}   
 Using the star structure on $H(U_q(sl(2,\CC)_{\RR}))^{\otimes n}\otimes U_q(sl(2,\CC)_{\RR})^{\otimes p}$, we first show that:  
\begin{eqnarray}  
\stackrel{ I}{\mathfrak B}{}\!\!(i)^{\star}  =  \sum_j v_{\tilde{ I}}^{-1} \; \stackrel{\tilde{ I}}{W} \; S^{-1}(\stackrel{\tilde{ I}}{x}_{j}) \; \stackrel{\tilde{ I}}{\mathfrak B}\!\!(i) \; \stackrel{\tilde{ I}}{y}_{j} \; \stackrel{\tilde{ I}}{\mu} \; \stackrel{\tilde{ I}}{W}{}\!\!^{-1} \;\; .  
\end{eqnarray}  
In order to prove this, we introduce the permutation $P_{12}$ and we have  
\begin{eqnarray*}  
\stackrel{ I}{\mathfrak B}{}\!\!(i)^{\star} & = & tr_2(P_{12} \; \stackrel{ I}{\LL}{}\!\!_1^{(-)-1 \star} \; \stackrel{ I}{\LL}{}\!\!_2^{(+) \star})  \\  
 & = &  tr_2(P_{12} \; \stackrel{ I}{W}_1 \; \stackrel{I}{\mu}{}\!\!_1^{-1}  \; \stackrel{\tilde{I}}{\LL}{}\!\!_1^{(-)-1} \; \stackrel{ I}{\mu}_1 \; \stackrel{I}{W}{}\!\!_1^{-1} \; \stackrel{ I}{W}_2 \; \stackrel{\tilde{ I}}{\LL}{}\!\!_2^{(+)} \; \stackrel{ I}{W}{}\!\!_2^{-1}) \\  
 & = & \stackrel{ I}{W}_1 \; tr_2(P_{12} \; \stackrel{\tilde{ I}}{\LL}{}\!\!_1^{(-)-1} \; \stackrel{\tilde{ I}}{\LL}{}\!\!_2^{(+)} \; \stackrel{ I}{\mu}{}\!\!_2^{-1}) \; \stackrel{ I}{\mu}_1 \; \stackrel{ I}{W}{}\!\!_1^{-1} \\  
 & = & \stackrel{ I}{W}_1 \; tr_2(P_{12} \; S^{-1}(\stackrel{\tilde{ I}}{x}_{j2} ) \; \stackrel{\tilde{ I}}{\LL}{}\!\!_2^{(+)} \; \stackrel{\tilde{ I} \tilde{ I}}{\RR} \;\stackrel{\tilde{ I}}{\LL}{}\!\!_1^{(-)-1} \; \stackrel{\tilde{ I}}{y}_{j1}  \; \stackrel{ I}{\mu}{}\!\!_2^{-1}) \; \stackrel{ I}{\mu}_1 \; \stackrel{ I}{W}{}\!\!_1^{-1} \\  
 & = & v_{\tilde{ I}}^{-1} \; \stackrel{\tilde{ I}}{W} \; S^{-1}(\stackrel{\tilde{ I}}{x}_{j}) \; \stackrel{\tilde{ I}}{\mathfrak B}\!\!(i) \; \stackrel{\tilde{ I}}{y}_{j} \; \stackrel{\tilde{ I}}{\mu} \; \stackrel{\tilde{ I}}{W}{}\!\!^{-1} \;\; .  
\end{eqnarray*}  
Then, it is easy to compute that the star acts on  $\stackrel{ I}{\mathfrak K}\!\!(i)$ and on its inverse as:  
\begin{eqnarray}  
\stackrel{ I}{\mathfrak K}\!\!(i)^{\star} \; = \; \stackrel{\tilde{ I}}{W}{}\!\!(i)\;  \stackrel{\tilde{ I}}{\mathfrak K}\!\!(i) \; \stackrel{\tilde{ I}}{W}{}\!\!(i)^{-1} \;,\;\;\;\; \stackrel{ I}{\mathfrak K}\!\!(i)^{-1 \star} \; = \; \stackrel{\tilde{ I}}{W}{}\!\!(i)\; \stackrel{\tilde{ I}}{\mu}{}^{-1} \; \stackrel{ I}{\mathfrak K}\!\!(i)^{-1} \; \stackrel{\tilde{ I}}{\mu} \; \stackrel{\tilde{ I}}{W}{}\!\!(i)^{-1} \;\;.  
\end{eqnarray}  
  
Finally, from the previous relations, we have  
\begin{eqnarray}  
(\stackrel{ I}{\mathfrak K}\!\!(i) \stackrel{ I}{\mathfrak B}\!\!(i) \stackrel{ I}{\mathfrak K}\!\!(i)^{-1})^{\star} & = & \left( tr_{23}(P_{12} P_{23} \stackrel{ I}{\mathfrak K}_3\!\!(i) \stackrel{ I}{\mathfrak B}_2\!\!(i) \stackrel{ I}{\mathfrak K}_1\!\!(i)^{-1}) \right)^{\star} \nonumber \\  
& = & tr_{23}( \stackrel{ I}{\mathfrak K}_1\!\!(i)^{-1\star} \stackrel{ I}{\mathfrak B}_2\!\!(i)^{\star} \stackrel{ I}{\mathfrak K}_3\!\!(i)^{\star}) \nonumber \\  
& = & tr_{23}(\stackrel{\tilde{ I}}{\mathfrak K}_1\!\!(i)^{-1} \stackrel{\tilde{ I}}{\mu}_1 \stackrel{\tilde{ I}}{W}{}\!\!_1^{-1} \stackrel{\tilde { I}}{v}{}\!\!_2^{-1} S^{-1}(\stackrel{\tilde{ I}}{x}_{2(j)}) \stackrel{\tilde { I}}{\mathfrak B}_2\!\!(i) \stackrel{\tilde{ I}}{y}_{2(j)} \stackrel{\tilde{ I}}{W}{}\!\!_3 \stackrel{\tilde{ I}}{\mathfrak K}_3\!\!(i)) \nonumber \\  
& = & \stackrel{\tilde{ I}}{v}{}\!\!^{-1} \stackrel{\tilde{ I}}{W} S^{-1}(\stackrel{\tilde{ I}}{x}_{(j)}) \stackrel{\tilde{ I}}{\mathfrak K}\!\!(i) \stackrel{\tilde{ I}}{x}_{(k)} S^{-1}(\stackrel{\tilde{ I}}{x}_{(l)}) \stackrel{\tilde { I}}{\mathfrak B}\!\!(i) \stackrel{\tilde{ I}}{y}_{(l)} \stackrel{\tilde{ I}}{y}_{(k)} \stackrel{\tilde{ I}}{\mathfrak K}\!\!(i)^{-1}  \stackrel{\tilde{ I}}{y}_{(j)}  \stackrel{\tilde{ I}}{\mu} \stackrel{\tilde{ I}}{W}{}\!\!^{-1} \nonumber \\  
& = & \stackrel{\tilde{ I}}{v}{}\!\!^{-1} \stackrel{\tilde{ I}}{W}  S^{-1}(\stackrel{\tilde{ I}}{x}_{(j)}) \left(  \stackrel{\tilde{ I}}{\mathfrak K}\!\!(i) \stackrel{\tilde { I}}{\mathfrak B}\!\!(i)  \stackrel{\tilde{ I}}{\mathfrak K}\!\!(i)^{-1} \right) \stackrel{\tilde{ I}}{y}_{(j)}  \stackrel{\tilde{ I}}{\mu} \stackrel{\tilde{ I}}{W}{}\!\!^{-1} \;\;.\nonumber  
\end{eqnarray}  
As a result, we have shown that the property holds true for the monodromies $\BB(i)$.  
The other cases are proved along the same lines.  
 $\Box$

\section*{III. Unitary Representations of the Moduli Algebra in the $SL(2,\CC)_{\RR}$ case.}   
\section*{III.1 Unitary representation of the graph algebra.}  
The reader is invited to read the appendix $A.1$ where the basic results on harmonic analysis are recalled. 
\begin{proposition} 
$H(U_q(sl(2,\CC)_{\RR}))$ admits a unitary representation acting on the space $Fun_{cc}(SL_q(2,\CC)_{\RR})$ endowed with the hermitian form  (\ref{L2hermitian}) and constructed as: 
\begin{eqnarray}  
  H(U_q(sl(2,\CC)_{\RR})) & \longrightarrow & End(Fun_{cc}(SL_q(2,\CC)_{\RR})), \nonumber \\  
   Pol(SL_q(2,\CC)_{\RR}) \ni f    & \mapsto &  m_f \;\;\;/\;\;\;m_f(g)  =  fg \;\;,\forall g \in  
Fun_{cc}(SL_q(2,\CC)_{\RR})\nonumber\\  
  U_q(sl(2,\CC)_{\RR})  \ni x  & \mapsto & \nabla_x \;\;\;/\;\;\; \nabla_x (g)  =  g_{(1)} \; \big< x, g_{(2)} \big > \;\; ,\forall g\in Fun_{cc}(SL_q(2,\CC)_{\RR}) .\nonumber  
\end{eqnarray}     
\end{proposition}   
\Proof Trivial to check. 
$\Box$ 
 
{}From the Alekseev isomorphism and the study of harmonic analysis on $SL_q(2,\CC)_{\RR}$, we obtain a simple description of unitary representations of the graph algebra ${\cal L}_{n,p}$:   
\begin{proposition}  
Let $\alpha_1,...,\alpha_p\in \SSP$ we denote by   
${\cal H}_{n,p}[\alpha]=Fun_{cc}(SL_q(2,\CC)_{\RR})^{\otimes n}\otimes 
{\VV}(\alpha)$ the pre-Hilbert space with sesquilinear form 
$<,>=(\otimes_{i=1}^n<,>_i )\otimes (\otimes_{j=1}^{p}<,>_{j+n})$ where $<,>_i$ 
is the $L^2$ hermitian form (\ref{L2hermitian}) on the $i$-th copy of    
$Fun_{cc}(SL_q(2,\CC)_{\RR})$ and $<,>_{n+j}$ is the hermitian form on 
$\stackrel{\alpha_j}{\VV}.$   ${\cal H}_{n,p}[\alpha]$ is endowed with a structure of 
${\cal L}_{n,p}$ module using the Alekseev isomorphism. This 
representation is unitary, in the sense that:  \begin{equation}  \forall a\in 
{\cal L}_{n,p}, \forall v,w\in {\cal H}_{n,p},\;\;\; <a^{\star}\rhd v, w>=<v, 
a\rhd w>.  \end{equation}  \end{proposition}   
\Proof   
This is a simple consequence of the previous proposition.  
$\Box$.    
  
We now come to the central part of our work: the construction of a  unitary representation of the   
moduli-algebra. We could have hoped to apply the same method as in proposition (\ref{prop:repofmoduli}), unfortunately this is not possible because there is no normalizable states in ${\cal H}_{n,p}$ or in its completion with respect to  
 $<,>$ which are invariant under the action of $U_q(sl(2,\CC)_{\RR})$ induced  from the algebra generated by  
$\stackrel{I}{C}.$ We cannot exclude the existence of a quantum analogue of Faddeev-Popov procedure to solve this problem but we were unable to proceed along this path. Instead we  will give explicit formulas for the action of  
$\stackrel{P}{\cal O}{}_{n,p}^{(\pm)}$ on the space of invariant vectors, with a suitable hermitian form, and verify that this representation is unitary. 
We will enlarge the representation space of the graph algebra as follows: we  will transfer the representation of the graph-algebra on the dual conjugate space   
 $\overline{{\cal H}_{n,p}}{[\alpha]}^{*}$ as follows: 
\begin{equation} 
\;\;\forall \overline{\phi} \in\overline{ {\cal H}_{n,p}[\alpha]}{}^{*}, \forall v \in {\VV}[\alpha],  
\forall a \in {\cal L}_{n,p}, (a\rhd \overline{\phi},\overline{v})= 
(\overline{\phi}, \overline{a^{\star}\rhd v})=\overline{(\phi,a^{\star}\rhd v) }. 
\end{equation} 
$\overline{ {\cal H}_{n,p}[\alpha]}{}^{*}$ is naturally endowed with a structure of $U_q(SL(2,\CC)_{\RR})$ module which admits, as we will see, invariant elements.  This method is similar in spirit to the concept of refined algebraic quantization program \cite{ALMMT}. 
In order not to complicate notations, we will prefer to work with the right module  
$ {\cal H}_{n,p}[\alpha]{}^{*}$ associated to the anti-representation: 
\begin{equation} 
\;\;\forall {\phi} \in {\cal H}_{n,p}[\alpha]{}^{*} , \forall v \in {\VV}[\alpha],  
\forall a \in {\cal L}_{n,p}, (\phi \lhd a,{v})= 
({\phi}, {a \rhd v}), 
\end{equation} 
this is of course completely equivalent. 
We will continue to denote by $\rho_{n,p}[\alpha]$ this antirepresentation.

The construction of a unitary representation of the moduli algebra is exposed through elementary steps: the  p-punctured sphere, the genus-n surface and finally the general case.

 \section*{III.2 Unitary Representation of the moduli algebra }  
  
\subsection*{III.2.1. The moduli algebra of a p-punctured sphere}  
This first subsection is devoted to a precise description of unitary  representations  
of the moduli-algebra on a sphere with p punctures associated to unitary irreducible  representations  
${\alpha}_1, \cdots , {\alpha}_p.$

Let us begin with some particular examples. The three first cases (i.e. $p=1$, $2$ or $3$) are singular in the sense that the representation of the moduli algebra is one-dimensional or zero-dimensional.   
 
In the case of the one-punctured sphere, the moduli algebra has the following structure: 
\begin{itemize} 
\item $M_q(S^2,SL(2,\CC)_{\RR};\Pi)=\{0\}$ if $\Pi$ is not the trivial representation  
\item $M_q(S^2,SL(2,\CC)_{\RR};\stackrel{I}{\Pi})=\CC$ where $I=(0,0)$. 
\end{itemize} 
As a result  
 the representation is non zero-dimensional if and only if the representation associated to the puncture is the trivial representation. In this case the corresponding one dimensional representation $H_{0,1}$ is trivially unitarizable. 
  
In the case of the two-punctured sphere, the moduli algebra has the following structure: 
\begin{itemize} 
\item $M_q(S^2,SL(2,\CC)_{\RR};\stackrel{\alpha_1}{\Pi},\stackrel{\alpha_2}{\Pi})=\{0\}$ if 
 $\stackrel{\alpha_1}{\Pi}$ is not equivalent to  $\stackrel{\alpha_2}{\Pi}$   
\item $M_q(S^2,SL(2,\CC)_{\RR};\stackrel{\alpha_1}{\Pi},\stackrel{\alpha_1}{\Pi})=\CC$. 
\end{itemize} 
 
As a result  
the representation is non-zero dimensional if and only if the representations associated to the  
two punctures are the same, i.e. $\alpha_1 = \alpha_2 $. The associated module  is  
$(  {{\VV}(\alpha_1,\alpha_1)}{}^{*}){}^{U_q(sl(2,\CC)_{\RR})}$ generated by  
 $\stackrel{\alpha}{{\omega}}_{0,2}$ where $\alpha=(\alpha_1,\alpha_1)$, such that  
\begin{eqnarray}  
<\stackrel{\alpha}{\omega}_{0,2} \; , \; \stackrel{I}{{e}}_i\!\!(\alpha_1) \otimes  \stackrel{J}{{e}}_j\!\!(\alpha_1)> \; = \Psi^{0}_{\alpha_1 \alpha_1} \big(\stackrel{I}{e}_i\!\!(\alpha_1) \otimes  \stackrel{J}{e}_j\!\!(\alpha_1)\big)\;\;. 
\end{eqnarray}    
 
We can endow this module with a  Hilbert structure $<.\vert.>$ such that $<\stackrel{\alpha}{{\omega}}_{0,2} \vert \stackrel{\alpha}{{\omega}}_{0,2}>=1$. This is a unitary one-dimensional module $H_{0,2}(\alpha)$ of the moduli algebra because of the property 
 $\overline{\vartheta_{I\Pi}}=\vartheta_{\tilde{I}\Pi}$ for unitary representations $\Pi$, and $I\in \SSF.$ \\

In the case of three punctures, the moduli algebra has the following structure: 
\begin{itemize} 
\item $M_q(S^2,SL(2,\CC)_{\RR};\stackrel{\alpha_1}{\Pi},\stackrel{\alpha_2}{\Pi},\stackrel{\alpha_3}{\Pi} )=\CC$ if 
$m_{\alpha_1}+m_{\alpha_2}+m_{\alpha_3}\in \ZZ,$ 
\item $M_q(S^2,SL(2,\CC)_{\RR};\stackrel{\alpha_1}{\Pi},\stackrel{\alpha_2}{\Pi},\stackrel{\alpha_3}{\Pi})=\{0\}$ in the other cases. 
\end{itemize} 
In the first case, let us denote by $\alpha=(\alpha_1,\alpha_2,\alpha_3).$ 
 The associated module is  
$( { {\VV}(\alpha)}{}^{*}){}^{U_q(sl(2,\CC)_{\RR})}$ generated by  
 $\stackrel{\alpha}{\omega}_{0,3}$ such that  
\begin{eqnarray}  
<\stackrel{\alpha}{\omega}_{0,3} \; , \; \stackrel{I}{{e}}_i\!\!(\alpha_1) \otimes  \stackrel{J}{{e}}_j\!\!(\alpha_2)\otimes  \stackrel{K}{{e}}_k\!\!(\alpha_3)> \; ={\big( \Psi^{0}_{\alpha_3\alpha_3}\Psi^{\alpha_3}_{\alpha_1 \alpha_2}\big)\big(\stackrel{I}{e}_i\!\!(\alpha_1) \otimes  \stackrel{J}{e}_j\!\!(\alpha_2)\otimes  \stackrel{K}{e}_k\!\!(\alpha_3)\big)}\;\;.  
\end{eqnarray}    
We can endow this module with a  Hilbert structure $<.\vert.>$ such that $<\stackrel{\alpha}{{\omega}}_{0,3} \vert \stackrel{\alpha}{{\omega}}_{0,3}>=1$. This is a unitary one-dimensional module $H_{0,3}(\alpha)$ of the moduli algebra for the same reasons as before.

\begin{definition} \label{defofPsi}
Let $\alpha=(\alpha_1,...,\alpha_n)\in {\mathbb S}^n, \beta=(\beta_3,....,\beta_n)\in {\mathbb S}^{n-2},  
\kappa\in {\mathbb S},$ 
we define  
$\Psi_{\alpha}^\kappa(\beta)\in  
Hom_{U_q(sl(2,\CC))_{\RR}}(\VV(\alpha), \stackrel{\kappa}{\VV})$ as 
\begin{equation} 
\Psi_{\alpha}^\kappa(\beta)= 
\Psi^{\kappa}_{\beta_n\alpha_n} 
\Psi^{\beta_n}_{\beta_{n-1}\alpha_{n-1}}\cdots 
\Psi^{\beta_{4}}_{\beta_{3}\alpha_{3}}\Psi^{\beta_{3}}_{\alpha_{1}\alpha_{2}}. 
\end{equation} 
\end{definition}

\begin{definition} 
For any  family   $\beta = (\beta_3, \cdots , \beta_{p-1}) \in {\mathbb S}^{ p-3}$ we define  
 the linear form  $\stackrel{\alpha}{{\omega}}_{0,p}[\beta]\in {{\VV}(\alpha)}^{*}$ by: 
\begin{equation}  
 \stackrel{\alpha}{\omega}_{0,p}[\beta] = 
\Psi^{0}_{\alpha}(\beta_3, \cdots , \beta_{p-1},\alpha_p). 
\end{equation} 
\end{definition}

 These linear forms satisfy the constraints  
\begin{eqnarray}  
 \; \stackrel{\alpha}{{\omega}}_{0,p}\!\![\beta]\lhd \stackrel{I}{C}{}\!\!^{(\pm)}_{0,p} \ 
 \;& = & 
 \; \stackrel{\alpha}{{\omega}}_{0,p}\!\![\beta] \;,\;\;\;\; 
\; \\
\stackrel{\alpha}{{\omega}}_{0,p}\!\![\beta] \; \lhd tr_{q}(\stackrel{ I}{\MM}\!\!(n+i)) 
& = & \; \vartheta_{I \alpha_i}  \stackrel{\alpha}{{\omega}}_{0,p}\!\![\beta] \;\;,\; \forall { I} \in \SSF, \forall i=1,...,p \;\;. \label{omegabetaareinvariants} 
\end{eqnarray}  
In particular  $\stackrel{\alpha}{{\omega}}_{0,p}\!\![\beta]$ are  invariant elements of  
$ {{\VV}(\alpha)}^{*}.$ 
It will be convenient to denote  
$\stackrel{\alpha}{\omega}_{0,p}[\beta]^{I}_{a}=  
<\stackrel{\alpha}{\omega}_{0,p}[\beta],\bigotimes_{i=1}^{p} \stackrel{I_i}{{e}}_{a_i}\!\!(\alpha_i)>.$ 
 
We will now smear $\stackrel{\alpha}{{\omega}}_{0,p}\!\![\beta]$ with a function $f$ sufficiently regular, in  
order to obtain ``wave packet.'' We have chosen functions which are analytic in the spirit of the Paley-Wiener theorem.  
 
In the sequel we will use the following notations:  
we  define  ${\cal S}=(\onehalf \ZZ)^2$ and for  $\beta\in  {\mathbb S}^k,$ we will denote by  
$\rho_{\beta}=(\rho_{\beta_1},...,\rho_{\beta_k})$ and  
$m_{\beta}= 
(m_{\beta_{1}},..., m_{\beta_{k}}).$ Reciprocally  $m\in (\onehalf\ZZ)^k,\rho\in \CC^k$ is associated to a unique element   $\beta(m,\rho)\in {\mathbb S}^k.$  
If $f$ is a complex valued function on $ {\mathbb S}^k$ we will denote by 
 $f_{m}$ the function on $\CC^k$ defined by $f_m(\rho)=f(\beta(m,\rho)).$

\begin{definition} 
We define $\AA^{(k)}$ to be the set of functions $f$ on ${\mathbb S}^{k}$ such that  
\begin{itemize} 
\item $f(\beta_1, ...,\beta_{r-1},\underline{\beta_r},\beta_{r+1},...,\beta_{k})=f(\beta_1, ...,\beta_{r-1},\beta_r,\beta_{r+1},...,\beta_{k}),\;\;\forall r=1,...,k,$ 
\item  
$f_m$ is a non zero function for only a finite number of $m\in \onehalf\ZZ^k,$ 
\item 
$f_m$ is a Laurent series  in the variables  $q^{i\rho_{\beta_1}},...,q^{i\rho_{\beta_k}},$ convergent for all $\rho_{\beta_1},..., \rho_{\beta_k}\in \CC.$ 
\end{itemize} 
 \end{definition}

For $\beta\in {\mathbb S}^{p-3}$ we define  
\begin{equation} 
\hat{\Xi}_{0,p}[\alpha,\beta]^{-1}=e^{i\pi(\alpha_1+...+\alpha_p)}\zeta(\alpha_1,\alpha_2,\beta_3) 
\prod_{j=3}^{p-2}\zeta(\beta_j,\alpha_j,\beta_{j+1}) 
 \zeta(\beta_{p-1},\alpha_{p-1},\alpha_p) 
\prod_{j=3}^{p-1}{\nu_1(d_{\beta})}. 
\end{equation} 
For $I=(I_1,...,I_p)$ we denote 
 $\nu^I(\alpha)=\prod_{j=1}^p \nu^{(I_j)}(\alpha_j).$ 
 
\begin{lemma} \label{omegaxinu}
$\stackrel{\alpha}{\omega}_{0,p}[\beta]{}^{I}_{a}\hat{\Xi}_{0,p}[\alpha,\beta]\nu^{I}(\alpha)$ is equal to  
$P[\alpha,\beta]/Q[\beta]$ where $P[\alpha,\beta]$ is a Laurent polynomial in the variables  
$q^{i\rho_{\alpha_1}},...,q^{i\rho_{\alpha_{p}}},q^{i\rho_{\beta_3}},...,q^{i\rho_{\beta_{p-1}}}$ and 
 $Q[\beta]=\prod_{j=3}^{p-1}(i\rho_{\beta_j}-B_j)_{2B_j+1}$ with  
$B_j \in \onehalf\NN.$ 
Let $M$ be the set of $J \in (\onehalf \NN)^{p-3}$ such that the inequalities:  $Y(J_3,I_1,I_2)=1,..., Y(J_{p-2},I_{p-2},J_{p-1})=1,  Y(I_{p-1}, J_{p-1}, I_p)=1$ hold. 
We have $B_j=max \{J_j, J \in M\}.$  
\end{lemma} 
 
\Proof 
Trivial application of the proposition (\ref{polesofreducedelement}) of the Appendix A.2. 
$\Box$ 
 
For $\beta\in {\mathbb S}^{p-3}$ we define  
$\xi[\beta]=\prod_{j=3}^{p-1}\xi(2\beta_j^l+1),$   
${\cal P}[\beta]=\prod_{j=3}^{p-1}{\cal P}(\beta_j)$,  where $\xi$ and the Plancherel weight  ${\cal P}$ are defined in the appendix A.1, as well as $\Xi_{0,p}[\alpha,\beta]=\hat{\Xi}_{0,p}[\alpha,\beta]\xi[\beta].$ 
 
\begin{proposition} 
We can define a subset of  ${{\VV}(\alpha)}^{*}$  by: 
\begin{equation}  
H_{0,p}(\alpha)=\{\stackrel{\alpha}{\omega}_{0,p}\!\!(f)=\int_{\SSP^{(p-3)}} d\beta {\cal P}[\beta] f(\beta)\Xi_{0,p}[\alpha,\beta] \stackrel{\alpha}{{\omega}}_{0,p}\!\![\beta], \mbox{with}\;\; f \in \AA^{(p-3)}\}. 
\end{equation} 
The map $\AA^{(p-3)}\rightarrow H_{0,p}(\alpha)$ which sends $f$ to $\stackrel{\alpha}{\omega}_{0,p}\!\!(f)$ is an injection. 
\end{proposition} 
\Proof 
In order to show that $\stackrel{\alpha}{\omega}_{0,p}\!\!(f)$ is well defined it is sufficient to show that \newline 
$\xi[\beta]\Xi_{0,p}[\alpha,\beta] \stackrel{\alpha}{{\omega}}_{0,p}[\beta]{}^{I}_{a}$ is a Laurent series  in  
$q^{i\rho_{\beta_3}},...,q^{i\rho_{\beta_{p-1}}}$, which is the case because  
$\xi[\beta]$ cancels the simple poles of $\Xi_{0,p}[\alpha,\beta] \stackrel{\alpha}{{\omega}}_{0,p}[\beta]^{I}_{a}.$ 
The sum over $m_{\beta}$ is finite because of the condition on $f$. 

We now prove injectivity of this map. 
Assume that  $\stackrel{\alpha}{\omega}_{0,p}\!\!(f)$ is zero, we would therefore have $ \stackrel{\alpha}{\omega}_{0,p}\!\!(f)(av)=0$ for all   
$a\in U_q(sl(2,\CC)_{\RR})^{\otimes p}$ and $v\in {{\VV}[\alpha]}.$ 
If $c$ is a central element of $U_q(sl(2,\CC)_{\RR})$ we denote by $c(\beta)$ its value on the module $\stackrel{\beta}{\VV}.$ 
In particular if $c_3,...,c_{p-1}$ are elements of the center of  
$U_q(sl(2,\CC)_{\RR})$, we take $a=\Delta(c_3)...\Delta^{(p-2)}(c_p)$ where  
$\Delta^{(k)}$ are the iterated coproducts, and using the intertwiner property we obtain that $ \stackrel{\alpha}{\omega}_{0,p}\!\!(fg(c_3,...,c_{p-1}))=0$ where  
$g(c_3,...,c_{p-1})(\beta)=\prod_{j=3}^{p-1}c_j(\beta_j).$ 
By a similar argument as the one used in \cite{BR3}  
(proof of Th 12 (Plancherel Theorem)) , we obtain that f has to satisfy the identity: 
$f(\beta)\Xi_{0,p}[\alpha,\beta] \stackrel{\alpha}{{\omega}}_{0,p}\!\![\beta]=0.$ 
It remains to show that this implies $f=0$. This follows from a similar argument exposed  in \cite{BR4} (Th 4) which uses the asymptotics of the reduced elements. 
$\Box$

We define, for $\alpha \in {\mathbb S}_P^{p}, \beta\in {\mathbb S}_P^{p-3}$, 
\begin{eqnarray}
 { M}_{0,p}[\alpha,\beta]=M(\alpha_1,\alpha_2,\beta_3) 
\prod_{j=3}^{p-2} M(\beta_j,\alpha_j,\beta_{j+1}) 
 M(\beta_{p-1},\alpha_{p-1},\alpha_p)
\end{eqnarray}
where $M$ is defined in the appendix A.2, and we denote for 
 $\alpha \in {\mathbb S}_P^{p}, \beta\in {\mathbb S}_P^{p-3}$ 
 \begin{equation} 
\Upsilon_{0,p}[\alpha,\beta]=\frac{\vert \Xi_{0,p}[\alpha,\beta] \vert^2}{{M}_{0,p}[\alpha,\beta]}\label{Upsilon}. \end{equation} 
 
\begin{lemma} \label{analyticityofupsilon0p}
$\Upsilon_{0,p}[\alpha,\beta]$ is an analytic function of the real variables $\rho_{\beta_3}, ...,\rho_{\beta_{p-1}}$ and therefore admits an analytic continuation for  $\rho_{\beta_3}, ..., \rho_{\beta_{p-1}} \in {\CC}^{p-3},$ i.e
$\beta_3,...,\beta_{p-1} \in {\cal S}^{p-3}.$
\end{lemma} 
\Proof 
We can extend $M$ to ${\mathbb S}^3$ as follows \cite{BR4}: 
\begin{eqnarray} 
M(\alpha,\beta,\gamma)= 
\psi_1(\alpha,\beta,\gamma) 
\frac{(q^{-1}-q)q^{m_\alpha+m_\beta+m_\gamma}} 
{\nu_1(d_{\alpha^l})\nu_1(d_{\underline{\alpha^r}}) 
\nu_1(d_{\beta^l})\nu_1(d_{\underline{\beta}^r})\nu_1(d_{\gamma^l})\nu_1(d_{\underline{\gamma^r}})} 
\Theta(\alpha,\beta,\gamma) 
\end{eqnarray} 
where 
\begin{eqnarray} 
\hskip-1cm&& \Theta(\alpha,\beta,\gamma)= 
\frac{\theta(2\alpha^r+1) \theta(2\beta^r+1) \theta(2\gamma^r+1)} 
{\theta(\alpha^r\!+\!\beta^r\!+\!\gamma^r\!+\!2)\theta(\underline{\alpha}^r\!+\!\beta^r\!+\!\gamma^r+2) 
 \theta(\alpha^r\!+\!\underline{\beta}^r\!+\!\gamma^r\!+\!2)  \theta(\alpha^r\!+\!\beta^r\!+\!\underline{\gamma}^r\!+\!2)}\nonumber\\ 
\hskip-1cm&&\psi_1(\alpha,\beta,\gamma)=\frac{ 
\varphi_{(2\alpha^r,-2m_{\alpha}-1)} 
\varphi_{(2\beta^r,-2m_{\beta}-1)} 
\varphi_{(2\gamma^r,-2m_{\gamma}-1)} 
}{ 
\varphi_{(i\rho_{\alpha+\beta+\gamma},m_{\alpha+\beta+\gamma})}\varphi_{(i\rho_{\underline{\alpha}+\beta+\gamma},m_{\underline{\alpha}+\beta+\gamma})}\varphi_{(i\rho_{\alpha+\underline{\beta}+\gamma},m_{\alpha+\underline{\beta}+\gamma})}\varphi_{(i\rho_{\alpha+\beta+\underline{\gamma}},m_{\alpha+\beta+\underline{\gamma}})}}.\nonumber 
\end{eqnarray} 
 
Moreover $\vert \Xi_{0,p}[\alpha,\beta] \vert^2$ can be extended to $\beta_3, ..., \beta_{p-1} \in {\mathbb S}^{p-3}$ by  
\begin{equation} 
\vert \Xi_{0,p}[\alpha,\beta] \vert^2=\frac{\prod_{j=3}^{p-1}(-1)^{2m_{\beta_j}}q^{4im_{\beta_j}\rho_{\beta_j}}}{{\psi}_2{}_{(\alpha_1,\alpha_2,\beta_3)}\prod_{j=3}^{p-2}{\psi}_2{}_{(\beta_j,\alpha_j,\beta_{j+1})}{\psi}_2{}_{(\beta_{p-1},\alpha_{p-1},\alpha_{p})}}\Xi_{0,p}[\alpha,\beta]^2 
\end{equation} 
where 
\begin{eqnarray} 
\hskip-1cm&&\psi_2{}_{(\alpha,\beta,\gamma)}=\frac{q^{\frac{i}{2}(\rho_{\alpha+\beta+\gamma}\vert m_{\alpha+\beta+\gamma}\vert+ 
\rho_{\underline{\alpha}+\beta+\gamma}\vert m_{\underline{\alpha}+\beta+\gamma}\vert+ 
\rho_{\alpha+\underline{\beta}+\gamma}\vert m_{\alpha+\underline{\beta}+\gamma}\vert+ 
\rho_{\alpha+\beta+\underline{\gamma}}\vert m_{\alpha+\beta+\underline{\gamma}}\vert)}}{ 
\varphi_{(-i\rho_{\alpha+\beta+\gamma},\vert m_{\alpha+\beta+\gamma}\vert)} 
\varphi_{(-i\rho_{\underline{\alpha}+\beta+\gamma},\vert m_{\underline{\alpha}+\beta+\gamma}\vert)} 
\varphi_{(-i\rho_{\alpha+\underline{\beta}+\gamma},\vert m_{\alpha+\underline{\beta}+\gamma}\vert)} 
\varphi_{(-i\rho_{\alpha+\beta+\underline{\gamma}},\vert m_{\alpha+\beta+\underline{\gamma}}\vert)}}.\nonumber 
\end{eqnarray} 
Due to the following obvious properties,  
\begin{eqnarray*} 
&&\varphi_{(2\beta^r,-2m_{\beta}-1)}^2=(-1)^{2m_{\beta}+1}\\ 
&&\varphi_{(i\rho_{\alpha+\beta+\gamma}, m_{\alpha+\beta+\gamma})}\varphi_{(-i\rho_{\alpha+\beta+\gamma},\vert m_{\alpha+\beta+\gamma}\vert)}=e^{i\frac{\pi}{2}(\vert m_{\alpha+\beta+\gamma}\vert-m_{\alpha+\beta+\gamma})}\\ 
&&(\nu_1(d_{\beta^l})\nu_1(d_{\underline{\beta^r}}))^2=(d_{\beta^l})_1(d_{\underline{\beta^r}})_1 
\end{eqnarray*} 
and the fact that $\Xi_{0,p}[\alpha,\beta]^2({\Theta}{}_{(\alpha_1,\alpha_2,\beta_3)}\prod_{j=3}^{p-2}{\Theta}{}_{(\beta_j,\alpha_j,\beta_{j+1})}{\Theta}{}_{(\beta_{p-1},\alpha_{p-1},\alpha_{p})})^{-1}$ is analytic in the complex variables $\rho_{\beta_3},...,\rho_{\beta_{p-1}},$ $\Upsilon_{0,p}[\alpha,\beta]$ is an analytic function of the real variables  $\rho_{\beta_3}, ...,\rho_{\beta_{p-1}}.$ We will still use the notation $\Upsilon_{0,p}[\alpha,\beta]$ to denote its analytic continuation for  $\rho_{\beta_3}, ...,\rho_{\beta_{p-1}}\in \CC.$  
$\Box$\\ 
 
Remark: Note that $\Theta(\alpha,\beta,\gamma)$ is left invariant  under permutation of $\alpha,\beta,\gamma$ and that $\Theta(\alpha,\beta,\gamma)$ is left invariant under the following shifts: 
$\Theta(\alpha+s,\beta,\gamma)=\Theta(\alpha,\beta,\gamma)$ for $s\in \ZZ^2,$ and  
$\Theta(\alpha,\beta,\gamma)=\Theta(\alpha+(i\frac{\pi}{\ln q}, 
i\frac{\pi}{\ln q}),\beta,\gamma).$ 
 
\begin{proposition} 
The space $H_{0,p}(\alpha)$  is endowed with the following pre-Hilbert space structure: 
\begin{eqnarray}  
<\stackrel{\alpha}{\omega}_{0,p}\!\!(f) \; \vert \; \stackrel{\alpha}{\omega}_{0,p}\!\!(g)> =  
\int_{\mathbb S_P^{p-3}} d\beta {\cal P}[\beta]\; \overline{f(\beta)}\Upsilon_{0,p}[\alpha,\beta] \; g(\beta)\;\;. \label{scalarproduct0P} 
\end{eqnarray}  
\end{proposition} 
\Proof 
We use injectivity of the map $f\mapsto \stackrel{\alpha}{\omega}_{0,p}\!\!(f)$ to show that this hermitian product is unambiguously defined. 
The convergence of the integrals in the real variables $\rho_{\beta_3}, ...,\rho_{\beta_{p-1}}$ is ensured by the analyticity of the integrand and the convergence of the sums in $m_{\beta_3}, ...,m_{\beta_{p-1}}$ comes from the fact that the wave packets have finite support in these discrete variables.\\ 
The positivity of this hermitian product is due to formulas (\ref{Upsilon})(\ref{defM}). Showing it is definite is trivial. 
$\Box$\\ 
 
The major theorem of this section is the result that $H_{0,p}(\alpha)$ is a right unitary module of the moduli algebra $M_q(S^2, \stackrel{\alpha}{\Pi}).$ This is a non trivial result which is divided in the following steps. 
In Lemma 1 we  compute the action of an element of $M_q(S^2, \stackrel{\alpha}{\Pi})$ on  
$ \stackrel{\alpha}{\omega}_{0,p}\!\![\beta].$ In particular if $\beta$ is in  
$\SSP$, the result is a linear combination of  
$ \stackrel{\alpha}{\omega}_{0,p}\!\![\gamma],$ with $\gamma$ in ${\mathbb S}.$ This comes from the fact that the observables are constructed with intertwiners of finite dimensional representations of $U_q(sl(2,\CC)_{\RR})$ and that the tensor product of finite dimensional representations and principal representations decomposes as a finite direct sum of infinite dimensional representations, non unitary in general, of ${\mathbb S}$ type as in  
(\ref{tensoroffiniteandinfinite}). 
As a result the action of an element $a$ of $M_q(S^2, \stackrel{\alpha}{\Pi})$ on $ \stackrel{\alpha}{\omega}_{0,p}\!\![f]$ can be defined after the use of a change of integration in the complex plane, thanks to Cauchy theorem. The proof of unitarity of this representation is reduced to properties of the kernel $\Upsilon_{0,p}$ under shifts.

\begin{proposition} \label{actionofOinthepuncturecase}
Let $\beta\in{\mathbb S}^{p-3}$,  the action of   
$\stackrel{P}{\cal O}{}\!\!^{(\pm)}_{0,p}\in {\cal L}_{0,p}^{inv}$ on $\stackrel{\alpha}{\omega}_{0,p}[\beta]$ is given by: 
\begin{equation}  
\stackrel{\alpha}{\omega}_{0,p}[\beta] \lhd \stackrel{P}{\cal O}{}\!\!^{(\pm)}_{0,p} \; = \; \sum_{s\in{\cal  S}^{p-3} } \stackrel{P}{K}{}\!\!_{0,p}^{(\pm)} 
 \left( \begin{array}{c} \!\!\alpha\!\! \\ \!\! \beta,s \!\!\end{array} \right) 
 \stackrel{\alpha}{\omega}_{0,p}\!\! \lbrack \beta+s \rbrack \;\;, \label{actionofOon omega} 
\end{equation}  
where: 
 
1. the functions  $\stackrel{ P}{K}{}\!\!_{0,p}^{(\pm)} \left( \begin{array}{c} \alpha \\\beta,s \end{array} \right)$ are non zero only for a finite set  of $s\in {\cal S}^{p-3}$ according to  the selection rules imposed by the palette P. Moreover any  element  $s$ of this finite set satisfy  
$s^l, s^r\in {\ZZ}^{p-3}.$ 
 
2. the functions  $\stackrel{ P}{K}{}\!\!_{0,p}^{(\pm)} \left( \begin{array}{c}  \!\!\alpha \!\! \\ \!\!\beta,s \!\! \end{array} \right)$ belong to  
$\frac{\Xi_{0,p}[\alpha,\beta+s]}{\Xi_{0,p}[\alpha,\beta]} 
\CC(q^{i\rho_{\alpha_1}},...,q^{i\rho_{\alpha_p}}, 
q^{i\rho_{\beta_3}},...,q^{i\rho_{\beta_{p-3}}} )$, and more precisely we have: 
$\stackrel{ P}{K}{}\!\!_{0,p}^{(\pm)} \left( \begin{array}{c}  \!\!\alpha \!\! \\ \!\!\beta,s \!\! \end{array} \right)\frac{\Xi_{0,p}[\alpha,\beta]}{\Xi_{0,p}[\alpha,\beta+s]}=\frac{P[\alpha,\beta]}{\prod_{j=3}^{p-1}Q_j(q^{i\rho_{\beta_j}})}$ 
where $P[\alpha,\beta]\in\CC(q^{i\rho_{\alpha_1}},...,q^{i\rho_{\alpha_p}}) 
[q^{i\rho_{\beta_3}}, q^{-i\rho_{\beta_{3}}},..., q^{i\rho_{\beta_{p-1}}}, q^{-i\rho_{\beta_{p-1}}}]$ and $Q_j$ is a polynomial with zeroes in $\{q^{n}, n\in \onehalf \ZZ\}$. 
\end{proposition}

\begin{figure} 
\psfrag{i1}{$$} 
\psfrag{i2}{$$} 
\psfrag{i3}{$$} 
\psfrag{i4}{$$} 
\psfrag{al1}{$\alpha_1$}  
\psfrag{al2}{$\alpha_2$}  
\psfrag{al3}{$\alpha_3$}  
\psfrag{al4}{$\alpha_4$}  
\psfrag{be}{$\beta$}  
\psfrag{n1}[][]{$N_1\;\;$}  
\psfrag{n2}{$N_2$}  
\psfrag{n3}{$N_3$} 
\psfrag{n4}{$N_4$} 
\psfrag{u3}[][]{$U_3\;\;$} 
\psfrag{u4}[][]{$U_4\;\;$} 
\psfrag{t3}{$T_3$} 
\psfrag{t4}{$T_4$} 
\psfrag{w}{$W$} 
\psfrag{0}{$0$} 
\centering 
\scalebox{0.8}[0.8]{\includegraphics[scale=0.5]{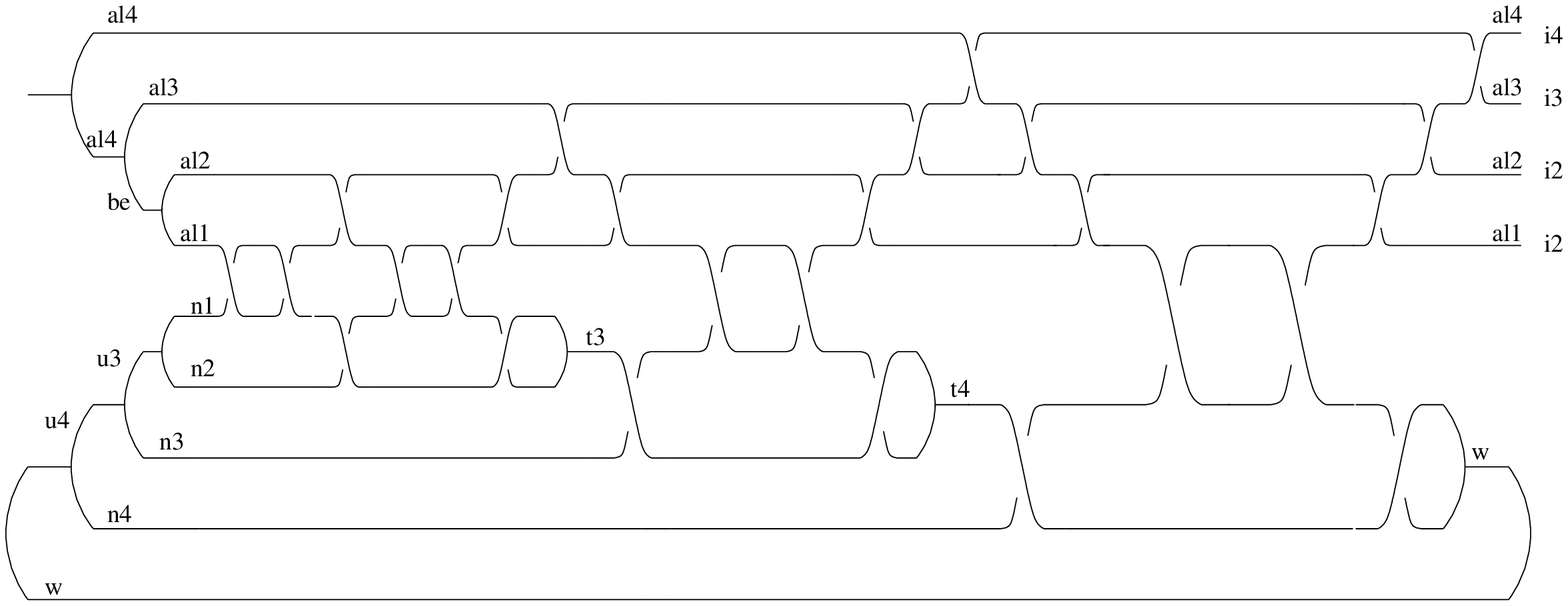}}
\caption{Expression of $\stackrel{P}{\cal F}{}\!\!_{0,4}^{(+)}$.}  
\label{fig:calF} 
\end{figure}

\Proof 
\begin{equation} 
(\stackrel{\alpha}{\omega}_{0,p}\!\!\lbrack \beta \rbrack \lhd \stackrel{P}{\cal O}{}\!\!^{(\pm)}_{0,p}) \stackrel{I}{e}_{a}\!\!(\alpha)  =  
 \stackrel{ P}{\cal F}{}\!\!_{0,p}^{(\pm)} \left( \begin{array}{c}  \!\!\alpha \!\! \\  \!\!\beta  \!\!\end{array} \right)\stackrel{I}{e}_{a}\!\!(\alpha) ,  
\end{equation}  
 
where $\stackrel{P}{\cal F}{}\!\!_{0,p}^{(\pm)} \left( \begin{array}{c}  \!\!\alpha  \!\!\\  \!\!\beta \!\! \end{array} \right)$ are   elements of  
 $\Hom_{U_q(sl(2,\CC)_{\RR})} 
(\stackrel{\alpha}{\VV}, \CC).$  The picture for 
$\stackrel{P}{\cal F}{}\!\!_{0,p}^{(+)} \left( \begin{array}{c}  \!\!\alpha  \!\!\\  \!\!\beta \!\! \end{array} \right)$ is shown in  figure \ref{fig:calF}, whereas the picture for $\stackrel{P}{\cal F}{}\!\!_{0,p}^{(-)} \left( \begin{array}{c}  \!\!\alpha  \!\!\\  \!\!\beta \!\! \end{array} \right)$ is the same after having turned overcrossing colored  by couples of finite dimensional representations into  the corresponding undercrossing. The picture for  $p>4$ punctures is a straighforward generalization.

When $\alpha_1,...,\alpha_n,\beta_3,...,\beta_{p-3}$ are fixed elements of $\SSF,$  the non zero elements of the family  
$\{\stackrel{\alpha}{\omega}_{0,p}[\beta+s], s\in {\cal S}^{p-3}\} ,$ form a   basis of   $\Hom_{U_q(sl(2,\CC)_{\RR})} 
(\stackrel{\alpha}{\VV}, \CC).$ 
As a result we obtain  
 
\begin{equation}  
\stackrel{P}{\cal F}{}\!\!_{0,p}^{(\pm)} \left( \begin{array}{c}  \!\!\alpha  \!\!\\  \!\!\beta  \!\!\end{array} \right) 
= \; \sum_{s\in{\cal  S}^{p-3} } \stackrel{P}{K}{}\!\!_{0,p}^{(\pm)} 
 \left( \begin{array}{c} \!\!\alpha\!\! \\ \!\!\beta,s\!\! \end{array} \right) 
 \stackrel{\alpha}{\omega}_{0,p}\!\! \lbrack \beta+s \rbrack \;\;,  
\end{equation} 
 
where  $\stackrel{P}{K}{}\!\!_{0,p}^{(\pm)} 
 \left( \begin{array}{c}  \!\!\alpha \!\! \\  \!\!\beta,s  \!\!\end{array} \right)$ can be computed as: 
\begin{equation} 
 \stackrel{P}{K}{}\!\!_{0,p}^{(\pm)} 
 \left( \begin{array}{c}  \!\!\alpha  \!\!\\  \!\!\beta,s \!\! \end{array} \right)= 
\stackrel{P}{\cal F}{}\!\!_{0,p}^{(\pm)} \left( \begin{array}{c} \!\! \alpha \!\! \\  \!\!\beta \!\! \end{array} \right) \stackrel{\alpha}{\eta}\!\! \lbrack \beta+s \rbrack 
\end{equation} 
with  
$\stackrel{\alpha}{\eta}\!\! \lbrack \beta \rbrack= 
\Phi_{\beta_{3}}^{\alpha_{1}\alpha_{2}} 
\Phi_{\beta_{4}}^{\beta_{3}\alpha_{3}} 
\cdots 
\Phi_{\alpha_{p}}^{\beta_{p-1}\alpha_{p-1}} 
\Phi_{0}^{\alpha_p\alpha_p} 
.$ \\
 
\begin{figure} 
\psfrag{al1}{$\alpha_1$}  
\psfrag{al2}{$\alpha_2$}  
\psfrag{al3}{$\alpha_3$}  
\psfrag{al4}{$\alpha_4$}  
\psfrag{be}{$\beta$}  
\psfrag{bes}{$\beta+s$}  
\psfrag{n1}[][]{$N_1\;\;$}  
\psfrag{n2}{$N_2$}  
\psfrag{n3}{$N_3$} 
\psfrag{n4}{$N_4$} 
\psfrag{u3}[][]{$U_3\;\;$} 
\psfrag{u4}[][]{$U_4\;\;$} 
\psfrag{t3}{$T_3$} 
\psfrag{t4}{$T_4$} 
\psfrag{w}{$W$} 
\psfrag{0}{$0$} 
\centering 
\scalebox{0.8}[0.8]{\includegraphics[scale=0.5]{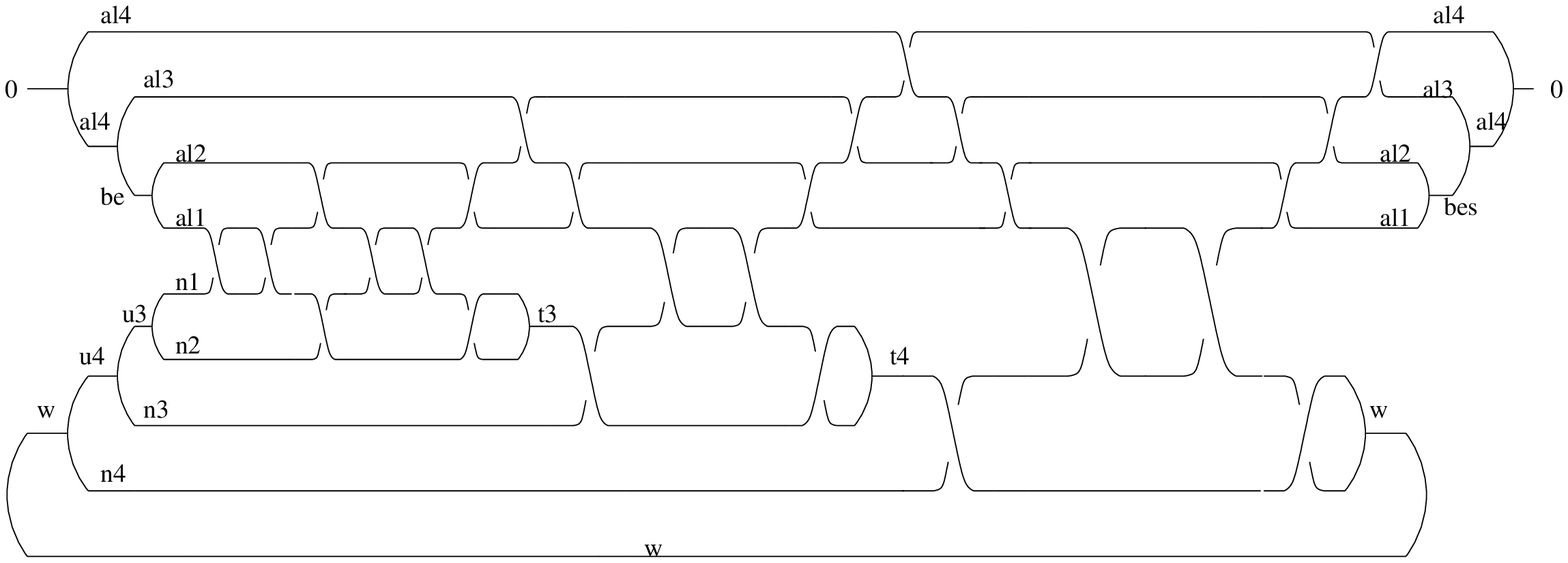}} 
\caption{Expression of  $\stackrel{P}{K}{}\!\!_{0,4}^{(+)}$.}  
\label{fig:valueofK} 
\end{figure}

$ \stackrel{P}{K}{}\!\!_{0,p}^{(+)} 
 \left( \begin{array}{c}  \!\!\alpha  \!\!\\  \!\!\beta,s \!\! \end{array} \right)$ is therefore represented by the picture of figure \ref{fig:valueofK}. The same comments for the figure representing $\stackrel{P}{\cal F}{}\!\!_{0,p}^{(\pm)} \left( \begin{array}{c}  \!\!\alpha  \!\!\\  \!\!\beta  \!\!\end{array} \right)$ apply here. 
 
Note that the selection rules in the finite dimensional case, implies that  if  
$\stackrel{\alpha}{\omega}_{0,p}[\beta]\not=0$ then the possible non zero elements $\stackrel{\alpha}{\omega}_{0,p}[\beta+s]$ are those  for $s$ satisfying $s^l, s^r \in \ZZ^{p-3}.$

Using the property   (\ref{factorizationof3j})  of factorization of finite dimensional intertwiners, 
we obtain that 
\begin{eqnarray} \label{factorisation}
 \stackrel{P}{K}{}\!\!_{0,p}^{(\pm)} 
 \left( \begin{array}{c}  \!\!\alpha  \!\!\\  \!\!\beta,s \!\! \end{array} \right)=  
\stackrel{P^l}{K}{}\!\!_{0,p}^{l(\pm)} 
 \left( \begin{array}{c} \!\! \alpha^l \!\! \\ \!\! \beta^l,s^l  \!\!\end{array} \right)  
\stackrel{P^r}{K}{}\!\!_{0,p}^{r(\pm)} 
 \left( \begin{array}{c} \!\! \alpha^r  \!\!\\  \!\!\beta^r,s^r \!\! \end{array} \right)
\end{eqnarray} 
where $\stackrel{P^l}{K}{}\!\!_{0,p}^{l(\pm)} 
\left( \begin{array}{c} \!\! \alpha^l  \!\!\\  \!\!\beta^l,s^l \!\! \end{array} \right),  
\stackrel{P^r}{K}{}\!\!_{0,p}^{r(\pm)} 
 \left( \begin{array}{c} \!\! \alpha^r \!\! \\ \!\! \beta^r,s^r \!\! \end{array} \right)$ are computed (\ref{expressionofK0pintermof6j}) in the appendix (B.1) and expressed in terms of $6j(0)$ coefficients.  
As a result it is straightforward to define a continuation of $ \stackrel{P}{K}{}\!\!_{0,p}^{(\pm)} 
 \left( \begin{array}{c}  \!\!\alpha  \!\!\\  \!\!\beta,s \!\! \end{array} \right)$  for $\alpha_1,...,\alpha_p, \beta_3,...,\beta_{p-1}\in {\mathbb S}, s\in {\cal S}^{p-3} $ maintaining (\ref{factorisation}) and  
by  replacing where needed $6j(0)$ by $6j(1)$ or  $6j(3)$ in (\ref{expressionofK0pintermof6j}). 
 
It can be checked, from this definition, that $\frac{ \Xi_{0,p}[\alpha,\beta]}{ \Xi_{0,p}[\alpha,\beta+s]}\stackrel{P}{K}{}\!\!_{0,p}^{(\pm)} \left( \begin{array}{c} \!\! \alpha \!\! \\  \!\!\beta, s \!\! \end{array} \right)$  is a rational function in $q^{i\rho_{\alpha_1}},...,q^{i\rho_{\alpha_p}}, 
q^{i\rho_{\beta_3}},...,q^{i\rho_{\beta_{p-3}}}.$

We recall that $\stackrel{\alpha}{\omega}_{0,p}[\beta]{}^{I}_{a}\Xi_{0,p}[\alpha,\beta]\nu^{I}(\alpha)$ is an element of 
$ \CC(q^{i\rho_{\alpha_1}},...,q^{i\rho_{\alpha_p}}, 
q^{i\rho_{\beta_3}},...,q^{i\rho_{\beta_{p-3}}})$. 
{}From this property we can deduce the more general result that 
$\stackrel{P}{\cal F}{}\!\!_{0,p}^{(\pm)} \left( \begin{array}{c} \!\!\alpha\!\! \\ \!\!\beta\!\! \end{array} \right)\stackrel{I}{e}_{a}\!\!(\alpha)  \Xi_{0,p}[\alpha,\beta] 
\nu^{I}(\alpha)$ is also an element of $ \CC(q^{i\rho_{\alpha_1}},...,q^{i\rho_{\alpha_p}}, 
q^{i\rho_{\beta_3}},...,q^{i\rho_{\beta_{p-3}}}).$ 
The property in $q^{i\rho_{\beta_3}},...,q^{i\rho_{\beta_{p-3}}}$ is a direct consequence of the definition of $ \Xi_{0,p}[\alpha,\beta]$, the only non trivial fact is to show that this also holds for $q^{i\rho_{\alpha_1}},...,q^{i\rho_{\alpha_p}}.$ 
This comes from the structure of the graph in the figure \ref{fig:calF}, the expression of  
 $\stackrel{I\alpha}{\RR}$ in terms of the coefficients $\Lambda^{BC}_{AD}(\alpha)$, and the fact that  $\frac{{\cal N}^{(A)}(\alpha)}{{\cal N}^{(D)}(\alpha)}\Lambda^{BC}_{AD}(\alpha)$ is a Laurent polynomial in $q^{i\rho_\alpha}$.

As a result we obtain that: 
\begin{eqnarray}  
\hskip-1cm&&\Xi_{0,p}[\alpha,\beta] \nu^{I}(\alpha)\stackrel{P}{\cal F}{}\!\!_{0,p}^{(\pm)} \left( \begin{array}{c}\!\! \alpha  \!\! \\ \!\!\beta \!\! \end{array} \right){}^{I}_{a} \; -  \sum_{ s\in {\cal S}^{p-3}}\frac{ \Xi_{0,p}[\alpha,\beta]}{ \Xi_{0,p}[\alpha,\beta+s]}\stackrel{P}{K}{}\!\!_{0,p}^{(\pm)} \left( \begin{array}{c} \!\! \alpha \!\! \\  \!\!\beta, s \!\! \end{array} \right) \big(\stackrel{\alpha}{\omega}_{0,p}\!\! \lbrack \beta+s \rbrack \big){}^{I}_{a}\Xi[\alpha,\beta+s] \nu^{I}(\alpha) \;, \nonumber 
\end{eqnarray}  
 is an element of  $ \CC(q^{i\rho_{\alpha_1}},...,q^{i\rho_{\alpha_p}}, 
q^{i\rho_{\beta_3}},...,q^{i\rho_{\beta_{p-1}}})$ which vanishes for an infinite number of sufficiently large $i\rho_{\alpha_1}, ..., i\rho_{\alpha_p},i\rho_{\beta_3},...,i\rho_{\beta_{p-1}}\in \onehalf \ZZ^+.$ As a result this rational function is nul and  we obtain relation (\ref{actionofOon omega}).$\Box$ 
 
Remark.  If $a\in {\cal L }_{0,p}^{inv},$  $a=\sum_{P} \lambda_{P} \stackrel{P}{\cal O}{}\!\!^{(\pm)}_{0,p}$ we define \\
$\stackrel{ a}{K}{}\!\!_{0,p}^{(\pm)} \left( \begin{array}{c} \alpha \\\beta,s \end{array} \right)= \sum_{P}\lambda_P\stackrel{ P}{K}{}\!\!_{0,p}^{(\pm)} \left( \begin{array}{c} \alpha \\\beta,s \end{array} \right).$

We will now endow $H_{0,p}(\alpha)$ with a structure of right ${\cal L}_{0,p}^{inv}$ module in the following sense: 
 
\begin{proposition} 
 $ {\cal L}_{0,p}^{inv}$  acts on $\VV(\alpha)^{*}$ with  
$\rho_{0,p}[\alpha]$ and  leaves  the space 
 $(\VV(\alpha)^{*})^{U_q(sl(2,\CC)_{\RR})}$ invariant. However the  subspace  $H_{0,p}(\alpha)$ is in general not invariant.  
We define the domain of $a\in {\cal L}_{0,p}^{inv} $ associated to $ \rho_{0,p}[\alpha]$ to be the subspace $D(a)\subset \AA^{(p-3)}$ defined as: 

f belongs to $D(a)$ if and only if for all $s\in {\cal S}^{p-3}$ the functions
  
$\beta\mapsto  \stackrel{ a}{K}{}\!\!_{0,p}^{(\pm)} \left( \begin{array}{c} \alpha \\\beta,s \end{array} \right)\frac{\Xi_{0,p}[\alpha,\beta]{\cal P}[\beta]}{\Xi_{0,p}[\alpha,\beta+s]{\cal P}[\beta+s]}f(\beta)$ are elements of  $\AA^{(p-3)}.$ 
The action of $a$ on an element $\stackrel{\alpha}{\omega}_{0,p}\!\!(f)\in D(a)\subset H_{0,p}(\alpha)$ belongs to $H_{0,p}(\alpha)$, and we have: 
\begin{equation}  
 \stackrel{\alpha}{\omega}_{0,p}\!\!(f) \lhd a = 
 \stackrel{\alpha}{\omega}_{0,p}\!\!(f\lhd a)  \;\;,\nonumber   
\end{equation}  
with  
\begin{equation}  
(f \lhd a)(\beta) = 
\sum_{s\in {\cal S}^{p-3}} f(\beta-s) \; \stackrel{a}{K}{}\!\!^{(\pm)}_{0,p} \left( \begin{array}{c} \alpha \\ \beta-s ,s \end{array} \right) 
\frac{\Xi_{0,p}[\alpha,\beta-s]{\cal P}[\beta-s]}{\Xi_{0,p}[\alpha,\beta]{\cal P}[\beta]}.   
\end{equation}  
 
In general the  domain $D(a)$ is not equal to $\AA^{(p-3)},$ but it always contains $Q_a[\beta]\AA^{(p-3)}$ where 
 $Q_a[\beta]=\prod_{j=3}^{p-3}((2\beta_j^l+n_j)_{k_j})^{p_j} 
((2\beta_j^r+n_j')_{k_j'})^{p_j'}$ for some   $n_j, n_j'$  integers and  $k_j,k_j',p_j,p_j'$  non negative   
integers depending on $a.$ 
\end{proposition} 
\Proof 
This property follows from  the invariance of the sum of integrals under the change  
$\beta\mapsto \beta-s$, which amounts to replace $m_\beta$ by  $m_\beta-m_s$ and $\rho_\beta$ by  $\rho_\beta-\rho_s$ with $\rho_s\in i\ZZ$  given by $i\rho_s=s^l+s^r+1.$ The former is a change of index in a sum whereas the latter is implied by Cauchy Theorem if the function considered is analytic, which follows from the  definition of $D(a)$. 
$\Box$

It is clear from the properties (\ref{omegabetaareinvariants}), that the anti-representation $\rho_{0,p}[\alpha]$ on $H_{0,p}[\alpha]$ decends to the quotient, and defines an anti-representation $\tilde{\rho}_{0,p}[\alpha]$ of 
 $M_q(S^2,SL(2,\CC)_{\RR};\alpha).$  
 
Our main result is that ${\tilde\rho}_{0,p}[\alpha]$ is unitary. 
 
\begin{theorem} \label{theorem1} 
The anti-representation ${\tilde\rho}_{0,p}[\alpha]$ of  $M_q(S^2,SL(2,\CC)_{\RR};\alpha)$ is unitary: 
\begin{equation} 
\forall a\in M_q(S^2,SL(2,\CC)_{\RR},[\alpha]), 
 \forall v\in D(a^{\star}), \forall w\in D(a), 
 <v\lhd a^{\star}\vert w>=<v\vert w\lhd a>.   
\end{equation} 
where $<.\vert .>$ is the positive sesquilinear form defined by (\ref{scalarproduct0P}). 
\end{theorem} 
 \Proof 
The first step amounts to extend  to ${\mathbb S}^{p-3}$ the function  
$\beta\mapsto \Upsilon_{0,p}[\alpha,\beta]$ entering in the definition of  $<.\vert.>$. 
For this task we use lemma \ref{analyticityofupsilon0p}. As a result, the extension of  
$\Upsilon_{0,p}[\alpha,\beta]$, denoted by the same notation, is an entire  function in the variables  
$\rho_{\beta_3},...,\rho_{\beta_{p-1}}.$ 
We then compute: 
\begin{eqnarray} 
&&<\stackrel{\alpha}{\omega}_{0,p}(f)\vert \stackrel{\alpha}{\omega}_{0,p}(g)\lhd \stackrel{P}{\cal O}{}\!\!^{(+)}_{0,p} >= \nonumber\\ 
&&\sum_{s}\int_{\SSP^{p-3}} 
d\beta {\cal P}(\beta+s)\overline{f(\beta)} 
\Upsilon_{0,p}[\alpha,\beta]g(\beta+s) 
 \stackrel{P}{K}{}\!\!^{(+)}_{0,p} \left( \begin{array}{c} \alpha \\ \beta+s ,-s \end{array} \right) 
\frac{\Xi_{0,p}[\alpha,\beta+s]}{\Xi_{0,p}[\alpha,\beta]}. \nonumber 
\end{eqnarray} 
On the other hand we have: 
\begin{eqnarray*} 
&&<\stackrel{\alpha}{\omega}_{0,p}(f)\lhd (\stackrel{P}{\cal O}{}\!\!^{(+)}_{0,p})^{\star} \vert \stackrel{\alpha}{\omega}_{0,p}(g) >=<\stackrel{\alpha}{\omega}_{0,p}(f)\lhd \stackrel{\tilde P}{\cal O}{}\!\!^{(-)}_{0,p} \vert \stackrel{\alpha}{\omega}_{0,p}(g) >=\\ 
&&=\sum_{s}\int_{\SSP^{p-3}} 
d\beta {\cal P}(\beta+s)\overline{f(\beta+s)} 
\Upsilon_{0,p}[\alpha,\beta]g(\beta) 
\overline{ \stackrel{\tilde P}{K}{}\!\!^{(-)}_{0,p} \left( \begin{array}{c} \alpha \\ \beta+s ,-s \end{array} \right)} 
\frac{\overline{\Xi_{0,p}[\alpha,\beta+s]}}{\overline{\Xi_{0,p}[\alpha,\beta]}}\\ 
&&=\sum_{s}\int_{\SSP^{p-3}} 
d\beta {\cal P}(\beta+s)\overline{f(\beta+s)} 
\Upsilon_{0,p}[\alpha,\beta]g(\beta) 
 \stackrel{\tilde P}{K}{}\!\!^{(-)}_{0,p} \left( \begin{array}{c} \overline{\alpha} \\ \overline{\beta}+s ,-s \end{array} \right) 
\frac{\overline{\Xi_{0,p}[\alpha,\beta+s]}}{\overline{\Xi_{0,p}[\alpha,\beta]}}. 
\end{eqnarray*} 
Using Cauchy theorem to shift $\rho_{\beta}$ in $\rho_{\beta}+\rho_s$ and simultaneously changing the index of the sum over the dumb variable $m_{\beta}$ and $s$, the last expression is also equal to: 
\begin{equation} 
\sum_{s}\int_{\SSP^{p-3}}d\beta {\cal P}[\beta]\overline{f(\beta)}\Upsilon_{0,p}[\alpha,\beta+s]g(\beta+s) 
 \stackrel{\tilde P}{K}{}\!\!^{(-)}_{0,p} \left( \begin{array}{c} \overline{\alpha} \\ \overline{\beta} ,-\tilde{s} \end{array} \right) 
\frac{\overline{\Xi_{0,p}[\alpha,\beta]}}{\overline{\Xi_{0,p}[\alpha,\beta+s]}}. \nonumber
\end{equation} 
As a result, proving unitarity is equivalent to showing the relation: 
\begin{eqnarray} 
\Upsilon_{0,p}[\alpha,\beta] 
 \stackrel{P}{K}{}\!\!^{(+)}_{0,p} \left( \begin{array}{c} \alpha \\ \beta+s ,-s \end{array} \right) 
\frac{\Xi_{0,p}[\alpha,\beta+s]}{\Xi[\alpha,\beta]}{\cal P}[\beta+s] =\nonumber \\
\Upsilon_{0,p}[\alpha,\beta+s] \stackrel{\tilde P}{K}{}\!\!^{(-)}_{0,p} \left( \begin{array}{c} \overline{\alpha} \\ \overline{\beta} ,-\tilde{s} \end{array} \right) 
\frac{\overline{\Xi_{0,p}[\alpha,\beta]}}{\overline{\Xi_{0,p}[\alpha,\beta+s]}}{\cal P}[\beta], 
\end{eqnarray} 
when $\beta\in \SSP^{p-3}.$ 
We now make use of the symmetries of the relations satisfied by $\stackrel{P}{K}{}\!\!^{(+)}_{0,p} \left( \begin{array}{c} \alpha \\ \beta ,s \end{array} \right)$ proved in the proposition (\ref{prop:symmetriesofK0p}) of the appendix A.3. 
\begin{equation} 
\hskip-0.5cm \stackrel{\tilde P}{K}{}\!\!^{(-)}_{0,p} \left( \begin{array}{c} \overline{\alpha} \\ \overline{\beta} ,-\tilde{s} \end{array} \right)= 
 \stackrel{\tilde P}{K}{}\!\!^{(-)}_{0,p} \left( \begin{array}{c} \tilde{\underline{\alpha}} \\ \tilde{\underline{\beta}} ,-\tilde{s} \end{array} \right)=\stackrel{ P}{K}{}\!\!^{(+)}_{0,p} \left( \begin{array}{c} {\underline{\alpha}} \\ \underline{\beta}-s ,{s} \end{array} \right)=\psi_{0,p}[\alpha,\beta,s]\stackrel{ P}{K}{}\!\!^{(+)}_{0,p} \left( \begin{array}{c} \alpha \\ \beta+s ,-s \end{array} \right)\nonumber 
\end{equation} 
As a result the unitarity condition is equivalent to prove the following quasi-invariance under imaginary integer shifts: 
\begin{equation} 
\frac{\Upsilon_{0,p}[\alpha,\beta]}{\vert\Xi_{0,p}[\alpha,\beta]\vert^2{\cal P}(\beta)}= 
\frac{\Upsilon_{0,p}[\alpha,\beta+s]}{\vert\Xi_{0,p}[\alpha,\beta+s]\vert^2{\cal P}(\beta+s)}\psi_{0,p}[\alpha,\beta,s] 
\end{equation} 
which, due to the explicit expression of $\Upsilon_{0,p}[\alpha,\beta],$ reduces to  
\begin{equation} 
\frac{\psi_1{}_{(\alpha_1,\alpha_2,\beta_3+s_3)}\prod_{j=3}^{p-2}\psi_1{}_{(\beta_j+s_{j},\alpha_j,\beta_{j+1}+s_{j+1})}\psi_1{}_{(\alpha_{p-1},\beta_{p-1}+s_{p-1},\alpha_p)}}{\psi_1{}_{(\alpha_1,\alpha_2,\beta_3)}\prod_{j=3}^{p-2}\psi_1{}_{(\beta_j,\alpha_j,\beta_{j+1})}\psi_1{}_{(\alpha_{p-1},\beta_{p-1},\alpha_p)}}= \psi_{0,p}[\alpha,\beta,s] 
\end{equation} 
which is a trivial fact. 
$\Box$

	\subsubsection*{III.2.2 The moduli algebra of a genus-n surface}  
In this subsection, we will construct a  unitary representation of   the moduli algebra on a  surface of arbitrary genus n. The graph algebra  
${\cal L}_{n,0}$ is isomorphic to  
$H(U_q(sl(2,\CC)_{\RR}))^{\otimes n}$  and acts  on  
$Fun_{cc}(SL(2,\CC)_{\RR})^{\otimes n}$. In order to find invariant vectors, we will apply the  technique  developped in  the p-punctures case: we will transfer this action on the dual space  
$(Fun_{cc}(SL(2,\CC)_{\RR})^{\otimes n})^{*}$ and extract a subspace of invariant elements. 
We use the notations of the end of appendix A.1. 
Let $\alpha\in {\mathbb S}^n,$ we define an element $\prod_{k=n}^1  
\stackrel{\alpha_k}{\GG}(k)\in (U_q(sl(2,\CC)_{\RR})^*)^{\otimes n}\otimes \otimes_{k=n}^1\End(\stackrel{\alpha_k}{\VV}),$ where  
 $\stackrel{\alpha_k}{\GG}(k)=p_k(\stackrel{\alpha_k}{\GG})$ where $p_k$ is the inclusion of $U_q(sl(2,\CC)_{\RR})^*$ in the $k-th$ copy of  
$(U_q(sl(2,\CC)_{\RR})^*)^{n}.$

{}From this element  we construct an  element of  $(Fun_{cc}(SL(2,\CC)_{\RR})^{\otimes n})^{*}\otimes \otimes_{k=n}^1\End(\stackrel{\alpha_k}{\VV})$ defined as  
$\iota^{\otimes n}(\otimes_{k=n}^1 \stackrel{\alpha_k }{\GG})=\prod_{k=n}^1\iota(\stackrel{\alpha_k}{\GG}(k))$ where $\iota$ is recalled in the appendix A.1 
The action of the lattice algebra on $\iota(\stackrel{\alpha_k}{\GG}(k))$ is easily computed by dualization: 
\begin{eqnarray} 
\hskip-1cm\iota(\stackrel{\alpha_k }{\GG}(k)) \lhd \stackrel{I}{\LL}{}\!\!^{(\pm)}(k)  =   \iota (\stackrel{\alpha_k }{\GG}(k)) \stackrel{I \alpha_k}{\RR}{}\!\!^{(\pm)-1} &,&  \iota(\stackrel{\alpha_k }{\GG}(k)) \lhd \stackrel{I}{\LL}{}\!\!^{(\pm)}(j)  =  \iota(\stackrel{\alpha_k }{\GG}(k)) \; \forall j\not= k,\nonumber\\ 
\iota(\stackrel{\alpha_k}{\GG}(k)) \lhd \stackrel{I}{\tilde{\LL}}{}\!\!^{(\pm)}(k)  =    \stackrel{I\alpha_k}{\RR}{}\!\!^{(\pm)} \iota(\stackrel{\alpha_k }{\GG}( k)) &,& 
\iota(\stackrel{\alpha_k }{\GG}(k)) \lhd \stackrel{I}{\tilde{\LL}}{}\!\!^{(\pm)}(j)  =  \iota(\stackrel{\alpha_k }{\GG}(k)) \; \forall j\not=k,\nonumber\\  
\iota(\stackrel{\alpha_k}{\GG}(k)) \lhd \stackrel{I}{\GG}(k)   =  \sum_{\alpha_k'\in{\mathbb S}(\alpha_k,I)} \Phi^{\alpha_k I}_{\alpha_k'}  \iota(\stackrel{\alpha_k' }{\GG}(k)) \Psi_{\alpha_k I}^{\alpha_k'}&,& 
\iota(\stackrel{\alpha_k }{\GG}(k)) \lhd \stackrel{I}{\GG}(j)   = \iota(\stackrel{\alpha_k }{\GG}(k)) \iota(\stackrel{I}{\GG}(j)) \;\forall j\not=k.\nonumber\\ 
\end{eqnarray}

 Before studying the general case, it is interesting to  give a detailed exposition of the genus one case which contains all the ideas without being too cumbersome.  
 
Given $\alpha \in {\mathbb S}$, we can define the character  
${\omega}_{1,0}(\alpha)\in (Fun_{cc}(SL_q(2,\CC)))^{*}$ of this representation by (\cite{BR3}):  
\begin{eqnarray}  
{\omega}_{1,0}(\alpha) = \sum_{ABC} \Clebphi{A}{B}{C}{j}{l}{n} \Clebpsi{B}{A}{C}{m}{i}{n} \Lambda^{BC}_A (\alpha) \iota(\stackrel{A}{k}{}\!\!^i_j \otimes \stackrel{B}{E}{}\!\!^m_l )\;\;.  
\end{eqnarray}  
 
The algebra ${\cal L}_{1,0}^{inv}$ acts on the right of $(Fun_{cc}(SL_q(2,\CC)))^{*}$  
with $\rho_{1,0}$ and we obviously have: 
\begin{eqnarray} 
\omega_{1,0}(\alpha) 
\lhd \stackrel{I}{C}_{1,0}={\omega}_{1,0}(\alpha).
\end{eqnarray}

\begin{figure} 
\psfrag{al}[][]{$\alpha+s$}  
\psfrag{alp}[][]{$\alpha$}  
\psfrag{i}{$I$}  
\psfrag{j}{$J$}  
\psfrag{w}{$W$}  
\centering 
\includegraphics[scale=0.5]{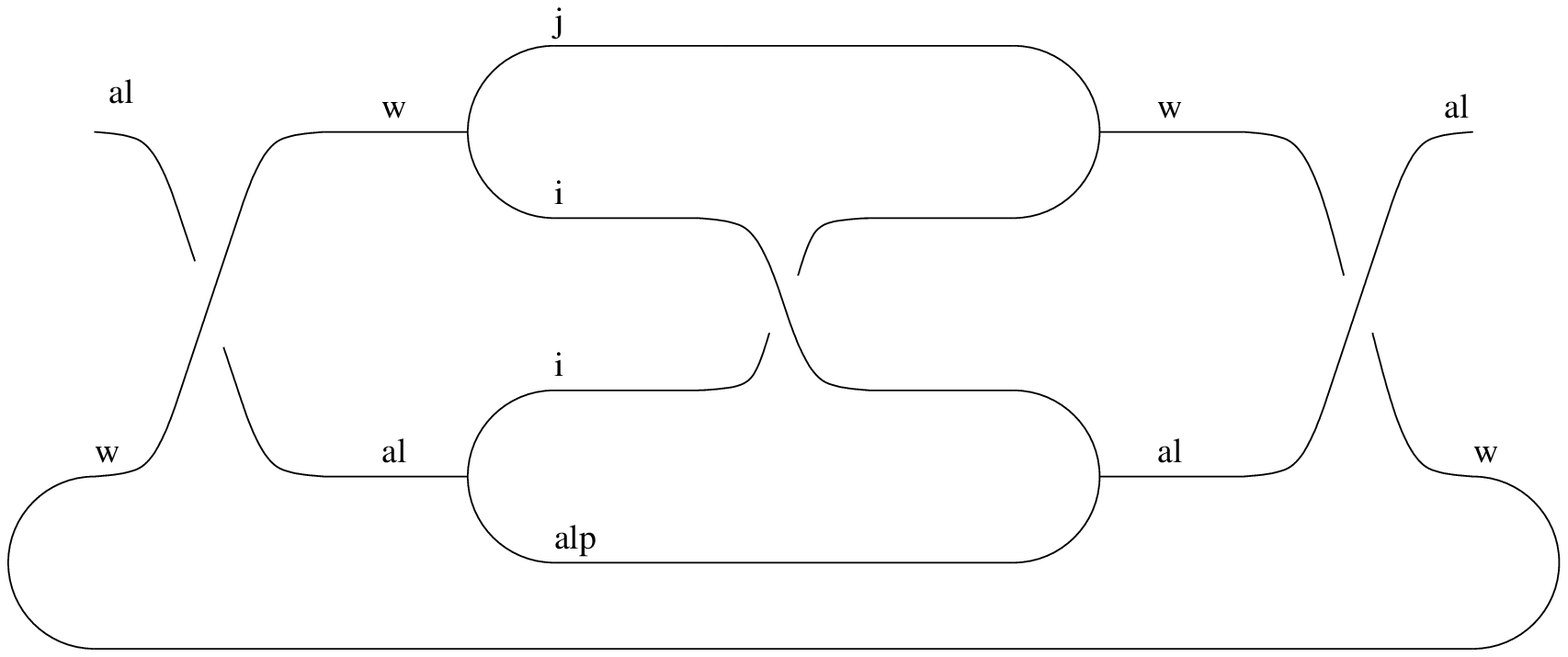} 
\caption{Expression of $\stackrel{P}{K}{}\!\!_{1,0}$.} 
\label{fig:observableoftorus} 
\end{figure}

\begin{lemma} 
The  element $ \stackrel{IJW}{\cal O}{}\!\!_{1,0}$ acts on the invariant  
element  ${\omega}_{1,0}(\alpha)$ as follows: 
\begin{equation} 
{\omega}_{1,0}(\alpha)\lhd\stackrel{IJW}{\cal O}{}\!\!_{1,0}  =  
\; \sum_{s\in{\cal S}} \; \Lambda^{WJ}_{II}[\alpha,s] 
{\omega}_{1,0}(\alpha+s) 
\end{equation} 
where $ \Lambda^{WJ}_{II}[\alpha,s]$ is given in the definition  
(\ref{factorizationlambda}) . 
\end{lemma} 
\Proof 
Computing the action of $\stackrel{IJW}{\cal O}{}\!\!_{1,0}$ on ${\omega}_{1,0}(\alpha)$ amounts to evaluate the intertwiner given by the graph (figure \ref{fig:observableoftorus}) 
which is equal  to  $ \Lambda^{WJ}_{II}[\alpha,s].$ 
$\Box$

\begin{proposition} 
We  define a subset of  $(Fun_{cc}(SL_q(2,\CC)))^{*}$  by: 
 
\begin{equation}  
H_{1,0}=\{ {\omega}_{1,0}(f)=\int_{\SSP} d\alpha {\cal P}(\alpha) f(\alpha)  {\omega}_{1,0}(\alpha), \;\; f \in \AA^{}\}. 
\end{equation}

 $\rho_{1,0}$ defines a right action of ${\cal L}_{1,0}^{inv}$ on $(Fun_{cc}(SL_q(2,\CC)))^{*}$ which in general does not leave $H_{1,0}$ invariant.
To the  element $\stackrel{ IJW}{\cal O}{}\!\!_{1,0}$ in  ${\cal L}_{1,0}^{inv}$ we define its domain $D(\stackrel{ IJW}{\cal O}{}\!\!_{1,0})\subset \AA$ as:
$f$ belongs to $D(\stackrel{ IJW}{\cal O}{}\!\!_{1,0})$ if and only if $\alpha\mapsto  
\Lambda^{WJ}_{II}[\alpha,s]
\frac{{\cal P}(\alpha)}{{\cal P}(\alpha+s)}f(\alpha)$ is an element of $\AA$ for all $s\in {\cal S}.$
We therefore have:
\begin{equation} 
{\omega}_{1,0}(f)\lhd\stackrel{ IJW}{\cal O}{}\!\!_{1,0}= 
{\omega}_{1,0}(f\lhd \stackrel{IJW}{\cal O}{}\!\!_{1,0}) 
\end{equation} 
with 
 
\begin{equation} 
(f\lhd \stackrel{ IJW}{\cal O}{}\!\!_{1,0})(\alpha)= 
 \sum_{s\in {\cal S}} f(\alpha+s) 
 \Lambda^{WJ}_{II}[\alpha+s, -s]\frac{{\cal P}(\alpha+s)}{{\cal P}(\alpha)}\;\;.\label{actiononchif} 
\end{equation}  
The map $\AA \rightarrow H_{1,0}$ which sends $f$ to ${\omega}_{1,0}(f)$ is an injection. 
We can therefore endow $H_{1,0}$ with the following pre-Hilbert structure 
\begin{equation} 
<{\omega}_{1,0}(f)\vert{\omega}_{1,0}(g)>=\int_{\SSP}d\alpha {\cal P}(\alpha)^2\overline{f(\alpha)}g(\alpha). 
\end{equation} 
The righ action ${\rho}_{1,0}$ acting on  $H_{1,0}$ descends to the quotient 
$M_q(\Sigma_{1,0},SL(2,\CC)_{\RR}))$ defines a right action 
 $\tilde{\rho}_{1,0}$, which  is unitary. 
\end{proposition} 
\Proof 
It is important to note that $<{\omega}_{1,0}(\alpha),\phi>$ where $\phi\in Fun_{cc}(SL_q(2,\CC)_{\RR})$ is a Laurent polynomial in $q^{i\rho_{\alpha}}$ because $\Lambda^{BC}_{AA}(\alpha)$ is itself a Laurent polynomial in   $q^{i\rho_{\alpha}}.$ As a result ${\omega}_{1,0} (f)$ is well defined for $f\in \AA$ and  
the proof of  the relation (\ref{actiononchif}) follows from Cauchy theorem. 

The proof of the injection is a straighforward consequence of the proof of Theorem 12 of \cite{BR3}.
Indeed if we have ${\omega}_{1,0}(f)=0$, it implies that 
$\int_{\SSP}d\alpha {\cal P}(\alpha)  \Lambda_A^{BC}(\alpha)f(\alpha)=0,$ for every $A,B,C.$
The proof of Plancherel theorem 12 of \cite{BR3} precisely shows that under weaker condition on $f$ than $f\in\AA$ the last relation implies $f=0.$

To prove the unitarity, we compute  
\begin{eqnarray*} 
<{\omega}_{1,0}(f) \vert {\omega}_{1,0}(g) \lhd \stackrel{IJW}{\cal O}{}\!\!_{1,0} > =  \sum_{s\in {\cal S}} \int_{\SSP} d\alpha {\cal P}(\alpha)
{\cal P}(\alpha+s) \overline{f(\alpha)}  \Lambda^{WJ}_{II}[\alpha+s, -s] g(\alpha  +s).
\end{eqnarray*} 
On the other hand we have 
\begin{eqnarray} 
<{\omega}_{1,0}(f) \lhd \stackrel{IJW}{\cal O}{}\!\!_{1,0}^{\star} \vert {\omega}_{1,0}(g) > &=&  \sum_{s\in {\cal S}} \int_{\SSP}d\alpha {\cal P}(\alpha)
{\cal P}(\alpha+s)\overline{f(\alpha+s)}  \overline{\Lambda^{\tilde{W}\tilde{J}}_{\tilde{I}\tilde{I}}[\alpha+s, -s]} g(\alpha) \nonumber \\ 
&=&  \sum_{s\in {\cal S}} \int_{\SSP} d\alpha {\cal P}(\alpha)
{\cal P}(\alpha+s)\overline{f(\alpha+s)}  \Lambda^{\tilde{W}\tilde{J}}_{\tilde{I}\tilde{I}}[\overline{\alpha}+s, -s] g(\alpha). \nonumber 
\end{eqnarray} 
Using Cauchy theorem to shift $\rho_{\beta}$ and changing the index of the sum, the last expression is equal to 
\begin{eqnarray*} 
 \sum_{s\in {\cal S}} \int_{\SSP} d\alpha{\cal P}(\alpha)
{\cal P}(\alpha+s) \overline{f(\alpha)}  \Lambda^{\tilde{W}\tilde{J}}_{\tilde{I}\tilde{I}}[\overline{\alpha}, -\tilde{s}] g(\alpha+s). 
\end{eqnarray*} 
As a result, proving the unitarity is equivalent to showing the relation: 
\begin{eqnarray*} 
\Lambda^{\tilde{W}\tilde{J}}_{\tilde{I}\tilde{I}}[\overline{\alpha}, -\tilde{s}] = \Lambda^{{W}{J}}_{{I}{I}}[\alpha+s, -s] , 
\end{eqnarray*} 
which is consequence of the equality:  
$\Lambda^{BC}_{AD}(\alpha)=\Lambda^{BC}_{AD}(\underline{\alpha})$. 
$\Box$ 
 
The simplest non trivial observables are the Wilson loops $\stackrel{I}{W}_A$ and $\stackrel{I}{W}_B$ around the handles taken in the representation  
$I\in \SSF$.  
It is  straightforward, from the previous proposition, to compute the action of these observables:  
\begin{eqnarray}  
(f \lhd \stackrel{I}{W}_A)(\alpha)  & = &  
\sum_{s^l=-I^l,s^r=-I^r}^{s^l=I^l,s^r=I^r} 
  \frac{v_{\alpha+s}}{v_{\alpha}} f(\alpha+s)\\  
(f  \lhd  \stackrel{I}{W}_B)(\alpha)  & = & \vartheta_{I\alpha}  f(\alpha).  
\end{eqnarray}  
{}From the representation of the Heisenberg double, $W_A$ acts by multiplication in the real space and by  finite difference in the Fourier space  whereas $W_B$ acts by left and right derivations  in the real space  and by multiplication in  
the Fourier space.  
 
This closes the construction for the genus one. The generalization to the genus n is similar in spirit  
although the technical details are much more involved. This is what we will now develop. We will assume that $n\geq 2.$ 
 
Let $\alpha,\beta,\gamma \in {\mathbb S}$ and let $A \in \End(\stackrel{\alpha}{\VV})$ of finite dimensional corank, we can define $(A \otimes 1)\Phi^{\alpha\beta}_{\gamma} \in Hom(\stackrel{\gamma}{\VV},\stackrel{\alpha}{\VV}\otimes\stackrel{\beta}{\VV})$ as follows  
\begin{equation} 
(A \otimes 1)\Phi^{\alpha\beta}_{\gamma}(v)=\sum_{IJij}<\stackrel{I}{e}_i{}^{*}\otimes\stackrel{J}{e}_j{}^{*},\Phi^{\alpha\beta}_{\gamma}(v)>(A\stackrel{I}{e}_{i})\otimes\stackrel{J}{e}_{j} 
\end{equation} 
where the sums are finite. A similar conclusion holds true for $(1 \otimes B)\Phi^{\alpha\beta}_{\gamma}$ with $corank (B) <+\infty.$ 
 
\begin{definition} \label{defofP}
Let $\lambda=(\lambda_1,...,\lambda_n)$ element of  ${\mathbb S}^n$, and  
$\sigma=(\sigma_3,...,\sigma_{n}), \tau=(\tau_3,...,\tau_{n})$ elements of  
${\mathbb S}^{n-2}$ and $\kappa \in {\mathbb S}.$ We define for $2 \leq n$ 
\begin{eqnarray}  
{\wp}_2 & = & \Psi_{\lambda_2\lambda_1}^{\sigma_{3}} \; \iota(\stackrel{\lambda_2}{\GG}{}\!\!(2))  
 \; \stackrel{\lambda_2}{\mu}{}\!\!^{-1} \; \stackrel{\lambda_2\lambda_1}{\RR'} \; \stackrel{\lambda_2}{\mu} \; \iota(\stackrel{\lambda_1}{\GG}{}\!\!(1))   \stackrel{\lambda_2 \lambda_1}{\RR}{}^{\!\!(-)} \; 
 \; \Phi^{\lambda_2\lambda_1}_{\tau_{3}}, \nonumber\\ 
&&\in (Fun_{cc}(SL_q(2,\CC)_{\RR})^{\otimes n})^{*} \otimes \Hom(\stackrel{\tau_{3}}{\VV},\stackrel{\sigma_{3}}{\VV}), 
  \nonumber \\  
{\wp}_{i} & = & \Psi_{\lambda_i\sigma_i}^{\sigma_{i+1}} \; \iota(\stackrel{\lambda_i}{\GG}{}\!\!(i))  
 \; \stackrel{\lambda_i}{\mu}{}\!\!^{-1} \; \stackrel{\lambda_i\sigma_i}{\RR'} \; \stackrel{\lambda_i}{\mu} \; {\wp}_{i-1}  \stackrel{\lambda_i\sigma_i }{\RR}{}^{\!\!(-)} \; 
 \; \Phi^{\lambda_i\tau_i}_{\tau_{i+1}}, \nonumber\\ 
&&\in (Fun_{cc}(SL_q(2,\CC)_{\RR})^{\otimes n})^{*} \otimes \Hom(\stackrel{\tau_{i+1}}{\VV},\stackrel{\sigma_{i+1}}{\VV}),\;\; 
\forall i=3,..., n\;\;,
\end{eqnarray}  
with the convention  $\tau_{n+1}=\sigma_{n+1}=\kappa$.

We can introduce  ${\omega}_{n,0}[\kappa,\lambda,\sigma,\tau]\in (Fun_{cc}(SL_q(2,\CC)_{\RR})^{\otimes n})^*$  as being: 
\begin{eqnarray}  
{\omega}_{n,0}(\kappa,\lambda,\sigma,\tau) = tr_{\kappa} \Big({\wp}_n\Big) \;\;. 
\end{eqnarray}  
${\omega}_{n,0}[\kappa,\lambda,\sigma,\tau]$ is well defined because $<{\omega}_{n,0}[\kappa,\lambda,\sigma,\tau] , f_n \otimes...\otimes f_1 >=tr_{\kappa}(<\wp_n,f_n \otimes...\otimes f_1 >)$ and $<\wp_n,f_n \otimes...\otimes f_1 >$ is, by construction, of finite rank and finite corank. 
\end{definition}

\begin{proposition} 
${\omega}_{n,0}[\kappa,\lambda,\sigma,\tau]$ are invariant vectors: 
${\omega}_{n,0}[\kappa,\lambda,\sigma,\tau]\lhd \stackrel{I}{C}_{n,0}= 
{\omega}_{n,0}[\kappa,\lambda,\sigma,\tau].$ 
\end{proposition} 
\Proof  
Using  
\begin{eqnarray}  
  \iota(\stackrel{\lambda_i}{\GG}(i)) \lhd  \stackrel{ I}{C}{}\!\!_{n,0}^{(\pm)}\!^{-1} \; = 
 \stackrel{{I} \lambda_i}{\RR}{}\!\!^{(\pm)-1} 
\;\iota( \stackrel{\lambda_i}{\GG}(i)) \; \; 
 \; \stackrel{ I}{\mu} \; \stackrel{{I} \lambda_i}{\RR}{}\!\!^{(\pm)} \; \stackrel{ I}{\mu}{}\!\!^{-1} \;, \nonumber 
\end{eqnarray}  
it is  immediate to show, by recursion, that  
\begin{eqnarray}  
\wp_i \lhd \stackrel{ I}{C}{}\!\!_{n,0}^{(\pm)}\!^{-1}  \; = \;  \stackrel{{ I} \sigma_{i+1}}{\RR}{}\!\!^{(\pm)-1} \;\; 
\wp_i \; 
 \stackrel{ I}{\mu} \; \stackrel{{ I} \tau_{i+1}}{\RR}{}\!\!^{(\pm)} \; \stackrel{ I}{\mu}{}\!\!^{-1} \;\;\;\;\;\;\forall i=2,..., n-1.  \nonumber 
\end{eqnarray}  
and then  
\begin{eqnarray}  
\wp_n \lhd \stackrel{ I}{C}{}\!\!_{n,0}^{(\pm)}\!^{-1}  \; = \;  \stackrel{{ I} \kappa}{\RR}{}\!\!^{(\pm)-1} \;\; 
\wp_n \; 
 \stackrel{ I}{\mu} \; \stackrel{{ I} \kappa}{\RR}{}\!\!^{(\pm)} \; \stackrel{ I}{\mu}{}\!\!^{-1} \;  \nonumber 
\end{eqnarray}  
which allows us to conclude. $\Box$  
 
We will denote for $n\geq 2,$ 
 
\begin{eqnarray} 
\hat{\Xi}_{n,0}[\kappa,\lambda,\sigma,\tau]^{-1} 
&=&e^{2i\pi(\lambda_1+...+\lambda_n-\kappa)}\nu_1(d_{\kappa})^2 
\zeta(\lambda_1,\lambda_2,\tau_3)\prod_{j=3}^{n} 
\zeta(\lambda_j,\tau_j,\tau_{j+1})\times\nonumber\\ 
&\times&\zeta(\lambda_1,\lambda_2,\sigma_3)\prod_{j=3}^{n} 
\zeta(\lambda_j,\sigma_j,\sigma_{j+1}) 
\prod_{j=3}^{n}\nu_1(d_{\sigma_j})\nu_1(d_{\tau_j}). 
\end{eqnarray}

We have the analogue of the lemma (\ref{omegaxinu}) : 
\begin{lemma} 
Let $f\in Fun_{cc}(SL_q(2,\CC))^{\otimes n}$,  for every  $x\in \kappa\cup\lambda\cup\sigma\cup\tau$ there exists $I_x\in\onehalf \NN$, such that  
$$\hat{\Xi}_{n,0}[\kappa,\lambda,\sigma,\tau] 
 \prod_{x\in \kappa\cup\lambda\cup\sigma\cup\tau} (\nu^{(I_x)}(x))^2 
<{\omega}_{n,0}[\kappa,\lambda,\tau,\sigma],f>$$ is a Laurent polynomial in the variables 
 $(q^{i\rho_x}), x\in\kappa\cup\lambda\cup\sigma\cup\tau$. 
\end{lemma} 
\Proof 
The proof is analogous to the related lemma of the p-punctures case. $\Box$ 
 
We will denote  
$\Xi_{n,0}[\kappa,\lambda,\sigma,\tau]= 
\hat{\Xi}_{n,0}[\kappa,\lambda,\sigma,\tau] \xi[\kappa]\xi[\lambda]\xi[\tau]\xi[\sigma].$

\begin{proposition} 
We can define a subset of   $(Fun_{cc}(SL_q(2,\CC))^{\otimes n})^{\star}$ by: 
\begin{equation} 
H_{n,0}=\{{\omega}_{n,0}(f), f\in{\AA^{(3n-3)}}\} 
 \end{equation}                                                                   
where  
\begin{equation} 
{\omega}_{n,0}(f)=\int_{\SSP^{3n-3}} 
d\kappa d\lambda d\sigma d\tau\;\;\; 
{\omega}_{n,0}[\kappa,\lambda,\sigma,\tau]\Xi_{n,0}[\kappa,\lambda,\sigma,\tau]  
 {\cal P}[\kappa,\lambda,\sigma,\tau]f(\kappa,\lambda,\sigma,\tau). 
\end{equation} 
The map $\AA^{(3n-3)}\rightarrow H_{n,0}$ which sends $f$ to  
${\omega}_{n,0}(f)$ is an injection. 
\end{proposition} 
\Proof
The proof of the injection is similar to $p$ punctures case.
$\Box$

We define for $n\geq 2$, 
\begin{eqnarray} 
{ M}_{n,0}[\kappa, \lambda, \sigma, \tau]&=& 
M(\lambda_1,\lambda_2,\tau_3)\prod_{j=3}^{n} 
M(\lambda_j,\tau_j,\tau_{j+1})\times\nonumber\\ 
&\times&M(\lambda_1,\lambda_2,\sigma_3)\prod_{j=3}^{n} 
M(\lambda_j,\sigma_j,\sigma_{j+1})
\end{eqnarray} 
with $\tau_{n+1}=\sigma_{n+1}=\kappa.$

We define for $(\kappa, \lambda, \sigma, \tau)\in\SSP^{3n-3}$ the function 
$$\Upsilon_{n,0}[\kappa, \lambda, \sigma, \tau]= 
\frac{\vert\Xi_{n,0}[\kappa, \lambda, \sigma, \tau]\vert^2} 
{M_{n,0}[\kappa, \lambda, \sigma, \tau]} \frac{{\cal P}(\lambda)}{{\cal P}(\kappa)}.$$ 
It can be checked that this function is analytic in $\rho_x\in \RR$ for  
$x\in\kappa\cup\lambda\cup\sigma\cup\tau,$ and we will still denote by  
$\Upsilon_{n,0}[\kappa, \lambda, \sigma, \tau]$ the analytic continuation to ${\mathbb S}^{3n-3}.$

\begin{proposition} 
The space $H_{n,0}$ is endowed with a structure of pre-Hilbert space as follows: 
\begin{equation} 
<{\omega}_{n,0}(f) \vert {\omega}_{n,0}(g)>= 
\int_{\SSP^{3n-3}}d\kappa d\lambda d\sigma d\tau {\cal P}(\kappa,\lambda,\sigma,\tau) 
 \overline{f[\kappa, \lambda, \sigma, \tau]} 
\Upsilon_{n,0}[\kappa, \lambda, \sigma, \tau]  
g[\kappa, \lambda, \sigma, \tau]. 
\label{scalarproductn0} 
\end{equation} 
\end{proposition} 
\Proof 
Same proof as in the p-puncture case: use the injectivity of the map to show that the scalar product is defined unambiguously and the fact that $\Upsilon_{n,0}$ is analytic. 
$\Box$

\begin{figure}
\psfrag{I1}{$I_1$}
\psfrag{I2}{$I_2$}
\psfrag{I3}{$I_3$}
\psfrag{J1}{$J_1$}
\psfrag{J2}{$J_2$}
\psfrag{J3}{$J_3$}
\psfrag{K1}{$K_1$}
\psfrag{K2}{$K_2$}
\psfrag{K3}{$K_3$}
\psfrag{L1}{$L_1$}
\psfrag{L2}{$L_2$}
\psfrag{L3}{$L_3$}
\psfrag{U3}{$U_3$}
\psfrag{T3}{$T_3$}
\psfrag{W}{$W$}
\psfrag{la1+l1}{$\lambda_1+\ell_1$}
\psfrag{la1}{$\lambda_1$}
\psfrag{la3}{$\lambda_3$}
\psfrag{la2}{$\lambda_2$} 
\psfrag{la2+l2}{$\lambda_2+\ell_2$} 
\psfrag{la3+l3}{$\lambda_3+\ell_3$} 
\psfrag{l1}{$\lambda_1$}
\centering
\scalebox{0.8}[0.8]{\includegraphics[scale=0.8]{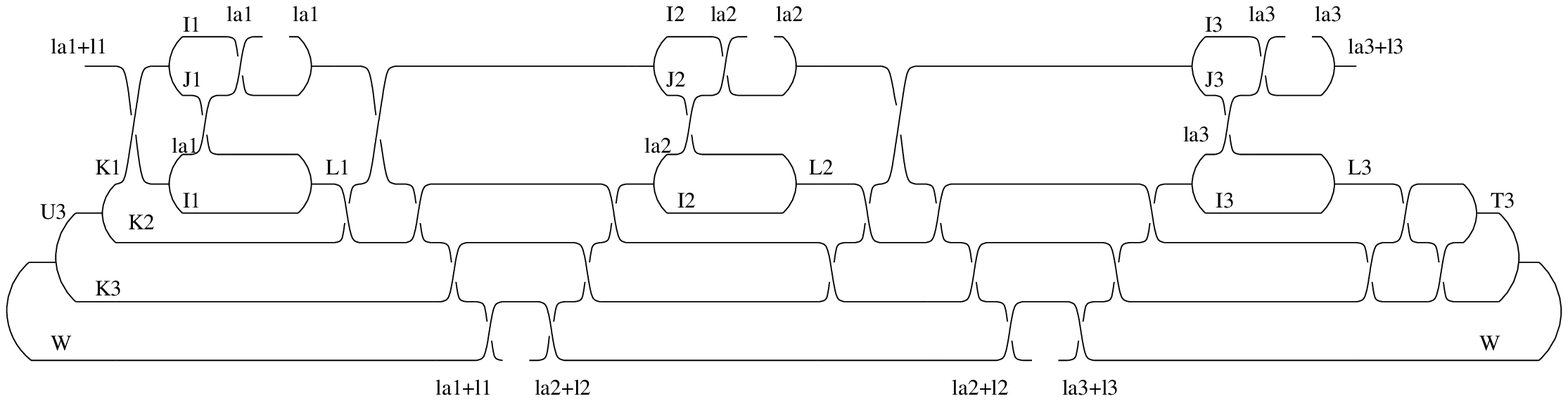}}
\caption{Expression of $\stackrel{P}{F}{}\!\!_{3,0}^{(+)}$.} 
\label{fig:valueofFforgenus3}
\end{figure}

\begin{proposition} 
Let $(\kappa,\lambda,\sigma,\tau)\in {\mathbb S}^{3n-3}$,  the action of   
$\stackrel{P}{\cal O}{}\!\!^{(\pm)}_{n,0}\in {\cal L}_{0,p}^{inv}$ on ${\omega}_{n,0}[\kappa,\lambda,\sigma,\tau]$ 
 is given by: 
\begin{equation}  
{\omega}_{n,0}[\kappa,\lambda,\sigma,\tau] \lhd 
 \stackrel{P}{\cal O}{}\!\!^{(\pm)}_{n,0} \; = \; 
 \sum_{(k,\ell,s,t)\in{\cal  S}^{3n-3} } \stackrel{P}{K}{}\!\!_{n,0}^{(\pm)} 
 \left( \begin{array}{c} \!\!\kappa;k\!\! \\ \!\! \lambda,\sigma,\tau; \ell,s,t \!\!\end{array} \right) 
 {\omega}_{n,0}[\kappa+k,\lambda+\ell,\sigma+s,\tau+t] \label{actionofOon omegagenusn} 
\end{equation}  
where: 
 
1. the functions  $\stackrel{ P}{K}{}\!\!_{n,0}^{(\pm)}\left( \begin{array}{c} \!\!\kappa;k\!\! \\ \!\! \lambda,\sigma,\tau; \ell,s,t \!\!\end{array} \right)$ are non zero only for a finite set  of $(k,\ell,s,t)\in {\mathbb S}^{3n-3}$ according to  the selection rules imposed by the palette P. Moreover any  element  $(k,\ell,s,t)$ of this finite set satisfy  
$(s^l,t^l),(s^r,t^r) \in {\ZZ}^{2n-4}.$ 
 
2. the function $\stackrel{ P}{K}{}\!\!_{n,0}^{(\pm)}\left( \begin{array}{c} \!\!\kappa;k\!\! \\ \!\! \lambda,\sigma,\tau; \ell,s,t \!\!\end{array} \right)$  belongs to  $\frac{\Xi_{n,0}[\kappa+k,\lambda+\ell,\sigma+s,\tau+t]}{\Xi_{n,0}[\kappa,\lambda,\sigma,\tau]} 
\CC(q^{i\rho_{x}})_{x\in \kappa\cup\lambda\cup\tau\cup\sigma}.$ 
\end{proposition} 
\Proof
Using the expression of the observables and their action on $(Fun_{cc}(SL_q(2,\CC))^{\otimes n})^*$, we have for $f\in Fun_{cc}(SL_q(2,\CC)^{\otimes n})$, 
\begin{equation}
(\bigotimes_{i=1}^n \iota(\stackrel{\lambda_i}{\GG}(i)) \lhd \stackrel{P}{\cal O}{}\!\!^{(\pm)}_{n,0})(f)  = 
 \sum_{\ell_1,\cdots,\ell_n\in {\cal S}} tr_{\lambda_1+\ell_1,\cdots,\lambda_n+\ell_n} (\stackrel{ P}{F}{}\!\!_{n,0}^{(\pm)} (\lambda,\ell) \bigotimes_{i=1}^n \iota(\stackrel{\lambda_i+\ell_i}{\GG}(i))(f)) , 
\end{equation} 

where $\stackrel{P}{F}{}\!\!_{n,0}^{(\pm)}(\lambda,\ell) $ are   elements of 
 $End_{U_q(sl(2,\CC)_{\RR})}(\bigotimes_{i=1}^n \stackrel{\lambda_i+\ell_i}{\VV} \otimes  \stackrel{\lambda_i}{\VV})$.  
Note that the sum over $\ell$ does not exist in the $p$-puncture case. It now appears because the expression of the  observable contains  multiplication by $\stackrel{I}{\GG}$ (with
$ I\in \SSF$), which therefore  corresponds, after acting on ${\omega}_{n,0}[\kappa,\lambda,\sigma,\tau]$,  to  tensor  representations of the principal series with finite dimensional representations. The picture for
$\stackrel{P}{F}{}\!\!_{n,0}^{(+)}(\lambda,l)$ is shown in  figure \ref{fig:valueofFforgenus3} (for the case $n=3$), whereas the picture for $\stackrel{P}{F}{}\!\!_{n,0}^{(-)}(\lambda,l)$ is the same after having turned overcrossing colored  by couples of finite dimensional representations into  the corresponding undercrossing. The picture for arbitrary genus $n$ is a straighforward generalization.

\begin{figure}
\psfrag{I1}{$I_1$}
\psfrag{I2}{$I_2$}
\psfrag{I3}{$I_3$}
\psfrag{J1}{$J_1$}
\psfrag{J2}{$J_2$}
\psfrag{J3}{$J_3$}
\psfrag{K1}{$K_1$}
\psfrag{K2}{$K_2$}
\psfrag{K3}{$K_3$}
\psfrag{L1}{$L_1$}
\psfrag{L2}{$L_2$}
\psfrag{L3}{$L_3$}
\psfrag{U3}{$U_3$}
\psfrag{T3}{$T_3$}
\psfrag{W}{$W$}
\psfrag{l1}{$\lambda_1$} 
\psfrag{l2}{$\lambda_2$} 
\psfrag{l3}{$\lambda_3$}
\psfrag{l1p}{$\lambda_1\!+\!\ell_1$} 
\psfrag{l2p}{$\lambda_2\!+\!\ell_2$} 
\psfrag{l3p}{$\lambda_3\!+\!\ell_3$}
\psfrag{k}{$\kappa$}
\psfrag{kp}{$\kappa\!+\!k$}
\psfrag{sp}{$\sigma\!+\!s$}
\psfrag{s}{$\sigma$}
\psfrag{sp}{$\sigma\!+\!s$}
\psfrag{t}{$\tau$}
\psfrag{tp}{$\tau\!+\!t$}
\centering
\scalebox{0.8}[0.8]{\includegraphics[scale=0.8]{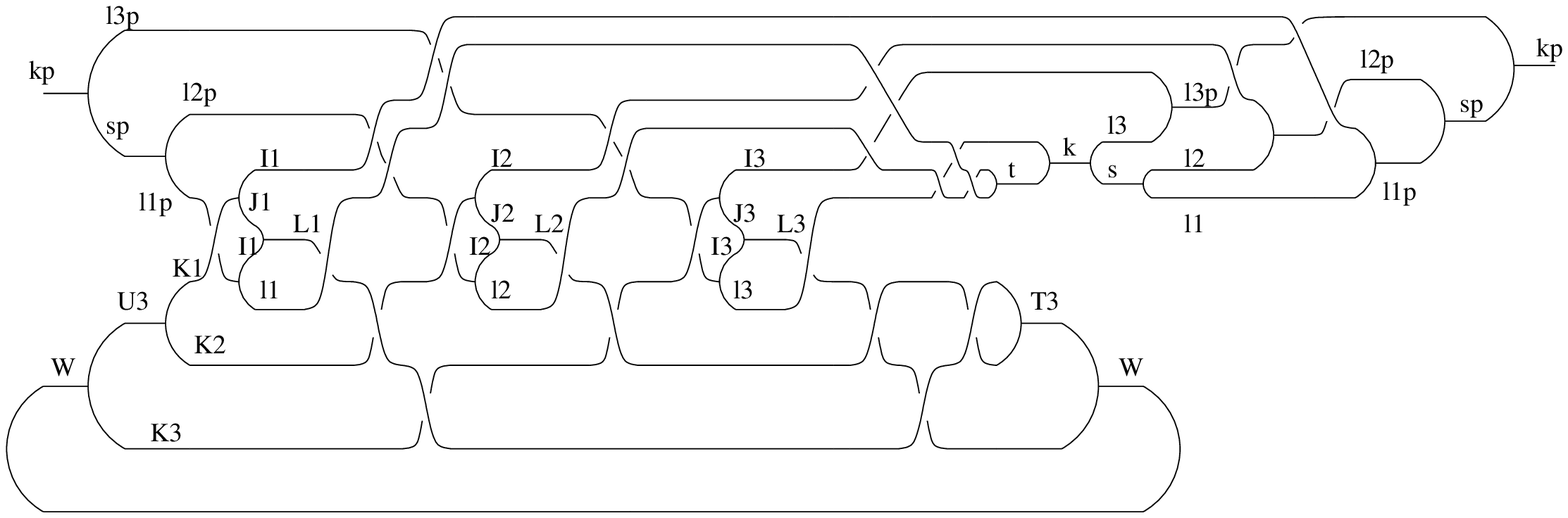}}
\caption{Expression of  $\stackrel{P}{K}{}\!\!_{3,0}^{(+)}$.} 
\label{fig:valueofKforgenus3}
\end{figure}

We will use the  same method as in the proposition (\ref{actionofOinthepuncturecase}),i.e we first define and prove everything in the case where 
$(\kappa,\lambda, \sigma,\tau)$ belongs to  $\SSF^{3n-3}$ and then we use the continuation method to extend it to  $\SSP^{3n-3}.$ We have
\begin{eqnarray}
(\omega_{n,0}[\kappa,\lambda,\sigma,\tau] \lhd \stackrel{P}{\cal O}{}\!\!^{(\pm)}_{n,0})(f)  = \sum_\ell \stackrel{P}{\cal F}{}\!\!_{n,0}^{(\pm)} \left( \begin{array}{c}  \!\!\kappa  \!\!\\  \!\!\lambda,\sigma,\tau;\ell  \!\!\end{array} \right)
\end{eqnarray}
where $\stackrel{P}{\cal F}{}\!\!_{n,0}^{(\pm)} \left( \begin{array}{c}  \!\!\kappa  \!\!\\  \!\!\lambda,\sigma,\tau;l  \!\!\end{array} \right)$ are elements of $(Fun_{cc}(SL_q(2,\CC))^{\otimes n})^*$ and is  a linear combination of the  elements $\omega_{n,0}[\kappa+k,\lambda+\ell,\sigma+s,\tau+t]$ as follows:
\begin{eqnarray}
\stackrel{P}{\cal F}{}\!\!_{n,0}^{(\pm)} \left( \begin{array}{c}  \!\!\kappa  \!\!\\  \!\!\lambda,\sigma,\tau;\ell  \!\!\end{array} \right) = \sum_{k,s,t} \stackrel{P}{ K}{}\!\!_{n,0}^{(\pm)} \left( \begin{array}{c}  \!\!\kappa;k  \!\!\\  \!\!\lambda,\sigma,\tau;\ell,s,t  \!\!\end{array} \right)\omega_{n,0}[\kappa+k,\lambda+\ell,\sigma+s,\tau+t]\;.
\end{eqnarray}
$ \stackrel{P}{K}{}\!\!_{n,0}^{(+)}
 \left( \begin{array}{c}  \!\! \kappa;k  \!\!\\  \!\!\lambda,\sigma,\tau;\ell,s,t  \!\! \end{array} \right)$ is represented by the picture \ref{fig:valueofKforgenus3}, and the same comments for the figure representing $\stackrel{P}{ F}{}\!\!_{n,0}^{(\pm)} (\lambda,\ell)$ apply here.

Note that the selection rules in the finite dimensional case, implies that  if 
${\omega}_{n,0}[\kappa,\lambda,\sigma,\tau]\not=0$ then the possible non zero elements ${\omega}_{n,0}[\kappa+k,\lambda+l,\sigma+s,\tau+t]$ are those  for $s$ and $t$ satisfying $s^l, s^r, t^l, t^r \in \ZZ^{n-2}.$

Using the property  (\ref{factorizationof3j}) of factorization of finite dimensional intertwiners,
we obtain that
\begin{equation}
 \stackrel{P}{K}{}\!\!_{n,0}^{(\pm)}
 \left( \begin{array}{c}  \!\!\kappa;k  \!\!\\  \!\!\lambda,\sigma,\tau;\ell,s,t \!\! \end{array} \right)= 
\stackrel{P^l}{K}{}\!\!_{n,0}^{l(\pm)}
 \left( \begin{array}{c} \!\!  \kappa^l;k^l  \!\!\\  \!\!\lambda^l,\sigma^l,\tau^l;\ell^l,s^l,t^l  \!\!\end{array} \right) 
\stackrel{P^r}{K}{}\!\!_{n,0}^{r(\pm)}
 \left( \begin{array}{c} \!\! \kappa^r;k^r  \!\!\\  \!\!\lambda^r,\sigma^r,\tau^r;\ell^r,s^r,t^r  \!\! \end{array} \right)
\end{equation} 
where the functions $\stackrel{P^l}{K}{}\!\!_{n,0}^{l(\pm)}$
, 
$\stackrel{P^r}{K}{}\!\!_{n,0}^{r(\pm)}$ are computed in the proposition (\ref{prop:expressionofKforgenus}) and expressed in terms of $6j$ symbols. 
The definition of the continuation of  $\stackrel{P^l}{K}{}\!\!_{n,0}^{l(\pm)}$ and $\stackrel{P^r}{K}{}\!\!_{n,0}^{r(\pm)}$ precisely uses these  expressions where $6j(0)$ are replaced by $6j(1)$ and $6j(3)$ where needed.
$\Box$

If $a\in {\cal L}_{n,0}^{inv}$ we define $\stackrel{ a}{K}{}\!\!_{n,0}^{(\pm)}\left( \begin{array}{c} \!\!\kappa;k\!\! \\ \!\! \lambda,\sigma,\tau; \ell,s,t \!\!\end{array} \right)$ by linearity as in the $p$-puncture case. 
  
We will now endow $H_{n,0}$ with a structure of right ${\cal L}_{n,0}^{inv}$ module in the following 
 sense: 
 
\begin{proposition} 
 $ {\cal L}_{n,0}^{inv}$  acts on $(Fun_{cc}(SL_q(2,\CC))^{\otimes n})^{*}$ with  
$\rho_{n,0}$ and  leaves  $(Fun_{cc}(SL_q(2,\CC))^{\otimes n})^{*} )^{inv}$ invariant. The subspace  
 $H_{n,0}$ is in general not invariant.  
We define the domain of $a\in {\cal L}_{n,0}^{inv} $ associated to $ \rho_{n,0} $  
to be the subspace $D(a)\subset \AA^{(3n-3)}$ defined as: 
f belongs to $D(a)$ if and only if for all $(k,\ell,s,t)\in {\cal S}^{3n-3}$ the functions  
$$(\kappa,\lambda,\sigma,\tau)\mapsto  \stackrel{a}{K}{}\!\!_{n,0}^{(\pm)}\left( \begin{array}{c} \!\!\kappa;k\!\! \\ \!\! \lambda,\sigma,\tau; \ell,s,t \!\!\end{array} \right) 
 \frac{(\Xi_{n,0}{\cal P})[\kappa,\lambda,\sigma,\tau]f(\kappa,\lambda,\sigma,\tau)} 
{(\Xi_{n,0}{\cal P})[\kappa+k,\lambda+\ell,\sigma+s,\tau+t]}$$ are elements of  $\AA^{(3n-3)}.$ 
The action of $a$ on an element ${\omega}_{n,0}(f)\in D(a)\subset H_{n,0}$ belongs to $H_{n,0}$, and we have: 
\begin{equation}  
 {\omega}_{n,0}(f) \lhd a = 
 {\omega}_{n,0}(f\lhd a)  \;\;,\nonumber   
\end{equation}  
with  
\begin{eqnarray}  
&&\hskip-1cm(f \lhd a)(\kappa,\lambda,\sigma,\tau) =\nonumber\\ 
&&\hskip-1cm\sum_{(k,\ell,s,t)\in {\cal S}^{3n-3}} \!\!\!\! 
 \frac{ \stackrel{a}{K}{}\!\!_{n,0}^{(\pm)}\left( \begin{array}{c} \!\!\kappa-\!k;k\!\! \\ \!\! \lambda\!-\!\ell,\sigma\!-\!s,\tau\!-\!t; \ell,s,t \!\!\end{array} \right) 
 f[\kappa\!-\!k,\lambda\!-\!\ell,\sigma\!-\!s,\tau\!-\!t]} 
{(\Xi_{n,0}{\cal P})[\kappa\!-\!k,\lambda\!-\!\ell,\sigma\!-\!s,\tau\!-\!t]^{-1}(\Xi_{n,0}{\cal P})[\kappa,\lambda,\sigma,\tau]} . 
\end{eqnarray}  
\end{proposition}

\begin{theorem}  \label{theorem2}
The representation ${\tilde\rho}_{n,0}$ of  $M_q(\Sigma 
,SL(2,\CC)_{\RR})$ is unitary: 
\begin{equation} 
\forall a\in M_q(\Sigma,SL(2,\CC)_{\RR}), 
 \forall v\in D(a^{\star}), \forall w\in D(a), 
 <v\lhd a^{\star}\vert w>=<v\vert w\lhd a>.   
\end{equation} 
where $<.\vert .>$ is the positive sesquilinear form defined by (\ref{scalarproductn0}). 
\end{theorem}  
\Proof 
Using Cauchy theorem for the integration in $\rho_x$ and reindexing the summation on $m_x$ for $x\in \lambda\cup\sigma\cup\tau\cup\kappa$ the  proof of unitarity of the representation reduces to the identity:
\begin{eqnarray} 
&&\Upsilon_{n,0}(\kappa,\lambda,\sigma,\tau)\stackrel{ P}{K}{}\!\!_{n,0}^{(\pm)}\left( \begin{array}{c} \!\!\kappa+\!k;-k\!\! \\ \!\! \lambda\!\!+\!\!\ell,\sigma\!\!+\!\!s,\tau\!\!+\!\!t; -\ell,-s,-t \!\!\end{array} \right)\frac{(\Xi_{n,0}{\cal P})(\kappa\!\!+\!\!k,\lambda\!\!+\!\!\ell,\sigma\!\!+\!\!s,\tau\!\!+\!\!t)}{\Xi_{n,0}(\kappa,\lambda,\sigma,\tau)}=\nonumber\\ 
&&\Upsilon_{n,0}(\kappa\!\!+\!\!k,\lambda\!\!+\!\!\ell,\sigma\!\!+\!\!s,\tau\!\!+\!\!t)\stackrel{ \tilde{P}}{K}{}\!\!_{n,0}^{(\mp)}\left( \begin{array}{c} \!\!\overline{\kappa};-\tilde{k}\!\! \\ \!\! \overline{\lambda},\overline{\sigma},\overline{\tau}; -\tilde{\ell},-\tilde{s},-\tilde{t} \!\!\end{array} \right) 
\frac{\overline{(\Xi_{n,0}{\cal P})(\kappa,\lambda,\sigma,\tau)}}{\overline{\Xi_{n,0}(\kappa\!\!+\!\!k,\lambda\!\!+\!\!\ell,\sigma\!\!+\!\!s,\tau\!\!+\!\!t)}} .
\end{eqnarray} 
when $(\kappa,\lambda,\sigma,\tau)\in\SSP^{3n-3}.$

In order to show this relation, we
make use of the following identities which are proved in the appendix:
\begin{eqnarray}
&&\stackrel{ \tilde{P}}{K}{}\!\!_{n,0}^{(-)}\left( \begin{array}{c} \!\!\overline{\kappa};-\tilde{k}\!\! \\ \!\! \overline{\lambda},\overline{\sigma},\overline{\tau}; -\tilde{\ell},-\tilde{s},-\tilde{t} \!\!\end{array} \right)=
\stackrel{ \tilde{P}}{K}{}\!\!_{n,0}^{(-)}\left( \begin{array}{c} \!\!\underline{\tilde\kappa};-\tilde{k}\!\! \\ \!\! \underline{\tilde\lambda},
\underline{\tilde\sigma},\underline{\tilde\tau}; -\tilde{\ell},-\tilde{s},-\tilde{t} \!\!\end{array} \right)
= \nonumber\\
&&
=\frac{[d_{\lambda}]}{[d_{\lambda+\ell}]} \frac{[d_{\kappa + k}]}{[d_{\kappa}]}\stackrel{P}{K}{}\!\!_{n,0}^{(+)}\left( \begin{array}{c} \!\!
\underline{\kappa}-k;{k}\!\! \\ \!\! \underline{\lambda}-\ell,
\underline{\sigma}-s,\underline{\tau}-t; {\ell},{s},{t} \!\!\end{array} \right)
= \nonumber\\
&&
=\psi_{n,0}(\kappa,\lambda,\sigma,\tau;k,\ell,s,t) \frac{[d_{\lambda}]}{[d_{\lambda+\ell}]} \frac{[d_{\kappa + k}]}{[d_{\kappa}]} \stackrel{P}{K}{}\!\!_{n,0}^{(+)}\left( \begin{array}{c} \!\!
{\kappa}+k;-{k}\!\! \\ \!\! {\lambda}+\ell,
{\sigma}+s,{\tau}+t; -{\ell},-{s},-{t} \!\!\end{array} \right).\nonumber
\end{eqnarray}
As a result showing unitarity is reduced to showing the quasi-invariance under shifts:
\begin{equation}
\frac{\Upsilon_{n,0}[\kappa,\lambda,\sigma,\tau]}
{\vert\Xi_{n,0}[\kappa,\lambda,\sigma,\tau]\vert^2
{\cal P}(\lambda)^2 {\cal P}(\sigma,\tau)}=\frac{\psi_{n,0}(\kappa,\lambda,\sigma,\tau,;k,\ell,s,t)\Upsilon_{n,0}[\kappa+k,\lambda+\ell,\sigma+s,\tau+t]}
{\vert\Xi_{n,0}[\kappa+k,\lambda+\ell,\sigma+s,\tau+t]\vert^2
{\cal P}(\lambda+\ell)^2  {\cal P}(\sigma+s,\tau+t) }.
\end{equation}
This is a direct consequence of the fact that  $\Upsilon_{n,0}$ is expressed in terms of the function $\Theta$ which satisfies:
$\Theta(\alpha+s,\beta+s,\gamma)=\Theta(\alpha,\beta,\gamma)$ for $s\in{\cal S}.$
$\Box$

\subsection*{III.3. The moduli algebra for the general case}  
  
This subsection generalizes the previous ones: we construct a unitary representation of the moduli algebra on a punctured surface of arbitrary genus $n$. The graph algebra ${\cal L}_{n,p}$ is isomorphic to $H(U_q(sl(2,\CC)_{\RR}))^{\otimes n} \otimes U_q(sl(2,\CC)_{\RR})^{\otimes p}$ and acts on ${\cal H}_{n,p}(\alpha) = Fun_{cc}(SL(2,\CC)_{\RR})^{\otimes n} \otimes \VV(\alpha)$ where $\alpha=(\alpha_1,\cdots,\alpha_p)\in \SSP^p$ denotes the representations assigned to the punctures. 

Before expressing a theorem for the general case similar to theorems \ref{theorem1} and \ref{theorem2}, it is interesting to study the representation of the moduli algebra on the one punctured torus. 

\begin{figure}  
\psfrag{lap}{$\lambda + \ell$}  
\psfrag{la}{$\lambda$}  
\psfrag{al}{$\alpha$}  
\psfrag{W}{$W$}  
\psfrag{K}{$K$}
\psfrag{N}{$N$}
\psfrag{I}{$I$}
\psfrag{J}{$J$}
\psfrag{L}{$L$}
\centering 
\includegraphics[scale=0.7]{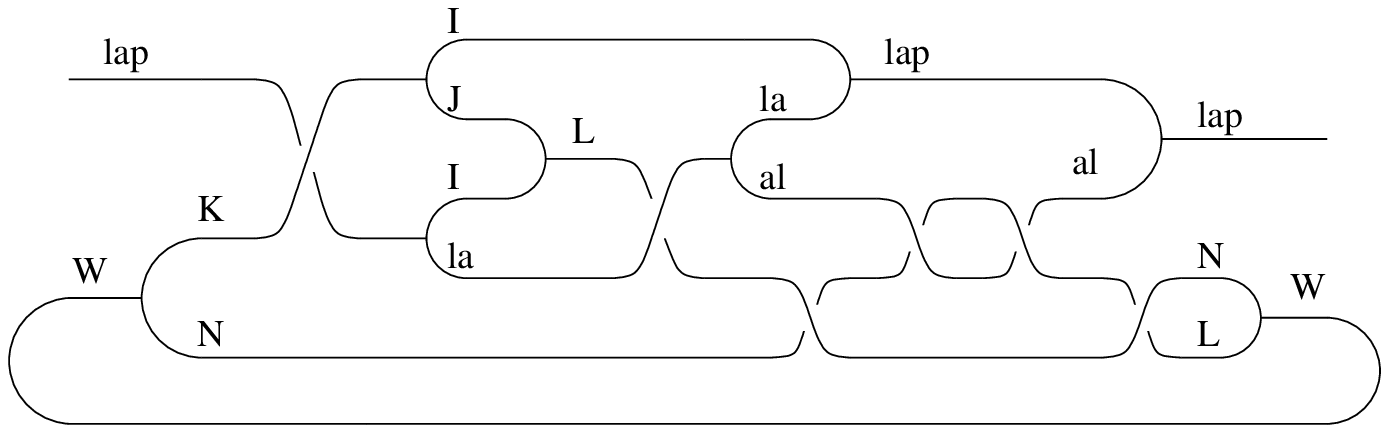} 
\caption{Expression of $\stackrel{P}{K}{}\!\!^{(+)}_{1,1}$.}  
\label{genus1punc1} 
\end{figure}

Given $\alpha, \lambda \in \mathbb{S}$, we can define $\stackrel{\alpha}{\omega}_{1,1}(\lambda) \in (Fun_{cc}(SL_q(2,\CC)_{\RR}) \otimes \VV(\alpha))^{*}$ by:
\begin{eqnarray}
<\stackrel{\alpha}{\omega}_{1,1}(\lambda) , f \otimes v> = tr_{\lambda}(<\stackrel{\lambda}{\GG},f> \Psi^{\lambda}_{\alpha \lambda}(v \otimes id)) 
\end{eqnarray}
where $f \in Fun_{cc}(SL_q(2,\CC)_{\RR})$ and $v \in \VV(\alpha)$.
The algebra ${\cal{L}}_{1,1}^{inv}$ acts on the right of $(Fun_{cc}(SL_q(2,\CC)_{\RR}) \otimes \VV(\alpha))^{*}$ with $\rho_{1,1}$ and we have:

\begin{eqnarray}
\stackrel{\alpha}{\omega}_{1,1}(\lambda) \lhd \stackrel{I}{C}_{1,1} & = & \stackrel{\alpha}{\omega}_{1,1}(\lambda) \\
\stackrel{\alpha}{\omega}_{1,1}(\lambda) \lhd tr_q(\stackrel{I}{M}) & = & \vartheta_{I \alpha} \stackrel{\alpha}{\omega}_{1,1}(\lambda)\;.
\end{eqnarray}
An observable is given by a palette $P=(I,J,N;K,L,W) \in \SSF^6$ as follows:
\begin{eqnarray}
\stackrel{P}{\cal O}{}\!\!_{1,1}^{(\pm)} = \frac{v_K^{1/2}}{v_I^{1/2} v_J^{1/2}}tr_{W}(\stackrel{W}{\mu} \Psi^{W}_{KN} \stackrel{(I,J)}{\theta}(K,L) \stackrel{NL}{\RR}{}\!\!^{(\pm)} \stackrel{N}{M}\stackrel{NL}{\RR}{}\!\!^{(\pm)-1} \Phi^{LN}_{W}) \nonumber
\end{eqnarray}
where $\stackrel{(I,J)}{\theta}(K,L) = \Psi^{K}_{JI} \stackrel{J}{B} \stackrel{JI}{\RR}{}\!\!' \stackrel{I}{A} \stackrel{JI}{\RR}{}\!\!^{(-)} \Phi^{JI}_L$.
After a direct computation, we can show that the action of this observable on $\stackrel{\alpha}{\omega}_{1,1}(\lambda)$ is given by:
\begin{eqnarray}
\stackrel{\alpha}{\omega}_{1,1}(\lambda) \lhd \stackrel{P}{\cal O}{}\!\!_{1,1}^{(\pm)} = \sum_{\ell \in \cal S} \stackrel{P}{K}{}\!\!_{1,1}^{(\pm)} \left( \begin{array}{c} \alpha \\ \lambda;\ell \end{array} \right) \stackrel{\alpha}{\omega}_{1,1}(\lambda +\ell)
\end{eqnarray}
where $\stackrel{P}{K}{}\!\!_{1,1}^{(+)} \left( \begin{array}{c} \alpha \\ \lambda;\ell \end{array} \right)$ is given by the graph (figure \ref{genus1punc1}). The expression for $\stackrel{P}{K}{}\!\!_{1,1}^{(-)} \left( \begin{array}{c} \alpha \\ \lambda;\ell \end{array} \right)$ is given by the same graph after having turned overcrossing colored by couples of finite dimensional representations into the corresponding undercrossing.

For $\alpha,\lambda \in \mathbb S$, we define
\begin{eqnarray}
\Xi_{1,1}[\alpha,\lambda]^{-1} & =& \hat{\Xi}_{1,1}[\alpha,\lambda]  \xi(\lambda)^2 \label{Xi11} \\
\hat{\Xi}_{1,1}[\alpha,\lambda] &=& e^{i\pi \lambda} \zeta(\lambda,\lambda,\alpha) \nu_1(d_{\lambda})
\end{eqnarray}

For $\alpha,\lambda \in \SSP$, we define ${ M}_{1,1}[\alpha,\lambda]=M(\alpha,\lambda,\lambda)$ where $M$ is defined in the appendix A.2 and we denote
\begin{eqnarray}
\Upsilon_{1,1}[\alpha,\lambda] = \frac{\vert \Xi_{1,1}(\alpha,\lambda) \vert^2}{{ M}_{1,1}[\alpha,\lambda]}.
\end{eqnarray}
We can show, as in the lemma \ref{analyticityofupsilon0p}, that $\Upsilon_{1,1}[\alpha,\lambda]$ is an analytic function of the real variable $\rho_{\lambda}$ and therefore admits an analytic continuation for $\lambda \in {\mathbb S}$.

\begin{proposition}
We define a subset of $(Fun_{cc}(SL_q(2,\CC)_{\RR})\otimes \VV(\alpha))^*$ by:
\begin{eqnarray}
H_{1,1}(\alpha) = \{\stackrel{\alpha}{\omega}_{1,1}(f) = \int_{\SSF} d\lambda \; {\cal P}(\lambda) \Xi_{1,1}[\alpha,\lambda] f(\lambda) \stackrel{\alpha}{\omega}_{1,1}(\lambda) , \;\; f \in \AA \}.
\end{eqnarray}
For the elements $f$ belonging to the domain of
 $\stackrel{P}{\cal O}{}\!\!_{1,1}^{(\pm)}$ we have :
\begin{eqnarray}
\stackrel{\alpha}{\omega}_{1,1}(f) \lhd \stackrel{P}{\cal O}{}\!\!_{1,1}^{(\pm)} = \stackrel{\alpha}{\omega}_{1,1}(f \lhd \stackrel{P}{\cal O}{}\!\!_{1,1}^{(\pm)})
\end{eqnarray}
where
\begin{eqnarray}
(f \lhd \stackrel{P}{\cal O}{}\!\!_{1,1}^{(\pm)})(\lambda) = \sum_{\ell \in \cal S} \stackrel{P}{K}{}\!\!_{1,1}^{(\pm)} \left( \begin{array}{c} \alpha \\ \lambda +\ell;-\ell \end{array} \right) \frac{{\cal P}(\lambda + \ell)   \Xi_{1,1}[\alpha,\lambda + \ell]}{{\cal P}(\lambda)   \Xi_{1,1}[\alpha,\lambda]} f(\lambda + \ell) \label{rho11}
\end{eqnarray}
The map $\AA \longrightarrow H_{1,1}(\alpha)$ is an injection. We can therefore endow $H_{1,1}(\alpha)$ with the following pre-Hilbert structure
\begin{eqnarray}
<\stackrel{\alpha}{\omega}_{1,1}(f) \vert \stackrel{\alpha}{\omega}_{1,1}(g)> = \int_{\SSP} d\lambda {\cal P}({\lambda}) \overline{f(\lambda)} \Upsilon_{1,1}[\alpha,\lambda] g(\lambda)\;\;.
\end{eqnarray}
 $\tilde{\rho}_{1,1}$ is an antirepresentation of 
$M_q(\Sigma_{1},SL(2,\CC)_{\RR}; \alpha)$  which is   unitary.
\end{proposition}
\Proof
It is easy to show that  $\stackrel{\alpha}{\omega}_{1,1}(f)$ is well defined and the expression (\ref{rho11}) of the action of $\tilde{\rho}_{1,1}$ follows from  Cauchy theorem. Using the usual method of proof,  unitarity is equivalent to the identity:
\begin{eqnarray}
\Upsilon_{1,1}[\alpha,\lambda] \stackrel{P}{K}{}\!\!_{1,1}^{(\pm)} \left( \begin{array}{c} \alpha \\ \lambda +\ell;-\ell \end{array} \right) \frac{\Xi_{1,1}[\alpha,\lambda+\ell]}{\Xi_{1,1}[\alpha,\lambda]} {\cal{P}}(\lambda+\ell) = \nonumber \\
\Upsilon_{1,1}[\alpha,\lambda+\ell] \stackrel{\tilde{P}}{K}{}\!\!_{1,1}^{(\mp)} \left( \begin{array}{c} \overline{\alpha} \\ \overline{\lambda};-\tilde{\ell} \end{array} \right) \frac{\overline{\Xi_{1,1}[\alpha,\lambda]}}{\overline{\Xi_{1,1}[\alpha,\lambda+\ell]}} \overline{{\cal{P}(\lambda)}}
\end{eqnarray}
when $(\alpha,\lambda) \in \SSP^2$.

In order to show this relation, we make use of the following identities which are proved in the appendix:
\begin{eqnarray}
\stackrel{\tilde{P}}{K}{}\!\!_{1,1}^{(\mp)} \left( \begin{array}{c} \overline{\alpha} \\ \overline{\lambda};-\tilde{\ell} \end{array} \right) & = & \stackrel{{P}}{K}{}\!\!_{1,1}^{(\pm)} \left( \begin{array}{c} \underline{\alpha} \\ \underline{\lambda}-\ell;{\ell} \end{array} \right)\\
&=& \psi_{1,1}(\alpha,\lambda;\ell) \stackrel{P}{K}{}\!\!_{1,1}^{(\pm)} \left( \begin{array}{c} {\alpha} \\ {\lambda}+\ell;-{\ell} \end{array} \right)\;\;.
\end{eqnarray}
As a result, showing unitarity reduces to the relation:
\begin{eqnarray}
\frac{\Upsilon_{1,1}[\alpha,\lambda]}{\vert\Xi_{1,1}[\alpha,\lambda]\vert^2 {\cal P}(\lambda)} =  \psi_{1,1}(\alpha,\lambda;\ell) \frac{\Upsilon_{1,1}[\alpha,\lambda+\ell]}{\vert\Xi_{1,1}[\alpha,\lambda+\ell]\vert^2 {\cal P}(\lambda+\ell)}
\end{eqnarray}
which holds. $\Box$

This proposition closes the construction for the torus with one puncture. 
All the tools are now ready to construct a right unitary module of the moduli algebra $M_q(\Sigma_n,SL(2,\CC)_{\RR};\alpha)$ .For this reason, we will just describe the representation of the moduli algebra without giving the technical details. 
 
\begin{proposition}
Let $\alpha=(\alpha_1,\cdots,\alpha_p) \in {\mathbb S}^p$, $\lambda=(\lambda_1,\cdots,\lambda_n) \in  {\mathbb{S}}^n$, $\beta=(\beta_3, \cdots,\beta_p) \in {\mathbb{S}}^{p-2}$,\;$\sigma=(\sigma_3,\cdots,\sigma_{n+1}), \tau=(\tau_3,\cdots,\tau_{n+1})$ elements of ${\mathbb S}^{n-1}$ and $\delta \in \mathbb S$. We define $\stackrel{\alpha}{\omega}_{n,p}(\beta,\lambda,\sigma,\tau,\delta) \in (Fun_{cc}(SL_q(2,\CC)_{\RR})^{\otimes n} \otimes \VV(\alpha))^*$ by
\begin{eqnarray}
<\stackrel{\alpha}{\omega}_{n,p}(\beta,\lambda,\sigma,\tau,\delta),\phi \otimes v> = 
tr_{\sigma_{n+1}} \left( <\wp_n(\lambda,\sigma,\tau),\phi> \Psi^{\tau_{n+1}}_{\delta \;\sigma_{n+1}} \left(\Psi^{\delta}_{\alpha}(\beta) \right) (v) \right)
\end{eqnarray}
where $\phi \in  Fun_{cc}(SL_q(2,\CC)_{\RR})^{\otimes n}$ and $v \in  \VV(\alpha)$.

 $\Psi^{\delta}_{\alpha}(\beta) \in Hom_{U_q(sl(2,\CC)_{\RR})}(\VV(\alpha),\stackrel{\delta}{\VV})$ has been introduced in the definition \ref{defofPsi} and $\wp_n(\lambda,\sigma,\tau) \in (Fun_{cc}(SL_q(2,\CC)_{\RR})^{\otimes n})^* \otimes Hom(\stackrel{\tau_{n+1}}{\VV},\stackrel{\sigma_{n+1}}{\VV})$ has been introduced in the definition \ref{defofP}.

$\stackrel{\alpha}{\omega}_{n,p}(\beta,\lambda,\sigma,\tau,\delta)$ are invariant vectors:
\begin{eqnarray}
\stackrel{\alpha}{\omega}_{n,p}(\beta,\lambda,\sigma,\tau,\delta) \lhd \stackrel{I}{C}_{n,p} = \stackrel{\alpha}{\omega}_{n,p}(\beta,\lambda,\sigma,\tau,\delta)\;\;,
\end{eqnarray}
and they satisfy the constraints:
\begin{eqnarray}
\stackrel{\alpha}{\omega}_{n,p}(\beta,\lambda,\sigma,\tau,\delta) \lhd tr_{q}(\stackrel{I}{\MM}(n+i)) = \vartheta_{I\alpha_i} \stackrel{\alpha}{\omega}_{n,p}(\beta,\lambda,\sigma,\tau,\delta) \;,\; \forall i =1,...,p\;\;.
\end{eqnarray}
\end{proposition}
\Proof
The proof of the invariance is a direct consequence of the factorization \\
$\stackrel{I}{C}{}\!\!_{n,p}^{(\pm)} = \stackrel{I}{C}{}\!\!_{n,0}^{(\pm)} \stackrel{I}{C}{}\!\!_{0,p}^{(\pm)}$. The action of $ tr_{q}(\stackrel{I}{\MM}(n+i))$ is obtained immediately. $\Box$\\

We will define:
\begin{eqnarray*}
\hat{\Xi}_{n,p}[\alpha,\beta,\lambda,\sigma,\tau,\delta]^{-1}  =  \hat{\Xi}_n[\lambda,\sigma,\tau]^{-1} \hat{\Xi}_c[\sigma_{n+1},\tau_{n+1},\delta]^{-1}
\hat{\Xi}_p[\alpha,\beta,\delta]^{-1}\;\;,
\end{eqnarray*}
where
\begin{eqnarray*}
\hat{\Xi}_n[\lambda,\sigma,\tau]^{-1} & = & e^{i\pi(2\lambda_1+...+2\lambda_n - \sigma_{n+1} -\tau_{n+1})} \zeta(\lambda_1,\lambda_2,\sigma_3) \prod_{j=3}^n \nu_1(d_{\sigma_j}) \zeta(\lambda_j,\sigma_j,\sigma_{j+1}) \nonumber \\
&& \times \zeta(\lambda_1,\lambda_2,\tau_3) \prod_{j=3}^n \nu_1(d_{\tau_j}) \zeta(\lambda_j,\tau_j,\tau_{j+1}) \\
\hat{\Xi}_p[\alpha,\beta,\delta]^{-1} & = & e^{i\pi(\alpha_1+...+\alpha_n)}  \zeta(\alpha_1,\alpha_2,\beta_3) \prod_{j=3}^{p-1}  \nu_1(d_{\beta_j}) \zeta(\alpha_j,\beta_j,\beta_{j+1}) \nonumber \\
&&\times \nu_1(d_{\delta}) \zeta(\alpha_p,\beta_p,\delta) \\
 \hat{\Xi}_c[\sigma_{n+1},\tau_{n+1},\delta]^{-1} & = & e^{i\pi(\sigma_{n+1} +\delta-\tau_{n+1})} \nu_1(d_{\tau_{n+1}}) \zeta(\sigma_{n+1},\tau_{n+1},\delta)\;.
\end{eqnarray*}
{}From these functions, we also define 
\begin{eqnarray*}
{\Xi}_{n,p}[\alpha,\beta,\lambda,\sigma,\tau,\delta]  =  \hat{\Xi}_{n,p}[\alpha,\beta,\lambda,\sigma,\tau,\delta] \xi(\beta) \xi(\lambda) \xi(\sigma) \xi(\tau) \xi(\delta)\;\;.
\end{eqnarray*}

We will also denote ${ M}_{n,p}[\alpha,\beta,\lambda,\sigma,\tau,\delta] = 
{ M}_{n}[\lambda,\sigma,\tau] { M}_{c}[\sigma_{n+1},\tau_{n+1},\delta] { M}_{p}[\alpha,\beta,\delta]$ with:
\begin{eqnarray*}
{ M}_{n}[\lambda,\sigma,\tau] & = & M(\lambda_1,\lambda_2,\sigma_3) M(\lambda_1,\lambda_2,\tau_3) \prod_{j=3}^n (M(\lambda_i,\sigma_i,\sigma_{i+1}) M(\lambda_i,\tau_i,\tau_{i+1})) \\
{ M}_{p}[\alpha,\beta,\delta] & = & M(\alpha_1,\alpha_2,\beta_3) M(\alpha_p,\beta_p,\delta) \prod_{j=3}^n M(\alpha_i,\beta_i,\beta_{i+1}) \\
{ M}_{c}[\sigma_{n+1},\tau_{n+1},\delta] & = & M(\sigma_{n+1},\tau_{n+1},\delta).
\end{eqnarray*}

Finally, for $\alpha={\alpha_1,\cdots,\alpha_p} \in {\mathbb{S}}_P^p$, $\lambda=(\lambda_1,\cdots,\lambda_n) \in  {\mathbb{S}}_P^n$, $\beta=(\beta_3, \cdots,\beta_p) \in {\mathbb{S}}_P^{p-2}$,\;$\sigma=(\sigma_3,\cdots,\sigma_{n+1}), \tau=(\tau_3,\cdots,\tau_{n+1})$ elements of ${\mathbb S}_P^{n-1}$ and $\delta \in \SSP$, we denote:
\begin{eqnarray}
\Upsilon_{n,p}[\alpha,\beta,\lambda,\sigma,\tau,\delta] = \frac{\vert \Xi_{n,p}[\alpha,\beta,\lambda,\sigma,\tau,\delta] \vert^2}{{ M}_{n,p}[\alpha,\beta,\lambda,\sigma,\tau,\delta]} \frac{{\cal P}(\lambda)}{{\cal P}(\tau_{n+1})}\;\;.
\end{eqnarray}
We can show that $\Upsilon_{n,p}[\alpha,\beta,\lambda,\sigma,\tau,\delta]$ is analytic in $\rho_x$ for $x \in \beta \cup \lambda \cup \sigma \cup \tau \cup \delta$ and we will still denote by $\Upsilon_{n,p}[\alpha,\beta,\lambda,\sigma,\tau,\delta]$ the analytic continuation to  $\alpha,\beta,\lambda,\sigma,\tau,\delta \in {\mathbb S}^{3n+p-3}$.

\begin{theorem}
We define a subset of $(Fun_{cc}(SL_q(2,\CC)_{\RR})^{\otimes n})^* \otimes \VV(\alpha)^*$ by:
\begin{eqnarray}
H_{n,p}(\alpha)=\{\stackrel{\alpha}{\omega}_{n,p}(f)=\int_{\SSP^{3n+p-3}} d\beta d\lambda d\sigma d\tau d\delta \;  {\cal P}(\beta,\lambda,\sigma,\tau,\delta) \Xi_{n,p}[\alpha,\beta,\lambda,\sigma,\tau,\delta] \nonumber \\
f(\beta,\lambda,\sigma,\tau,\delta) \stackrel{\alpha}{\omega}_{n,p}(\beta,\lambda,\sigma,\tau,\delta), f \in \AA \}
\end{eqnarray}
The map $\AA^{3n+p-3} \longrightarrow H_{n,p}(\alpha)$ which sends $f$ to $\stackrel{\alpha}{\omega}_{n,p}(f)$ is an injection. 

The algebra ${\cal L}_{n,p}^{inv}$ acts on the right of $ (Fun_{cc}(SL_q(2,\CC)_{\RR})^{\otimes n})^* \otimes \VV(\alpha)^*$ with $\rho_{n,p}$ which descends to a right action $\tilde{\rho}_{n,p}$ on $H_{n,p}$ by:
\begin{eqnarray}
\stackrel{\alpha}{\omega}_{n,p}(f) \lhd \stackrel{P}{\cal O}_{n,p} = \stackrel{\alpha}{\omega}_{n,p}(f \lhd \stackrel{P}{\cal O}_{n,p})\;.
\end{eqnarray}
The subspace $H_{n,p}(\alpha)$ is in general not invariant. We define the domain of $a \in {\cal L}_{n,p}^{inv}$ to be the subspace $D(a)$.

We can endow $H_{n,p}(\alpha)$ with a pre-Hilbert structure as follows:
\begin{eqnarray}
<\stackrel{\alpha}{\omega}_{n,p}(f) \vert \stackrel{\alpha}{\omega}_{n,p}(f)> = \int_{\SSP^{3n+p-3}} d\beta d\lambda d\sigma d\tau d\delta \;  {\cal P}(\beta,\lambda,\sigma,\tau,\delta) \Upsilon_{n,p}[\alpha,\beta,\lambda,\sigma,\tau,\delta] \nonumber \\
\overline{f(\beta,\lambda,\sigma,\tau,\delta)}g(\beta,\lambda,\sigma,\tau,\delta) \;\;.
\end{eqnarray}

The representation $\tilde{\rho}_{n,p}$ of $M_q(\Sigma,SL(2,\CC)_{\RR})$ is unitary:
\begin{eqnarray}
\forall a \in M_q(\Sigma,SL(2,\CC)_{\RR}), \forall v \in D(a^{\star}), \forall w \in D(a), <v \lhd a^{\star} \vert w> = <v \vert w \lhd a>.
\end{eqnarray}
\end{theorem}

\Proof
We perform the proof similarly to the proof of the theorems \ref{theorem1} and \ref{theorem2}.

First, using usual continuation arguments, we compute the action of $\stackrel{P}{\cal O}{}\!\!^{(\pm)}_{n,p}$ on $\stackrel{\alpha}{\omega}_{n,p}[\beta,\lambda,\sigma,\tau,\delta]$ when $\alpha={\alpha_1,\cdots,\alpha_p} \in {\mathbb{S}}^p$, $\lambda=(\lambda_1,\cdots,\lambda_n) \in  {\mathbb{S}}^n$, $\beta=(\beta_3, \cdots,\beta_p) \in {\mathbb{S}}^{p-2}$,\;$\sigma=(\sigma_3,\cdots,\sigma_{n+1}), \tau=(\tau_3,\cdots,\tau_{n+1})$ elements of ${\mathbb S}^{n-1}$ and $\delta \in \mathbb S$. We show that:
\begin{eqnarray}
\stackrel{\alpha}{\omega}_{n,p}[\beta,\lambda,\sigma,\tau,\delta] \lhd \stackrel{P}{\cal O}{}\!\!^{(\pm)}_{n,p} = \sum_{b,\ell,s,t,d \in \cal S} \stackrel{P}{K}{}\!\!_{n,p}^{(\pm)} \left(\begin{array}{c} \alpha \\ \beta,\lambda,\sigma,\tau,\delta;b,\ell,s,t,d \end{array} \right) \nonumber \\
\stackrel{\alpha}{\omega}_{n,p}[\beta+b,\lambda+\ell,\sigma+s,\tau+t,\delta+d]
\end{eqnarray}
The functions $\stackrel{P}{K}{}\!\!_{n,p}^{(\pm)} \left(\begin{array}{c} \alpha \\ \beta,\lambda,\sigma,\tau,\delta;b,\ell,s,t,d \end{array} \right)$, defined by the graph in picture (fig \ref{genus2punctures2}), are non zero only for a set of $b,\ell,s,t,d \in \cal S$ according to the selection rules imposed by the palette $P$.

If $a \in {\cal L}_{n,p}^{inv}$, we define $\stackrel{a}{K}{}\!\!_{n,p}^{(\pm)} \left(\begin{array}{c} \alpha \\ \beta,\lambda,\sigma,\tau,\delta;b,\ell,s,t,d \end{array} \right)$ by linearity as in the previous cases.
 The subspace $H_{n,p}(\alpha)$  is in general not left invariant by the action of ${\cal L}_{n,p}^{inv}$. So, we define the domain $D(a)$ of $a$ such that: $f$ belongs to $D(a)$ if and only if for all $b,\ell,s,t,d \in \cal S$ the functions
\begin{eqnarray}
(\alpha, \beta,\lambda,\sigma,\tau,\delta) \mapsto \stackrel{a}{K}{}\!\!_{n,p}^{(\pm)} \left(\begin{array}{c} \alpha \\ \beta,\lambda,\sigma,\tau,\delta;b,\ell,s,t,d \end{array} \right) f(\beta,\lambda,\sigma,\tau,\delta)\nonumber \\
 \frac{\Xi_{n,p}[\alpha, \beta,\lambda,\sigma,\tau,\delta] {\cal P}[\beta,\lambda,\sigma,\tau,\delta] }{\Xi_{n,p}[\alpha, \beta+b,\lambda+\ell,\sigma+s,\tau+t],\delta+d {\cal P}[\beta+b,\lambda+\ell,\sigma+s,\tau+t,\delta+d]}
\end{eqnarray}
are elements of $\AA^{(3n+p-3)}$. The action $\rho_{n,p}$ of ${\cal L}_{n,p}^{inv}$ on $(Fun_{cc}(SL_q(2,\CC)_{\RR})^{\otimes n}\otimes {\VV}(\alpha))^*$ descends to an action $\tilde{\rho}_{n,p}$ on $H_{n,p}[\alpha]$ defined by:
\begin{eqnarray}
\stackrel{\alpha}{\omega}_{n,p}(f) \lhd a = \stackrel{\alpha}{\omega}_{n,p}(f \lhd a)\;,
\end{eqnarray}
with
\begin{eqnarray}
&&(f \lhd a)(\beta,\lambda,\sigma,\tau,\delta) = \sum_{b,\ell,s,t,d \in \cal S} 
\stackrel{a}{K}{}\!\!_{n,p}^{(\pm)} \left(\begin{array}{c} \alpha \\ \beta-b,\lambda-\ell,\sigma-s,\tau-t,\delta-d;b,\ell,s,t,d \end{array} \right) \nonumber\\
&&\frac{\Xi_{n,p}[\alpha, \beta-b,\lambda-\ell,\sigma-s,\tau-t,\delta-d] {\cal P}[\beta-b,\lambda-\ell,\sigma-s,\tau-t,\delta-d] }{\Xi_{n,p}[\alpha, \beta,\lambda,\sigma,\tau,\delta] {\cal P}[\beta,\lambda,\sigma,\tau,\delta] } \nonumber \\
&&f(\beta-b,\lambda-\ell,\sigma-s,\tau-t,\delta-d).
\end{eqnarray}

Finally, using Cauchy theorem, unitarity reduces to the identity:
\begin{eqnarray}
&&\Upsilon_{n,p}[\alpha,\beta,\lambda,\sigma,\tau,\delta] \stackrel{P}{K}{}\!\!_{n,p}^{(\pm)} \left(\begin{array}{c} \alpha \\ \beta+b,\lambda+\ell,\sigma+s,\tau+t,\delta+d;-b,-\ell,-s,-t,-d \end{array} \right) \nonumber \\
&&\frac{\Xi_{n,p}[\alpha, \beta+b,\lambda+\ell,\sigma+s,\tau+t,\delta+d] {\cal P}[\beta+b,\lambda+\ell,\sigma+s,\tau+t,\delta+d] }{\Xi_{n,p}[\alpha, \beta,\lambda,\sigma,\tau,\delta] } = \nonumber \\
&&\Upsilon_{n,p}[\alpha,\beta+b,\lambda+\ell,\sigma+s,\tau+t,\delta+d] \stackrel{P}{\tilde{K}}{}\!\!_{n,p}^{(\mp)} \left(\begin{array}{c} \overline{\alpha} \\\overline{\beta},\overline{\lambda},\overline{\sigma},\overline{\tau},\overline{\delta};-\tilde{b},-\tilde{\ell},-\tilde{s},-\tilde{t},-\tilde{d} \end{array} \right) \nonumber \\
&&\frac{\overline{\Xi_{n,p}[\alpha, \beta,\lambda,\sigma,\tau,\delta]} \overline{{\cal P}[\beta,\lambda,\sigma,\tau,\delta]} }{\overline{\Xi_{n,p}[\alpha, \beta+b,\lambda+\ell,\sigma+s,\tau+t,\delta+d]}}
\end{eqnarray}
where $\alpha,\beta,\lambda,\sigma,\tau \in \SSP$.
In order to show this relation, we make use of the identities related in the proposition \ref{symnp} given in the appendix and the proof of unitarity reduces to the relation
\begin{eqnarray*}
&&\frac{\Upsilon_{n,p}[\alpha,\beta,\lambda,\sigma,\tau,\delta]}{\vert \Xi_{n,p}[\alpha, \beta,\lambda,\sigma,\tau,\delta] \vert ^2 {\cal P}[\beta,\lambda,\sigma,\tau]} \frac{{\cal P}(\tau_{n+1})}{{\cal P}(\lambda)} \psi_{n,p}^{-1}(\alpha, \beta,\lambda,\sigma,\tau,\delta;b,\ell,s,t,d)= \nonumber \\
&&\frac{\Upsilon_{n,p}[\alpha,\beta+b,\lambda+\ell,\sigma+s,\tau+t,\delta+d]}{\vert \Xi_{n,p}[\alpha, \beta+b,\lambda+\ell,\sigma+s,\tau+t,\delta+d] \vert ^2 {\cal P}[\beta+b,\lambda+\ell,\sigma+s,\tau+t,\delta+d]} \frac{{\cal P}(\tau_{n+1}+t_{n+1})}{{\cal P}(\lambda+\ell)} 
\end{eqnarray*}
which holds immediately. $\Box$

\section*{IV. Discussion and Conclusion} 

The major result of our work is the proof that there exists a unitary representation of the quantization of the moduli space of the flat $SL(2,\CC)_{\RR}$ connections on a punctured surface. This is a non trivial result which 
necessitates to integrate the formalism of combinatorial quantization and harmonic analysis on $SL_q(2,\CC)_{\RR}.$ However we have  not studied in details properties of this unitary representation. 
It would indeed be interesting to analyze the domains of definition of the operators $\rho_{n,p}({\cal O})$ and to study the possible extensions of $\rho_{n,p}({\cal O}).$ 
Related mathematical questions, which are answered positively in the compact group  case, are the following: 
\begin{itemize} 
\item is the representation  $\rho_{n,p}$ irreducible? 
 \item in this case,  is it the only irreducible representation up to equivalence? 
 \item Does this representation provides a unitary representation of the mapping class group? 
\end{itemize}

What remains also to be done is to relate  precisely the quantization of $SL(2,\CC)_{\RR}$ Chern-Simons theory to quantum gravity in de Sitter space. 

Discarding the problem of degenerate metrics, the two theories  are classically equivalent. As  pointed out in \cite{AT,Wi1}, three dimensional lorentzian gravity written in the first order formalism is  a gauge theory, specifically a Chern-Simons theory associated to the Lorentz group when the cosmological constant $\Lambda$ is positive. This equivalence was extensively studied (see for example \cite{Ca} and the references therein) and it is possible to relate the observables of Chern-Simons theory with the geometric parameters associated to the metric solution of Einstein equations. 
As usual we denote by $l_P$ the Planck length, $l_P=\hbar G$ and the cosmological constant $\Lambda$ is related to the cosmological length $l$ by $\Lambda=l^{-2}.$ These two length scales define a dimensionless constant $l_P/l$. The semiclassical regime is obtained when $\hbar$ approaches zero, and the relation between Chern-Simons $SL(2,\CC)_{\RR}$ and gravity, imposes $q=1-l_P/l+o(l_P/l).$
It is a central question to control the other terms of the expansion and the non perturbative corrections. This issue is not adressed here but could possibly   be done by comparing two expectation value of observables in 
$SL(2,\CC)_{\RR}$-Chern-Simons theory and in quantum gravity in de Sitter space. The construction of a unitary representation of the observables, provided in our work, is a step in this direction.  

In the rest of this discussion we will provide a relation between the mass and the spin of a particle and the parameters $(m,\rho)$ of the insertion of a principal representation.

 Let us first give  the metric of de Sitter space associated to massive and spinning particules. In a neighbourhood of a massive spinning particle of mass $m_p$ and spin $j$ , the metrics takes the form of Kerr-de Sitter solution (\cite{BBM,Pa}):
\begin{eqnarray*}  
ds^2 & = & -(8l_P M-\frac{r^2}{l^2}+\frac{(8l_P j)^2}{4r^2})dt^2 \\
&& +(8l_P M-\frac{r^2}{l^2}+\frac{(8l_P j)^2}{4r^2})^{-1}dr^2+
r^2(-\frac{8l_P j}{2r^2}dt+d\phi)^2 \\
& = & -\frac{(r^2+r_-^2)(r_+^2-r^2)}{r^2}dt^2 +\frac{r^2}{(r^2+r_-^2)(r_+^2-r^2)} dr^2 +r^2(d\phi+\frac{r_+r_-}{r^2}dt)^2
\end{eqnarray*}
where we have defined 
\begin{eqnarray*}
r_{+} = 2 l \sqrt{l_P M+ l_P \sqrt{M^2+\frac{j^2}{l^2}}} \;\;\;,\;\;\;
 r_{-} =  2l \sqrt{-l_P M+ l_P \sqrt{M^2+\frac{j^2}{l^2}}},
\end{eqnarray*}
where $8 l_P M=1-8l_P m_p$.

This metric has  a cosmological event horizon located in $r=r_+$. Following \cite{Pa}, this metric can be conveniently written as 
\begin{eqnarray*}
ds^2 = {\sinh}^2R\left( \frac{r_+dt}{l} - r_- d\phi \right)^2 - l^2 dR^2 + 
\cosh^2R \left( \frac{r_-dt}{l} + r_+ d\phi \right)^2
\end{eqnarray*}
with $r^2 = r_+^2\cosh^2 R + r_-^2 \sinh^2 R$. In this coordinate system, the cosmological horizon is located at $R = 0$; the exterior of the horizon is described for real $R$  and the interior for imaginary value of $R$. To this metric we can associate an orthonormal  cotriad $e_\mu^a$ with $g_{\mu\nu}=e_\mu^a e_\nu^b \eta_{ab}$ and its  spin connection $\omega_\mu{}^a_b.$ These data  define a flat $SL(2,\CC)_{\RR}$ connection 
\begin{eqnarray*}
A_\mu=\omega_\mu^a J_a+ \frac{1}{l} e_\mu^a P_a
\end{eqnarray*}
where $J_a, P_a$ are the generators of $so(3,1)$ and $\omega_\mu^a=\onehalf \epsilon_{abc}\omega_\mu{}^{ab}.$ The explicit value of the connection is given by Park (\cite{Pa}) in the spinorial representation and a trivial computation shows that:
\begin{eqnarray}
W_{cl}(A) = 2 \cosh \left( 2\pi \frac{ r_- - i r_+}{l} \right) \label{Wclassique}
\end{eqnarray}
where $W_{cl}(A)$ is the classical holonomy of the connection along  a circle centered around the world-line of the particle.

In the quantization of Chern-Simons theory that we provided,  a puncture is  colored with a principal unitary representation $\alpha=(m,\rho)$. The monodromy $\stackrel{I}{M}_q(A)$ around a puncture belongs to the center of the moduli algebra and it is defined by $\stackrel{I}{M}_q(A) = \vartheta_{I\alpha}$. In order to compare it to the classical case, one can trivially evaluate the associated holonomy in the {\it spinorial} representation $I=(1/2,0)$:
\begin{eqnarray}
W_q(A)=q^{m+i\rho}+q^{-(m+i\rho)}. \label{Wquantique}
\end{eqnarray}

At the semi classical level, the comparison between  (\ref{Wclassique}, \ref{Wquantique}), implies the relations:
\begin{eqnarray}
\mp 2\pi r_+ =  \rho l_P \;\;\;,\;\;\; \pm 2r_- =  m l_P\;.
\end{eqnarray}
which is equivalent to 
\begin{eqnarray}
32 \pi^2 l_P M= \frac{l_P^2}{l^2}(\rho^2-m^2)\;\;\;,\;\;\; 
16 \pi^2 j=\frac{1}{\pi^2} \frac{l_p^2}{l^2} \rho m\;\;.
\end{eqnarray}

{}From these  relations we note that  $r_-$ is quantized in units of $l_P$ whereas $r_+$ has a continuous spectrum. 
It is much more delicate to understand what  is the physical meaning of holonomies  around several punctures.
 In particular, the energy of such a system in de Sitter space is not, a priori, well defined in the absence of boundaries, even at the classical level. It should be a good chalenge to understand the classical and quantum behaviour of particules in dS space in the light of  Chern-Simons theory on a $p$-punctured sphere. We will give an analysis of this problem in a future work.

A comparison, similar to the analysis given above,  between the classical geometry  and the quantization in the Chern-Simons approach has been given in \cite{Ez} for the genus one case.  The generalization to the genus n case is open. Of particular interest is the construction of coherent states, lying in $H_{n,p}$ and  approaching a classical 3 metric. This subject is up to now still in its infancy.

\newpage

\section*{Appendix A: Quantum Lorentz group}

 \subsection*{A.1 Representations and harmonic analysis}   
In this work we have chosen $q\in \RR, 0<q<1.$  
 
For $x \in \CC$, we denote $\lbrack x \rbrack _q \; = \; \lbrack x \rbrack \; = \; \frac{q^x - q^{-x}}{q-q^{-1}}, d_x \;= \; 2x+1$, and $v_x^{1/4} \; = \; \exp (i \frac{\pi}{2} x) q^{-\frac{x(x+1)}{2}}.$ 
 
The square root of a complex number is defined  by:   
\begin{eqnarray}   
\forall x\in \CC, \sqrt{x}=\sqrt{\vert x \vert}e^{i \frac{Arg(x)}{2}},   
\mbox{where}\;\; x=\;\vert x \vert e^{i  Arg(x)}, Arg(x) \in ]-\pi,\pi],    
\end{eqnarray}   
For all complex number $z$ with non zero imaginary part,  we define $\epsilon(z)=\;\mbox{sign}\; (\Im(z)).$   
 
 We will define  the following basic functions: $\forall z \in \CC, \forall n \in {\mathbb Z},$   
\begin{eqnarray*}   
&&(z)_{\infty}=(q^{2z},q^2)_{\infty}=   
\prod_{k=0}^{+\infty}(1-q^{2z+2k})\;,\;\;\;\;\;\;\;\;\;\;\;\;   
(z)_{n}=\frac{(z)_{\infty}}{(z+n)_{\infty}},\\   
&&\nu_{\infty}(z)=\prod_{k=0}^{+\infty}\sqrt{1-q^{2z+2k}}   
\;,\;\;\;\;\;\;\;\;\;\;\;\;\;\;\;\;\;\;\;\;\;\;\;\;\;\;\;\;\;\;\;\;\nu_{n}(z)=\frac{\nu_{\infty}(z)}{\nu_{\infty}(z+n)}.   
\end{eqnarray*}   
Let us define the function $\xi,\theta:$ 
\begin{eqnarray} 
 \xi(z)&=&(z)_{\infty}(1-z)_{\infty},\xi(z)=\xi(z+i\frac{\pi}{\ln q})\\  
\theta (z) & = & \xi(z) q^{z^2-z} e^{i \pi z}/ (1)_{\infty}^2 ,  
\theta(z+1)=\theta(z).  
\end{eqnarray}     
With our choice of square root, we have $\overline{\nu_n(z)}= \nu_n(\overline{z}).$  
It is also convenient to introduce the following function $\varphi:\CC\times \ZZ \longrightarrow \{1,i,-i,-1\}$, defined by 
\begin{equation} 
\varphi(z,n)=\nu_n(z-n+1)\nu_{-n}(n-z)q^{-nz+\onehalf n(n-1)}. 
\end{equation}

We will    
recall in this appendix  fondamental results  on  ${U}_q(sl(2,\mathbb{C})_{\RR}).$ We will give a summary of      
the  harmonic analysis  on $SL_q(2,\CC)_{\RR}$, for a complete treatment see \cite{BR3,BR3'}.

 $U_q(su(2))$, for $q \in \rbrack 0, 1 \lbrack$, is defined as being the star Hopf algebra generated  by the elements  $J_{\pm}$, $q^{\pm J_z}$, and the relations:  
\begin{eqnarray}  
q^{\pm J_z}  q^{\mp J_z} \; = \; 1\;,\;\;\;\;  q^{J_z} J_{\pm} q^{-J_z} \; = \; q^{\pm 1} J_{\pm} \; , \;\;\;\; \lbrack J_+,\, J_- \rbrack \; = \; \frac{q^{2 J_z} \; - \; q^{-2J_z}}{q \; - \; q^{-1}} \;\;\; .  
\end{eqnarray}  
The coproduct is defined by   
\begin{eqnarray}  
\Delta (q^{\pm J_z}) \; = \; q^{\pm J_z} \; \otimes \; q^{\pm J_z} \; , \;\;\;\;\; \Delta J_{\pm} \; = \; q^{- J_z} \; \otimes \; J_{\pm} \; + \; J_{\pm} \; \otimes \; q^{J_z} \; ,  
\end{eqnarray}    
and the star structure is given by:  
\begin{eqnarray}  
(q^{J_z}){}^{\star} \; = \; q^{J_z} \; , \;\;\;\;\; J_{\pm}^{\star} \; = \; q^{\mp 1} \; J_{\mp} \; .  
\end{eqnarray}  
This Hopf algebra is a ribbon quasi-triangular Hopf algebra. The action of the generators on an orthonormal basis of an irreducible representation of spin $I$ is given by the following expressions:  
\begin{eqnarray}  
q^{J_z} \; \stackrel{I}{e}_m \; & = & \; q^{m} \; \stackrel{I}{e}_m \; , \\  
J_{\pm} \; \stackrel{I}{e}_m \; & = & \; q^{\pm \frac{1}{2}} \sqrt{\lbrack I \pm m +1 \rbrack \lbrack I \mp m \rbrack } \stackrel{I}{e}_{m \pm 1}\;\; .  
\end{eqnarray}  
 
The element $\mu$ is given by $\mu \; = \; q^{2J_z}$.   
  
For a representation $\pi \; = \; \stackrel{I}{\pi}$ of $U_q(su(2))$, we define the conjugate representation $\bar{\pi}(x) \; = \; \overline{\pi(S^{-1} x^{\star})}$ and we have $\bar{\pi}(x) \; = \; \stackrel{I}{w} \; \pi(x) \; \stackrel{I}{w}{}^{-1}$ with $\stackrel{I}{w}$ matrices whose components are  defined by $\stackrel{I}{w}{}^{\bar{m}}_n = w_{mn} = v_I^{1/2} e^{-i \pi m} q^m \delta_{m,-n}$.\\

${U}_q(sl(2,\mathbb{C})_{\RR})$  is defined to be the quantum double of    
 ${U}_q(su(2)).$ Therefore   ${U}_q(sl(2,\mathbb{C})_{\RR})=    
 {U}_q(su(2))\otimes  {U}_q(su(2))^{*}$ as     
a vector space, where $ {U}_q(su(2))^{*}$ denotes the restricted dual of    
${U}_q(su(2)),$ i.e the Hopf algebra spanned by the matrix elements of finite dimensional representations  of  ${U}_q(su(2)).$    
   
A basis of $ {U}_q(su(2))^{*}$ is the set of matrix elements in an orthonormal basis of irreducible unitary     
representations of $ {U}_q(su(2))$, which we will denote by    
 ${\buildrel {{}_{B}} \over  g}{}^i_j, B\in \halfinteger, i,j=-B,...,B.$   
   
It can be shown that $ {U}_q(su(2))^{*}$ is isomorphic, as a star Hopf algebra, to the quantum envelopping algebra $ {U}_q(an(2))$ where $an(2)$ is the  Lie algebra  of traceless  complex upper triangular 2$\times$2 matrices with real diagonal. 
   
\medskip  
   
${  U}_q(su(2))$ being a factorizable Hopf algebra, it is possible to give a nice    
generating family  of ${U}_q(su(2))$. Let us     
introduce,     
for each $I\in \halfinteger$ the elements    
 ${\buildrel I \over L}{}^{\!(\pm)}\in { End}(\CC^{d_I} )\otimes {U}_q(su(2))$    
 defined by    
 ${\buildrel I \over L}{}^{\!(\pm)}=({\buildrel I\over \pi} \otimes id)(R^{(\pm)}).$     
The matrix elements of ${\buildrel I \over L}{}^{\!(\pm)}$ when $I$ describes  $\halfinteger$ span the vector space  ${U}_q(su(2))$.   
      
The star Hopf algebra structure on ${U}_q(sl(2,\mathbb{C})_{\RR})$  is described in  details in \cite{BR3}.  Let us simply recall  that we have:    
\begin{eqnarray*}    
&&\hskip -1.2cm{\buildrel I \over L}{}^{(\pm)}{}^{i}_{j} {\buildrel J \over L}{}^{(\pm)}{}^{k}_{l}=    
\sum_{Kmn} \Clebphi{I}{J}{K}{i}{k}{m}\!{\buildrel K\over L}{}^{(\pm)}{}^{m}_{n}\!\Clebpsi{I}{J}{K}{j}{l}{n}\;,\;\;\;{\buildrel {IJ} \over {R}}_{12} \; {\buildrel I \over L}{}^{(+)}_1 {\buildrel J \over L}{}^{(-)}_2 =    
 {\buildrel J \over L}{}^{(-)}_2 {\buildrel I \over L}{}^{(+)}_1 \; {\buildrel {IJ} \over {R}}_{12},\label{Rll}\\   
&&\hskip -1.2cm{\buildrel I \over g}{}^{i}_{j} {\buildrel J \over g}{}^{k}_{l}=\sum_{Kmn}     
\Clebphi{I}{J}{K}{i}{k}{m}\!{\buildrel K \over g}{}^{m}_{n}\!\Clebpsi{I}{J}{K}{j}{l}{n}\;,\;\;\;\;\;\;\;\;\;\;\;\;\;\;\;\;\;    
{\buildrel {IJ} \over  R}{}^{(\pm)}_{12} {\buildrel I \over L}{}^{(\pm)}_1 {\buildrel J \over g}_2=    
{\buildrel J \over g}_2 {\buildrel I \over L}{}^{(\pm)}_1 {\buildrel {IJ} \over  R}{}^{(\pm)}_{12},\label{RLg}    
\end{eqnarray*}    
\vskip -0.4cm    
\begin{eqnarray*}    
&&\hskip -1cm\Delta({\buildrel I \over L}{}^{(\pm)}{}^a_b)=\sum_{c}{\buildrel I \over L}{}^{(\pm)}{}^c_b     
\otimes {\buildrel I \over L}{}^{(\pm)}{}^a_c \;,\;\;\;\;\;\;\;\;\;\;    
\Delta({\buildrel I \over g}{}^a_b)=\sum_{c}{\buildrel I \over    
g}{}^c_b \otimes {\buildrel I \over g}{}^a_c  ,\label{coproduitD}\\    
&&\hskip -1cm({\buildrel I \over L}{}^{(\pm)}{}^a_b)^{\star}= S^{-1}({\buildrel I \over L}{}^{(\mp)}{}^b_a)\;,\;\;\;\;\;\;\;\;\;\;\;\;\;\;\;\;\;\;\;\;\;\;\;({\buildrel I \over g}{}^a_b)^{\star}=S^{-1}({\buildrel I \over g}{}^b_a).\label{staralg}    
\end{eqnarray*}    
The center of $ {U}_q(sl(2,\mathbb{C})_{\mathbb{R}})$ is a polynomial algebra in two variables $\Omega_+, \Omega_-$ and we have $\Omega_{\pm}=   
tr({\buildrel \onehalf \over \mu}{}^{-1}{\buildrel \onehalf \over L}{}^{(\mp)}{}^{-1} {\buildrel \onehalf \over g}).$   
   
We will denote by $\mathbb{S}$ the set of couples $\alpha=(\alpha^l,\alpha^r)\in\CC^{2}$ such that  
 $m_{\alpha}=\alpha^l-\alpha^r\; \in \; \frac{1}{2}\ZZ.$ We will define $i\rho_{\alpha}=\alpha^l+\alpha^r+1 \;$. Reciprocally we will denote  
 $\alpha(m,\rho)\in \mathbb{S} $ the unique element associated to $m\in   \frac{1}{2}\ZZ$ and $\rho\in \CC.$ 
 
We will use the following definitions: 
 $\alpha \in \mathbb S$,  $v_{\alpha}^{1/4}=v_{\alpha^l}^{1/4} v_{\alpha^{r}}^{-1/4}, e^{i\pi \alpha}=e^{i\pi \alpha^l} e^{i\pi \alpha^r}.$ We define  $[d_{\alpha}]= [d_{\alpha^r}][d_{\alpha^l}]$ as well as  
$\nu_1(d_{\alpha})=\nu_1(d_{\alpha^r})\nu_1(d_{\alpha^l}).$ 
  We can extend $\varphi$ to  ${\mathbb S}\times {\ZZ\times\ZZ }$ as follows: 
$\varphi(\alpha,s)=\varphi(\alpha^l,s^l)\varphi(\alpha^r,s^r).$ For $\alpha=(\alpha_1,\cdots,\alpha_n) \in {\mathbb S}^n$ and $s=(s_1,\cdots,s_n) \in (\ZZ^2)^n$, we define: $\phi(\alpha,s)=\prod_{i=1}^n \phi(\alpha_i,s_i)$.

We distinguish two subsets of the previous one, $\mathbb{S}_P$ (resp. $\mathbb{S}_F$),  defined by $\rho_{\alpha} \; \in \RR$ (resp.$(\alpha^l,\alpha^r) \in \frac{1}{2} \ZZ^{+} \times \frac{1}{2} \ZZ^{+}$).

For $\alpha\in \SSF$ we define the vector space $\stackrel{\alpha}{\VV}=  
\bigoplus_{C=\vert \alpha^l-\alpha^r\vert}^{\alpha^l+\alpha^r} {\buildrel {C}\over V}$.  
  
For  $\alpha\in \mathbb{S}\setminus \SSF $ we define the vector space $\stackrel{\alpha}{\VV}=  
\bigoplus_{C, C-\vert \alpha^l-\alpha^r\vert\in \NN} {\buildrel {C}\over V}$.  
  
For $\alpha \in  \mathbb{S}$ we define a representation $\stackrel{\alpha}{\Pi}$ on the vector space   
$\stackrel{\alpha}{\VV}$ by the following expressions of the action of the generators of ${U}_q(sl(2,\mathbb{C})_{\mathbb{R}})$:  
  
\begin{eqnarray}    
{\buildrel {{}_{B}} \over  L}{}^{(\pm)}{}^i_j \;  {\buildrel {{}_{C}} \over e}_r &=    
&\sum_{n} {\buildrel {{}_{C}} \over e}_n\;\;    
 {\buildrel {{}_{BC}}\over  R}{}^{(\pm)}{}^{in}_{jr},\label{repl}\\    
{\buildrel {{}_{B}} \over  g}{}^i_j  \;  {\buildrel {{}_{C}} \over e}_r &=&     
\sum_{DEpk} {\buildrel {{}_{E}} \over e}_p\;\;     
\Clebphi{E}{B}{D}{p}{i}{k}\Clebpsi{B}{C}{D}{j}{r}{k} \Lambda^{BD}_{EC}(\alpha),\label{repg}    
\end{eqnarray}     
where $ {\buildrel {{}_{C}} \over e}_r, r=-C,...,C$ is an orthonormal basis of ${\buildrel {C}\over V}$ and   
 the complex numbers $\Lambda^{BD}_{EC}(\alpha )$ have been defined in    \cite{BR3}.   
It is a basic result that these coefficients can be expressed in  terms of $6j$ symbols \cite{BR3,BR4} as follows:    
\begin{eqnarray}    
&&\hskip -1cm\Lambda^{BC}_{AD}(\alpha)= \sum_{p} \!    
\sixjn{\alpha^l}{\alpha^r}{A}{B}{C}{\alpha^l+p}{\epsilon} \!\sixjn{\alpha^l}{\alpha^r}{D}{B}{C}{\alpha^l+p}{\epsilon}   
\frac{v_{\alpha^l+p}}{v_{\alpha^r}}     
\frac{v^{1/4}_{A}v^{1/4}_{D}}{v^{1/2}_{B}v^{1/2}_{C}}\label{formlamb6j}\\    
&&= \sum_{p} \!\sixjn{\alpha^r}{\alpha^l}{A}{B}{C}{\alpha^l+p}{\epsilon}\!    
\sixjn{\alpha^r}{\alpha^l}{D}{B}{C}{\alpha^l+p}{\epsilon} \frac{v_{\alpha^l}}{v_{\alpha^l+p}}     
\frac{v^{1/2}_{B}v^{1/2}_{C}}{v^{1/4}_{A}v^{1/4}_{D}},    
\end{eqnarray}   
with $\epsilon=0$ if $\alpha\in \SSF$ and $\epsilon=1$ in the other case.  
 
Let $A\in\onehalf \ZZ,\alpha\in {\mathbb S}$ we denote 
\begin{eqnarray} 
&&\nu^{(A)}(\alpha)=\nu_{2A+1}(\alpha^l+\alpha^r-A+1),\\ 
&&N^{(A)}(\alpha)=\frac{\nu_{\infty}(A+i\rho_{\alpha}+1)\nu_{\infty}(i\rho_{\alpha}-A)}{(m_\alpha+i\rho_\alpha)_{\infty}\nu_1(m_\alpha+i\rho_\alpha)} 
e^{-i\pi\frac{m_{\alpha}}{2}\epsilon({i\rho})}e^{i\pi A}q^{-\frac{m_{\alpha}}{2}}. 
\end{eqnarray} 
We have shown in \cite{BR3} that both  
$\frac{N^{(A)}(\alpha)}{N^{(D)}(\alpha)}\Lambda_{AD}^{BC}(\alpha)$ and  
$\frac{N^{(D)}(\alpha)}{N^{(A)}(\alpha)}\Lambda_{AD}^{BC}(\alpha)$ are Laurent polynomials in  
$q^{i\rho_{\alpha}}.$

A representation $\stackrel{\alpha}{\pi}$ associated to an element  $\alpha\in \mathbb{S}_P$ is said to belong to the principal series (infinite dimensional unitary representation) whereas a  representation $\stackrel{\alpha}{\pi}$ associated to an element  $\alpha\in \mathbb{S}_F$ is an irreducible finite dimensional representation.

The action of the center on the module  $\stackrel{\alpha}{\VV}$ is such that   
 $\stackrel{\alpha}{\Pi}(\Omega_{\pm})=\omega_{\pm}(\alpha) id$ where  
 $\omega_{+}(\alpha)=q^{2\alpha^l+1}+q^{-2\alpha^l{-1}},  
 \omega_{-}(\alpha)=q^{2\alpha^r+1}+q^{-2\alpha^r{-1}}.$   
  
 For   $\alpha=(\alpha^l,\alpha^r)\in \mathbb{S}$ we can define the  elements    
$\tilde{\alpha}, \overline{\alpha}$ and $\underline{\alpha }$ defined by    
 $\tilde{\alpha}=(\alpha^{r},\alpha^l),\overline{\alpha}= 
(\overline{\alpha^l},\overline{\alpha^r})$ and $\underline{\alpha }=(-\alpha^l-1,-\alpha^r-1).$

The two modules $\stackrel{\alpha}{\VV}$ and   
$\stackrel{\underline{\alpha}}{\VV}$ are equivalent and we have 
$\Lambda^{BC}_{AD}(\alpha)=\Lambda^{BC}_{AD}(\underline{\alpha}).$ 
Note also that $\stackrel{\alpha}{\VV}$ is a module equivalent to   
 $\stackrel{\beta}{\VV}$ if $\beta=(\alpha^l+i\frac{\pi}{\ln q}, \alpha^r+i\frac{\pi}{\ln q})$ because  
 $\Lambda^{BC}_{AD}(\alpha)$ depends only on $q^{2\alpha^l+1} $ and $q^{2\alpha^r+1}.$ As a result we can always assume that   
$\rho_\alpha\in ]+\frac{\pi}{\ln q}, -\frac{\pi}{\ln q}].$

We can endow $\stackrel{\alpha}{\VV}$ with a structure of pre-Hilbert space by defining the hermitian form $<.,.>$  such   
that the basis   $\{{\buildrel {{}_{C}} \over e}_r(\alpha), C-\vert m_{\alpha} \vert \in \halfinteger, r=-C,\dots,C\}$ of   
$\stackrel{\alpha}{\VV}$ is orthonormal.     
     
Representations of the principal series are unitary in the sense that   
$\forall v, w\in \stackrel{\alpha}{\VV},\;\; \forall a\in  {U}_q(sl(2,\mathbb{C})_{\mathbb{R}}),$ $ <a^{*}v,w>=<v, aw>,$  
 this last property being equivalent   to the relation:   
$\Lambda^{BC}_{AD}(\tilde{\alpha})={\overline {\Lambda^{BC}_{AD}(\alpha)}}.$      
\medskip   
   
When $\alpha\in \SSP$, we  will denote by $\stackrel{\alpha}{\mathbb{H}}$  the separable Hilbert space, completion of  $\stackrel{\alpha}{\VV}$ which Hilbertian basis is      
 $\{{\buildrel {{}_{C}} \over e}_r(\alpha), C-\vert m_{\alpha} \vert \in \halfinteger, r=-C,\dots,C\}.$\medskip

Let us now recall some basic facts about the algebra of functions on $SL_q(2,\CC)_{\RR}$ \cite{PW, BR3}. We will use the notations of \cite{BR3}.    
 The space of compact supported functions on the quantum    
 Lorentz group, denoted $Fun_c(SL_q(2,\CC)_{\RR})$ is,  by definition,     
$Fun(SU_q(2)') \otimes \left( \bigoplus_{ I \in \onehalf \ZZ^{+}} End(\CC^{d_I})\right).$ This is a $C^{*}$ algebra without unit. It contains the dense *-subalgebra \cite{BR1}    
 $Fun_{cc}(SL_q(2,\CC)_{\RR})=Pol(SU_q(2)')\otimes \left( \bigoplus_{ I \in \onehalf \ZZ^{+}} End(\CC^{d_I})\right)$ which is a multiplier Hopf algebra \cite{VD}, and which can be understood as being the quantization of the algebra generated by  polynomials functions on $SU(2)$ and compact supported functions on $AN(2).$ 
    
$(\kE{C}{m}{n}{D}{p}{q})_{C,D,m,n,p,q}$ is a vector  basis of  $Fun_{cc}(SL_q(2,\CC)_{\RR})$ which is defined, for example,  by duality from the generators of the envelopping algebra:     
\begin{eqnarray}    
&&<{\buildrel {{}_{A}} \over  L}{}^{(\pm)}{}^i_j \otimes {\buildrel {{}_{B}} \over  g}{}^k_l, \;\kE{C}{m}{n}{D}{p}{q}>=    
{\buildrel {{}_{AC}}\over  R}{}^{(\pm)}{}^{im}_{jn}\delta_{B,D}\delta^{k}_{q}\delta^{p}_{l}.    
\end{eqnarray}    
We can describe completely the structure of the multiplier Hopf algebra in this basis;    
\begin{eqnarray}    
&&\hskip -1.5cm\kE{A}{i}{j}{B}{k}{l} \cdot \kE{C}{m}{n}{D}{p}{q}= \sum_{Frs} \Clebphi{A}{C}{F}{i}{m}{r}\Clebpsi{A}{C}{F}{j}{n}{s} \delta^p_l     
\delta_{B,D} \kE{F}{r}{s}{B}{k}{q}\nonumber\\    
&&\hskip -1.5cm\Delta(\kE{A}{i}{j}{B}{k}{l})={\cal F}^{-1}_{23}\;    
(\!\!\!\!\sum_{\substack{{C,D,m,}\\{p,q,r,s}}}\!\!\!\!\Clebphi{C}{D}{B}{q}{s}{l}\!\!\Clebpsi{C}{D}{B}{p}{r}{k}\!\kE{A}{i}{m}{C}{p}{q} \!\otimes \!    
 \kE{A}{m}{j}{D}{r}{s}\;)\;{\cal F}_{23}\nonumber\\    
&&\hskip -1.5cm\epsilon(\kE{A}{i}{j}{B}{k}{l})=\delta^i_j \delta_{B,0}\;\;\;\;\;\;\;(\kE{A}{i}{j}{B}{k}{l})^{\star}=S^{-1}({\buildrel A \over k}{}^j_i) \! \otimes     
\! {\buildrel B \over E}{}^l_{k}\;\;\;\;\mbox{with}\;\;    
{\cal F}^{-1}_{12}=\sum_{J,x,y}\!\!{\buildrel J \over E}{}^x_y \! \otimes \! S^{-1}\!({\buildrel J \over k}{}^y_x). 
\end{eqnarray}    
The space of right and left invariant  linear forms (also called Haar measures) on $Fun_c(SL_q(2,\CC)_{\RR})$  is a vector space of dimension one and we will pick one element $h$, which is defined by:    
\begin{eqnarray}    
h(\kE{A}{i}{j}{B}{m}{l})= \delta_{A,0}({\buildrel {{}_{B}} \over \mu}{}^{-1})^m_l \qd{B}.\label{haar}    
\end{eqnarray}    
This Haar measure can be used to define an hermitian form on $Fun_c(SL_q(2,\CC)_{\RR}):$  
\begin{equation}  
\forall f,g \in Fun_c(SL_q(2,\CC)_{\RR}), \;\;\;<f,g>=h(f^{\star}g).\label{L2hermitian}  
\end{equation}  
Using the associated $L^2$ norm, $\vert \vert \; a \; \vert \vert_{L^2}=h(a^* a)^{\onehalf}$, we can complete the space  $Fun_c(SL_q(2,\CC)_{\RR})$   into the Hilbert space of $L^2$ functions on the quantum Lorentz group, denoted $L^2(SL_q(2,\CC)_{\RR})$.    
    
 $Fun_{cc}(SL_q(2,\CC)_{\RR})$ is a multiplier Hopf algebra with basis $(u_i)=((\kE{C}{m}{n}{D}{p}{r})_{C,D,m,n,p,r})$. The restricted dual of $Fun_{cc}(SL_q(2,\CC)_{\RR})$, denoted  ${\tilde   U}_q(sl(2,\CC)_{\RR}),$   is the vector space spanned by the dual basis   
 $(u^{i})=   
({\buildrel {{}_{C}}\over X }{}^{n}_m \otimes {\buildrel {{}_{D}}\over g}{}^r_p).$  It is also, by duality, a multiplier Hopf algebra and   ${  U}_q(sl(2,\CC)_{\RR} )$ is included as an algebra in the  multiplier algebra $M({\tilde   U}_q(sl(2,\CC)_{\RR} ))$.      
If $\Pi$ is the  principal representation  of ${  U}_q(sl(2,\CC)_{\RR} )$, acting on $\stackrel{\alpha}{\VV},$ it is possible \cite{BR1} to associate to it a unique representation ${\tilde \Pi}$ of   ${\tilde   U}_q(sl(2,\CC)_{\RR} ),$ acting on  
 $\stackrel{\alpha}{\VV}$, such that    
 ${\tilde \Pi}({\buildrel {{}_{C}}\over X }{}^{n}_m)({\buildrel {{}_{D}} \over e}_r)=\delta_{C,D}\delta^{n}_{r}{\buildrel {{}_{D}} \over e}_m.$   
    
We define for all $f$ element of  $Fun_{cc}(SL_q(2,\CC)_{\RR})$, the operator  $\Pi(f)=\sum_i {\tilde\Pi}(u^i)\; h(u_i \; f).$ It is easy to show that  $\Pi(f)$ is of finite rank and of finite corank.

The matrix elements of the representation $\stackrel{\alpha}{\Pi}$ are  the linear form on $U_q(sl(2,\CC)_{\RR})$ which expression is:  
\begin{eqnarray}  
\stackrel{\alpha}{\GG}{}^{Ai}_{Bj} = \sum_{MD} \Clebphi{A}{D}{M}{m}{r}{x} \Clebpsi{D}{B}{M}{s}{j}{x} \Lambda^{DM}_{AB} (\alpha) \stackrel{A}{k}{}^i_m \otimes \stackrel{D}{E}{}^s_r\;. \label{matrixelementofrepresentation} 
\end{eqnarray}  
 We have a natural inclusion $Fun_{cc}(SL_q(2,\CC)_{\RR}) \stackrel{\iota}{\rightarrow} (Fun_{cc}(SL_q(2,\CC)_{\RR}))^{*}$ defined thanks to the Haar measure $h$ on the quantum group (\cite{BR3}) by: if $f \in Fun(SL_q(2,\CC))$ we have $\iota(f)(a) = h(fa), \; \forall a \in Fun_{cc}(SL_q(2,\CC))$. $\iota$ extends naturally to $(U_q(sl(2,\CC)_{\RR})^*$ and we trivially have: 
$\stackrel{\alpha}{\Pi}(f)^{Ai}_{Bj}=<\iota(\stackrel{\alpha}{\GG}{}^{Ai}_{Bj}),f>. $

\medskip   
   
If $f$ is a function on ${\mathbb S}_P$, we will define   $f_m$ to be the function  
defined by $f_m(\rho)= f(\alpha(m,\rho)).$      
We will denote   
\begin{equation}  
 \int_{\SSP} d(\alpha) \; f(\alpha)=    
\!\!\!\sum_{m \in \onehalf \ZZ} \; \frac{2\pi}{\vert \ln q\vert }  
\int_{\frac{\pi}{\ln q}}^{\frac{-\pi}{\ln q}} \!\!\!d\rho\;\;    
\; f_m(\rho).   
\end{equation}

The Plancherel  formula can be written as:    
\begin{eqnarray*}     
\hskip -0.6cm \forall \psi\in Fun_{cc}(SL_q(2,\CC)_{\RR}),\;\; \vert \vert \; \psi \;    
 \vert \vert_{L^2}^2  = \; \int_{\SSP}d(\alpha)  {\cal P}(\alpha)  \;\;    
{\rm tr} ( \; \stackrel{\alpha}{\Pi}(\mu^{-1})  \stackrel{\alpha}{\Pi}(\psi)   
(\stackrel{\alpha}{\Pi}(\psi))^{\dagger})  
 \end{eqnarray*}    
where  we have denoted    
$    
{\cal P}(m,\rho)=    
(q-q^{-1})^2\;[m+i\rho][m-i\rho].$

\subsection*{A.2 Tensor Product and Clebsch-Gordan maps.}   
The aim of this section is to give explicit formulae for the intertwiners (Clebsch-Gordan maps) between the representation $\stackrel{\alpha}{\Pi} \otimes  \stackrel{\beta}{\Pi}$ and the representation $\stackrel{\gamma}{\Pi}$, in term of complex continuations of 6-j symbols of $U_q(su(2))$, where $\alpha, \beta, \gamma$  belong to ${\mathbb S}_P$ or to ${\mathbb S}_F$ .  
Let us recall that the tensor product of these representations decomposes as:  
  
\begin{eqnarray}  
\stackrel{\alpha}{\VV} \; \otimes \;  
 \stackrel{\beta}{\VV} & =  
 & \bigoplus_{\gamma^l= \vert \alpha^l - \beta^l \vert ,  
 \;\gamma^r = \vert \alpha^r - \beta^r \vert}  
^{\alpha^l + \beta^l, \gamma^r =\alpha^r+\beta^r}  
 \stackrel{\gamma}{\VV},\;\;\; \text{where}\;\;\;   
  \alpha, \beta  \in {\mathbb S}_F  \;\; , 
 \label{tensoroffiniteandfinite}\\  
\stackrel{\alpha}{\VV} \; \otimes \; \stackrel{\beta}{\VV} & = &   
\bigoplus_{m=-\alpha^l, n=-\alpha^r}^{m=\alpha^l, n=\alpha^r}  
 \stackrel{(m,n)+\beta}{\VV} \;\; , \;\; \alpha \in {\mathbb S}_F \; , \; \beta  \in {\mathbb S}  ,  
\label{tensoroffiniteandinfinite}\\   
\stackrel{\alpha}{\HH} \; \otimes \; \stackrel{\beta}{\HH} & = &   
\bigoplus_{m_\gamma \in J_{m_\alpha,m_\beta}} \int^{\oplus} d\rho_\gamma \;  
 \stackrel{\gamma}{\HH} \;\; , \;\; \alpha, \beta, \gamma \; \in {\mathbb S}_P \;\; \label{tensorofinfiniteandinfinite},    
\end{eqnarray}  
where $J_{m,n}= \lbrace p \in \frac{1}{2} \ZZ ,\; m+n+p \in \ZZ \rbrace$.   
 
Let $(\alpha,\beta,\gamma)\in {\mathbb S}^3,$   
the intertwiner $\Psi_{\alpha,\beta}^{\gamma}:  
\stackrel{\alpha}{\VV} \otimes \stackrel{\beta}{\VV} \longrightarrow \stackrel{\gamma}{\VV}$ is defined in \cite{BR4} by:  
\begin{eqnarray}  
\Psi_{\alpha,\beta}^{\gamma}(\stackrel{A}{e}_i (\alpha) \otimes \stackrel{B}{e}_j (\beta)) =   
\sum_{C,k} \stackrel{C}{e}_k (\gamma) \Clebpsi{\alpha}{\beta}{\gamma}{Ai}{Bj}{Ck}= 
 \sum_{C,k} \stackrel{C}{e}_k (\gamma) \Clebpsi{A}{B}{C}{i}{j}{k} \ElemRedpsi{C}{\alpha}{\beta}{\gamma}{A}{B} \;\label{sl2cintertwiner}   
\end{eqnarray}  
where the reduced elements $ \ElemRedpsi{C}{\alpha}{\beta}{\gamma}{A}{B}$ are defined by:  
\begin{eqnarray}  
 \ElemRedpsi{C}{\alpha}{\beta}{\gamma}{A}{B} \; = \;  
 \sqrt{[d_B]} e^{-i\pi B} 
 \sum_{p} 
 \sixj{\beta^l}{\alpha^l}{\gamma^l}{C}{\gamma^r}{\alpha^l+p}   
\sixj{\alpha^r}{\alpha^l}{A}{C}{B}{\alpha^l+p}  
\sixj{\gamma^r}{\alpha^l+p}{\beta^l}{B}{\beta^r}{\alpha^r} \nonumber \\  
\frac{v_{\alpha^r}^{1/2}}{v_{\alpha^l+p}^{1/2}} \frac{v_{\alpha^l}^{1/4}}{v_{\alpha^r}^{1/4}}  
 \frac{v_{\beta^l}^{1/4}}{v_{\beta^r}^{1/4}} \frac{v_{\gamma^r}^{1/4}}{v_{\gamma^l}^{1/4}} 
 \frac{v_B^{1/4} v_C^{1/4}}{v_A^{1/4}} 
q^{(\alpha^r+\beta^l-\gamma^r)} e^{i \pi (\alpha^r+\beta^l-\gamma^r)} \frac{\nu_1(d_{\gamma^r})}{\nu_1(d_{\alpha^r}) \nu_1(d_{\beta^l})}  \label{ElementReduitPsi} 
 \; ,  
\end{eqnarray}  
where $\nu_1$ is  defined in the appendix of \cite{BR4}, and is such that $\nu_1(x)^2=1-q^{2x},$ and the $6j$ coefficients are of type $0, 1$ or $3$ depending on the nature of  $\alpha,\beta,\gamma$ and their expressions are given in \cite{BR4}. 
 
Remarks: 
 
1. Note that from the analysis of proposition \ref{polesofreducedelement}, the only possible singularities of this expression appear when one of the numbers  
$2\alpha^l+1, 2\beta^l+1, 2\gamma^l+1$ is an integer. We shall denote such a configuration special. This is explained from the fact that generically $\dim  \; \Hom(\stackrel{\alpha}{\VV}\otimes \stackrel{\beta}{\VV}, \stackrel{\gamma}{\VV})=1$ except possibly for the case where $2\alpha^l+1, 2\beta^l+1, 2\gamma^l+1$ is an integer.  
A couple  $(\alpha,\beta)\in {\mathbb S}^2$ is said to be of finite type if $\alpha$ or $\beta$ belongs to $\SSF.$ In this case, if $\gamma\in{\mathbb S}, 
 N_{\alpha,\beta}^{\gamma}=\dim \; \Hom(\stackrel{\alpha}{\VV}\otimes \stackrel{\beta}{\VV}, \stackrel{\gamma}{\VV})$  is zero or one dimensional. We denote by ${\mathbb S}(\alpha,\beta)$ the set of elements $\gamma\in {\mathbb S}$ such that  $N_{\alpha,\beta}^{\gamma}=1.$ 
 
2. The formula of the reduced element  gives in particular  
for  the special configuration  where $\alpha=\beta\not=(-\onehalf,-\onehalf) , \gamma=0:$ 
 $\Psi_{\alpha,\alpha}^{0}:  
\stackrel{\alpha}{\VV} \otimes \stackrel{\alpha}{\VV} \longrightarrow \CC$ with:  
\begin{eqnarray}  
\Psi_{\alpha,\alpha}^{0}(\stackrel{A}{e}_i (\alpha) \otimes \stackrel{B}{e}_j (\alpha)) =   
\Clebpsi{\alpha}{\alpha}{0}{Ai}{Bj}{00}= 
 \Clebpsi{A}{B}{0}{i}{j}{0} \ElemRedpsi{0}{\alpha}{\alpha}{0}{A}{B} \;   
\end{eqnarray}  
where the reduced element $ \ElemRedpsi{0}{\alpha}{\alpha}{0}{A}{B}$ is defined as:  
\begin{eqnarray}  
 \ElemRedpsi{0}{\alpha}{\alpha}{0}{A}{B} \; = \;  
 Y^{(1)}_{(A;m_{\alpha})}\delta_{A,B}e^{-i\pi A}\sqrt{[d_A]}  
 \frac{q^{\alpha^r+\alpha^l}e^{i \pi \alpha}}{\nu_1(d_{\alpha}) } 
.\nonumber   
\end{eqnarray}

3.   Let $(\alpha,\beta,\gamma)\in {\mathbb S}^3,$  we can define  the  
intertwiner 
 \begin{equation} 
{\tilde  \Psi}_{\alpha,\beta}^{\gamma}= 
{ \Psi}_{\alpha,\beta}^{\gamma} 
q^{(-\alpha^r-\beta^l+\gamma^r)} 
e^{-i\pi(\alpha^r+\beta^l-\gamma^r)} 
 \frac{v_{\beta^r}^{1/4}v_{\gamma^l}^{1/4}} 
{v_{\beta^l}^{1/4}v_{\gamma^r}^{1/4}} 
\frac{\nu_1(d_{\alpha^r}) \nu_1(d_{\beta^l})}{\nu_1(d_{\gamma^r})} 
\end{equation} 
which  is the intertwiner that we used in our previous work \cite{BR4}. 
 
The new  normalisation is such that when $\alpha,\beta,\gamma \in \SSF$  the following proposition holds. 
 
\begin{proposition} \label{factorizationof3j}. 
Let $I\in \SSF,$ the module $\stackrel{I}{\VV}$ decomposes as $\stackrel{I}{\VV}=\stackrel{I^l}{V}\otimes \stackrel{I^r}{V}$ according to the isomorphism ${\mathfrak U}_q(sl(2,\CC)_{\RR})={\mathfrak U}_q(sl(2))\otimes_{R^{-1}}{\mathfrak U}_q(sl(2)).$ The basis $\stackrel{A}{e}_i(I)$ defined by (\ref{repl})(\ref{repg}) is expressed as  
\begin{equation}\stackrel{A}{e}_i(I)=\Clebphi{I^l}{I^r}{A}{t}{u}{i} 
v^{\frac{1}{4}}_{I^l}v^{\frac{1}{4}}_{I^r}v^{-\frac{1}{4}}_{A}\stackrel{I^l}{e}_t\otimes\stackrel{I^r}{e}_u.\end{equation} 
The expression of the  ${\mathfrak U}_q(sl(2,\CC)_{\RR})$ intertwiner $\Psi^{K}_{IJ}=\Psi^{K^l}_{I^lJ^l}\otimes\Psi^{K^r}_{I^rJ^r}\stackrel{I^rJ^l}{R}{}^{\!\!-1}$ in this basis is exactly given by the formulas (\ref{sl2cintertwiner})(\ref{ElementReduitPsi}). 
\end{proposition} 
\Proof 
Trivial computation. 
$\Box$ 
 
We will denote by  $(\stackrel{\alpha}{\VV})^{r*}$ the restricted dual of 
$ \stackrel{\alpha}{\VV}$ defined as $(\stackrel{\alpha}{\VV})^{r*}= 
\bigoplus_{C, C-\vert \alpha^l-\alpha^r\vert\in \NN}  
({\buildrel {C}\over V})^{*}.$ It is endowed with a structure of $U_q(sl(2,\CC)_{\RR})$ module and the two modules $(\stackrel{\alpha}{\VV})^{r*}$ and $\stackrel{\alpha}{\VV}$ are isomorphic from remark 2.

Similarly we can define for  $(\alpha, \beta, \gamma)\in {\mathbb S}^3$ and $(A,B,C)\in \onehalf\ZZ^+,$ 
  
\begin{eqnarray}  
  \ElemRedphi{A}{B}{\gamma}{\alpha}{\beta}{C} =  \sqrt{[d_B]} e^{-i\pi B} 
 \sum_{p} 
 \sixj{\beta^l}{\alpha^l}{\gamma^l}{C}{\gamma^r}{\alpha^l+p}   
\sixj{\alpha^r}{\alpha^l}{A}{C}{B}{\alpha^l+p}  
\sixj{\gamma^r}{\alpha^l+p}{\beta^l}{B}{\beta^r}{\alpha^r} \nonumber \\  
\frac{v_{\alpha^l+p}^{1/2}}{v_{\alpha^r}^{1/2}} \frac{v_{\alpha^r}^{1/4}}{v_{\alpha^l}^{1/4}}  
 \frac{v_{\beta^r}^{1/4}}{v_{\beta^l}^{1/4}} \frac{v_{\gamma^l}^{1/4}}{v_{\gamma^r}^{1/4}} \frac{v_A^{1/4}}{v_B^{1/4} v_C^{1/4}} 
 q^{(\alpha^r+\beta^l-\gamma^r)} e^{i \pi (\alpha^r+\beta^l-\gamma^r)} \frac{\nu_1(d_{\gamma^r})}{\nu_1(d_{\alpha^r}) \nu_1(d_{\beta^l})} . \label{ElementReduitPhi} 
\end{eqnarray}   
 
We endow $(\stackrel{\alpha}{\VV}{}^{r*} \; \otimes \; \stackrel{\beta}{\VV}{}^{r*})^{*}$ with the  structure of $U_q(sl(2,\CC)_{\RR})$ module such that the natural inclusion  
$\stackrel{\alpha}{\VV} \; \otimes \; \stackrel{\beta}{\VV}\hookrightarrow (\stackrel{\alpha}{\VV}{}^{r*} \; \otimes \; \stackrel{\beta}{\VV}{}^{r*})^{*}$ is an interwiner. 
As a result we can define for  $\alpha,\beta,\gamma\in  
{\mathbb S}$ the intertwiner  
$\Phi^{\alpha,\beta}_{\gamma}: \stackrel{\gamma}{\VV} \longrightarrow  
(\stackrel{\alpha}{\VV}{}^{r*} \; \otimes \; \stackrel{\beta}{\VV}{}^{r*})^*$ by:  
\begin{eqnarray}  
\Phi^{\alpha,\beta}_{\gamma}(\stackrel{C}{e}_k (\gamma))& =&  
\sum_{A,i;B,j}\stackrel{A}{e}_i (\alpha) \; \otimes \;\stackrel{B}{e}_j (\beta) 
\Clebphi{\alpha}{\beta}{\gamma}{Ai}{Bj}{Ck}\nonumber\\   
&=& 
\sum_{A,i; B,j} (\stackrel{A}{e}_i (\alpha) \; \otimes \;\stackrel{B}{e}_j (\beta) )  
 \Clebphi{A}{B}{C}{i}{j}{k}  \ElemRedphi{A}{B}{\gamma}{\alpha}{\beta}{C} 
\label{sumdefinitionPhi} . 
\end{eqnarray}     
  
Note that it is only in the   case where $(\alpha, \beta)$ is of finite type that the sum is finite, and in that case $\Phi^{\alpha,\beta}_{\gamma}$ is an intertwiner from $\stackrel{\gamma}{\VV}$ to  
$\stackrel{\alpha}{\VV}\otimes \stackrel{\beta}{\VV}.$

The normalization of the intertwiners has been chosen in order to have the following orthogonality relations: 
if $(\alpha,\beta)$ is of finite type, we have  
\begin{eqnarray} 
&&\Psi^{\gamma'}_{\alpha,\beta}\Phi^{\alpha,\beta}_{\gamma}= 
N_{\alpha,\beta}^{\gamma}\delta^{\gamma'}_{\gamma}  
id_{\stackrel{\gamma}{\VV}},\\ 
&&\sum_{\gamma\in {\mathbb S}(\alpha,\beta)} 
\Phi^{\alpha,\beta}_{\gamma}\Psi^{\gamma}_{\alpha,\beta}= 
id_{\stackrel{\alpha}{\VV}\otimes \stackrel{\beta}{\VV}}. 
\end{eqnarray} 
 
\begin{proposition} 
The reduced element satisfy the following symmetries: 
\begin{eqnarray} 
\hskip-1cm \ElemRedpsi{C}{\alpha}{\beta}{\gamma}{A}{B}&=&\ElemRedphi{B}{A}{\gamma}{\beta}{\alpha}{C}\\ 
\hskip-1cm\ElemRedpsi{C}{\alpha}{\beta}{\gamma}{A}{B}&=& 
\ElemRedphi{A}{B}{\tilde\gamma}{\tilde\alpha}{\tilde\beta}{C}\\ 
\hskip-1cm\ElemRedpsi{C}{\alpha}{\beta}{\gamma}{A}{B}&=& 
q^{\alpha^r+\alpha^l-\gamma^l-\gamma^r}e^{i\pi (C-A+\alpha-\gamma)} 
\frac{[d_A]^{\onehalf}\nu_1(d_{\gamma})}{[d_C]^{\onehalf}\nu_1(d_{\alpha})} 
\ElemRedpsi{A}{\tilde\gamma}{\beta}{\tilde\alpha}{C}{B} 
\end{eqnarray} 
\end{proposition} 
\Proof 
 Left to the reader.
$\Box$ 
 
The following proposition precises the nature of the singularities of $\Psi_{\alpha,\beta}^{\gamma}$ in the variables 
 $q^{i\rho_\alpha}, q^{i\rho_\beta},q^{i\rho_\gamma}.$  
Let $\alpha,\beta,\gamma \in \mathbb{S}$ we denote  
\begin{eqnarray} 
\varsigma (\alpha,\beta,\gamma) &=& 
\nu_{\vert m_\alpha+m_\beta+m_\gamma\vert} 
(\onehalf (i\rho_\alpha+i\rho_\beta+i\rho_\gamma+1-\vert m_\alpha+m_\beta+m_\gamma\vert))\\ 
\zeta(\alpha,\beta,\gamma)&=&\varsigma(\alpha,\beta,\gamma) 
\varsigma(\underline{\alpha},\beta,\gamma) 
\varsigma(\alpha,\underline{\beta},\gamma) 
\varsigma(\alpha,\beta,\underline{\gamma}). 
\end{eqnarray} 
Note that $\varsigma (\alpha,\beta,\gamma)= 
\varsigma (\tilde{\alpha},\tilde{\beta},\tilde{\gamma}).$ 
 
\begin{proposition}\label{polesofreducedelement} 
The reduced elements $\ElemRedpsi{C}{\alpha}{\beta}{\gamma}{A}{B}$ satisfies the property: 
\begin{equation} 
\ElemRedpsi{C}{\alpha}{\beta}{\gamma}{A}{B}e^{-i\pi(\alpha+\beta-\gamma)} 
=\frac{\zeta(\alpha,\beta,\gamma)\nu_1(2\gamma^l+1)\nu_1(2\gamma^r+1)} 
{\nu^{(A)}(\alpha)\nu^{(B)}(\beta)\nu^{(C)}(\gamma)} 
P^{ABC}_{m_{\alpha}m_{\beta}m_{\gamma}}(q^{i\rho_\alpha},q^{i\rho_{\beta}},q^{i\rho_\gamma}) 
\end{equation} 
where $P^{ABC}_{m_{\alpha}m_{\beta}m_{\gamma}}$  is a  polynomial. 
\end{proposition} 
 
\Proof 
This property comes from a careful study of the expression of the reduced element in term of 6j(1) and 6j(3) given by (\ref{ElementReduitPsi}). 
 First of all from the expression of the 6j, the square roots can only be of the form $\nu_1(z+n)$, where $n\in \ZZ$ and $z\in \{2\alpha^l+1, 2\beta^l+1,2\gamma^l+1,  
\epsilon_{\alpha}\alpha^l+\epsilon_{\beta}\beta^l+\epsilon_{\gamma}\gamma^l\}$ with $\epsilon_{\alpha}, \epsilon_{\beta}, \epsilon_{\gamma} \in \{1,-1\}$.
 
We first study the behaviour in $\beta$ and $\gamma.$ It is easy to see from the explicit definition of 
 the $6j(3)$ (given in \cite{BR4}) that the behaviour in $\beta$ and in $\gamma$ is of the form  
$$\frac{\nu_1(2\gamma^l+1)\nu_1(2\gamma^r+1)} 
{\nu^{(B)}(\beta)\nu^{(C)}(\gamma)}U(\alpha,\beta,\gamma) $$ with $U$ having no singularities in $\beta^l+n, \gamma^l+n.$ Using the first and second relation  of the previous proposition, we can exchange the role of $\alpha$ and $\beta$. This explains the behaviour 
 $\frac{\nu_1(2\gamma^l+1)\nu_1(2\gamma^r+1)}{\nu^{(A)}(\alpha)\nu^{(B)}(\beta)\nu^{(C)}(\gamma)}.$ 
 
We now analyze how  it depends on the variable $\rho_\alpha+\rho_\beta+\rho_\gamma.$ 
{}From the expression of the $6j$ given in (\cite{BR4}), we obtain that the reduced element can be expressed as  
$\frac{\nu_{\infty}(\alpha^r+\beta^r+\gamma^r+2)}{\nu_{\infty}(\alpha^l+\beta^l+\gamma^l+2)} 
R(\alpha,\beta,\gamma)$ where $R(\alpha,\beta,\gamma)$ has no singularities in  
$\alpha^l+\beta^l+\gamma^l+n $ as well as  
 $\frac{\nu_{\infty}(\alpha^l+\beta^l+\gamma^l+2)}{\nu_{\infty}(\alpha^r+\beta^r+\gamma^r+2)} 
T(\alpha,\beta,\gamma)$ where $T(\alpha,\beta,\gamma)$ has no singularities in  
$\alpha^l+\beta^l+\gamma^l+n.$ 
In order to prove this we just expand the $6j$ using its definition and  the trivial symmetries of  the $6j$. 
As a result  
when $m_{\alpha}+m_{\beta}+ m_{\gamma}\geq 0, \frac{\nu_{\infty}(\alpha^r+\beta^r+\gamma^r+2)}{\nu_{\infty}(\alpha^l+\beta^l+\gamma^l+2)}=\varsigma(\alpha,\beta,\gamma)$ and we obtain the announced result. In the case where $m_{\alpha}+m_{\beta} +m_{\gamma}<0$ we use the other behaviour. This analysis therefore implies that  the reduced element is equal to $ 
\varsigma({\alpha},{\beta},{\gamma})S(\alpha,\beta,\gamma)$ with $S$ regular in $\alpha^l+\beta^l+\gamma^l+n. $ 
The analysis for the combination  $\alpha^l+\beta^l-\gamma^l$ is similar. Indeed it is easy to show using only the definition of the 6j and a trivial symmetry that the behaviour is of the form  
$\frac{\nu_{\infty}(\alpha^r+\beta^r-\gamma^r+1)}{\nu_{\infty}(\alpha^l+\beta^l-\gamma^l+1)} 
U(\alpha,\beta,\gamma)$ with $U$ regular in $\alpha^l+\beta^l-\gamma^l+n. $ 
If $m_{\alpha}+m_{\beta} -m_{\gamma}>0$ then we obtain the announced behaviour 
 $\varsigma({\alpha},{\beta},\underline{\gamma})$, if not we use the first two relations of the previous formula to exchange $\alpha,\beta,\gamma$ with $\tilde\beta,\tilde\alpha,\tilde\gamma$, and we are back to the situation $m_{\tilde\alpha}+m_{\tilde\beta} -m_{\tilde\gamma}>0.$ 
The  other combination  $-\alpha^l+\beta^l+\gamma^l$ is  deduced from the previous analysis by using the third relation of the previous proposition. Finally the combination  $\alpha^l-\beta^l+\gamma^l$ is obtained using  the  exchange of the role of $ \alpha$ and $\beta$. 
This completes the proof. 
$\Box$ 
  
In \cite{BR4} we introduced a function $M:\SSP^3\rightarrow \RR^+$, as follows: 
\begin{eqnarray} 
&&M(\alpha,\beta,\gamma)= 
\vert\frac{(1-q^2)(1)_{\infty}^2 q^{2(m_{\alpha}+m_{\beta}+m_{\gamma})}} 
{(d_{\alpha^l})_1 (d_{\beta^l})_1(d_{\gamma^l})_1 
}\vert\times \nonumber \\ 
&&\times\vert\frac{\xi(d_{\alpha^r}, d_{\beta^r},d_{\gamma^r} )}{\xi(\alpha^r\!+\!\beta^r\!+\!\gamma^r\!+\!2,\underline{\alpha}^r\!+\!\beta^r\!+\!\gamma^r\!+\!2, \alpha^r\!+\!\underline{\beta}^r\!+\!\gamma^r\!+\!2,\alpha^r\!+\!\beta^r\!+\!\underline{\gamma}^r\!+\!2)}\vert.\label{defM} 
\end{eqnarray} 
This function is such that the intertwiner: 
\begin{eqnarray} 
&&{\tilde\Psi}_{\alpha\beta}:\stackrel{\alpha}{\VV}\otimes \stackrel{\beta}{\VV}\rightarrow  
\int^{\oplus}d \gamma \stackrel{\gamma}{\HH} \nonumber \\ 
&& v\otimes w\mapsto  
(\gamma\mapsto M(\alpha,\beta,\gamma)^{\frac{1}{2}}{\tilde\Psi}_{\alpha\beta}^{\gamma}(v\otimes w)) \nonumber
\end{eqnarray} 
is an isometry.

\subsection*{A.3 Miscellaneous properties}

\begin{proposition}  
Let $I =(I^l,I^r)\in \SSF$ indexing the  finite-dimensional representation  
 $\stackrel{I }{\Pi}$ of $U_q(sl(2,\CC)_{\RR})$. The star acts on the elements of $SL_q(2,\CC)_{\RR}$ as:  
\begin{eqnarray}  
\stackrel{I}{\GG}{}\!\!^{\star} & = & \stackrel{\tilde{I}}{W} \; \stackrel{\tilde{I }}{\GG} \; \stackrel{\tilde{I}}{W}{}\!\!^{-1} \;   
\end{eqnarray}  
where we have defined $\stackrel{I}{W}{}\!\!^{Aa}_{Bb} = e^{i \pi A} \; v_A^{-1/2} \; \stackrel{A}{w}_{ab} \; \delta^A_B$.  
\end{proposition}   
  
\Proof  {}From the expression of the matrix element of the representation (\ref{matrixelementofrepresentation}), we have  
\begin{eqnarray}  
\stackrel{I}{\GG}{}\!\!^{Ai}_{Bj} = \sum_{MD} \Clebphi{A}{D}{M}{m}{r}{x} \Clebpsi{D}{B}{M}{s}{j}{x} \Lambda^{DM}_{AB} ({I}) \stackrel{A}{k}{}\!\!^i_m \otimes \stackrel{D}{E}{}\!\!^s_r\;. \nonumber 
\end{eqnarray}  
Then the star acts as:  
\begin{eqnarray}  
\stackrel{I}{\GG}{}\!\!^{Ai\;\star}_{Bj} = \sum_{MD} \Clebphi{A}{D}{M}{m}{r}{x} \Clebpsi{D}{B}{M}{s}{j}{x} \overline{\Lambda^{DM}_{AB} ({I})} S^{-1}(\stackrel{A}{k}{}\!\!_i^m) \otimes \stackrel{D}{E}{}\!\!_s^r\; .  \nonumber 
\end{eqnarray}  
Using the relations  
\begin{eqnarray}  
\overline{\Lambda^{DM}_{AB} ({ I})} & = &  e^{i \pi (2D+2M-A-B)} \Lambda^{DM}_{AB} ({I})\;,\\  
S^{-1}(\stackrel{A}{k}{}\!\!_i^m) & = & \stackrel{A}{w}{}\!\!^{-1 \; mu} \; \stackrel{A}{k}{}\!\!_u^l \; \stackrel{A}{w}{}\!\!_{li} \;,  
\end{eqnarray}  
a straigtforward computation implies  
\begin{eqnarray}  
\stackrel{I}{\GG}{}\!\!^{Ai\;\star}_{Bj} \; = \; e^{i\pi A}v_A^{-1/2} \stackrel{A}{w}_{li} \; \stackrel{\tilde{I}}{\GG}{}\!\!^{Al}_{Bu} \;  e^{-i\pi B}v_B^{1/2} \stackrel{B}{w}{}\!\!^{-1 \;ui}\;\;.  \nonumber  
\end{eqnarray}  
$\Box$
 
\begin{proposition}  
The star acts on the elements of $U_q(sl(2,\CC)_{\RR})$ as:  
\begin{eqnarray}  
\stackrel{I}{\LL}{}\!\!^{(\pm) \; \star} = \stackrel{\tilde{I}}{W} \; \stackrel{\tilde{I}}{\LL}{}\!\!^{(\pm)} \; \stackrel{\tilde{I}}{W}{}\!\!^{-1} \; . \label{staronL} 
\end{eqnarray}  
\end{proposition}  
  
\Proof 
By definition, $\stackrel{I}{\LL}{}\!\!^{(\pm)}{}^{Aa}_{Bb}  =  (\stackrel{I}{\Pi}  \otimes  id)(\RR^{(\pm)})^{Aa}_{Bb}$, then  
\begin{eqnarray}  
\stackrel{I}{\LL}{}\!\!^{(\pm)}{}^{Aa \; \star}_{Bb} \; = \; \overline{\stackrel{I}{\Pi}(x_{i})^{Aa}_{Bb}} \; y_{i}^{\star} \;\; ,\nonumber   
\end{eqnarray}  
where we have denoted $\RR =\sum_{i}x_{i} \otimes y_{i}$. The previous  proposition and the relation  $\RR^{(\pm)\;\star \otimes \star} =  \RR^{(\pm) -1} = (S \otimes id) (\RR)$ imply (\ref{staronL}). A similar proof implies as well:  
\begin{eqnarray}  
\stackrel{I}{\LL}{}\!\!^{(\pm)-1 \; \star} = \stackrel{\tilde{I}}{W}  \stackrel{\tilde{I}}{\mu}{}\!\!^{-1}  \stackrel{\tilde{I}}{\LL}{}\!\!^{(\pm)-1}  \stackrel{\tilde{I}}{\mu}   \stackrel{\tilde{I}}{W}{}\!\!^{-1}\;. \nonumber
\end{eqnarray}  
$\Box$

It is also immediate to note that the R-matrix and the ribbon-element $\mu$ satisfy the relations:  
\begin{eqnarray}  
\overline{\stackrel{{  I} {  J}}{\RR}} & = & \stackrel{ \tilde{  I}}{W} \otimes \stackrel{ \tilde{  J}}{W} \; \stackrel{\tilde{  I} \tilde{   J}}{\RR} \; \stackrel{ \tilde{  I}}{W}{}\!\!^{-1} \otimes \stackrel{ \tilde{  J}}{W}{}\!\!^{-1} \;\; , \\  
\overline{\stackrel{  I}{\mu}} & = & \stackrel{ \tilde{  I}}{W} \; \stackrel{\tilde{  I}}{\mu} \; \stackrel{ \tilde{  I}}{W}{}^{-1} \;\; .  
\end{eqnarray}  
  
In order to define the star on the moduli algebra, we also need the following proposition:  
\begin{proposition}  
The complex conjugates of the Clebsch-Gordan maps of $U_q(sl(2, \CC)_{\RR})$ satisfy:    
\begin{eqnarray}  
\overline{\Phi^{  IJ}_{  K}} & = & \lambda^{\tilde{  I} \tilde{  J}}_{\tilde{  K}} \; (\stackrel{\tilde{  I}}{W} \otimes \stackrel{\tilde{  J}}{W}) \;  \stackrel{\tilde{  I} \tilde{  J}}{\RR} \; \Phi^{\tilde{  I}\tilde{  J}}_{\tilde{  K}} \;  \stackrel{\tilde{  K}}{W}{}\!\!^{-1} \;\; , \\  
\overline{\Psi_{  IJ}^{  K}} & = & \lambda^{\tilde{  I} \tilde{  J}-1}_{\tilde{  K}} \; \stackrel{\tilde{  K}}{W} \; \Psi_{\tilde{  I}\tilde{  J}}^{\tilde{  K}} \; \stackrel{\tilde{  I} \tilde{  J}}{\RR}{}\!\!^{-1} \; (\stackrel{\tilde{  I}}{W}{}\!\!^{-1} \otimes \stackrel{\tilde{  J}}{W}{}\!\!^{-1}) \;\; ,  
\end{eqnarray}  
where we have defined  
\begin{eqnarray}  
\lambda^{\tilde{  I} \tilde{  J}}_{\tilde{  K}} \; = \; e^{i \pi (I^{r} - I^l + J^{r} - J^l + K^l - K^{r})} \frac{v_{K^{r}}^{1/2}}{v_{I^{r}}^{1/2} v_{J^{r}}^{1/2}} \frac{v_{I^l}^{1/2} v_{J^l}^{1/2}}{v_{K^l}^{1/2}} \;\; , \label{valueoflambda} 
\end{eqnarray}  
for ${  I}, {  J}, {  K} \in \mathbb S_F$.  
\end{proposition}  
\Proof  We will give the proof of the first relation, the other one is  similar.  
$ \Phi^{  IJ}_{  K}$ is an intertwiner and then for $x \; \in \; U_q(sl(2,\CC)_{\RR})$ we have  
\begin{eqnarray}  
\stackrel{  I}{\Pi}(x_{(1)}) \otimes \stackrel{  J}{\Pi}(x_{(2)}) \; \Phi^{  IJ}_{  K} \; = \; \Phi^{  IJ}_{  K} \; \stackrel{  K}{\Pi}(x)\;\;. \nonumber 
\end{eqnarray}  
We take the complex-conjugate of the previous equation and an easy computation leads to:  
\begin{eqnarray}  
\overline{\stackrel{  I}{\Pi}(x_{(1)})} \otimes \overline{ \stackrel{  J}{\Pi}(x_{(2)}) } \; \overline{\Phi^{  IJ}_{  K}} & = & \overline{\Phi^{  IJ}_{  K}} \; \overline{\stackrel{  K}{\Pi}(x)} \nonumber\;\;,\\  
 \overline{\stackrel{  I}{\Pi}(S^{-1}(x_{(2)}^{\star}))} \otimes 
 \overline{ \stackrel{  J}{\Pi}(S^{-1}(x_{(1)}^{\star})) } \; \overline{\Phi^{  IJ}_{  K}} & = & \overline{\Phi^{  IJ}_{  K}} \; \overline{\stackrel{  K}{\Pi}(S^{-1}(x^{\star}))} \nonumber\;\;,\nonumber \\  
\stackrel{\tilde{  I}}{\Pi}(x_{(1)}) \otimes \stackrel{\tilde{  J}}{\Pi}(x_{(2)}) \; (\stackrel{\tilde{  I} \tilde{  J}}{\RR}{}\!\!^{-1} \stackrel{\tilde{  I}}{W}{}\!\!^{-1} \otimes \stackrel{\tilde{  J}}{W}{}\!\!^{-1} \; \overline{\Phi^{{  I}{  J}}_{{  K}}} \;  \stackrel{\tilde{  K}}{W}) & = & (\stackrel{\tilde{  I} \tilde{  J}}{\RR}{}\!\!^{-1} \stackrel{\tilde{  I}}{W}{}\!\!^{-1} \otimes \stackrel{\tilde{  J}}{W}{}\!\!^{-1} \; \overline{\Phi^{{  I}{  J}}_{{  K}}} \;  \stackrel{\tilde{  K}}{W}) \;\stackrel{\tilde{  K}}{\Pi}(x)\;\;. \nonumber  
\end{eqnarray}  
As $\Hom(\stackrel{  K}{V};\stackrel{  I}{V} \otimes \stackrel{  J}{V})$ is at most one dimensional, there exists $\lambda^{\tilde{  I} \tilde{  J}}_{\tilde{  K}} \in \CC$ such that:  
\begin{eqnarray}  
\overline{\Phi^{{  I}{  J}}_{{  K}}} \; = \; \lambda^{\tilde{  I} \tilde{  J}}_{\tilde{  K}} \stackrel{\tilde{  I}}{W} \otimes \stackrel{\tilde{  J}}{W} \; \stackrel{\tilde{  I} \tilde{  J}}{\RR} \;\Phi^{\tilde{  I}\tilde{  J}}_{\tilde{  K}} \;  \stackrel{\tilde{  K}}{W}{}\!\!^{-1}\;\;. \nonumber  
\end{eqnarray}  
Let us show that  $\lambda^{\tilde{  I} \tilde{  J}}_{\tilde{  K}}$ is given by the expression 
(\ref{valueoflambda}).  
We have  
\begin{eqnarray}  
\overline{<\stackrel{A}{e}{}\!\!^i \otimes \stackrel{B}{e}{}\!\!^j \vert \Phi^{{  I}{  J}}_{{  K}} \vert \stackrel{C}{e}{}\!\!_k >} \; = \; \overline{ \ElemRedphi{A}{B}{K}{I}{J}{C}} \; \Clebphi{A}{B}{C}{i}{j}{k}\;\;,  \nonumber  
\end{eqnarray}  
and a direct calculation shows that  
\begin{eqnarray}  
&&<\stackrel{A}{e}{}\!\!^i \otimes \stackrel{B}{e}{}\!\!^j \vert \stackrel{{  I}}{W} \otimes \stackrel{{  J}}{W} \; \stackrel{{  I} {  J}}{\RR} \; \Phi^{{  I} {  J}}_{{  K}} \;  \stackrel{{  K}}{W}{}\!\!^{-1}   \vert \stackrel{C}{e}{}\!\!_k > = \nonumber \\  
& = & \sum_{B'} \stackrel{I}{W}{}\!\!^{Ai}_{Al} \stackrel{J}{W}{}\!\!^{Bj}_{Bm} \stackrel{IJ}{R}{}\!\!^{Al \; Bm}_{an \; B'p} <\stackrel{A}{e}{}\!\!^n \otimes \stackrel{B'}{e}{}\!\!^p \vert \Phi^{{  I} {  J}}_{{  K}} \vert \stackrel{C}{e}{}\!\!_r> \stackrel{I}{W}{}\!\!^{-1 Cr}_{Ck} \nonumber \\ 
& = &  \sum_{B'} \Lambda^{AC}_{BB'}(J)  
\ElemRedphi{A}{B'}{K}{I}{J}{C}e^{i\pi(A+B-C)} \Clebphi{A}{B}{C}{i}{j}{k} \;\nonumber\text{\;\;(using the expressions of $R$ and $\Phi$)}\\  
&=& \sum_{B'MN} \sixj{J^r}{J^l}{B}{A}{C}{N} \sixj{J^l}{J^r}{B'}{A}{C}{N} \frac{v_{J^r}}{v_N} \frac{v_A^{1/2} v_C^{1/2}}{v_B^{1/4} v_{B'}^{1/4}}   
 e^{i\pi(A+B-C)} \frac{v_{K^r}^{1/4} v_{K^l}^{1/4}}{v_C^{1/4}}  \frac{v_A^{1/4}}{v_{I^r}^{1/4} v_{I^l}^{1/4}} \frac{v_{B'}^{1/4}}{v_{J^r}^{1/4} v_{J^l}^{1/4}}  \nonumber \\  
&  &\;\;\;\;\;\;\;\;\;\;\sixj{J^l}{J^r}{B'}{A}{C}{M}   
\sixj{J^r}{C}{M}{K^r}{I^l}{K^l} \sixj{J^r}{I^r}{K^r}{I^l}{N}{A} \frac{v_{I^r}^{1/2} v_N^{1/2}}{v_{K^r}^{1/2} v_A^{1/2}} \Clebphi{A}{B}{C}{i}{j}{k} \;\nonumber\\  
&=&\sum_M \sixj{J^l}{J^r}{B}{A}{C}{M} \sixj{J^l}{C}{M}{K^r}{I^l}{K^l} \sixj{J^r}{I^r}{K^r}{I^l}{M}{A} \nonumber \\ 
&&\;\;\;\;\;\;\;\;\;\;\frac{v_{J^r}^{1/2} }{v_M^{1/2}} \frac{v_C^{1/4} v_A^{1/4}}{v_B^{1/4}}  \frac{v_{J^r}^{1/4} v_{I^r}^{1/4} v_{K^l}^{1/4}}{v_{J^l}^{1/4} v_{I^l}^{1/4} v_{K^r}^{1/4}}e^{i\pi(A+B-C)}  
 \Clebphi{A}{B}{C}{i}{j}{k} \;\nonumber\text{\;\;(using orthogonality on $6j(0)$)}\\ 
& = & \sum_M \sixj{J^r}{J^l}{B}{A}{C}{M} \sixj{J^r}{C}{M}{K^l}{I^r}{K^r} \sixj{J^l}{I^l}{K^l}{I^r}{M}{A} \text{(using Yang-Baxter equation)}\nonumber \\ 
&&\;\;\;\;\;\;\;\;\;\; \frac{v_B^{1/4} v_C^{1/4}}{v_A^{1/4}} \frac{v_M^{1/2}}{v_{J^l}^{1/2}} \frac{v_{J^r}^{1/4} v_{I^r}^{1/4} v_{K^l}^{1/4}}{v_{J^l}^{1/4} v_{I^l}^{1/4} v_{K^r}^{1/4}}e^{i\pi(A+B-C)} \Clebphi{A}{B}{C}{i}{j}{k} \nonumber 
\end{eqnarray}  
As a result, from the defining relation of the reduced element given in the appendix A.2, one obtains the announced value (\ref{valueoflambda}) for $\lambda^{\tilde{  I} \tilde{  J}}_{\tilde{  K}}$. $\Box$ \\

The $6j$-symbols of $U_q(sl(2,\CC)_{\RR})$ when it includes at least a finite-dimensional representation are easily defined and  are  expressed as a product of two  continuations of $6j$-symbols of  $U_q(su(2))$.

\begin{definition}{\label{definition6j}} 
Let $I\in \SSF, \alpha,\beta,\gamma \in {\mathbb S},$ the space of intertwiners  
$\Hom(\stackrel{I}{\VV}\otimes\stackrel{\alpha}{\VV}\otimes \stackrel{\beta}{\VV}, \stackrel{\gamma}{\VV} )$ is a finite dimensional space of dimension at most $\dim \stackrel{I}{\VV}.$ The non zero elements   
$\Psi^{\gamma}_{\alpha'\beta} \Psi^{\alpha'}_{I\alpha}$ with $\alpha'\in {\mathbb S }(I,\alpha)$ form a basis of this space. The intertwiner $\Psi^{\gamma}_{I\gamma'} \Psi^{\gamma'}_{\alpha\beta}$ with 
 $\gamma'\in {\mathbb S }(I,\gamma)$ can be expressed in this basis and the components are the $6j$ coefficients of $U_q(sl(2,\CC)_{\RR})$ when it includes at least a finite-dimensional representation: 
 
\begin{equation} 
\Psi^{\gamma}_{I\gamma'} \Psi^{\gamma'}_{\alpha\beta}= 
\sum_{\alpha'\in {\mathbb S }(I,\alpha)} 
\sixj{\beta}{\gamma}{\alpha'}{I}{\alpha}{\gamma'} \Psi^{\gamma}_{\alpha'\beta} \Psi^{\alpha'}_{I\alpha}.\label{definition6jsl2c} 
\end{equation} 
 
\end{definition}

\begin{proposition}\label{factorisation6j} 
These  $6j$-symbols of $U_q(sl(2,\CC)_{\RR})$  are expressed as   a product of two  continuations of $6j$-symbols of $U_q(su(2))$:  
\begin{eqnarray} 
 \sixj{J}{L}{M}{I}{K}{N} & =   
& \sixj{J^l}{L^l}{M^l}{I^l}{K^l}{N^l}_{(0)}   
\sixj{J^r}{L^r}{M^r}{I^r}{K^r}{N^r}_{(0)} \label{6j0ofsl2c}\;\; \\  
 \sixj{\beta}{\beta'}{M}{I}{K}{\beta'} 
& = & \sixj{\beta^l}{\beta'{}^l}{M^l}{I^l}{K^l}{\beta''{}^l}_{(1)}   
\sixj{\beta^r}{\beta'{}^r}{M^r}{I^r}{K^r}{\beta''{}^r}_{(1)} \;\;\label{6j1ofsl2c} \\  
\sixj{\beta}{\gamma}{\alpha'}{I}{\alpha}{\gamma'} & =   
& \sixj{\beta^l}{\gamma^l}{\alpha'{}^l}{I^l}{\alpha^l}{\gamma'{}^l}_{(3)}   
\sixj{\beta^r}{\gamma^r}{\alpha'{}^r}{I^r}{\alpha^r}{\gamma'{}^r}_{(3)}\;\; \label{6j3ofsl2c} 
\end{eqnarray}  
where $I,J,K,L,M,N \in \mathbb S_F$ and $\alpha, \beta, \gamma,\alpha', \beta',\gamma', \beta'',  \in {\mathbb S}$ with the constraint that $\rho_{x}=\rho_{x'}$ with $x=\alpha, \beta, \beta', \gamma$. 
 \end{proposition}

\Proof  
The first relation (\ref{6j0ofsl2c}) is a trivial application of  proposition  
\ref{factorizationof3j}. 
The two other relations follows from standard continuation arguments. $\Box$

\begin{proposition}  
Let $\stackrel{\alpha}{\Pi}$ be an irreducible unitary representation of   
$U_q(sl(2,\CC)_{\RR}),$  labelled by the couple $\alpha\in \SSP.$  The   
explicit formula for $\vartheta_{I\alpha}$ where $I\in \SSF$ is  
\begin{equation}  
\vartheta_{I\alpha}=\frac{[(2I^l+1)(2\alpha^l+1)]}{[2\alpha^l+1]}\frac{[(2I^r+1)(2\alpha^r+1)]}{[2\alpha^r+1]}.  
\end{equation}  
\end{proposition}  
\Proof  
  
\begin{eqnarray}  
tr_{I} (\stackrel{  I}{\mu} \; \stackrel{  I \alpha}{\RR}  \; \stackrel{\alpha   I}{\RR})^{Cc}_{Cc} & = & \sum_{A} \; \stackrel{  I}{\mu}{}\!\!^{Aa}_{Aa} \; \stackrel{  I \alpha}{\RR}{}\!\!^{Aa \; Cc}_{Ab \; Cd} \; \stackrel{ \alpha   I}{\RR}{}\!\!^{Cd \; Ab}_{Cc \; Aa} \nonumber \\  
& = & \sum_{AFG} \stackrel{  A}{\mu}{}\!\!^{a}_{a} \Clebphi{C}{A}{F}{c}{a}{f} \Clebpsi{A}{C}{F}{b}{d}{f} \Lambda^{AF}_{CD}(\alpha) \nonumber \\  
& & \;\;\;\;\;\;\;\;\;\;\;\;\;\; \Clebphi{A}{C}{G}{b}{d}{g} \Clebpsi{C}{A}{G}{c}{a}{g} \Lambda^{DG}_{A}(I) \nonumber \\  
& = & \sum_{AF}  \frac{[ d_F ]}{[d_C]}\Lambda^{AF}_C(\alpha) \Lambda^{CF}_A(I) \;\;.\nonumber   
\end{eqnarray}  

{}From the linear relation (101) of \cite{BR3}, we deduce the following relation (when one takes $A=D$ and sums over $C$):  
\begin{eqnarray}  
\sum_{MN} \Lambda_A^{MN}(\alpha) \Lambda_M^{AN}(I) [d_N]=  
 \frac{[(2I^l+1)(2\alpha^l+1)]}{[2\alpha^l+1]}  
\sum_C [d_C]\Lambda_A^{I^rC}(\alpha) \frac{v_A^{1/2} v_{I^r}^{1/2}}{v_C^{1/2}} \;\;. \nonumber 
\end{eqnarray}  
As a result, we have  
\begin{eqnarray}  
tr_{  I} (\stackrel{  I}{\mu} \; \stackrel{  I \alpha}{\RR}  \; \stackrel{\alpha   I}{\RR})^{Cc}_{Cc} \; =  
  \frac{[(2I^l+1)(2\alpha^l+1)]}{[2\alpha^l+1]}  
 \frac{1}{[d_C]} \sum_A [d_A] \Lambda^{I^rA}_{C}(\alpha) \frac{v_C^{1/2} v_{I^r}^{1/2}}{v_A^{1/2}}\;\;.  \nonumber
\end{eqnarray}  
{}From the lemma 6 of \cite{BR3}, we finally have  
\begin{eqnarray}  
\frac{1}{[d_C]} \sum_A [d_A] \Lambda^{I^rA}_{C}(\alpha) \frac{v_C^{1/2} v_{I^r}^{1/2}}{v_A^{1/2}} \; =  \frac{[(2I^r+1)(2\alpha^r+1)]}{[2\alpha^r+1]}\;\;.  
\end{eqnarray}  \nonumber
 $\Box$

\begin{definition} 
\label{factorizationlambda} 
Let $A,B,C,D$ elements of $\SSF$,  $\alpha \in {\mathbb S},$ and  
$s\in {\cal S}$ 
we define  
\begin{eqnarray} 
\Lambda^{BC}_{AD}(\alpha,s)&=&  
\sum_{t\in {\cal S}} \!    
\sixj{\alpha+s}{\alpha}{A}{B}{C}{\alpha+t} 
 \!\sixj{\alpha+s}{\alpha}{D}{B}{C}{\alpha+t} 
\frac{v_{\alpha+t}}{v_{\alpha}}     
\frac{v^{1/4}_{A}v^{1/4}_{D}}{v^{1/2}_{B}v^{1/2}_{C}}\\ 
&=&\Lambda^{B^lC^l}_{A^lD^l}(\alpha^l+s^l,\alpha^l) 
\Lambda^{B^rC^r}_{A^rD^r}(\alpha^r,\alpha^r+s^r). 
\end{eqnarray} 
\end{definition}

\section*{Appendix B: Graphical proofs}  
\begin{figure}  
\psfrag{Ai}{$Ai$}  
\psfrag{Bj}{$Bj$}  
\psfrag{Ck}{$Ck$}  
\psfrag{Dl}{$Dl$}  
\psfrag{Phi}[][]{\;\;\;\;\;\;\;\;\;$\Clebphi{\alpha}{\beta}{\gamma}{Ai}{Bj}{Ck}$}  
\psfrag{Psi}{$\Clebpsi{\alpha}{\beta}{\gamma}{Ai}{Bj}{Ck}$}  
\psfrag{al}{$\alpha$}  
\psfrag{be}{$\beta$} 
\psfrag{ga}{$\gamma$} 
\psfrag{R-1}{$\stackrel{\alpha\beta}{\RR}{}^{-1}{}^{Ai\;Bj}_{Dl\;Ck}$} 
\psfrag{R}{$\stackrel{\alpha\beta}{\RR}{}^{'}{}^{Ai\;Bj}_{Dl\;Ck}$} 
\psfrag{W}{$\stackrel{\alpha}{\mathbb W}{}^{Ai}_{Bj}$} 
\psfrag{W-1}{$\stackrel{\alpha}{\mathbb W}{}^{-1}{}^{Ai}_{Bj}$} 
\centering 
\includegraphics[scale=0.7]{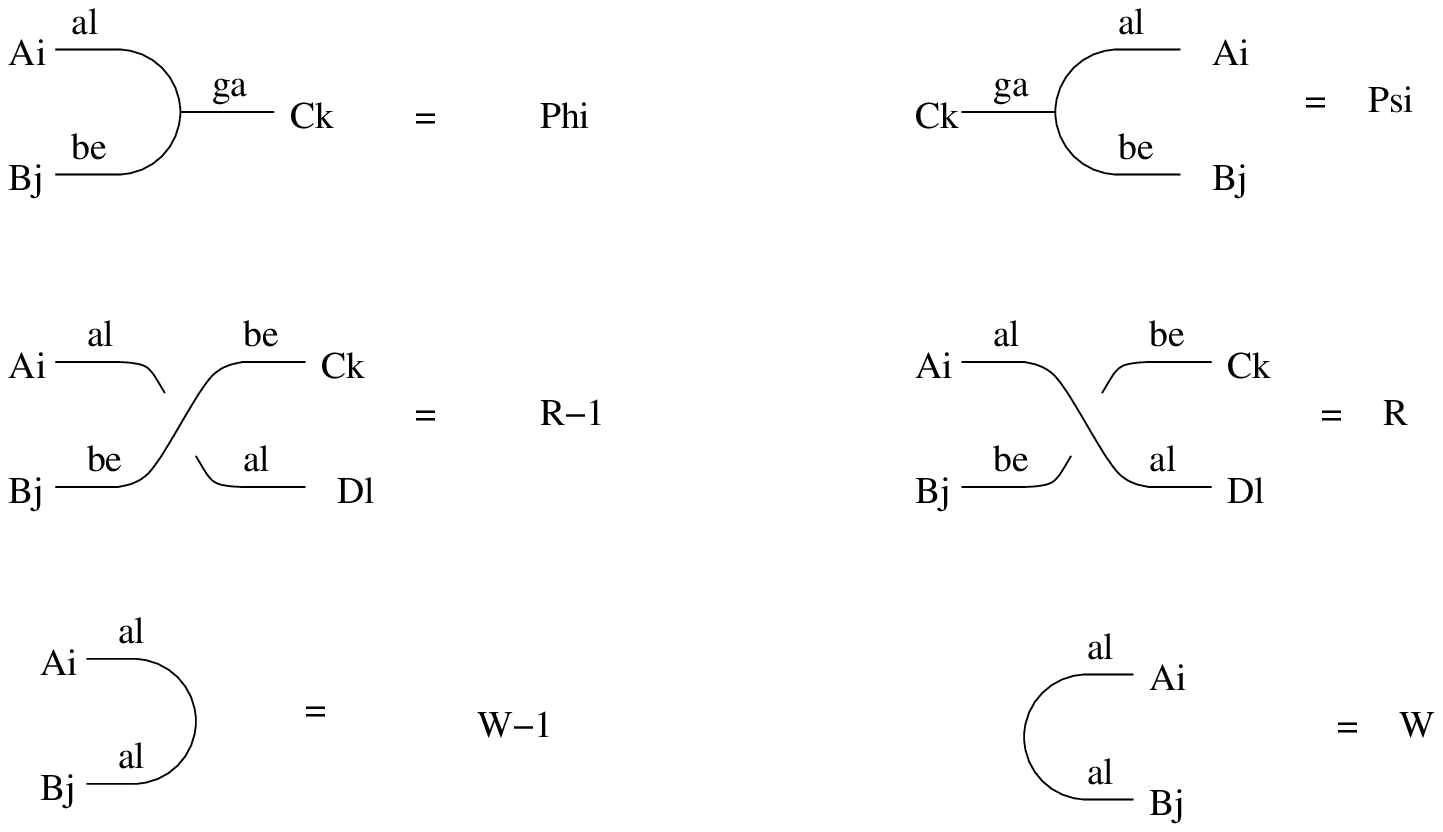} 
\caption{Pictorial representation.}  
\label{fig:picturalrepresentation} 
\end{figure}

This appendix is devoted to technical parts of the article that are proved using graphical methods. 
 We need first to introduce our conventions on  the graphical description of  intertwiners  of $U_q(sl(2,\CC)_{\RR})$ which are summarized in figure  
\ref{fig:picturalrepresentation}.  
  
\subsection*{B.1 The p-punctured sphere}  
We will present in this subsection an  explicit expression of the function $\stackrel{P}{K}{}\!\!_{0,p}^{(\pm)} \left(\begin{array}{c} \alpha \\ \beta,s \end{array} \right)$ in terms of $6j$ symbols. 
We have used a pictorial representation of  this function in the case $p=4$. The case of $p$-punctures is a straighforward generalisation. This enables us to give an explicit expression for the general case which is described in the following  proposition. 
We need to introduce new functions at this stage. 
If $P$ is the palette $P=(IJNKLUTW)$, with  
$I=J=K=L=\emptyset$ in the  p-punctured sphere case, we denote  
$P(i)$ the quintuplet  $(N_i, U_i, U_{i+1}, T_{i}, T_{i+1}).$ 
Let $Q=(N,U,U',T,T')\in \SSF^5$, $\alpha,\beta,\beta'\in {\mathbb S}, s, t, s',t'\in {\cal S}$, we define the function: 
 
\begin{eqnarray} 
 &&\hskip-0.8cm\stackrel{Q}{\cal K} 
{}^{(\pm)}\left(\begin{array}{c} \alpha \\ \beta,\beta' \end{array} 
\vline \begin{array}{c} t,t' \\ s,s' \end{array}  \right)= 
\sum_{a,b,c,d\in {\cal S}^4} \frac{v_{N}v_{\alpha}}{v_{\alpha+a}} 
 \frac{v_{\beta'+s'}^{1/2}}{v_{\beta'+c}^{1/2}}  
\big(\frac{v_{T'}^{1/2}}{v_{T}^{1/2}v_{N}^{1/2}} 
 \frac{v_{\beta'+b}^{1/2} v_{\beta'+d}^{1/2}}{v_{\beta'+c}^{1/2} v_{\beta'+t'}^{1/2}}\big)^{\pm1}\sixj{\beta'}{\beta'\!+\!t'}{\;\;\;U'}{N}{\;\;\;\;\;U}{\beta'\!+\!b}   \nonumber \\ 
&&\hskip0.3cm 
\sixj{\alpha}{\beta'\!+\!b}{\beta\!+\!t}{U}{\;\;\;\beta}{\;\beta'}  
\sixj{\alpha}{\beta'\!+\!b}{\;\beta\!+\!t}{T}{\;\beta \!+\! s}{\beta'\!+\!c}\sixj{\beta\!+\!s}{\beta'\!+\!d}{\;\alpha\!+\!a}{\;\;\;N}{\;\;\;\alpha}{\beta'\!+\!c}\sixj{\beta\!+\!s}{\beta'\!+\!d}{\;\;\alpha\!+\!a}{\;\;\;N}{\;\;\;\alpha}{\beta'\!+\!s'} \nonumber \\ 
&&\hskip0.3cm 
 \sixj{\beta'\!+\!s'}{\beta'\!+\!t'}{\;\;T'}{\;\;\;\;\;T}{\;\;\;\;\;N}{\beta'\!+\!d} 
\sixj{T}{\beta'\!+\!t'}{\beta'\!+\!d}{N}{\;\beta'\!+\!c}{\beta'\!+\!b}\;\;. 
\end{eqnarray}

\begin{figure}  
\centering 
\includegraphics[scale=0.7]{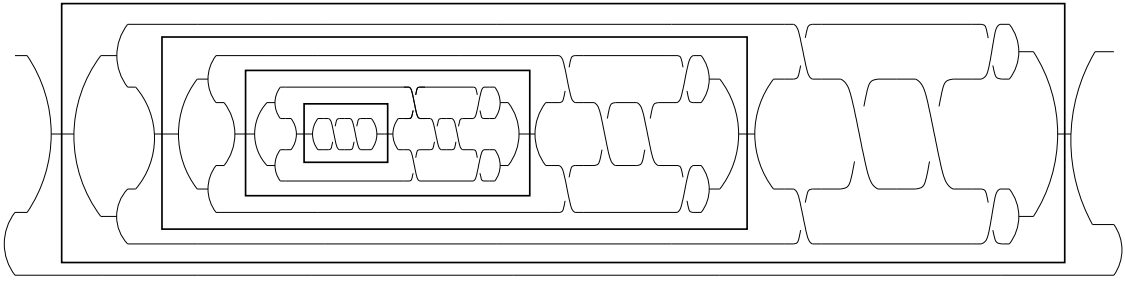} 
\caption{Iterative decomposition of $\stackrel{P}{K}{}\!\!_{0,p}^{(\pm)}$.}  
\label{fig:decomposition} 
\end{figure}

\begin{figure}  
\psfrag{al}{$\alpha$} 
\psfrag{be}{$\beta$} 
\psfrag{bep}{$\beta'$} 
\psfrag{n}{$N$} 
\psfrag{u}{$U$} 
\psfrag{up}{$U'$} 
\psfrag{T}{$T$} 
\psfrag{Tp}{$T'$} 
\psfrag{bes}{$\beta\!+\!s$} 
\psfrag{bepsp}{$\beta'+s'$} 
\psfrag{bet}[][]{$\beta+t$} 
\psfrag{beptp}[][]{$\beta'\!+\!t'$} 
\centering 
\includegraphics[scale=0.5]{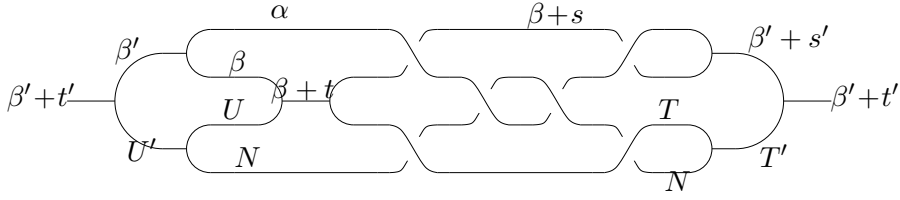} 
\caption{Elementary block in the $p$-punctured case.}  
\label{fig:elementaryblock} 
\end{figure}

\begin{proposition} 
 The following identity holds: 
\begin{eqnarray} \label{expressionofK0pintermof6j}
\stackrel{P}{K}{}\!\!_{0,p}^{(\pm)}\left(\begin{array}{c} \alpha \\ \beta,s \end{array} \right) \;\; = \;\;  [d_W]  
\sum_{{t\in {\cal S}^{p+1}}\atop {t_1=t_2, t_{p+1}=(0,0)} } 
 \prod_{i=1}^{p} \stackrel{P(i)}{\cal K}{}^{\!\!\!(\pm)} 
\left(\begin{array}{c} \alpha_i \\ \beta_i,\beta_{i+1} \end{array}\vline  \begin{array}{c} t_i, t_{i+1} \\ s_i,s_{i+1} \end{array}  \right)\;\; , 
\end{eqnarray} 
where $t \in {\cal S}^{p-1}$ and  we have used the following conventions: $s_1=s_2=(0,0)$, $U_1=T_1=(0,0)$, $U_2=T_2=N_1$, $\beta_1=(0,0)$, $\beta_2=\alpha_1$, $\beta_p=\alpha_p$, $\beta_{p+1}=(0,0)$, $s_p=(0,0)$ and $U_{p+1}=T_{p+1}=W$.   
 \end{proposition}  
 \Proof 
To prove this identity, we first prove it in the case where $\alpha_1,...,\alpha_p,\beta_3,...,\beta_{p-1}\in \SSF.$ 
$\stackrel{P}{K}{}\!\!_{0,p}^{(\pm)}\left(\begin{array}{c} \alpha \\ \beta,s \end{array} \right)$ is represented by the graph in figure \ref{fig:valueofK}, this graph is recasted in the graph shown in picture \ref{fig:decomposition} 
 after having used orthogonality relations: we have made  a summation on the coloring  $\beta_i+t_i$  of the lines crossing a box. 
This graph clearly shows an iterative structure, where the generic element is represented by the picture \ref{fig:elementaryblock}.  
The value of this elementary graph is $\stackrel{P(i)}{\cal K}{}^{\!\!\!(\pm)} 
\left(\begin{array}{c} \alpha_i \\ \beta_i,\beta_{i+1} \end{array}\vline  \begin{array}{c} t_i, t_{i+1} \\ s_i,s_{i+1} \end{array}  \right)$  
and is easily expressed in terms of $6j$ coefficients. 
The proof of the proposition when $\alpha_1,...,\alpha_p,\beta_3,...,\beta_{p-1}\in 
 {\mathbb S}$ is straighforward from continuation argument.$\Box$
  
A $U_q(sl(2,\CC)_{\RR})$ palette $P$ is equivalent to a couple of  two  
$U_q(sl(2))$ palettes denoted $(P^l, P^r).$ It is trivial to show that  
$ 
\stackrel{(P^l,0)}{K}{}\!\!_{0,p}^{(\pm)}\left(\begin{array}{c} \alpha \\ \beta,s \end{array} \right)$ (resp.$ 
\stackrel{(0,P^r)}{K}{}\!\!_{0,p}^{(\pm)}\left(\begin{array}{c} \alpha \\ \beta,s \end{array} \right)$ ) depends only on the variables $\alpha^l, \beta^l,s^l$ 
(resp. $\alpha^r, \beta^r,s^r$). 
We can therefore define $\stackrel{P^l}{K}{}\!\!_{0,p}^{l(\pm)}\left(\begin{array}{c} \alpha^l \\ \beta^l,s^l \end{array} \right) 
=\stackrel{(P^l,0)}{K}{}\!\!_{0,p}^{(\pm)}\left(\begin{array}{c} \alpha \\ \beta,s \end{array} \right)$ and similarly for the right variables. 
{}From the proposition (\ref{factorisation6j}) the following factorisation property is satisfied: 
\begin{equation} 
\stackrel{P}{K}{}\!\!_{0,p}^{(\pm)}\left(\begin{array}{c} \alpha \\ \beta,s \end{array} \right)=\stackrel{P^l}{K}{}\!\!_{0,p}^{l(\pm)}\left(\begin{array}{c} \alpha^l\\ \beta^l ,s^l \end{array} \right)\stackrel{P^r}{K}{}\!\!_{0,p}^{r(\pm)}\left(\begin{array}{c} \alpha^r \\ \beta^r ,s^r \end{array} \right). 
\end{equation} 
 
The proof of the unitarity of the representation of the moduli algebra uses as central tools the following proposition. 
 
\begin{proposition}\label{prop:symmetriesofK0p} 
The functions ${\stackrel{P}{K}{}\!\!_{0,p}^{(\pm)}\left(\begin{array}{c} \alpha \\ \beta,s \end{array} \right)}$ satisfy the symmetries: 
\begin{enumerate} 
\item $\overline{\stackrel{P}{K}{}\!\!_{0,p}^{(\pm)}\left(\begin{array}{c} \alpha \\ \beta,s \end{array} \right)}=\stackrel{P}{K}{}\!\!_{0,p}^{(\pm)}\left(\begin{array}{c} \overline{\alpha} \\ \overline{\beta} ,s \end{array} \right)$ 
\item ${\stackrel{P}{K}{}\!\!_{0,p}^{(\pm)}\left(\begin{array}{c} \alpha \\ \beta,s \end{array} \right)}=\stackrel{\tilde P}{K}{}\!\!_{0,p}^{(\mp)}\left(\begin{array}{c} \tilde{\alpha} \\ \tilde{\beta}+\tilde{s} ,-\tilde{s} \end{array} \right)$ 
\item  ${\stackrel{P}{K}{}\!\!_{0,p}^{(\pm)}\left(\begin{array}{c} \underline{\alpha} \\ \underline{\beta}, -s \end{array} \right)}=\psi_{0,p}[\alpha,\beta,s] \stackrel{ P}{K}{}\!\!_{0,p}^{(\pm)}\left(\begin{array}{c} {\alpha} \\ {\beta}, s \end{array} \right)$ with $\psi_{0,p}[\alpha,\beta,s] \in \{+1,-1\}.$ 
\end{enumerate} 
\end{proposition} 
 
\Proof 
The first property is a simple consequence of the following facts: $q$ is real, $\overline{\nu_p(z)}= 
\nu_p(\overline{z})$, the explicit expression of  
$\stackrel{P}{K}{}\!\!_{0,p}^{(\pm)}\left(\begin{array}{c} \alpha \\ \beta,s \end{array} \right)$ in  
terms of $6j$ and  
$\overline{v_I^{1/2}v_\alpha^{1/2}v_{\alpha+p}^{-1/2}}= 
v_I^{1/2}v_{\overline{\alpha}}^{1/2}v_{\overline{\alpha}+p}^{-1/2},$ if $I\pm p\in \NN.$ 
 
\begin{figure}  
\psfrag{al4}{$\alpha_4$} 
\psfrag{al3}{$\alpha_3$} 
\psfrag{al2}{$\alpha_2$} 
\psfrag{al1}{$\alpha_1$} 
\psfrag{be}{$\beta$} 
\psfrag{bes}[][]{$\;\;\beta\!+\!s$} 
\psfrag{n1}[][]{$\!\!N_1$} 
\psfrag{n2}[][]{$\!\!N_2$} 
\psfrag{n3}[][]{$\!\!N_3$} 
\psfrag{n4}[][]{$\!\!N_4$} 
\psfrag{w}[][]{$\!\!W$} 
\psfrag{u3}[][]{$U_3$} 
\psfrag{t3}[][]{$\!\!T_3$} 
\psfrag{u4}[][]{$\!\!U_4$} 
\psfrag{t4}[][]{$\!\!T_4$} 
\psfrag{0}{$0$} 
\centering 
\scalebox{0.6}[0.6]{\includegraphics[scale=0.5]{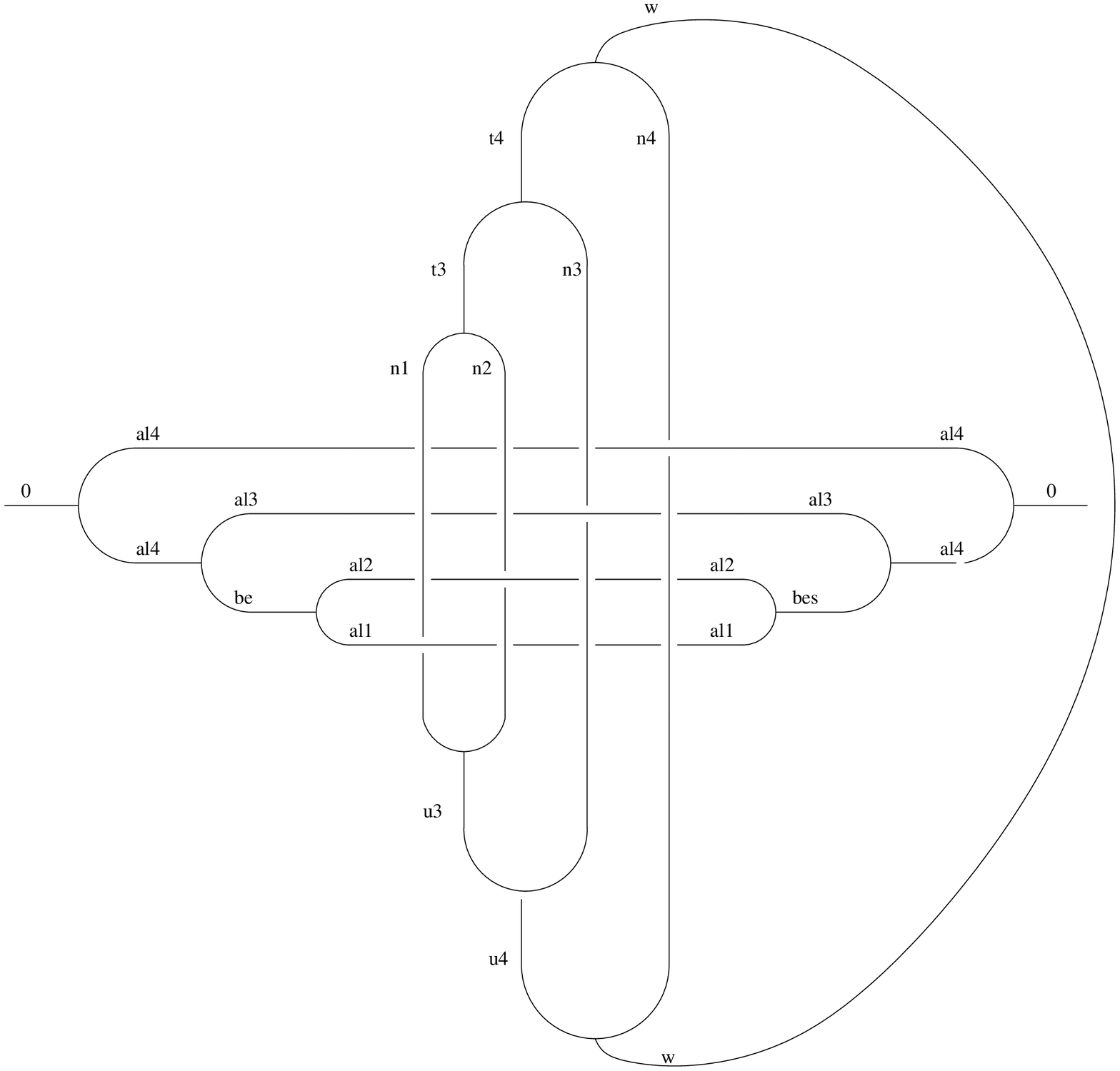}} 
\scalebox{0.6}[0.6]{\includegraphics[scale=0.5]{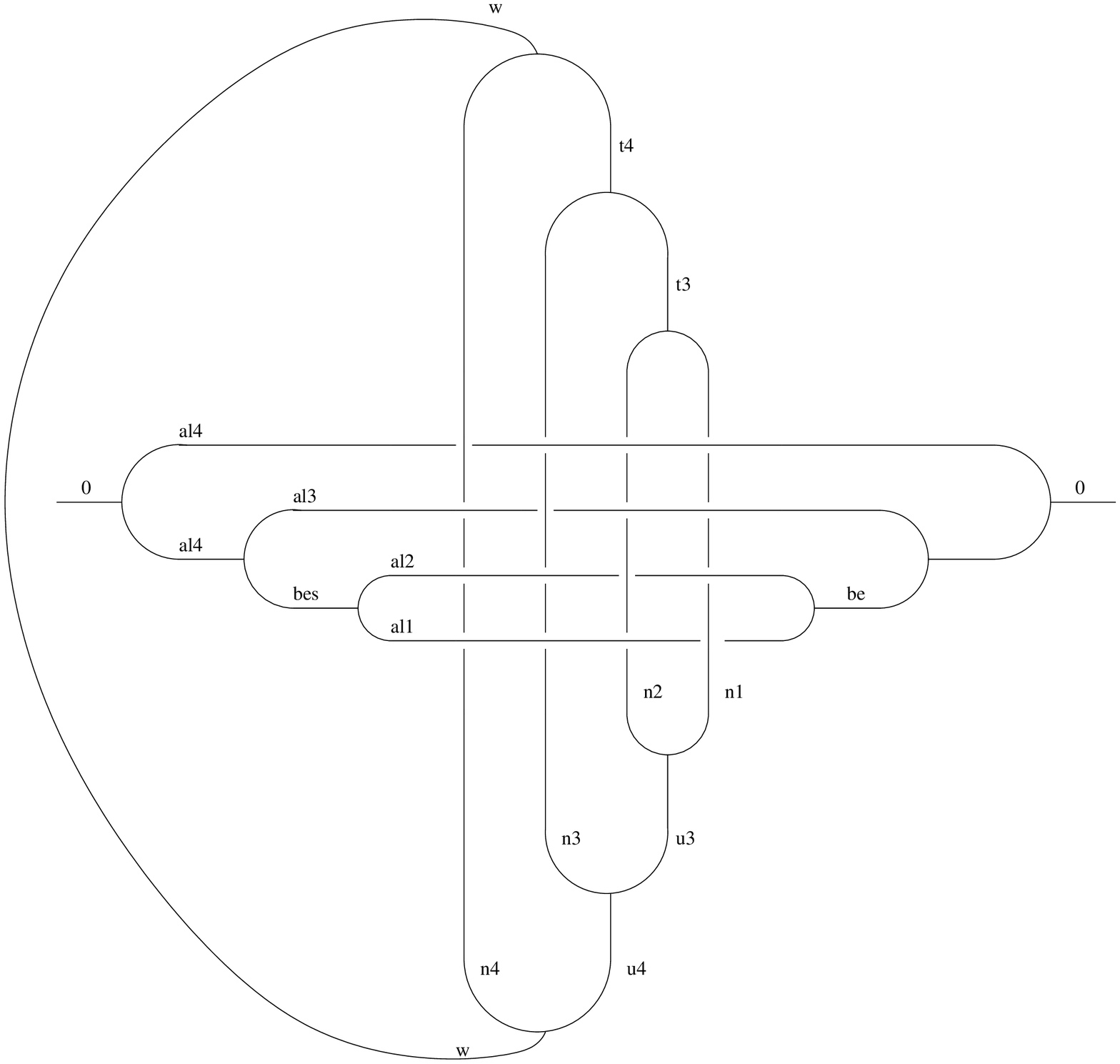}}       
\label{fig:valueofKplus}
\caption{Equivalent expressions of $\stackrel{P}{K}{}\!\!_{0,4}^{(+)}$ after trivial topological moves.}
\end{figure}

\begin{figure}  
\psfrag{al4}{$\widetilde{\alpha}_4$} 
\psfrag{al3}{$\widetilde{\alpha}_3$} 
\psfrag{al2}{$\widetilde{\alpha}_2$} 
\psfrag{al1}{$\widetilde{\alpha}_1$} 
\psfrag{bes}{$\widetilde{\beta}$} 
\psfrag{be}[][]{$\;\;\widetilde{\beta\!+\!s}$} 
\psfrag{n1}[][]{$\!\!\widetilde{N}_1$} 
\psfrag{n2}[][]{$\!\!\widetilde{N}_2$} 
\psfrag{n3}[][]{$\!\!\widetilde{N}_3$} 
\psfrag{n4}[][]{$\!\!\widetilde{N}_4$} 
\psfrag{w}[][]{$\!\!\widetilde{W}$} 
\psfrag{u3}[][]{$\widetilde{U}_3$} 
\psfrag{t3}[][]{$\!\!\widetilde{T}_3$} 
\psfrag{u4}[][]{$\!\!\widetilde{U}_4$} 
\psfrag{t4}[][]{$\!\!\widetilde{T}_4$} 
\psfrag{0}{$0$} 
\centering 
\scalebox{0.6}[0.6]{\includegraphics[scale=0.5]{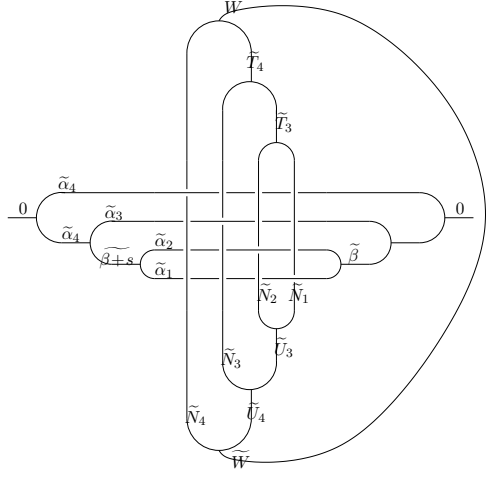}} 
\caption{Expression of $\stackrel{P}{K}{}\!\!_{0,4}^{(+)}$ after trivial manipulations.}  
\label{fig:valueofKminustilde} 
\end{figure}

To prove the second relation, we first notice that it is equivalent to prove the relation: 
\begin{eqnarray} 
&&{\stackrel{P}{K}{}\!\!_{0,p}^{(\pm)}\left(\begin{array}{c} \alpha \\ \beta,s \end{array} \right)}\frac{\Xi_{0,p}[\alpha,\beta]}{\Xi_{0,p}[\alpha,\beta+s]}= 
{\stackrel{\tilde{P}}{K}{}\!\!_{0,p}^{(\mp)}\left(\begin{array}{c} \tilde{\alpha} \\ \tilde{\beta}+\tilde{s},-\tilde{s} \end{array} \right)} 
\frac{\Xi_{0,p}[\tilde{\alpha},\tilde{\beta}+\tilde{s}]}{\Xi_{0,p}[\tilde{\alpha}, 
\tilde{\beta}]} 
\frac{\Xi_{0,p}[\alpha,\beta]\Xi_{0,p}[\tilde{\alpha},\tilde{\beta}]} 
{\Xi_{0,p}[\alpha,\beta+s]\Xi_{0,p}[\tilde{\alpha},\tilde{\beta}+\tilde{s}]}\nonumber 
\\ 
&&={\stackrel{\tilde{P}}{K}{}\!\!_{0,p}^{(\mp)}\left(\begin{array}{c} \tilde{\alpha} \\ \tilde{\beta}+\tilde{s},-\tilde{s} 
 \end{array} \right)} 
\frac{\Xi_{0,p}[\tilde{\alpha},\tilde{\beta}+\tilde{s}]}{\Xi_{0,p}[\tilde{\alpha}, 
\tilde{\beta}]} 
\frac{\Xi_{0,p}[{\alpha}, 
{\beta}]^2}{\Xi_{0,p}[{\alpha},{\beta}+{s}]^2} 
\end{eqnarray} 
which is an identity on rational functions in  
$q^{i\rho_{\alpha_1}},...,q^{i\rho_{\alpha_p}}, 
q^{i\rho_{\beta_3}},...,q^{i\rho_{\beta_{p-3}}}$. 
As a result, using the standard continuation argument, it is sufficient to show the second relation when $\alpha_1,...,\alpha_p,\beta_3,...,\beta_{p-3}\in \SSF.$ $\stackrel{P}{K}{}\!\!_{0,p}^{(+)}\left(\begin{array}{c} \alpha \\ \beta,s \end{array} \right)$ is represented by the picture 
 (\ref{fig:valueofKplus}) which is the same, after topological moves, as the left graph   
 (figure \ref{fig:valueofK}). This graph, after a flip, is equal to the right graph (figure \ref{fig:valueofK}).  
The value of this graph is equal to the value of the graph pictured in figure (\ref{fig:valueofKminustilde}),  
because of the property $\stackrel{IJ}{\RR}{}^{(+)}=\stackrel{\tilde{I}\tilde{J}}{\RR}{}^{(-)}.$ 
This last graph is equal to  $\stackrel{\tilde{P}}{K}{}\!\!_{0,p}^{(-)}\left(\begin{array}{c} \tilde{\alpha} \\  
\tilde{\beta}+\tilde{s},-\tilde{s} \end{array} \right).$ 
This ends the proof of the second property.

The third property follows from a detailed computation making use of the explicit expression of $\stackrel{P}{K}{}\!\!_{0,p}^{(+)}\left(\begin{array}{c} \alpha \\ \beta,s \end{array} \right)$ in terms of $6j.$ 
Using the symmetries of the $6j$ proved in \cite{BR4}, a direct computation show that: 
\begin{equation} 
\stackrel{Q}{\cal K} 
{}^{(\pm)}\left(\begin{array}{c}\underline{ \alpha} \\ \underline{\beta},\underline{\beta'} \end{array} 
\vline \begin{array}{c} -t,-t' \\ -s,-s' \end{array}  \right)= e^{i\pi(U-U'+T-T'+s-s')} 
\psi_{0,p}(\alpha,\beta,\beta',s,s') 
\stackrel{Q}{\cal K} 
{}^{(\pm)}\left(\begin{array}{c} \alpha \\ \beta,\beta' \end{array} 
\vline \begin{array}{c} t,t' \\ s,s' \end{array}  \right) 
\end{equation} 
with 
\begin{equation} 
\psi(\alpha,\beta,\beta',s,s')=\frac{ 
\varphi_{(\alpha+\beta+\beta',s+s') }   
\varphi_{(\alpha+\beta-\beta',s-s')}}    
{\varphi_{(-\alpha+\beta+\beta',-s+s')}    
\varphi_{(\alpha-\beta+\beta',-s-s')}}.   
\end{equation} 
It is easy to show that the selection rules imply that $\psi(\alpha,\beta,\beta',s,s')\in\{+1,-1\}.$ 
As a result we obtain the third property with  
\begin{equation} 
\psi[\alpha,\beta,s]= 
\prod_{i=1}^p\psi_{0,p}(\alpha_i,\beta_i,\beta_{i+1},s_i,s_{i+1}).  
\end{equation}

\subsection*{B.2 The genus-n surface}  
 Like the previous subsection, we will present in this one an  explicit expression of the function $\stackrel{P}{K}{}\!\!_{n,0}^{(\pm)} \left(\begin{array}{c} \kappa;k \\ \lambda, \sigma, \tau ; \ell,s,t \end{array} \right)$ in terms of $6j$ symbols.
We have used a pictorial representation of  this function in the case $n=3$ (the case $n=2$ being degenerate) . The case of arbitrary genus $n$ is a straighforward generalisation. 
We need to introduce new functions at this stage.\\

If $P$ is the palette $P=(IJNKLUTW)$, with 
$N=\emptyset$ for the genus-$n$ case, we denote 
$P(i)$ the 8-uplet  $(I_i,J_i, K_i, L_i, U_{i}, U_{i+1}, T_{i}, T_{i+1})$.
If  $I'=(I'_3,\cdots,I'_{n+1}) \in  \SSF^{n-2}$, we denote $II'(i) =(I_i,I'_i,I'_{i+1})$.
Let $Q=(I,J,K,L,U,U',T,T')\in \SSF^8$, $Q'=(I,J,K)\in \SSF^3$, $\lambda, \sigma, \sigma',\tau,\tau' \in {\mathbb S}$ and  $\ell,s,s',t,t',a,b,b',c,c'\in {\cal S}$, we define the functions:

\begin{eqnarray}
&&\hskip-0.8cm\stackrel{Q}{\cal K}
{}^{(\pm)}\left(\begin{array}{c} \tau' \\ \lambda,\tau \end{array}
\vline \begin{array}{c} b',c' \\ \ell,a,b,c \end{array}  \right)  =  
\sum_{x,y,z \in {\cal S}, X,Y \in \SSF} \left(\frac{v_{K}^{1/2}v_{T'}^{1/2}}{v_{X}^{1/2}v_{L}^{1/2}}\right)^{\pm 1} \frac{v_{X}^{1/2}v_{I}^{1/2}v_{L}^{1/2}}{v_{K}^{1/2}v_{Y}^{1/2}v_{L}^{1/2}} \frac{v_{\lambda+y}^{3/2}}{v_{\lambda+z}^{1/2}v_{\lambda+l} v_{\lambda+a}^{1/2}} \frac{v_{\tau'+c'}^{1/2}}{v_{\tau+x}^{1/2}} \nonumber \\
&&\hskip0.3cm \sixj{\lambda\!+\!l}{\tau'\!+\!c'}{\tau\!+\!x}{U'}{\tau\!+\!t}{\tau'\!+\!t'}
\sixj{\tau\!+\!t}{\tau\!+\!x}{U'}{K}{U}{\tau\!+\!c}
\sixj{\tau\!+\!b}{\tau\!+\!x}{X}{K}{T}{\tau\!+\!c} \nonumber \\
&&\hskip0.3cm\sixj{\lambda\!+\!a}{\lambda\!+\!z}{T'}{T}{L}{\lambda\!+\!y}
\sixj{\lambda\!+\!l}{\lambda\!+\!z}{X}{T}{K}{\lambda\!+\!y}
\sixj{\tau\!+\!b}{\tau'\!+\!c'}{\lambda\!+\!z}{T'}{\lambda\!+\!a}{\tau'\!+\!b'}  \nonumber \\
&&\hskip0.3cm \sixj{\tau'\!+\!c'}{\tau\!+\!b}{\lambda\!+\!z}{X}{\lambda\!+\!l}{\tau\!+\!x} 
\sixj{\lambda}{\lambda\!+\!y}{Y}{K}{I}{\lambda\!+\!l}
\sixj{I}{Y}{L}{I}{J}{K}
\sixj{\lambda}{\lambda\!+\!y}{Y}{L}{I}{\lambda\!+\!a}  \label{def1}
\end{eqnarray}
\begin{eqnarray}
&&\hskip-0.8cm \stackrel{Q'}{\cal N}
{}^{(1)}\left(\begin{array}{c} \tau' \\ \lambda,\tau \end{array}
\vline \begin{array}{c} t',b' \\ \ell,t,a,b \end{array}  \right)  = 
\sum_{x\in {\cal S}} \frac{v_{\tau}^{1/2} v_{\lambda}^{1/2} v_{K}^{1/2} v_{\tau+b}^{1/2}}{v_{\tau'}^{1/2} v_{\tau+x}^{1/2} v_{J}^{1/2}} \sixj{\tau'\!+\!b'}{\tau\!+\!b}{\lambda\!+\!a}{I}{\lambda}{\tau\!+\!x} \nonumber \\
&&\hskip0.3cm\sixj{\lambda}{\tau'\!+\!b'}{\tau\!+\!x}{K}{\tau}{\tau'} \sixj{\tau}{\tau\!+\!x}{K}{I}{J}{\tau\!+\!b}  \label{def2}\\
&&\hskip-0.8cm\stackrel{Q'}{\cal N}
{}^{(2)}\left(\begin{array}{c} \sigma' \\ \lambda,\sigma \end{array}
\vline \begin{array}{c}  s'\\ \ell,s \end{array}  \right)  =  
\sum_{x \in {\cal S}} \frac{v_{\lambda+x}^{1/2}}{v_{\lambda+l}^{1/2} v_J^{1/2}} \sixj{\sigma}{\sigma'\!+\!s'}{\lambda\!+\!x}{K}{\lambda\!+\!a}{\sigma'} \nonumber \\
&&\hskip0.3cm \sixj{\lambda\!+\!a}{\lambda\!+\!x}{K}{J}{I}{\lambda\!+\!l} \sixj{\sigma'\!+\!s'}{\lambda\!+\!l}{\sigma\!+\!s}{J}{\sigma}{\lambda\!+\!x}
 \label{def3} \end{eqnarray}

\begin{proposition}\label{prop:expressionofKforgenus}
 The following identity holds:
\begin{eqnarray}
\stackrel{P}{K}{}\!\!_{n,0}^{(\pm)}\left(\begin{array}{c} \kappa;k \\ \lambda,\sigma,\tau;\ell,s,t \end{array} \right) &  =  & \frac{[d_W]}{[d_{\kappa}]} \sum_{a,b,I'}  \stackrel{P}{G}{}\!\!_{n,0}^{(\pm)}\left(\begin{array}{c} \kappa;k \\ \lambda,\tau;\ell,a,b \end{array} \right) \nonumber \\
&&\stackrel{II'}{N}{}\!\!_{n,0}^{(1)}\left(\begin{array}{c} \kappa;k \\ \lambda,\tau;\ell,t,a,b \end{array} \right)
\stackrel{II'}{N}{}\!\!_{n,0}^{(2)}\left(\begin{array}{c} \kappa;k \\ \lambda,\sigma;\ell,s \end{array} \right)
\end{eqnarray}
where we have defined the functions
\begin{eqnarray}
\stackrel{P}{G}{}\!\!_{n,0}^{(\pm)}\left(\begin{array}{c} \kappa;k \\ \lambda,\tau;\ell,a,b \end{array} \right) & = & \sum_c \prod_{i=1}^n \stackrel{P(i)}{\cal K}{}^{(\pm)}\left(\begin{array}{c} \tau_{i+1} \\ \lambda_i,\tau_i \end{array}
\vline \begin{array}{c} b_{i+1},c_{i+1} \\ \ell_i,a_i,b_i,c_i \end{array}  \right) \\
\stackrel{II'}{N}{}\!\!_{n,0}^{(1)}\left(\begin{array}{c} \kappa;k \\ \lambda,\tau;\ell,t,a,b \end{array} \right) & = & \prod_{i=1}^{n} \stackrel{II'(i)}{\cal N}{}^{(1)}\left(\begin{array}{c} \tau_{i+1} \\ \lambda_i,\tau_i \end{array}
\vline \begin{array}{c} t_{i+1},b_{i+1} \\ \ell_i,t_i,a_i,b_i \end{array}  \right) \\
\stackrel{II'}{N}{}\!\!_{n,0}^{(2)}\left(\begin{array}{c} \kappa;k \\ \lambda,\sigma;\ell,s \end{array} \right) & = & \prod_{i=1}^{n} \stackrel{II'(i)}{\cal N}{}^{(2)}\left(\begin{array}{c} \sigma_{i+1} \\ \lambda_i,\sigma_i \end{array}
\vline \begin{array}{c}  s_{i+1}\\ \ell_i,s_i \end{array}  \right)\;\;.
\end{eqnarray}
In the sums, $a=(a_1,\cdots,a_n)\in {\cal S}^{n}$, $b=(b_2,\cdots,b_{n+1}) \in {\cal S}^{n}$,  $c=(c_2,\cdots,c_{n+1}) \in {\cal S}^{n-2}$ and we impose: $\tau_{n+1}=\sigma_{n+1}=\kappa$, $\tau_2=\sigma_2=\lambda_1$, $\tau_1=\sigma_1=(0,0)$, $U_1=T_1=(0,0)$, $U_2=K_2$, $T_2=L_2$,  $I'_2=I_1$ and $U_{n+1}=T_{n+1}=W$.
\end{proposition}

\begin{figure} 
\centering
\includegraphics[scale=0.6]{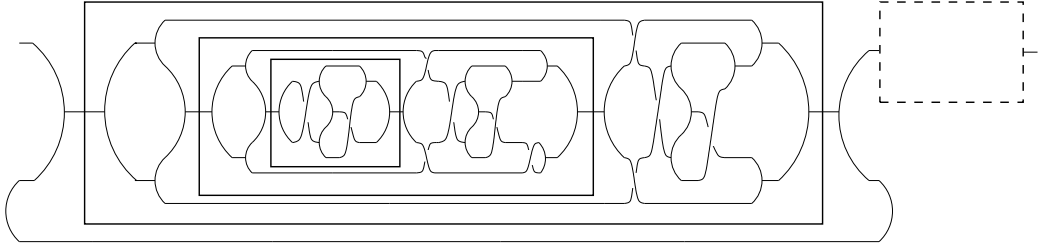}
\caption{The first part of the decomposition.} 
\label{fig:decompositiongenus3}
\end{figure}

\begin{figure} 
\centering
\includegraphics[scale=0.5]{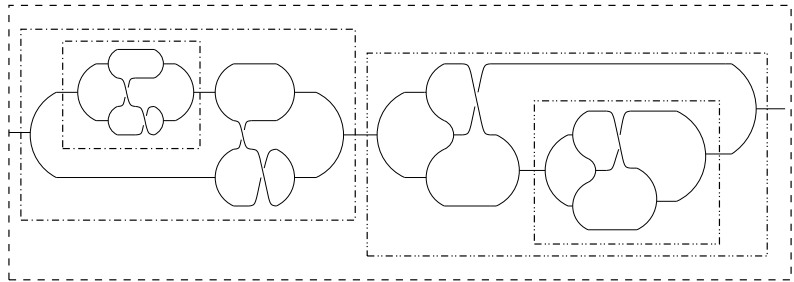}
\caption{The second and third part of the decomposition. This picture represents the dashed box of the previous picture.} 
\label{fig:normalisationgenus3}
\end{figure}

\begin{figure} 
\psfrag{la}[][]{$\lambda\!+\!\ell$}
\psfrag{si}[][]{$\sigma\!+\!s$}
\psfrag{ta}[][]{$\tau\!+\!t$}
\psfrag{sip}[][]{$\sigma'\!+\!s'$}
\psfrag{tap}[][]{$\tau'\!+\!t'$}
\psfrag{tp}[][]{$\tau'\!+\!t'$}
\psfrag{la+l}[][]{$\lambda$}
\psfrag{si+s}[][]{$\sigma$}
\psfrag{sip+sp}[][]{$\sigma'$}
\psfrag{t}[][]{$\tau+t$}
\psfrag{la+a}[][]{$\lambda\!+\!a$}
\psfrag{t+b}[][]{$\tau\!+\!b$}
\psfrag{t+c}[][]{$\tau\!+\!c$}
\psfrag{tp+bp}[][]{$\tau'\!+\!b'$}
\psfrag{tp+cp}[][]{$\tau'\!+\!c'$}
\psfrag{tap+bp}[][]{$\tau'\!+\!b'$}
\psfrag{tap+tp}[][]{$\tau'$}
\psfrag{ta+t}[][]{$\tau$}
\psfrag{c}[][]{$c$}
\psfrag{cp}[][]{$c'$}
\psfrag{U}[][]{$U$}
\psfrag{Up}[][]{$U'$}
\psfrag{T}[][]{$T$}
\psfrag{Tp}[][]{$T'$}
\psfrag{K}[][]{$K$}
\psfrag{L}[][]{$L$}
\psfrag{I}[][]{$I$}
\psfrag{J}[][]{$J$}
\psfrag{N}[][]{$J$}
\psfrag{M}[][]{$K$}
\centering
\scalebox{0.8}[0.8]{\includegraphics[scale=1]{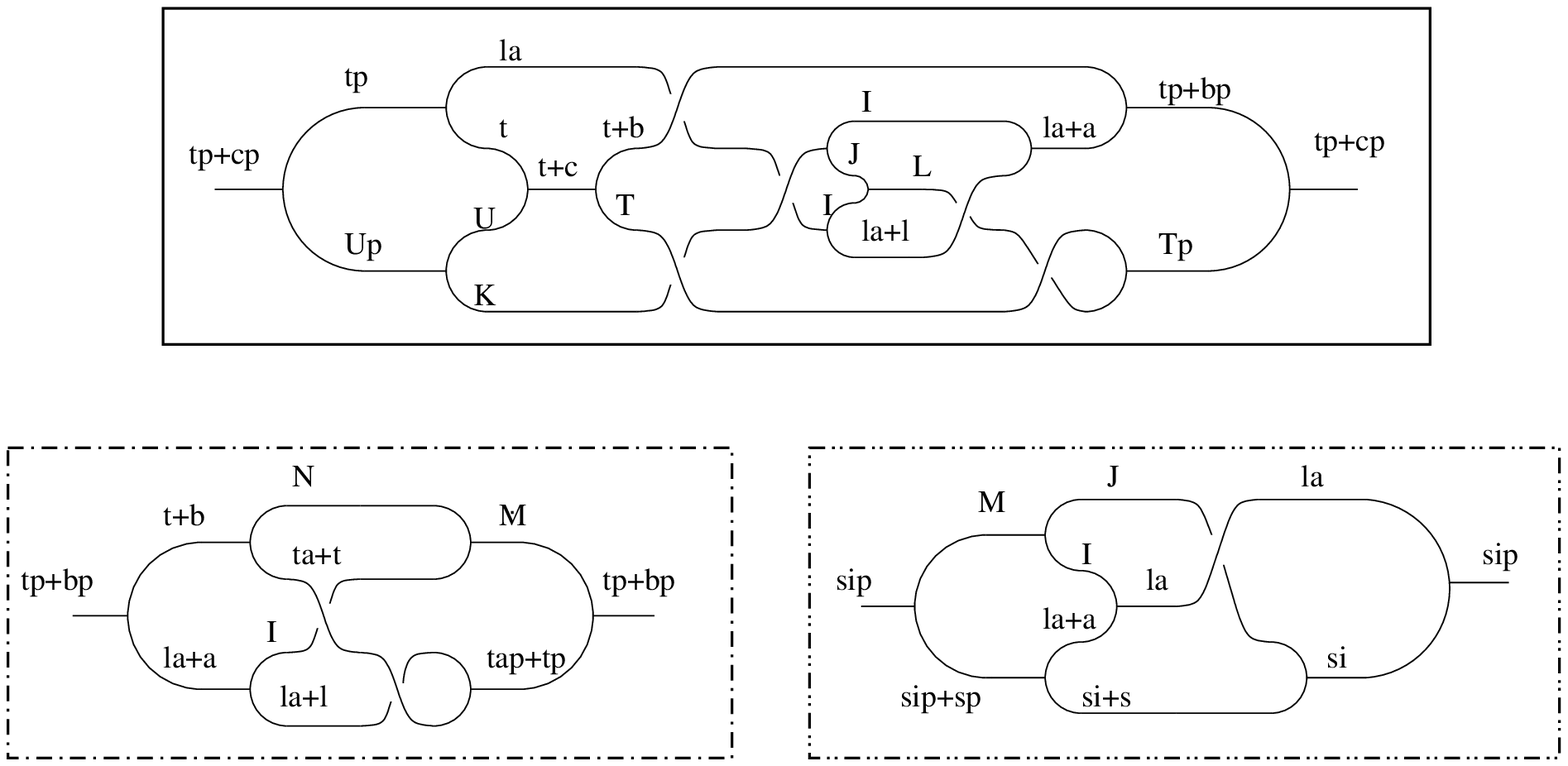}}
\caption{Elementary blocks in the genus $n$ case.} 
\label{fig:elementaryblockforgenus3}
\end{figure}

 \Proof
To prove this identity, we first prove it in the case where $\kappa,\lambda,\sigma,\tau \in \SSF^{3n-3}.$
$\stackrel{P}{K}{}\!\!_{n,0}^{(\pm)}\left(\begin{array}{c} \kappa;k \\ \lambda,\sigma,\tau;\ell,s,t \end{array} \right)$ is represented by the graph in figure \ref{fig:valueofKforgenus3}. This graph is recasted in the graph shown in pictures \ref{fig:decompositiongenus3},\ref{fig:normalisationgenus3}
 after having used topological moves and orthogonality relations: we have made  a summation on all the coloring of the lines crossing a box.
This graph clearly decomposes into three parts and each part shows an iterative structure whose generic elements are represented by the picture \ref{fig:elementaryblockforgenus3}. 

The value of each elementary graph is respectively $\stackrel{P(i)}{\cal K}{}^{(\pm)}\left(\begin{array}{c} \tau_{i+1} \\ \lambda_i,\tau_i \end{array}
\vline \begin{array}{c} b_{i+1},c_{i+1} \\ \ell_i,a_i,b_i,c_i \end{array}  \right)$, $\stackrel{II'(i)}{\cal N}{}^{(1)}\left(\begin{array}{c} \tau_{i+1} \\ \lambda_i,\tau_i \end{array}\vline \begin{array}{c} t_{i+1},b_{i+1} \\ \ell_i,t_i,a_i,b_i \end{array}  \right)$ and  $\stackrel{II'(i)}{\cal N}{}^{(2)}\left(\begin{array}{c} \sigma_{i+1} \\ \lambda_i,\sigma_i \end{array} \vline \begin{array}{c}  s_{i+1}\\ \ell_i,s_i \end{array}  \right)$.
It is easily expressed in terms of $6j$ coefficients which explains the definitions (\ref{def1},\ref{def2},\ref{def3}).
The proof of the proposition when $\kappa,\lambda,\sigma,\tau \in {\mathbb S}^{n-3}$ is straighforward from the continuation argument.$\Box$

Using the same argument and the same definitions as in the $p$-punctured case, the following factorisation property is also satisfied:
\begin{equation}
\stackrel{P}{K}{}\!\!_{n,0}^{(\pm)}\left(\begin{array}{c} \kappa;k \\ \lambda,\sigma,\tau;\ell,s,t \end{array} \right)=\stackrel{P^l}{K}{}\!\!_{n,0}^{l(\pm)}\left(\begin{array}{c} \kappa^l;k^l\\ \lambda^l,\sigma^l,\tau^l;\ell^l,s^l,t^l \end{array} \right)\stackrel{P^r}{K}{}\!\!_{0,p}^{r(\pm)}\left(\begin{array}{c}  \kappa^r;k^r \\   \lambda^r,\sigma^r,\tau^r;\ell^r,s^r,t^r\end{array} \right)\;\;.
\end{equation}\

The proof of unitarity of the representation of the moduli algebra uses as central tools the following proposition.

\begin{proposition}\label{prop:symmetriesofK}
The functions $\stackrel{P}{K}{}\!\!_{n,0}^{(\pm)}\left(\begin{array}{c} \kappa,k \\ \lambda,\sigma,\tau;\ell,s,t \end{array} \right)$ satisfy the symmetries:
\begin{enumerate}
\item $\overline{\stackrel{P}{K}{}\!\!_{n,0}^{(\pm)}\left(\begin{array}{c} \kappa,k \\ \lambda,\sigma,\tau;\ell,s,t \end{array} \right) }=\stackrel{P}{K}{}\!\!_{n,0}^{(\pm)}\left(\begin{array}{c} \overline{\kappa},k \\ \overline{\lambda},\overline{\sigma},\overline{\tau};\ell,s,t \end{array} \right)$

\item $\stackrel{P}{K}{}\!\!_{n,0}^{(\pm)}\left(\begin{array}{c} \kappa,k \\ \lambda,\sigma,\tau;\ell,s,t \end{array} \right) = \frac{[d_{\lambda}]}{[d_{\lambda+\ell}]} \frac{[d_{\kappa +k}]}{[d_{\kappa}]} \stackrel{\tilde{P}}{K}{}\!\!_{n,0}^{(\mp)}\left(\begin{array}{c} \tilde{\kappa}+\tilde{k},-\tilde{k} \\ \tilde{\lambda}+\tilde{\ell},\tilde{\sigma}+\tilde{s},\tilde{\tau}+\tilde{t};-\tilde{\ell},-\tilde{s},-\tilde{t} \end{array} \right)$

\item  $\stackrel{P}{K}{}\!\!_{n,0}^{(\pm)}\left(\begin{array}{c} \underline{\kappa};-k \\ \underline{\lambda},\underline{\sigma},\underline{\tau};-\ell,-s,-t \end{array} \right) = \psi_{n,0}[ \kappa, \lambda,\sigma,\tau;k,\ell,s,t] \stackrel{P}{K}{}\!\!_{n,0}^{(\pm)}\left(\begin{array}{c} \kappa;k \\ \lambda,\sigma,\tau;\ell,s,t \end{array} \right)$ \\
 with  $\psi_{n,0}[ \kappa, \lambda,\sigma,\tau;k,\ell,s,t] \in \{+1,-1\}$.
\end{enumerate}
\end{proposition}

\Proof
The proof of the first property is the same as the $p$-punctured case. The third property is also proved along similar lines. The only property which is drastically different from the p-puncture case is the second one. Indeed, contrary to the $p$-puncture case, it cannot be reduced to topological moves on graphs.  
To prove the second relation, we first use the standard continuation argument so that it is sufficient to show the same relation when $\kappa,\lambda,\sigma,\tau \in \SSF^{3n-3}$.

\begin{figure} 
\psfrag{lap}[][]{$\lambda_i\!+\!\ell_i$}
\psfrag{la}[][]{$\lambda_i$}
\psfrag{S}[][]{$S_i$}
\psfrag{I}[][]{$I_i$}
\psfrag{J}[][]{$J_i$}
\centering
\scalebox{0.8}[0.8]{\includegraphics[scale=1]{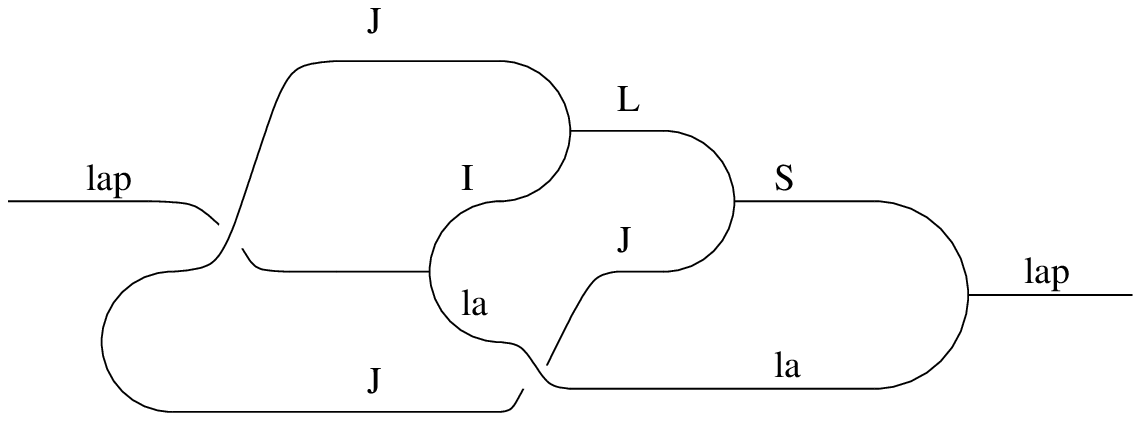}}
\caption{Definition of $\stackrel{S(i)}{\cal E}{}\!\!(\lambda_i,\ell_i)$.} 
\label{fig:definitionforsymetry1}
\end{figure}

\begin{figure} 
\psfrag{la1p}[][]{$\lambda_1\!+\!\ell_1$}
\psfrag{la2p}[][]{$\lambda_2\!+\!\ell_2$}
\psfrag{la1}[][]{$\lambda_1$}
\psfrag{la2}[][]{$\lambda_2$}
\psfrag{k}[][]{$\kappa$}
\psfrag{kp}[][]{$\kappa\!+\!k$}
\psfrag{S1}[][]{$S_1$}
\psfrag{S2}[][]{$S_2$}
\psfrag{I1}[][]{$I_1$}
\psfrag{I2}[][]{$I_2$}
\psfrag{M}[][]{$M$}
\centering
\scalebox{0.8}[0.8]{\includegraphics[scale=1]{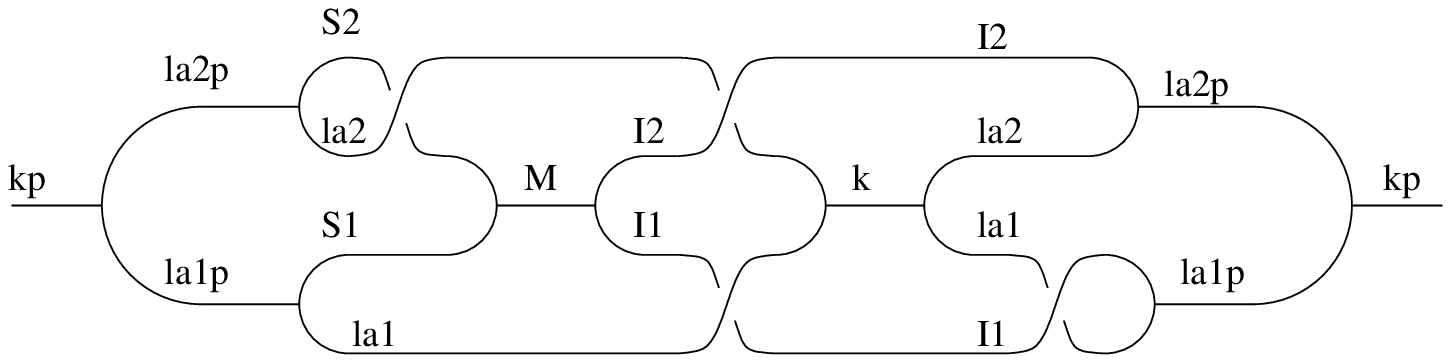}}
\caption{Definition of $\stackrel{P,S,M}{\cal C}
$.}
\label{fig:definitionforsymetry2}
\end{figure}

\begin{figure} 
\psfrag{S1}[][]{$S_1$}
\psfrag{S2}[][]{$S_2$}
\psfrag{I1}[][]{$I_1$}
\psfrag{I2}[][]{$I_2$}
\psfrag{J1}[][]{$J_1$}
\psfrag{J2}[][]{$J_2$}
\psfrag{K1}[][]{$K_1$}
\psfrag{K2}[][]{$K_2$}
\psfrag{L1}[][]{$L_1$}
\psfrag{L2}[][]{$L_2$}
\psfrag{M}[][]{$M$}
\psfrag{W}[][]{$W$}
\centering
\scalebox{0.8}[0.8]{\includegraphics[scale=1]{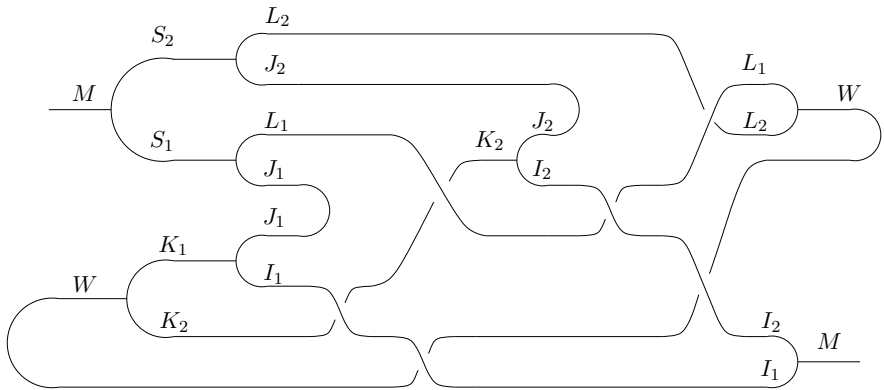}}
\caption{Definition of $\stackrel{P,S,M}{\cal D}{}\!\!^{(+)}$. The expression of $\stackrel{P,S,M}{\cal D}{}\!\!^{(-)}$ is obtained after having turned all overcrossing into undercrossing but not those which permute the representations ($L_1$,$K_2$) and ($L_1$,$L_2$).} 
\label{fig:definitionforsymetry3}
\end{figure}

The idea of this proof is to decompose the previous graph in different units which satisfy  the required property. We will perform it in the case $n=2$. This proof can be generalised for an arbitrary genus $n$. Let $P=(I_1,I_2,J_1,J_2,K_1,K_2,L_1,L_2,W)$ the palette associated to the genus $n=2$; we also introduce $S=(S_1,S_2) \in \SSF^2$ and $M\in\SSF$; finally we denote the quadruplet $PS(i)=(I_i,J_i,L_i,S_i)$, for $i\in\{1,2\}$. Let us define new functions $\stackrel{S(i)}{\cal E}{}\!\!(\lambda_i;\ell_i)$, $\stackrel{P,S,M}{\cal C}{}\!\! \left(\begin{array}{c} \kappa;k \\ \lambda;\ell \end{array} \right)$ and $\stackrel{P,S,M}{\cal D}{}\!\!^{(\pm)}$ respectively by the graphic expressions given in the figures \ref{fig:definitionforsymetry1}, \ref{fig:definitionforsymetry2}, \ref{fig:definitionforsymetry3}.

\begin{proposition}\label{prop:symetryrelation}
The previous functions satisfy the properties:
\begin{enumerate}
\item $\stackrel{S}{\cal E}{}\!\!(\lambda;\ell) = \stackrel{\tilde{S}}{\cal E}{}\!\!(\tilde{\lambda}+\tilde{\ell};-\tilde{\ell})$ for any $S \in \SSF^4$, $\lambda \in \mathbb{S}$ and $\ell \in {\cal S}$
\item $\sum_M  \stackrel{P,S,M}{\cal C}{}\!\! \left(\begin{array}{c} \kappa;k \\ \lambda;\ell \end{array} \right) \stackrel{P,S,M}{\cal D}{}\!\!^{(\pm)} = \sum_M \stackrel{\tilde{P},\tilde{S},M}{\cal C}{}\!\! \left(\begin{array}{c} \tilde{\kappa}+\tilde{k},-\tilde{k} \\ \tilde{\lambda}+\tilde{\ell};-\tilde{\ell} \end{array} \right) \stackrel{\tilde{P},\tilde{S},M}{\cal D}{}\!\!^{(\mp)}$ \\
for any $P,S \in \SSF$, $\lambda \in \mathbb{S}^2$, $\kappa \in \mathbb{S}$, $\ell \in {\cal S}^2$ and $k \in {\cal S}$ .
\end{enumerate}
\end{proposition}

\Proof
The first point is straigthforward when we evaluate the explicit expression of the graph \ref{fig:definitionforsymetry1}. Indeed, by a direct manipulation, we show that:
\begin{eqnarray}
\stackrel{PS(j)}{\cal E}{}\!\!(\lambda_j;\ell_j) = \frac{v_{I_j}^{1/2}}{v_{S_j}^{1/2}} e^{i\pi(I_j-L_j)} \frac{[d_{L_j}]^{1/2}}{[d_{I_j}]^{1/2}} \Lambda^{J_jL_j}_{I_j S_j}(\lambda_j+\ell_j,\lambda_j) \;\;, \nonumber
\end{eqnarray}
where $\Lambda^{J_iL_i}_{I_iS_i}(\lambda_i+\ell_i,\lambda_i)$ have already been introduced in the appendix A.1. and satisfies the required property.

\begin{figure} 
\psfrag{la1p}[][]{$\lambda_1\!+\!\ell_1$}
\psfrag{la2p}[][]{$\lambda_2\!+\!\ell_2$}
\psfrag{la1}[][]{$\lambda_1$}
\psfrag{la2}[][]{$\lambda_2$}
\psfrag{k}[][]{$\kappa$}
\psfrag{kp}[][]{$\kappa\!+\!k$}
\psfrag{S1}[][]{$S_1$}
\psfrag{S2}[][]{$S_2$}
\psfrag{I1}[][]{$I_1$}
\psfrag{I2}[][]{$I_2$}
\psfrag{M}[][]{$M$}
\psfrag{d/d}[][]{$\frac{e^{2i\pi l} [d_{\lambda}]}{[d_{\lambda+l}]}$}
\psfrag{v}[][]{$\frac{v_{\lambda_1+l_1}^{1/2}}{v_{\lambda_1}^{1/2}v_{I_1}^{1/2}} \frac{v_{\lambda_2+l_2}^{1/2}}{v_{\lambda_2}^{1/2}v_{S_2}^{1/2}}$}
\psfrag{d1/2}[][]{$\frac{e^{i\pi l} [d_{\lambda}]^{1/2}}{[d_{\lambda+l}]^{1/2}}$}
\psfrag{1/v}[][]{$v_{I_1}^{-1}v_{S_2}^{-1}$}
\centering
\scalebox{0.8}[0.8]{\includegraphics[scale=1]{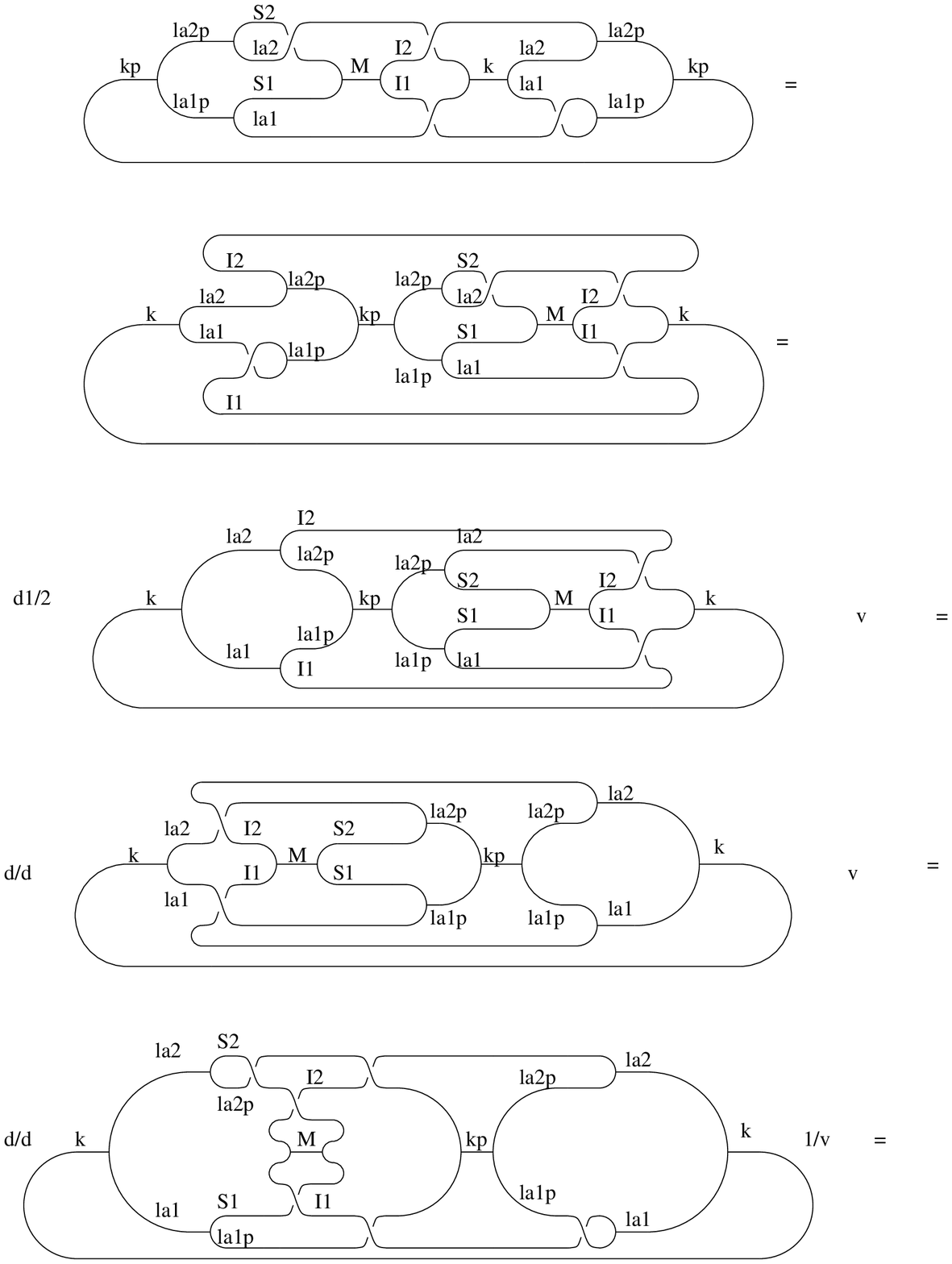}}
\caption{This sequence  of graphs shows the main lines of proof of the symmetry relation \ref{symetryrelation}. All the representations are supposed to be finite. To obtain the final result, we have to introduce two orthogonality relations between $(I_1,I_2)$ and $(S_1,S_2)$ in the last picture. The obtained graph decomposes in a graph similar to the first one and a residual graph whose  value is given in  \ref{symetryrelation}.}
\label{fig:proofofsymetrygenus2}
\end{figure}

The second point is more technical and it is proved in two steps. On one hand, after moves on the graph \ref{fig:definitionforsymetry2}, described in figure  \ref{fig:proofofsymetrygenus2}, we obtain the identity:
\begin{eqnarray} \label{symetryrelation}
&&\hskip5cm\stackrel{P,S,M}{\cal C}{}\!\! \left(\begin{array}{c} \kappa;k \\ \lambda;\ell \end{array} \right) = \nonumber \\
 &&\sum_N \frac{v_{I_2}^{1/2}}{v_{I_1}^{3/2} v_{S_2}^{1/2}} \frac{e^{i\pi(N-M)}}{v_M^{1/4} v_N^{1/4}} \frac{[d_M]^{1/2}}{[d_N]^{1/2}} \Lambda^{S_2S_1}_{MN}(I_1,I_2) \stackrel{\tilde{P},\tilde{S},\tilde{M}}{\cal C}{}\!\! \left(\begin{array}{c} \tilde{\kappa}+\tilde{k};-\tilde{k} \\ \tilde{\lambda}+\tilde{\ell};-\tilde{\ell} \end{array} \right)\frac{[d_{\lambda_1}][d_{\lambda_2}]}{[d_{\lambda_1+\ell_1}][d_{\lambda_2+\ell_2}]} \frac{[d_{\kappa + k}]}{[d_{\kappa}]}. \nonumber
\end{eqnarray}
On the other hand, we can evaluate the expression of $\stackrel{P,S,M}{\cal D}{}\!\!^{(\pm)}$ in term of $6j$ coefficients 
\begin{eqnarray}
&&\hskip-0.8cm\stackrel{P,S,M}{\cal D}{}\!\!^{(+)}  =  \frac{v_{J_1}^{1/2} v_{J_2}^{1/2}}{v_{S_2}^{1/2} v_{L_1}^{1/2} v_{I_1}^{1/2} v_{K_2}^{1/2}} e^{i\pi(I_1+I_2-K_1-K_2)} \frac{[d_{K_1}]^{1/2} [d_{K_2}]^{1/2}}{[d_{I_1}]^{1/2} [d_{I_2}]^{1/2}} \nonumber \\
&&\hskip0.3cm\sum_X \sixj{X}{L_2}{S_1}{L_1}{J_1}{W}
\sixj{L_2}{J_2}{S_2}{M}{S_1}{X}
\sixj{J_1}{X}{W}{K_2}{K_1}{I_1}
\sixj{J_2}{M}{X}{I_1}{K_2}{I_2} \\
&&\hskip-0.8cm\stackrel{P,S,M}{\cal D}{}\!\!^{(-)}  =  \frac{v_{K_1} v_{I_1}^{1/2} v_{S_2}^{1/2} v_{K_2}^{1/2}}{v_{L_1}^{1/2} v_{J_2}^{1/2} v_{J_1}^{1/2}} e^{i\pi(I_1+J_2+M-K_2-2W)} \frac{[d_W][d_{K_2}]^{1/2}}{[d_{I_1}]^{1/2} [d_{J_2}]^{1/2} [d_{M}]^{1/2}} \nonumber \\
&&\hskip0.3cm\sum_X \sixj{K_2}{X}{L_2}{L_1}{W}{K_1}
\sixj{X}{S_1}{I_1}{J_1}{K_1}{L_1}
\sixj{S_1}{S_2}{M}{I_2}{I_1}{X}
\sixj{L_2}{K_2}{X}{I_2}{S_2}{J_2}
\end{eqnarray}
and we can  easily show that:
\begin{eqnarray}
\sum_M \stackrel{P,S,M}{\cal D}{}\!\!^{(\pm)} \frac{v_{I_2}^{1/2}}{v_{I_1}^{1/2}} \frac{e^{i\pi(N-M)}}{v_M^{1/4} v_N^{1/4}} \frac{[d_M]^{1/2}}{[d_N]^{1/2}} \Lambda^{S_2S_1}_{MN}(I_1,I_2) = \stackrel{\tilde{P},\tilde{S},N}{\cal D}{}\!\!^{(\mp)}\;\;.
\end{eqnarray}
As a consequence, the second relation holds. $\Box$

\begin{figure} 
\psfrag{la1p}[][]{$\lambda_1\!+\!\ell_1$}
\psfrag{la2p}[][]{$\lambda_2\!+\!\ell_2$}
\psfrag{la1}[][]{$\lambda_1$}
\psfrag{la2}[][]{$\lambda_2$}
\psfrag{k}[][]{$\kappa$}
\psfrag{kp}[][]{$\kappa\!+\!k$}
\psfrag{S1}[][]{$S_1$}
\psfrag{S2}[][]{$S_2$}
\psfrag{I1}[][]{$I_1$}
\psfrag{I2}[][]{$I_2$}
\psfrag{M}[][]{$M$}
\psfrag{E1}[][]{$\stackrel{PS(1)}{{\cal E}}(\lambda_1,\ell_1)$}
\psfrag{E2}[][]{$\stackrel{PS(2)}{{\cal E}}(\lambda_2,\ell_2)$}
\psfrag{D}[][]{$\;\;\;\stackrel{P,S,M}{{\cal D}}{}\!\!^{(\pm)}$}
\centering
\scalebox{0.6}[0.6]{\includegraphics[scale=1]{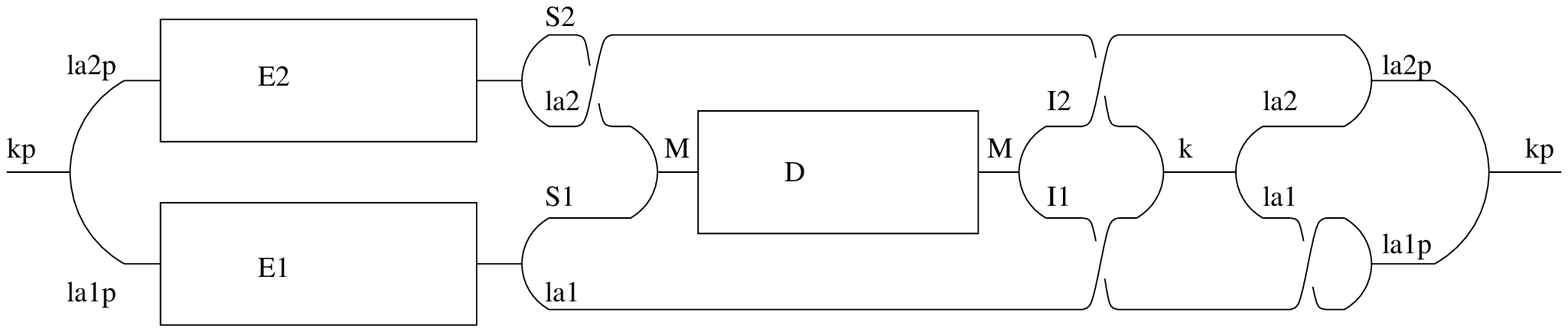}}
\caption{Decomposition of $\stackrel{P}{K}{}\!\!_{2,0}^{(\pm)}$.}
\label{structuregenus2}
\end{figure}

We are now ready to perform the proof of the second symmetry relation of the proposition \ref{prop:symmetriesofK}. The graphical expression of $ \stackrel{P}{K}{}\!\!_{2,0}^{(\pm)}\left(\begin{array}{c} \kappa,k \\ \lambda;\ell \end{array} \right)$ is obtained directly from the picture \ref{fig:valueofKforgenus3}. 
After some trivial topological moves, we introduce orthogonality relations and we show that (figure \ref{structuregenus2}):
\begin{eqnarray}
\stackrel{P}{K}{}\!\!_{2,0}^{(\pm)} \left(\begin{array}{c} \kappa,k \\ {\lambda};{\ell} \end{array} \right) =\sum_{S_1,S_2,M} \stackrel{PS(1)}{\cal E}{}\!\!(\lambda_1,\ell_1) \stackrel{PS(2)}{\cal E}{}\!\!(\lambda_2,\ell_2) \stackrel{P,S,M}{\cal C}{}\!\! \left(\begin{array}{c} \kappa,k \\ \lambda;\ell \end{array} \right) \stackrel{P,S,M}{\cal D}{}\!\!^{(\pm)} \;\;.
\end{eqnarray}
Finally, the symmetry relation is a direct consequence of the proposition \ref{prop:symetryrelation}. The structure of the proof is similar for the general case $n>2$. $\Box$

\newpage

\subsection*{B.1 The p-punctured genus-n surface}
In this appendix, we will give the technical details concerning the proof of unitarity in the case of a p-punctured genus-n surface. In the following, we will extensively study the one puncture torus and we will give the properties for the general case without giving the cumbersome proves.

The function $\stackrel{P}{K}{}\!\!_{1,1}^{(\pm)} \left(\begin{array}{c} \alpha \\ \lambda;\ell\end{array} \right)$, defined by the graph (\ref{genus1punc1}), satisfy the following symmetries:
\begin{enumerate}
\item $\overline{\stackrel{P}{K}{}\!\!_{1,1}^{(\pm)} \left(\begin{array}{c} \alpha \\ \lambda;\ell\end{array} \right)}=\stackrel{P}{K}{}\!\!_{1,1}^{(\pm)} \left(\begin{array}{c} \overline{\alpha} \\ \overline{\lambda};\ell\end{array} \right)$  
\item $\stackrel{P}{K}{}\!\!_{1,1}^{(\pm)} \left(\begin{array}{c} \alpha \\ \lambda;\ell\end{array} \right) = \stackrel{\tilde{P}}{K}{}\!\!_{1,1}^{(\mp)} \left(\begin{array}{c} \tilde{\alpha} \\ \tilde{\lambda}+\tilde{\ell};-\tilde{\ell}\end{array} \right)$ \label{sym11}
\item $\stackrel{P}{K}{}\!\!_{1,1}^{(\pm)} \left(\begin{array}{c} \underline{\alpha} \\ \underline{\lambda};-\ell\end{array} \right) = \psi_{1,1}(\alpha,\lambda,\ell) \stackrel{P}{K}{}\!\!_{1,1}^{(\pm)} \left(\begin{array}{c} {\alpha} \\{\lambda};\ell \end{array} \right)$, \\
with $\psi_{1,1}(\alpha,\lambda,\ell) \in \{+1,-1\}$.
\end{enumerate}
The first and the third properties are proved in the same spirit as in the previous cases. The proof of the second property cannot be reduced to topological moves on graphs. This property is proved along similar lines as we proved the genus 2 case. 

\begin{figure} 
\psfrag{lap}[][]{$\lambda\!+\!\ell$}
\psfrag{la}[][]{$\lambda$}
\psfrag{al}[][]{$\alpha$}
\psfrag{S}[][]{$S$}
\psfrag{I}[][]{$I$}
\psfrag{M}[][]{$M$}
\psfrag{J}[][]{$J$}
\psfrag{K}[][]{$K$}
\psfrag{L}[][]{$L$}
\psfrag{N}[][]{$N$}
\psfrag{W}[][]{$W$}
\centering
\scalebox{0.6}[0.6]{\includegraphics[scale=1]{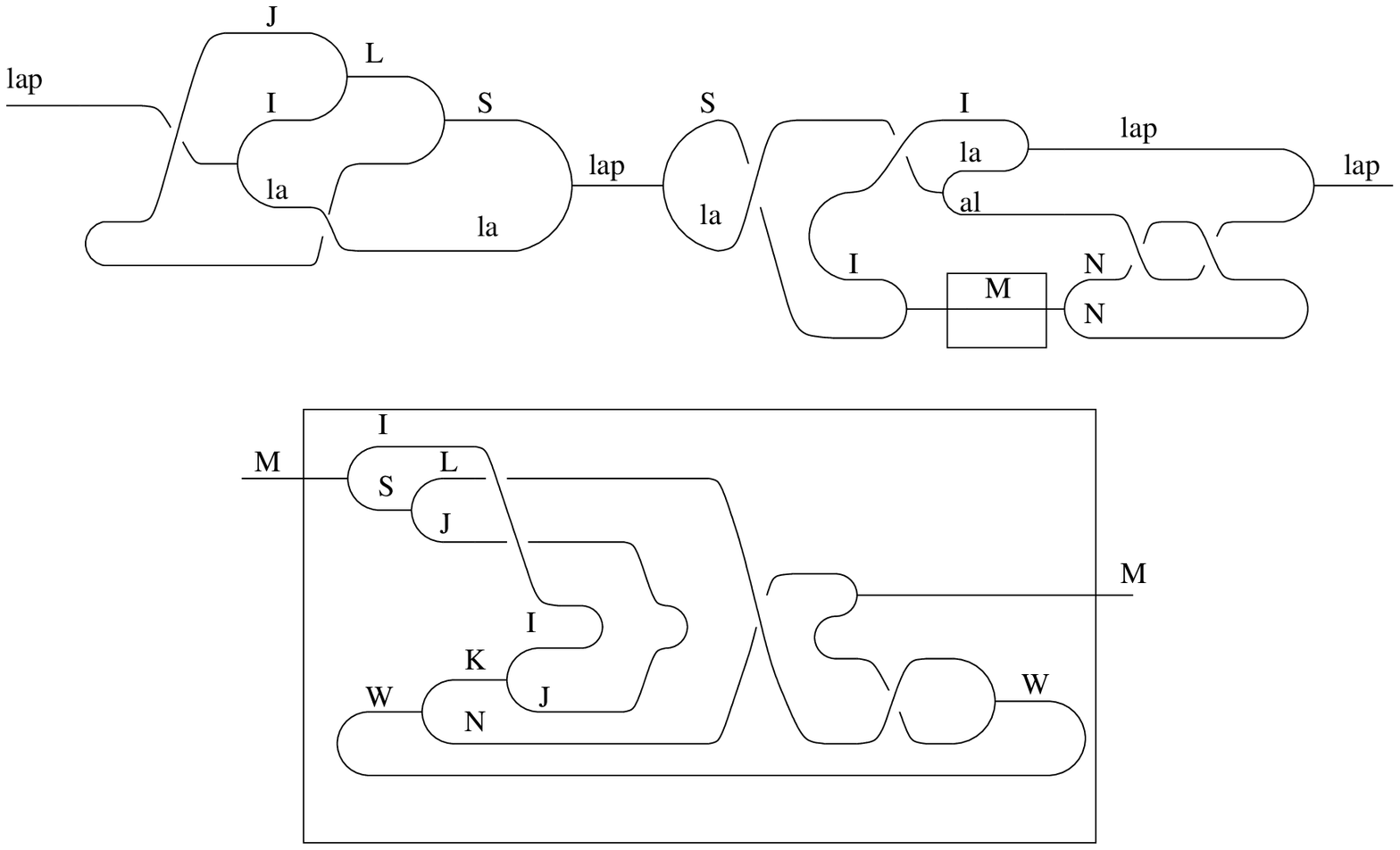}}
\caption{Expression of $\stackrel{P}{K}{}\!\!^{(+)}_{1,1}$ after topological moves and introducing orthogonality relations. The first unit on the top is $\stackrel{PS}{\cal E}(\lambda;\ell)$; the second on the top is $\stackrel{P,S,M}{\cal C}$; the last one on the box is $\stackrel{P,S,M}{\cal D}{}\!\!^{(+)} $.  }
\label{decomposition11}
\end{figure}

After topological moves, we decompose the graph in different units, by using orthogonality relations, as follows (figure \ref{decomposition11}):
\begin{eqnarray}
\stackrel{P}{K}{}\!\!_{1,1}^{(\pm)} \left(\begin{array}{c} \alpha \\ \lambda;\ell\end{array} \right) = \sum_{S,M \in \SSF} \stackrel{PS}{\cal E}(\lambda;\ell)\stackrel{P,S,M}{\cal C}\!\!\left( \begin{array}{c} \alpha \\ \lambda;\ell \end{array} \right) \stackrel{P,S,M}{\cal D}{}\!\!^{(\pm)}
\end{eqnarray}
where $P=(I,J,N,K,L,W)\in \SSF^6$ is the palette and $PS=(I,J,L,S) \in \SSF^4$.We show that these functions satisfies the similar properties as in the proposition \ref{prop:symetryrelation}. As a result, $\stackrel{P}{K}{}\!\!_{1,1}^{(\pm)} \left(\begin{array}{c} \alpha \\ \lambda;\ell\end{array} \right)$ satisfy the required symmetry relation. This closes the proof of unitarity for the one-punctured torus.

\begin{figure} 
\psfrag{la1p}[][]{$\lambda_1\!+\!\ell_1$}
\psfrag{la2p}[][]{$\lambda_2\!+\!\ell_2$}
\psfrag{la1}[][]{$\lambda_1$}
\psfrag{la2}[][]{$\lambda_2$}
\psfrag{al1}[][]{$\alpha_1$}
\psfrag{al2}[][]{$\alpha_2$}
\psfrag{tau}[][]{$\tau$}
\psfrag{sig}[][]{$\sigma$}
\psfrag{taup}[][]{$\tau+t$}
\psfrag{sigp}[][]{$\sigma+s$}
\psfrag{del}[][]{$\delta$}
\psfrag{delp}[][]{$\delta+d$}
\psfrag{I1}[][]{$I_1$}
\psfrag{I2}[][]{$I_2$}
\psfrag{W}[][]{$W$}
\psfrag{J1}[][]{$J_1$}
\psfrag{J2}[][]{$J_2$}
\psfrag{K1}[][]{$K_1$}
\psfrag{K2}[][]{$K_2$}
\psfrag{L1}[][]{$L_1$}
\psfrag{L2}[][]{$L_2$}
\psfrag{N1}[][]{$N_1$}
\psfrag{N2}[][]{$N_2$}
\psfrag{U3}[][]{$U_3$}
\psfrag{U4}[][]{$U_4$}
\psfrag{T3}[][]{$T_3$}
\psfrag{T4}[][]{$T_4$}
\centering
\scalebox{0.6}[0.6]{\includegraphics[scale=1]{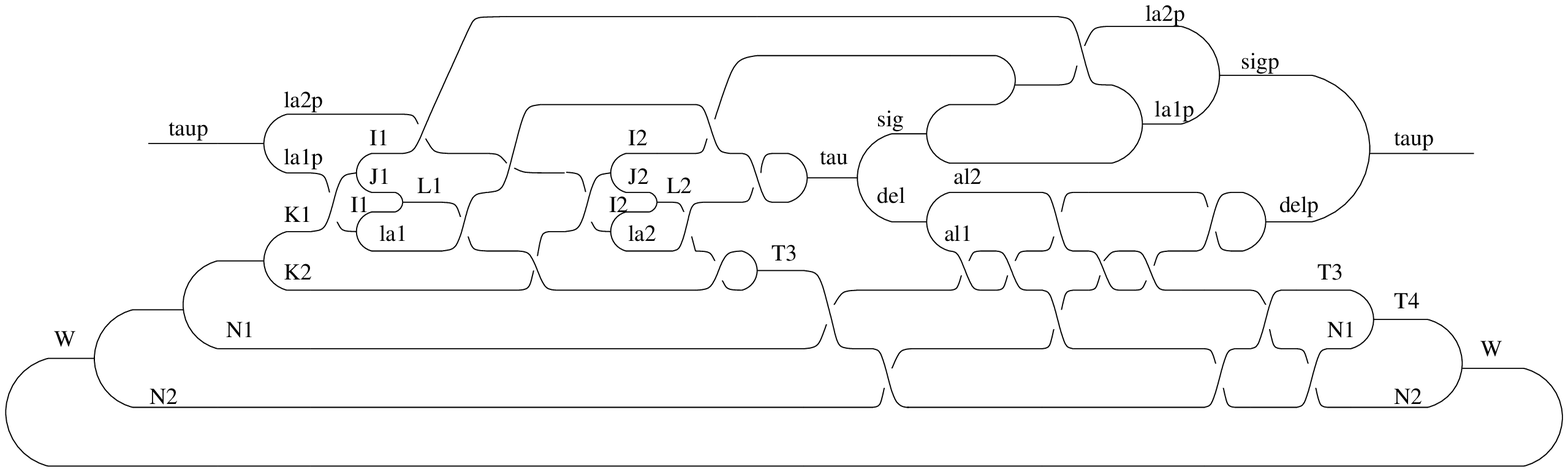}}
\caption{Expression of $\stackrel{P}{K}{}\!\!_{2,2}^{(+)}$.}
\label{genus2punctures2}
\end{figure}

For the general case, the function $\stackrel{P}{K}{}\!\!_{n,p}^{(\pm)} \left(\begin{array}{c} \alpha \\ \beta,\lambda,\sigma,\tau,\delta;b,\ell,s,t,d \end{array} \right)$ is obtained from  the graph in figure \ref{genus2punctures2} where we can notice an obvious iterative structure. 

This function satisfy the usual symmetry relations which are central to prove unitarity of the moduli algebra: 
\begin{proposition} \label{symnp}
The function $\stackrel{P}{K}{}\!\!_{n,p}^{(\pm)} \left(\begin{array}{c} \alpha \\ \beta,\lambda,\sigma,\tau,\delta;b,\ell,s,t,d \end{array} \right)$ satisfy the relations:
\begin{enumerate}
\item $\overline{\stackrel{P}{K}{}\!\!_{n,p}^{(\pm)} \left(\begin{array}{c} \alpha \\ \beta,\lambda,\sigma,\tau,\delta;b,\ell,,s,t,d\end{array} \right)} = \stackrel{P}{K}{}\!\!_{n,p}^{(\pm)} \left(\begin{array}{c} \overline{\alpha} \\ \overline{\beta},\overline{\lambda},\overline{\sigma},\overline{\tau},\overline{\delta};b,\ell,,s,t,d\end{array} \right)$
\item $\stackrel{P}{K}{}\!\!_{n,p}^{(\pm)} \left(\begin{array}{c} \alpha \\ \beta,\lambda,\sigma,\tau,\delta;b,\ell,,s,t,d\end{array} \right) = \frac{[d_{\lambda}]}{[d_{\lambda+\ell}]} \frac{[d_{\tau_{n+1} +t_{n+1}}]}{[d_{\tau_{n+1}}]} \\
\times \stackrel{\tilde{P}}{K}{}\!\!_{n,p}^{(\mp)} \left(\begin{array}{c} \tilde{\alpha} \\ \tilde{\beta}+\tilde{b},\tilde{\lambda}+\tilde{\ell},\tilde{\sigma}+\tilde{s},\tilde{\tau}+\tilde{t},\tilde{\delta}+\tilde{d};-\tilde{b},-\tilde{\ell},-\tilde{s},-\tilde{t},-\tilde{d},\end{array} \right)$ 
\item $\stackrel{P}{K}{}\!\!_{n,p}^{(\pm)} \left(\begin{array}{c} \underline{\alpha} \\ \underline{\beta},\underline{\lambda}, \underline{\sigma}, \underline{\tau}, \underline{\delta},;-b,-\ell,-s,-t,-d\end{array} \right) = \psi_{n,p}(\alpha,\beta,\lambda,\sigma,\tau,\delta,b,\ell,s,t,d) \\
 \times \stackrel{P}{K}{}\!\!_{n,p}^{(\pm)} \left(\begin{array}{c} {\alpha} \\\beta,{\lambda},\sigma,\tau,\delta;b, \ell,s,t,d \end{array} \right)$, 
with $\psi_{n,p}(\alpha,\beta,\lambda,\sigma,\tau,\delta,b,\ell,s,t,d) \in \{+1,-1\}$.
\end{enumerate}
\end{proposition}
This proposition is proved along the same lines as the corresponding proposition for the genus-n case. $\Box$

\bibliographystyle{unsrt}

\end{document}